\documentclass[iop,numberedappendix,twocolappendix]{emulateapj}

\usepackage[usenames]{color}
\usepackage[T1]{fontenc}
\usepackage{cases}

\newcommand{\bfr}{\mbox{\boldmath $r$}}

\newcommand{\bfx}{\mbox{\boldmath $x$}}

\newcommand{\bfd}{\mbox{\boldmath $d$}}

\slugcomment{Accepted for publication in {\it the Astrophysical Journal}}
\shorttitle{NUCLEOSYNTHESIS WITH CHARGED MASSIVE PARTICLE}
\shortauthors{KUSAKABE, KIM, CHEOUN ET AL.}

\begin{document}

\title{REVISED BIG BANG NUCLEOSYNTHESIS WITH LONG-LIVED NEGATIVELY CHARGED MASSIVE PARTICLES:  UPDATED RECOMBINATION RATES, PRIMORDIAL $^9$B\lowercase{e} NUCLEOSYNTHESIS, AND IMPACT OF NEW $^6$L\lowercase{i} LIMITS}

\author{Motohiko Kusakabe\altaffilmark{1,2}, K. S. Kim\altaffilmark{1}, Myung-Ki Cheoun\altaffilmark{2}, Toshitaka Kajino\altaffilmark{3,4}, Yasushi Kino\altaffilmark{5}, \\Grant. J. Mathews\altaffilmark{6}}

\altaffiltext{1}{School of Liberal Arts and Science, Korea Aerospace University, Goyang 412-791, Korea\\
{\tt motohiko@kau.ac.kr, kyungsik@kau.ac.kr}}
\altaffiltext{2}{Department of Physics, Soongsil University, Seoul 156-743, Korea\\
{\tt cheoun@ssu.ac.kr}}
\altaffiltext{3}{National Astronomical Observatory of Japan, 2-21-1 
Osawa, Mitaka, Tokyo 181-8588, Japan\\
{\tt kajino@nao.ac.jp}}
\altaffiltext{4}{Department of Astronomy, Graduate School of Science, 
University of Tokyo, 7-3-1 Hongo, Bunkyo-ku, Tokyo 113-0033, Japan\\
}
\altaffiltext{5}{Department of Chemistry, Tohoku University, Sendai 980-8578, Japan\\
{\tt y.k@m.tohoku.ac.jp}}
\altaffiltext{6}{Center for Astrophysics, Department of Physics, University of Notre Dame, Notre Dame, IN 46556, U.S.A.\\
{\tt gmathews@nd.edu}}

\begin{abstract}
We extensively reanalyze effects of a long-lived negatively charged massive particle, $X^-$, on big bang nucleosynthesis (BBN).  The BBN model with an $X^-$ particle was originally  motivated by the discrepancy between $^{6,7}$Li abundances predicted in standard BBN model and those inferred from observations of metal-poor stars.  In this model $^7$Be is destroyed via the recombination with an $X^-$ particle followed by radiative proton capture.  We calculate precise rates for the radiative recombinations of $^7$Be, $^7$Li, $^9$Be, and $^4$He with $X^-$.  In nonresonant rates we take into account respective partial waves of scattering states and respective bound states.  The finite sizes of nuclear charge distributions cause deviations in wave functions from those of point-charge nuclei.  For a heavy $X^-$ mass, $m_X\gtrsim 100$ GeV, the $d$-wave $\rightarrow$ 2P transition  is most important for $^7$Li and $^{7,9}$Be, unlike recombination with electrons.  Our new nonresonant rate of the $^7$Be recombination for $m_X=1000$ GeV is more than 6 times larger than the existing rate.  Moreover, we suggest a new important reaction for $^9$Be production: the recombination of $^7$Li and $X^-$ followed by deuteron capture.  We derive binding energies of $X$-nuclei along with reaction rates and $Q$-values.  We then calculate BBN and find that the amount of $^7$Be destruction depends significantly on the charge distribution of $^7$Be.  Finally, updated constraints on the initial abundance and the lifetime of the $X^-$ are derived in the context of revised upper limits to the primordial $^6$Li abundance.  Parameter regions for the solution to the $^7$Li problem are revised, and the primordial $^9$Be abundances is revised.
\end{abstract}


\section{INTRODUCTION}\label{sec1}

Standard big bang nucleosynthesis (SBBN)  is an important probe of the early universe.  This model explains the primordial light element abundances inferred from astronomical observations except for the $^7$Li abundance.  Additional nonstandard effects during big bang nucleosynthesis (BBN) may  be required  to explain the $^7$Li discrepancy.  However, such models are strongly constrained from the consistency in the other elemental abundances.  In this paper we re-examine in detail one intriguing solution to the $^7$Li problem, that due to a late-decaying negatively charged particle (possibly the stau as the next to lightest supersymmetric particle) denoted as the $X^-$. In previous work \citep{Kusakabe:2007fv} we showed that both decrease in $^7$Li and increase in $^6$Li abundances are possible in this model.  Recently, however, the primordial $^6$Li abundance has been revised downward  \citep{Lind:2013iza}, and there is now only an upper limit.  Hence, it is necessary to re-evaluate the $X^-$ solution in light of these new measurements.  We show that this remains a viable model for $^7$Li reduction without violating the new $^6$Li upper limit.

\subsection{Primordial Li  Observations}
The primordial lithium abundance is inferred from spectroscopic measurements of metal-poor stars (MPSs).  These stars have a roughly constant abundance ratio, $^7$Li/H$=(1-2) \times 10^{-10}$, as a
function of metallicity~\citep{Spite:1982dd,Ryan:1999vr,Melendez:2004ni,Asplund:2005yt,bon2007,shi2007,Aoki:2009ce,Hernandez:2009gn,Sbordone:2010zi,Monaco:2010mm,Monaco:2011sd,Mucciarelli:2011ts,Aoki:2012wb,Aoki2012b}.  The SBBN
model, however,  predicts a value that is higher by about a factor of $3-4$ [e.g., $^7$Li/H=$5.24 \times 10^{-10}$~\citep{Coc:2011az}] than the observational value when one
uses the baryon-to-photon ratio determined in the $\Lambda$CDM model from an analysis of the power spectrum of the cosmic
microwave background (CMB) radiation from the Wilkinson Microwave Anisotropy Probe \citep{Larson:2010gs, Hinshaw:2012fq} or  the Planck data \citep{Coc:2013eea}).   
This discrepancy suggests the need for a mechanism to reduce  the $^7$Li abundance during or after BBN.  Astrophysical processes such as the rotationally induced mixing \citep{Pinsonneault:1998nf,Pinsonneault:2001ub}, and the combination of atomic and turbulent diffusion~\citep{Richard:2004pj,Korn:2007cx,Lind:2009ta} might have reduced the $^7$Li abundance in stellar atmospheres although this possibility is constrained by the very narrow dispersion in observed Li abundances.

In previous work the $^6$Li/$^7$Li isotopic ratios for MPSs have also been measured and $^6$Li detections have been reported for the halo turnoff star HD 84937~\citep{smith93,smith98,Cayrel:1999kx}, the two Galactic  disk stars HD 68284 and HD 130551~\citep{Nissen:1999iq}, and other stars \citep{Asplund:2005yt,ino05,asp2008,gar2009,ste2010,ste2012}.  A large $^6$Li abundance of $^6$Li/H$\sim 6\times10^{-12}$ has then been suggested \citep{Asplund:2005yt}.  That abundance is $\sim$1000 times higher than the SBBN prediction, and is also significantly higher than the prediction from a  standard Galactic cosmic-ray nucleosynthesis model \citep[cf.][]{pra2006,pra2012}.  It has been noted for some time, however,  \citep{smith2001,Cayrel:2007te} that convective motion in stellar atmospheres could cause systematic asymmetries in the observed atomic line profiles and mimic the presence of $^6$Li~\citep{Cayrel:2007te}. Indeed, in a subsequent detailed analyses, \citet{Lind:2013iza}  found that most of the previous $^6$Li absorption feature  could be attributed to a combination of  3D turbulence and nonlocal thermal equilibrium (NLTE) effects in the model atmosphere. For the present purposes, therefore, we adopt the 2$\sigma$ from their G64-12 NLTE model with 5 parameters, corresponding to $^6$Li$/$H$ = (0.85 \pm 4.33) \times 10^{-12}$.    

Abundances of $^9$Be \citep{boe1999,Primas:2000ee,Tan:2008md,Smiljanic:2009dt,Ito:2009uv,Rich:2009gj} and B \citep{dun1997,gar1998,Primas:1998gp,cun2000} in MPSs have also been measured.  The observed abundances linearly scale with Fe abundances.  The linear relation between abundances of light elements and Fe is expected in  Galactic cosmic-ray nucleosynthesis models~\citep{ree1970,men1971,ree1974,pra2012}.  Any primordial abundances, on the other hand, should be observed as plateau abundances as in the Li case.  Be and B in the observed MPSs are not expected to be primordial.  Nonetheless, primordial abundances of Be and B may be found by future observations.  The strongest lower limit on the primordial Be abundance at present is log(Be/H)$<-14$ which has been derived from an observation of carbon-enhanced MPS BD+44$^\circ$493 of an iron abundance [Fe/H]$=-3.7$ \footnote{[A/B]$=\log(n_{\rm A}/n_{\rm B})-\log(n_{\rm A}/n_{\rm B})_\odot$, where $n_i$ is the number density of $i$ and the subscript $\odot$ indicates the solar value, for elements A and B.} with Subaru/HDS \citep{Ito:2009uv}.

\subsection{$X^-$ Solution}
As one of the solutions to the lithium problem, effects of negatively charged massive particles (CHAMPs or Cahn-Glashow particles) $X^-$~\citep{cahn:1981,Dimopoulos:1989hk,rujula90} during the BBN epoch have been studied \citep{Pospelov:2006sc,Kohri:2006cn,Cyburt:2006uv,Hamaguchi:2007mp,Bird:2007ge,Kusakabe:2007fu,Kusakabe:2007fv,Jedamzik:2007qk,Jedamzik:2007cp,Kamimura:2008fx,Pospelov:2007js,Kawasaki:2007xb,Jittoh:2007fr,Jittoh:2008eq,Jittoh:2010wh,Pospelov:2008ta,Khlopov:2007ic,Kawasaki:2008qe,Bailly:2008yy,Jedamzik:2009uy,Kamimura2010,Kusakabe:2010cb,Pospelov:2010hj,Kohri:2012gc,Cyburt:2012kp,Dapo2012}.  Constraints on supersymmetric models have been derived through  BBN calculations~\citep{Cyburt:2006uv,Kawasaki:2007xb,Jittoh:2007fr,Jittoh:2008eq,Jittoh:2010wh,Pradler:2007ar,Pradler:2007is,Kawasaki:2008qe,Bailly:2008yy}.  In addition, cosmological effects of fractionally charged massive particles (FCHAMPs) have been studied although the nucleosynthesis has not yet been studied \citep{Langacker:2011db}.  

Such long-lived CHAMPs and FCHAMPs which are also called heavy stable charged particles (HSCPs) appear in theories beyond the standard model, and have been searched in collider experiments.  Although the particles should leave characteristic tracks of long time-of-flights due to small velocities, and anomalous energy losses, they have never been detected.  The most stringent limit on scaler $\tau$ leptons (staus) has been derived using data collected with the Compact Muon Solenoid detector for $pp$ collisions at the Large Hadron Collider during the 2011 ($\sqrt[]{\mathstrut s}=7$ TeV, 5.0 fb$^{-1}$) and 2012 ($\sqrt[]{\mathstrut s}=8$ TeV, 18.8 fb$^{-1}$) data taking period.  The limit excludes stau mass below 500 GeV for the direct+indirect production model \citep{CMS2013JHEP}.  The limit on FCHAMPs with spin 1/2 that are neutral under $SU$(3)$_{C}$ and $SU$(2)$_L$ has also been derived from Compact Muon Solenoid searches.  It excludes the masses less than 310 GeV for charge number $q=2/3$, and masses less than 140 GeV for $q=1/3$ \citep{CMS:2012xi}.

The $X^-$ particles and nuclei $A$ can form new bound atomic systems ($A_X$ or $X$-nuclei) with binding energies $\sim O(0.1-1)$~MeV in the limit that the mass of $X^-$, $m_X$, is much larger than the nucleon mass \citep{cahn:1981,Kusakabe:2007fv}.  The $X$-nuclei are exotic chemical species with very heavy masses and chemical properties similar to normal atoms and ions.  The superheavy stable (long-lived) particles have been searched for in experiments, and multiple constraints on respective $X$-nuclei have been derived.  The spectroscopy of terrestrial water gives a limit on the number ratio of $X$/H$<10^{-28}-10^{-29}$ for $m_X=11-1100$ GeV \citep{Smith1982} while that of sea water gives the limits of $X$/H$<4\times 10^{-17}$ for $m_X=5-1500$ GeV \citep{Yamagata:1993jq} and $X$/H$<6\times 10^{-15}$ for $m_X=10^4-10^7$ GeV \citep{Verkerk:1991jf}.  Limits have been derived from analyses of other material, (1) $X$/(Na/23)$<5\times 10^{-12}$ for $m_X=10^2-10^5$ GeV \citep{Dick:1985wk}, (2) $X$/(C/12)$<2\times 10^{-15}$ for $m_X\leq 10^5$ GeV \citep{Tur1984}, and (3) $X$/(Pb/200)$<1.5\times 10^{-13}$ for $m_X\leq 10^5$ GeV \citep{Norman:1988fd}.  Furthermore, limits from analyses of H, Li, Be, B, C, O and F have been derived for $m_X= 10^2 -10^4$ GeV using commercial gases, lake and deep see water deuterium, plant $^{13}$C, commercial $^{18}$O, and reagent grade samples of  Li, Be, B, and F \citep{Hemmick:1989ns}.

If the $X^-$ particle exits during the BBN epoch, it opens new pathways of  atomic and nuclear reactions and affects the resultant nucleosynthesis ~\citep{Pospelov:2006sc,Kohri:2006cn,Cyburt:2006uv,Hamaguchi:2007mp,Bird:2007ge,Kusakabe:2007fu,Kusakabe:2007fv,Jedamzik:2007qk,Jedamzik:2007cp,Kamimura:2008fx,Pospelov:2007js,Kawasaki:2007xb,Jittoh:2007fr,Jittoh:2008eq,Jittoh:2010wh,Pospelov:2008ta,Khlopov:2007ic,Kawasaki:2008qe,Bailly:2008yy,Kamimura2010,Kusakabe:2010cb,Pospelov:2010hj,Kohri:2012gc,Cyburt:2012kp,Dapo2012}.  As the temperature of the universe decreases, positively charged nuclides gradually become electromagnetically bound to $X^-$'s.  Heavier nuclei with larger mass and charge numbers recombine earlier since their binding energies are larger \citep{cahn:1981,Kusakabe:2007fv}.  The formation of most $X$-nuclei proceeds through radiative recombination of nuclides $A$ and $X^-$ \citep{Dimopoulos:1989hk,rujula90}.  However, the $^7$Be$_X$ formation proceeds also through the non-radiative $^7$Be charge exchange reaction between a $^7$Be$^{3+}$
  ion and an $X^-$ \citep{Kusakabe:2013tra,2013PhRvD..88h9904K}.  The recombination of $^7$Be with $X^-$ occurs in a higher temperature environment than that of lighter nuclides does.  At  $^7$Be recombination, therefore, the thermal abundance of free electrons $e^-$ is still very high, and abundant $^7$Be$^{3+}$ ions can exist.  The charge exchange reaction then only affects the $^7$Be abundance.  
  
  Because of relatively small binding energies, the bound states cannot form until late in the BBN epoch.  At the low temperatures, nuclear reactions are already inefficient. Hence, the effect
of the $X^-$ particles is not large.  However, the $X^-$ particle can cause efficient production of $^6$Li~\citep{Pospelov:2006sc} with the weak destruction of $^7$Be~\citep{Bird:2007ge,Kusakabe:2007fu} depending on its abundance and lifetime \citep{Bird:2007ge,Kusakabe:2007fv,Kusakabe:2010cb}.

The $^6$Li abundance can significantly increase through the $X^-$-catalyzed transfer reaction $^4$He$_X$($d$, $X^-$)$^6$Li~\citep{Pospelov:2006sc}, where 1(2$,$3)4 signifies a reaction $1+2\rightarrow 3+4$.  The cross section of the reaction is six orders of magnitude larger than that of the radiative $^4$He($d$,$\gamma$)$^6$Li reaction through which $^6$Li is produced in SBBN model \citep{Hamaguchi:2007mp}.  Other transfer reactions such as $^4$He$_X$($t$,$X^-$)$^7$Li, $^4$He$_X$($^3$He,$X^-$)$^7$Be, and $^6$Li$_X$($p$,$X^-$)$^7$Be are also possible~\citep{Cyburt:2006uv}.  Their rates are, however, not so large as that of
the $^4$He$_X$($d$,$X^-$)$^6$Li since the former reactions involve a $\Delta
l=1$ angular momentum transfer  and consequently  a large hindrance of the
nuclear matrix element~\citep{Kamimura:2008fx}.

The most important reaction for a reduction of the primordial $^7$Li abundance~\footnote{$^7$Be produced during the BBN is transformed into $^7$Li by electron capture in the epoch of the recombination of $^7$Be and electron much later than the BBN epoch.  The primordial $^7$Li abundance is, therefore, the sum of abundances of $^7$Li and $^7$Be produced in BBN.  In SBBN with the baryon-to-photon ratio inferred from WMAP, $^7$Li is produced mostly as $^7$Be during the BBN.} is the resonant reaction $^7$Be$_X$($p$,$\gamma$)$^8$B$_X$
through the first atomic excited state of $^8$B$_X$~\citep{Bird:2007ge} and the atomic ground state of $^8$B$^\ast$($1^+$,0.770~MeV)$_X$, i.e., an atom
consisting of the $1^+$ nuclear excited state of $^8$B and an $X^-$~\citep{Kusakabe:2007fu}.  From a realistic estimate of binding energies of $X$-nuclei, however, the latter resonance has been found to be an inefficient pathway for $^7$Be$_X$ destruction~\citep{Kusakabe:2007fv}.

The $^8$Be$_X$+$p$ $\rightarrow ^9$B$_X^{\ast{\rm a}}
\rightarrow ^9$B$_X$+$\gamma$ reaction through the $^9$B$_X^{\ast{\rm a}}$
atomic excited state
of $^9$B$_X$~\citep{Kusakabe:2007fv} produces the $A=$9 $X$-nucleus so that it can possibly lead to the production of heavier nuclei.  This reaction, however, is not operative because of its large resonance energy~\citep{Kusakabe:2007fv}.

The resonant reaction
$^8$Be$_X$($n$, $X^-$)$^9$Be through the atomic ground state of
$^9$Be$^\ast$($1/2^+$, 1.684~MeV)$_X$, is another reaction producing mass number 9 nuclide~\citep{Pospelov:2007js}.  \citet{Kamimura:2008fx}, however,  adopted a realistic  root mean square charge
radius for $^8$Be of 3.39~fm, and found that $^9$Be$^*$($1/2^+$, 1.684~MeV)$_X$ is not a
resonance but a bound state located below the $^8$Be$_X$+$n$
threshold.  A subsequent four-body calculation
for an $\alpha+\alpha+n+X^-$ system confirmed that the
$^9$Be$^*$($1/2^+$, 1.684~MeV)$_X$ state is located below the
threshold~\citep{Kamimura2010}.  This was also confirmed by \citet{Cyburt:2012kp} using a three-body model.  The effect of the resonant reaction is, therefore, negligible.
The  detailed BBN calculations of \citet{Kusakabe:2007fv,Kusakabe:2010cb} precisely incorporate recombination reactions of nuclides and $X^-$ particles, nuclear reactions of $X$-nuclei, and their inverse reactions.  These calculations have also included  reaction rates estimated in a rigorous quantum few-body model \citep{Hamaguchi:2007mp,Kamimura:2008fx}.  The most realistic calculation \citep{Kusakabe:2010cb} shows no significant production of $^9$Be and heavier nuclides.

Reactions of neutral $X$-nuclei, i.e., $p_X$,
$d_X$ and $t_X$ can produce and destroy Li and Be~\citep{Jedamzik:2007qk,Jedamzik:2007cp}.  The rates for these reactions and the charge-exchange reactions 
$p_X$($\alpha$,$p$)$\alpha_X$, $d_X$($\alpha$,$d$)$\alpha_X$ and
$t_X$($\alpha$,$t$)$\alpha_X$ have been calculated in a rigorous quantum few-body model~\citep{Kamimura:2008fx}.  The cross sections for the charge-exchange reactions are much larger than those of the nuclear reactions so that the neutral $X$-nuclei $p_X$, $d_X$
and $t_X$ are quickly converted to $\alpha_X$ before they induce nuclear reactions.  The production and destruction of Li and Be is not significantly affected by the presence of neutral $X$-nuclei \citep{Kamimura:2008fx}.  This was confirmed in a detailed nuclear reaction network calculation \citep{Kusakabe:2010cb}.  It has been shown
in our previous work \citep{Kusakabe:2007fv,Kusakabe:2010cb} that concordance with the observational constraints on D, $^3$He, and $^4$He is maintained in the parameter region of $^7$Li reduction.

In this paper we present an extensive study on effects of a CHAMP, $X^-$, on BBN.  First, we study the effects of theoretical uncertainties in the nuclear charge distributions on the binding energies of nuclei and the $X^-$, reaction rates, and BBN.  Next, we derive the most precise radiative recombination rates for $^7$Be, $^7$Li, $^9$Be, and $^4$He with an $X^-$.  Finally, we suggest a new reaction for $^9$Be production, i.e. $^7$Li$_X$($d$, $X^-$)$^9$Be.  Based upon our updated BBN calculation, it is found that the amount of $^7$Be destruction depends significantly upon the assumed charge density for the  $^7$Be nucleus.  The most realistic constraints on the initial abundance and the lifetime of the $X^-$ are then derived, and the primordial $^9$Be abundance is also estimated.

In Sec. \ref{sec2}, models for the nuclear charge density are described.  In Sec.~\ref{sec3}, binding energies of the $X$-nuclei are calculated with both of a variational method and the integration of the Sch$\ddot{\rm o}$dinger equation, for different charge densities.  In Sec.~\ref{sec4}, reaction rates are calculated for the radiative proton capture of the $^7$Be$_X$($p$, $\gamma$)$^8$B$_X$ and $^8$Be$_X$($p$, $\gamma$)$^9$B$_X$ reactions.  Theoretical uncertainties in the rates due to the assumed charge density shapes are deduced.  In Sec.~\ref{sec5}, rates for the radiative recombination of $^7$Be, $^7$Li, $^9$Be, and $^4$He with $X^-$ particles are calculated.  Both nonresonant and resonant rates are derived.  The difference of the recombination rate for  $X^-$ particles compared to that  for electrons is shown.  In Sec.~\ref{sec6}, a new reaction for $^9$Be production is pointed out.  It is the radiative recombination of $^7$Li and an $X^-$ followed by  deuteron capture.  In Sec.~\ref{sec7}, the rates and $Q$-values for $\beta$-decays and nuclear 
reactions involving the $X^-$ particle are derived.  In Sec.~\ref{sec8}, a new reaction network calculation code is explained.  In Sec.~\ref{sec9}, we show the evolution of elemental abundances as a function of cosmic temperature, and derive the most realistic constraints on the initial abundance and the lifetime of the $X^-$.  
Parameter regions for the solution to the $^7$Li problem, and the prediction of primordial $^9$Be  are presented.  Sec. \ref{sec10} is devoted to a summary and conclusions.  In Appendix \ref{app1}, we comment on the electric dipole transitions of $X$-nuclei which change nuclear and atomic states simultaneously. \footnote{Throughout the paper, we use natural units, $\hbar=c=k_{\rm B}=1$, for the reduced Planck constant $\hbar$, the speed of light $c$, and the Boltzmann constant $k_{\rm B}$.  We use the usual notation
1(2$,$3)4 for a reaction $1+2\rightarrow 3+4$.}

\section{NUCLEAR CHARGE DENSITY}\label{sec2}

We assume that a negatively charged massive particle (CHAMP) with a single charge and spin zero, $X^-$,  exists during the BBN epoch.  We derive general constraints depending upon the mass of the $X^-$, i.e., $m_X$.  The mass is treated as one parameter.  Although the existence of very light CHAMPs is excluded by searches in collider experiments, their existence during the BBN epoch is also considered in this paper taking account of unknown mechanisms such as the time evolution of the mass of the $X^-$.  In order to estimate possible uncertainties in the binding energies of nuclei and $X^-$ particles which are associated with the nuclear charge density, we use three different shapes for  the charge density.  The first  is a Woods-Saxon shape:
\begin{equation}
 \rho_{\rm WS}(r^\prime)=\frac{ZeC_{\rm WS}}{1+\exp\left[\left(r^\prime-R\right)/a\right]},
\label{eq1}
\end{equation}
where
$r^\prime$ is the distance from the center of mass of the nucleus,
$Ze$ is the charge of the nucleus,
$R$ is the parameter characterizing the nuclear size,
$a$ is nuclear surface  diffuseness,
and $C_{\rm WS}$ is a normalization constant.
The $C_{\rm WS}$ value is fixed by the equation of charge conservation, $Ze=\int \rho_{\rm WS}~d\bfr^\prime$, and it is given by
\begin{equation}
 C_{\rm WS}=\left(4\pi \int_0^\infty \frac{{r^\prime}^2}{1+\exp\left[\left(r^\prime-R\right)/a\right]} dr^\prime\right)^{-1}.
\label{eq2}
\end{equation}
For a given value of diffuseness $a$, $R$ can be constrained so that the parameter set of ($a$, $R$) satisfies the  root-mean-square (RMS) charge radius $\langle r^2 \rangle_{\rm C}^{1/2}$ measured in nuclear experiments.

The potential between an $X^-$ and a nucleus $A$ ($XA$ potential) is calculated by folding the Coulomb potential with the charge density:
\begin{equation}
 V(r)=\int -\frac{e \rho(\bfr^\prime)}{x}~d\bfr^\prime,
\label{eq3}
\end{equation}
where
$\bfr$ is the position vector from an $X^-$ to the center of mass of $A$,
$\bfr^\prime$ is the position vector from the center of mass of $A$,
$\bfx=\bfr+\bfr^\prime$ is the displacement vector between the $X^-$ and the position, and $\rho(\bfr^\prime)$ is the charge density of the nucleus.  The charge density could be distorted from the density of normal nucleus by the potential of an $X^-$.  The distortion effect, however,  is relatively small because of the weak Coulomb potential.  Hence,  we neglect it in this study.  Under the assumption of a Woods-Saxon charge distribution $\rho_{\rm WS}(r^\prime)$, the potential reduces to the form
\begin{eqnarray}
 V_{\rm WS}(r)&=&-\frac{2\pi C_{\rm WS} Z e^2}{r} \int_0^\infty dr^\prime r^\prime
 \frac{(r+r^\prime)-|r-r^\prime|}{1+\exp[(r^\prime-R)/a]}~~.\nonumber\\
\label{eq4}
\end{eqnarray}

The second charge density adopted in this study is a Gaussian shape described by
\begin{equation}
\rho_{\rm G}(r^\prime)=\frac{Ze}{\pi^{3/2} b^3}\exp\left[-\left(\frac{r^\prime}{b}\right)^2\right]~,
\label{eq5}
\end{equation}
where
the range parameter is related to the RMS charge radius by $b=(2/3)^{1/2}\langle r^2_{\rm C} \rangle^{1/2}$.
The $XA$ potential is given by
\begin{equation}
 V_{\rm G}(r)=\int d\bfr^\prime \frac{-e\rho_{\rm G}(\bfr^\prime)}{x}=-\frac{Ze^2}{r} \mathrm{erf}\left(\frac{r}{b}\right),
\label{eq6}
\end{equation}
where $\mathrm{erf}(x)=2/\pi^{1/2} \int_0^x \exp(-t^2)~dt$ is the error function.

The third charge density is a square well given by
\begin{equation}
 \rho_{\rm w}(r^\prime)=\frac{3Ze}{4\pi r_0^3}H(r_0-r^\prime),
\label{eq7}
\end{equation}
where $H(x)$ is the Heaviside step function, and
the surface radius is related to the RMS charge radius by $r_0=(5/3)^{1/2}\langle r^2_{\rm C} \rangle^{1/2}$.
The $XA$ potential is then given by
\begin{eqnarray}
 V_{\rm w}(r)&=&-\frac{Ze^2}{2r_0} \left[3-\left(\frac{r}{r_0}\right)^2\right]~~~~~({\rm for}~r\le r_0) \nonumber\\
&&-\frac{Ze^2}{r}~~~~~({\rm for}~r>r_0).
\label{eq8}
\end{eqnarray}

\section{BINDING ENERGY}\label{sec3}

Binding energies and wave functions for bound states of nuclei $A+X^-$, i.e., denoted $X$-nuclei or $A_X$, are calculated for four different $X$-particle masses: $m_X$=1, 10, 100, and 1000 GeV.  We performed both numerical integrations of the Schr$\ddot{\rm o}$dinger equation with RADCAP~\citep{Bertulani:2003kr}\footnote{In the RADCAP code, there was an error in the numerical value of $\pi$, which was corrected.} and variational calculations with the Gaussian expansion method \citep{Hiyama:2003cu}.  It was confirmed that binding energies derived with the two methods generally agree with each others within $\sim$1 \%.

Table \ref{tab1} shows the adopted experimental RMS charge radii, and calculated binding energies of ground state (GS) $X$-nuclides for $m_X=100$ TeV.  This mass is chosen as one example in which the $X^-$ particle is much heavier than the lighter nuclei.  Hence,  the reduced mass of the $A+X^-$ system is given by $\mu=m_A m_X/(m_A+m_X)\rightarrow m_A$, where $m_A$ is the mass of nuclide $A$.  The second and the third columns show measured RMS charge radii and the associated reference, respectively.  Results for three different nuclear charge distributions, i.e., Gaussian (second column), homogeneous (third column), and Woods-Saxon (WS) with three values for the diffuseness parameter $a=0.45$ fm (WS45; fourth column), 0.4 fm (WS40; fifth), and 0.35 fm (WS35; sixth) are shown in the fourth to eighth columns, respectively.  We have chosen these three values for the diffuseness parameter $a$ since larger $a$ values do not lead to simultaneous solutions of $R$ that reproduce the  RMS charge radii for all nuclides.  

\placetable{tab1}

Binding energies of the first atomic excited states, $^8$B$_X^{\ast{\rm a}}$ and $^9$B$_X^{\ast{\rm a}}$, are also shown since they are important in resonant reactions through the atomic excited states that  result in  $^7$Be$_X$ destruction and  $^9$B$_X$ production.  The superscript $\ast{\rm a}$ indicates an atomic excited state, that  is different from a nuclear excited state indicated by a superscript $\ast$.  Binding energies of the first atomic excited states are, therefore, also calculated.    Binding energies for the Gaussian charge distribution are the largest. Those for a homogeneous distribution are the smallest, while  those for the WS distribution are intermediate.  In addition, with a larger diffuseness parameter, the binding energies are larger.  The reason for  this ordering of binding energies is as follows.  The five cases are arranged as: 1) Gaussian; 2) WS with a large $a$ value; 3) an intermediate $a$ value;  4) a  small $a$ value; and 5) the homogeneous distribution. 
These are listed in descending order of nuclear charge density at small radii $r$.  When the charge density at small $r$ is relatively large, the Coulomb potential between $A$ and $X^-$ is large.  
Then, large values for  the binding energies are derived.  It is noted that in all cases, the amplitudes of the Coulomb potentials are smaller than those for two point charges.  This is  because of the finite size of charge distribution of the nucleus $A$.  Binding energies are, therefore, smaller than those in the Bohr's atomic model.

Table \ref{tab2} shows calculated binding energies of GS $X$-nuclides and the first atomic excited states of  $^8$B$^{\ast{\rm a}}_X$ and $^9$B$^{\ast{\rm a}}_X$ in the WS40 model, for $m_X=1$ GeV (second column), 10 GeV (third column), 100 GeV (fourth column), and 1000 GeV (fifth column).  The WS charge distribution with a diffuseness parameter of a=0.4 fm is taken as our primary model in this paper.  When $m_X$ is larger, the reduced mass $\mu=m_A m_X/(m_A+m_X)$ is larger.  Binding energies are then larger.  However, the binding energies for $m_X=100$ GeV and 1000 GeV do not differ from each other since the reduce masses in both cases is already near the limiting value of $\mu=m_A m_X/(m_A+m_X)\rightarrow m_A$.

\placetable{tab2}

\begin{figure}
\begin{center}
\includegraphics[width=0.45\textwidth]{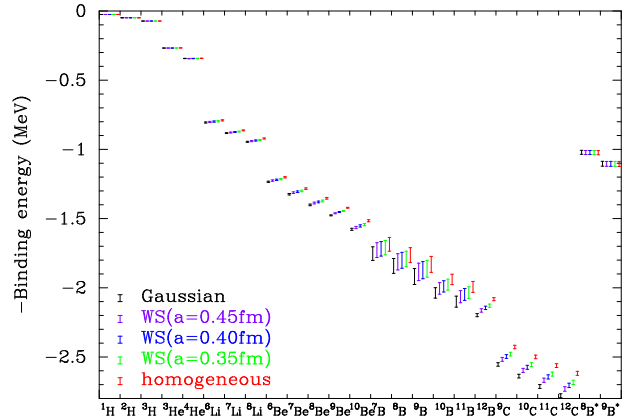}
\end{center}
\caption{Binding energies of nuclei and $X^-$ particles with $m_X=100$ TeV for different charge distributions.  These are  Gaussian (black lines), Woods-Saxon type with diffuseness parameters $a=0.45$ fm (purple lines), 0.40 fm (blue lines), and 0.35 fm (green lines), and a homogeneous well (red lines).  Error bars indicate uncertainties determined from uncertainties in the experimental RMS charge radii.  \label{fig1}}
\end{figure}

Figure \ref{fig1} shows binding energies of GS $X$-nuclides and the first atomic excited states of $^8$B$_X^{\ast{\rm a}}$ and $^9$B$_X^{\ast{\rm a}}$  in the five models of nuclear charge distribution for $m_X=100$ TeV.  As the nuclear mass increases, the nuclear charge number and the reduced mass become larger.  Therefore, heavier nuclei generally have larger binding energies.  Error bars indicate uncertainties originating from the experimental 1 $\sigma$  error in the  RMS charge radii.  Errors in binding energies of nuclides up to $^4$He are small, while those for heavier nuclides can be $\mathcal{O}$(0.1 MeV).  However, $Q$-values for most reactions involving $X$-nuclei heavier than $^4$He$_X$ are large, $\gtrsim 1$ MeV \citep[e.g.][]{Kusakabe:2007fv}.  


Effects of errors in binding energies on the rates of forward and inverse reactions are then small.  Two exceptions are $^7$Be$_X$($p$, $\gamma$)$^8$B$_X$ ($Q=0.64$ MeV) and $^8$Be$_X$($p$, $\gamma$)$^9$B$_X$ ($Q=0.33$ MeV).  These reactions are also exceptional because the resonant components in their reaction rates can be dominant.  For the reason described above, we adopted data calculated for the WS40 model, such as nuclear masses, reaction rates, coefficients for reverse reactions, and $Q$-values.  Only data for  the reactions $^7$Be$_X$($p$, $\gamma$)$^8$B$_X$ and $^8$Be$_X$($p$, $\gamma$)$^9$B$_X$ are calculated for three models of charge distribution, i.e., Gaussian, WS, and homogeneous types.  In the limit that the mass of the $X^-$ particle is much larger than that of light nuclides $\sim \mathcal{O}$(1 GeV), reaction rates of the radiative neutron capture are very small.  This is because the electric multipole moments approach  zero in this limit, and the electric matrix elements are very small.  This situation is similar to the case of the long-lived strongly interacting massive particle $X^0$ \citep{Kusakabe:2009jt}.  Then, we assume that rates of radiative neutron capture reactions are vanishingly small in this study.
This is   different from the assumption in \citet{Kusakabe:2007fv,Kusakabe:2010cb}.

\begin{figure}
\begin{center}
\includegraphics[width=0.45\textwidth]{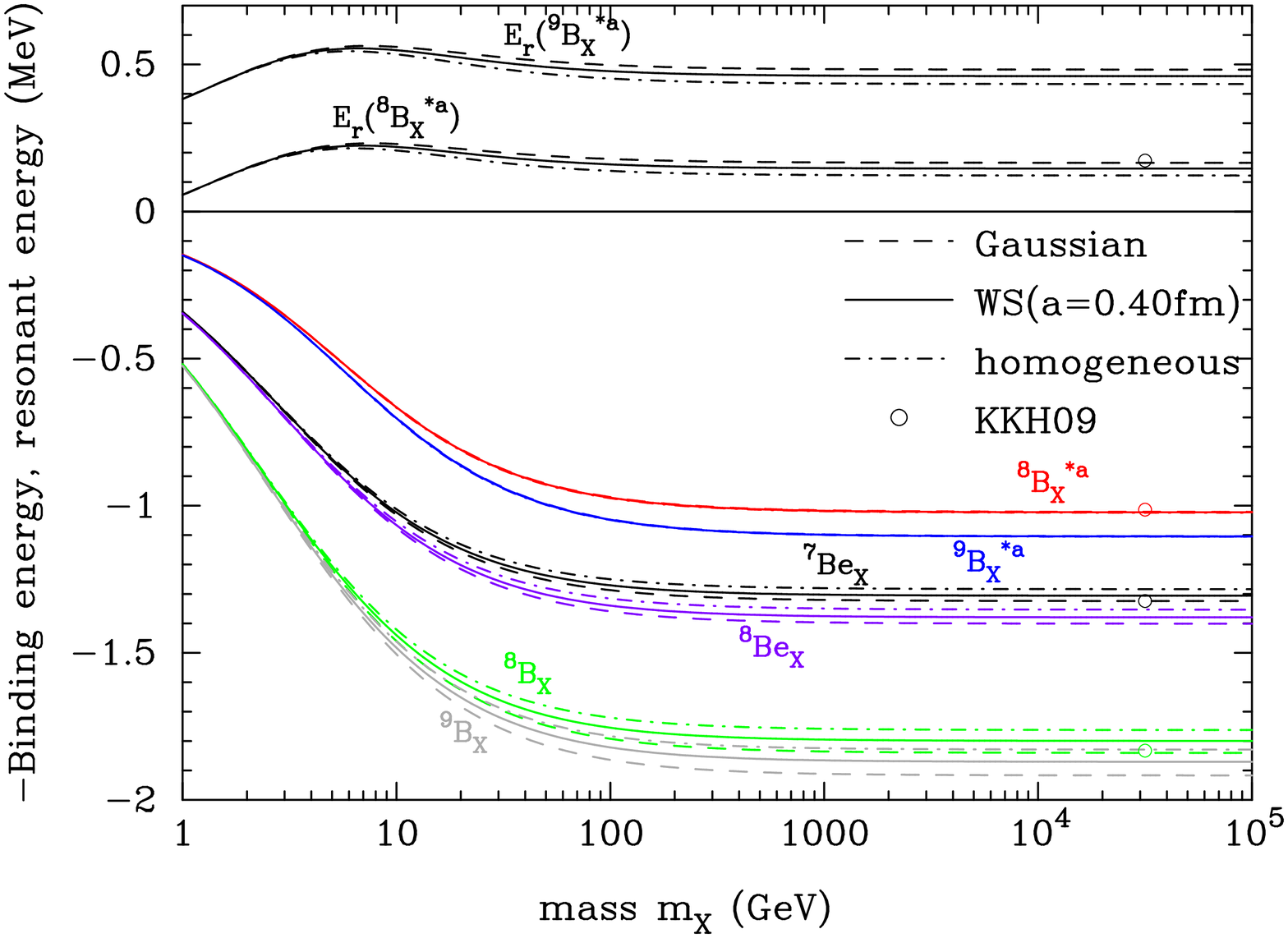}
\end{center}
\caption{Binding energies and resonance energies as a function of $m_X$.  The upper black lines show resonance energies in the reactions $^7$Be$_X$($p$, $\gamma$)$^8$B$_X$ and $^8$Be$_X$($p$, $\gamma$)$^9$B$_X$.  The lower lines show binding energies of $^7$Be$_X$ (black lines), $^8$Be$_X$ (purple lines), $^8$B$_X$ (green lines), $^9$B$_X$ (gray lines), and the first atomic excited states $^8$B$_X^{\ast{\rm a}}$ (red lines) and $^9$B$_X^{\ast{\rm a}}$ (blue lines).  Results for different nuclear charge distributions, i.e.  Gaussian (dashed lines), Woods-Saxon type with diffuseness parameter $a=0.40$ fm (solid lines), and homogeneous well (dot-dashed lines) are drawn.  Open circles show energy heights derived by a quantum many-body calculation \citep{Kamimura:2008fx} for $m_X=\infty$. \label{fig2}}
\end{figure}

Figure \ref{fig2} shows binding energies of GS $X$-nuclides and the first atomic excited states, $^8$B$_X^{\ast{\rm a}}$ and $^9$B$_X^{\ast{\rm a}}$, for nuclear charge distribution models of Gaussian (dashed lines), WS40 (solid lines), and homogeneous (dot-dashed lines) as a function of $m_X$.  Resonance energies $E_{\rm r}$ are also shown  for $^8$B$_X^{\ast{\rm a}}$ and $^9$B$_X^{\ast{\rm a}}$ measured relative to  the separation channels, $^7$Be$_X$+$p$ and $^8$Be$_X$+$p$.   Binding energies are larger when the value of $m_X$ is larger, and they approach  the asymptotic value in the limit of $\mu\rightarrow m_A$.  Maxima are observed in the curves of $E_{\rm r}$($^8$B$_X^{\ast{\rm a}}$) and $E_{\rm r}$($^9$B$_X^{\ast{\rm a}}$) at $m_X\lesssim 10$ GeV.  The resonance energies increase with increasing $m_X$ in the mass region of $m_X\lesssim 10$ GeV, while they are approximately saturated in the region of $m_X\gtrsim 10$ GeV.  Since rates of the resonant reactions are sensitive to the resonance energy heights, results of BBN including the existence of $X^-$ significantly depend on the mass $m_X$, as described below.  Open circles show binding energies of $E_{\rm B}$($^7$Be$_X$), $E_{\rm B}$($^8$B$_X$), $E_{\rm B}$($^8$B$_X^{\ast {\rm a}}$), and the resonance energy $E_{\rm r}$($^8$B$_X^{\ast {\rm a}}$) derived by a quantum many-body calculation for $m_X=\infty$ \citep{Kamimura:2008fx}.  The open circles are consistent with calculated values in the Gaussian model.


\begin{figure}
\begin{center}
\includegraphics[width=0.45\textwidth]{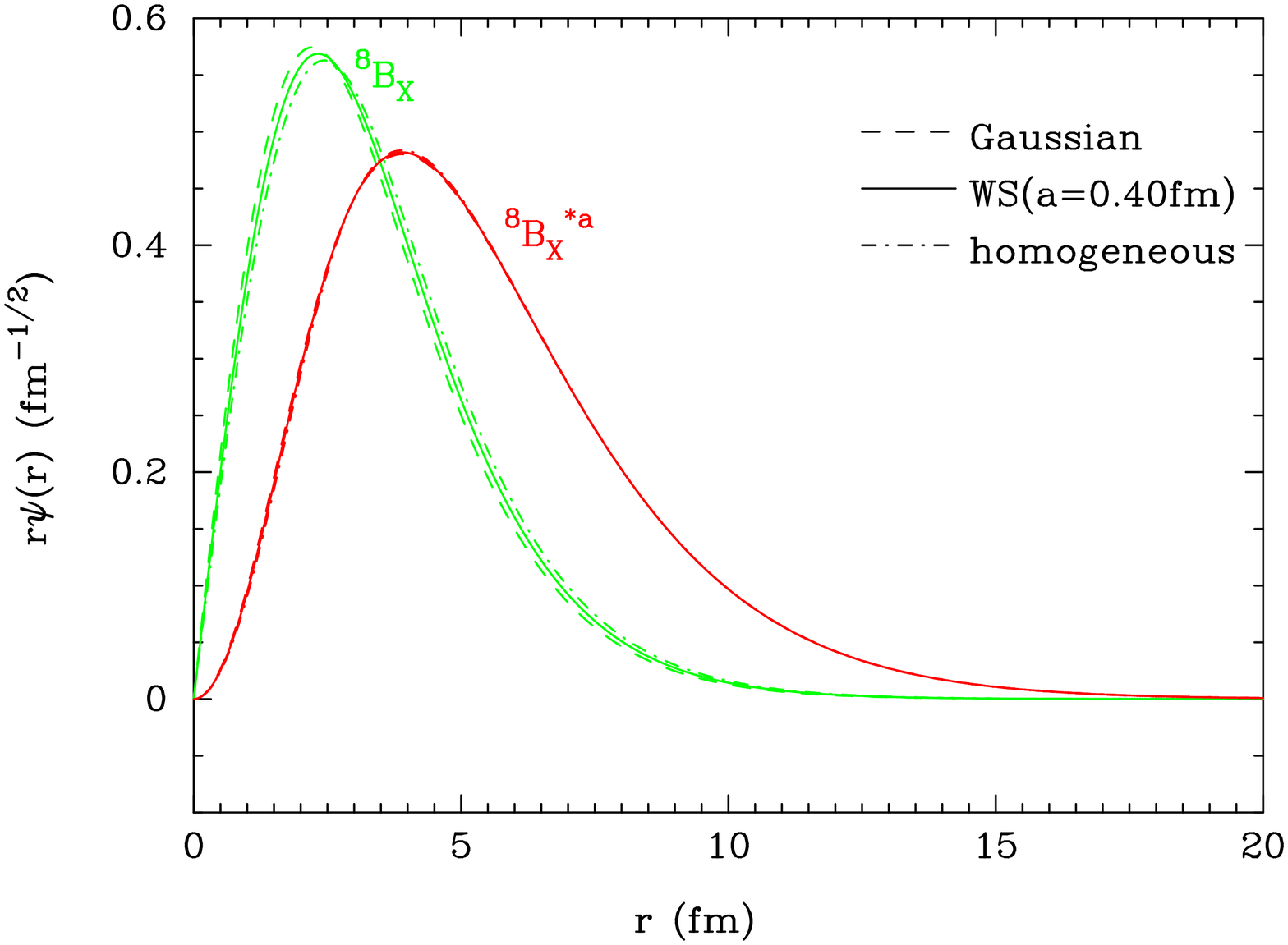}
\end{center}
\caption{Wave functions for the ground state of $^8$B$_X$ and the first atomic excited state $^8$B$_X^{\ast{\rm a}}$ as a function of radius $r$ for $m_X=1000$ GeV.  Lines are drawn for different nuclear charge distributions as labeled, i.e. Gaussian (dashed lines), Woods-Saxon type with diffuseness parameter $a=0.40$ fm (solid lines), and a homogeneous well (dot-dashed lines) .\label{fig3}}
\end{figure}

\begin{figure}
\begin{center}
\includegraphics[width=0.45\textwidth]{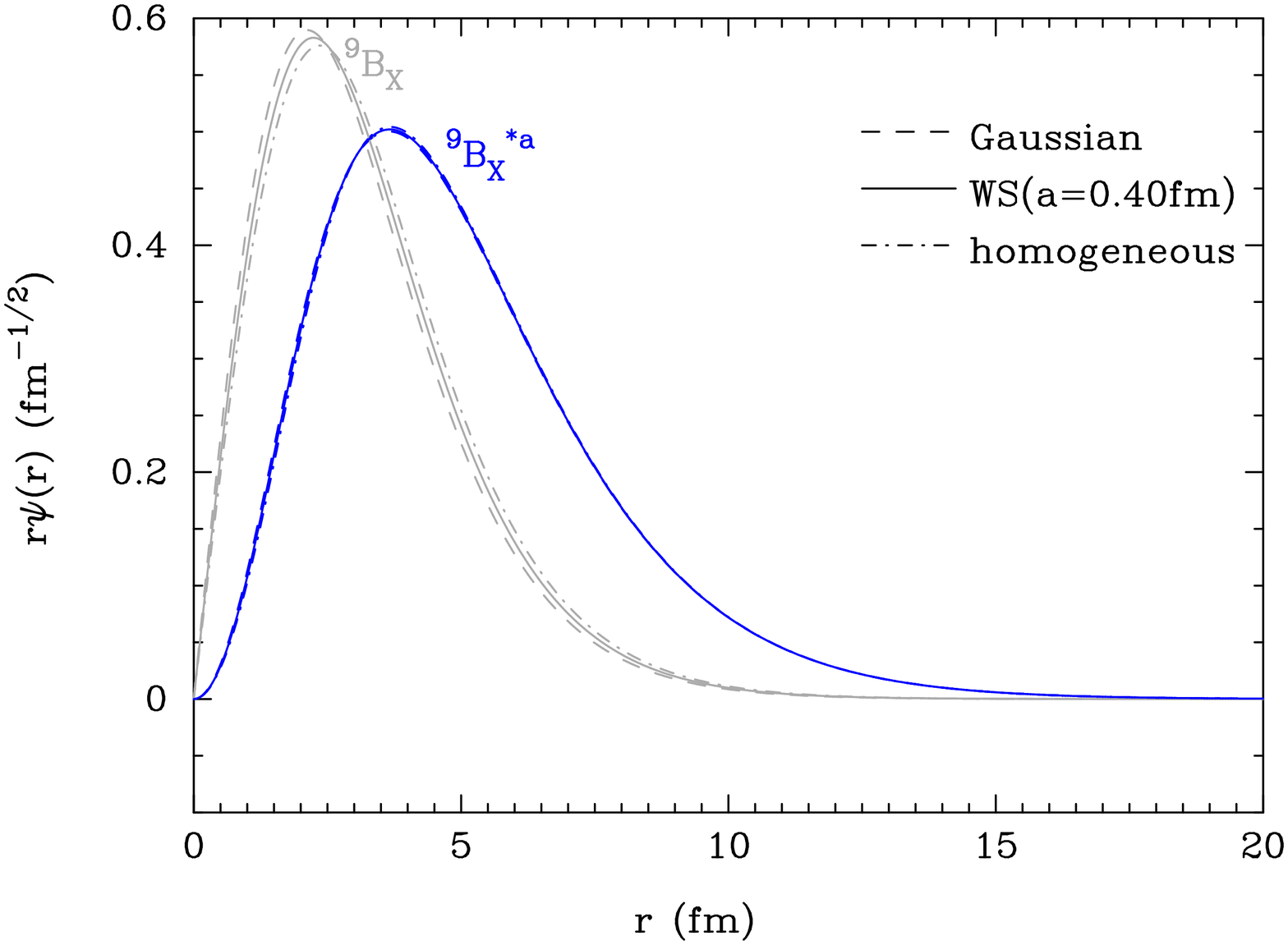}
\end{center}
\caption{Wave functions for the ground state of $^9$B$_X$ and the first atomic excited state $^9$B$_X^{\ast{\rm a}}$ as a function of radius $r$ for $m_X=1000$ GeV.  Lines are drawn for  different nuclear charge distributions as labeled, i.e. Gaussian (dashed lines), Woods-Saxon type with diffuseness parameter $a=0.40$ fm (solid lines), and a homogeneous well (dot-dashed lines).\label{fig4}}
\end{figure}

Figures \ref{fig3} and \ref{fig4} show wave functions of the GS and first atomic excited states of $^8$B$_X^{\ast{\rm a}}$ and $^9$B$_X^{\ast{\rm a}}$ for the case of $m_X=1000$ GeV with nuclear charge distribution models of Gaussian (dashed lines), WS40 (solid lines), and homogeneous (dot-dashed lines).  There are differences between the three lines for GS $^8$B$_X$ and $^9$B$_X$  although they are relatively small.  On the other hand, differences are hardly seen for the excited states.  Shapes of the charge distribution predominantly affects the  Coulomb potentials at small $r$ values.  When angular momentum exists such as in the $l=1$ excited states of $^8$B$_X^{\ast{\rm a}}$ and $^9$B$_X^{\ast{\rm a}}$, however, the effect of the centrifugal potential $l(l+1)/2\mu r^2$ is significant.  The effect of the nuclear charge distribution is, therefore, most important for GS $X$-nuclei whose amplitudes of wave functions at small $r$ is larger than those of the excited states.  
The Gaussian type has the largest Coulomb potential, the WS type has the second largest, and the homogeneous type the smallest.  Because of the Coulomb attractive force, the wave functions in the Gaussian model are located in a region of smaller $r$ than those in other models, while those in the homogeneous case are the most extended radially.


\section{RESONANT PROTON CAPTURE REACTIONS}\label{sec4}
Two important resonant reactions are
\begin{eqnarray}
^7{\rm Be}_X&+&p \rightarrow ^8{\rm B}_X^{\ast{\rm a}}(2\mathrm{P}) \rightarrow ^8{\rm B}_X+\gamma \nonumber\\
&&~~~[Q=m(^7{\rm Be}_X)+m(p)-m(^8{\rm B}_X)] \nonumber\\
^8{\rm Be}_X&+&p \rightarrow ^9{\rm B}_X^{\ast{\rm a}}(2\mathrm{P}) \rightarrow ^9{\rm B}_X+\gamma \nonumber\\
&&~~~[Q=m(^8{\rm Be}_X)+m(p)-m(^9{\rm B}_X)],
\label{eq9}
\end{eqnarray}
where (2P) indicates the atomic 2P state, and $m(A)$ and $m(A_X)$ are masses of nucleus $A$ and $X$-nucleus $A_X$, respectively.
Resonant rates for these radiative capture reactions can be calculated as follows.

The thermal reaction rate is derived as a function of temperature $T$ by numerically integrating the cross section over energy,
\begin{equation}
\langle \sigma v \rangle =\left(\frac{8}{\pi \mu}\right)^{1/2} \frac{1}{T^{3/2}} \int_0^\infty E \sigma(E) \exp\left(-\frac{E}{T}\right) dE,
\label{eq10}
\end{equation}
where
$E$ is the center of mass kinetic energy,
and $\sigma(E)$ is the reaction cross section as a function of $E$.

The thermal reaction rate for isolated and narrow resonances is given \citep{Angulo1999} by
\begin{eqnarray}
N_{\rm A} \langle \sigma v \rangle&=& N_{\rm A} \left(\frac{2\pi}{\mu}\right)^{3/2} \omega \gamma T^{-3/2} \exp\left(-E_{\rm r}/T \right) \nonumber\\
&=&1.5394\times 10^{11}~{\rm cm}^3 {\rm mol}^{-1} {\rm s}^{-1} A^{-3/2} \omega \gamma_{,{\rm MeV}} \nonumber\\
&&\times T_9^{-3/2} \exp(-11.605E_{\rm r,MeV}/T_9),
\label{eq11}
\end{eqnarray}
where
$N_{\rm A}$ is Avogadro's number,
$A$ is the reduced mass in  atomic mass units (amu) given by $A=A_1 A_2/(A_1+A_2)$ with $A_1$ and $A_2$ the masses of two interacting particles, 1 and 2, in amu,
$T_9=T/(10^9~{\rm K})$ is the temperature in units of $10^9$ K.
The parameter $\omega$ is a statistical factor defined by
\begin{equation}
\omega=(1+\delta_{12})\frac{(2J+1)}{(2I_1+1)(2I_2+1)},
\label{eq12}
\end{equation}
where
$I_i$ is the spin of the particle $i$,
$J$ is the spin of the resonance,
and $\delta_{12}$ is the Kronecker delta necessary to avoid a double counting of identical particles.  The quantity $\gamma$ is defined  by
\begin{equation}
\gamma \equiv \frac{\Gamma_{\rm i} \Gamma_{\rm f}}{\Gamma(E_{\rm r})},
\label{eq13}
\end{equation}
where $\Gamma_{\rm i}$ and $\Gamma_{\rm f}$ are the partial widths for the entrance and exit channels, respectively.  $\Gamma(E_{\rm r})$ is the total width for a resonance with resonance energy $E_{\rm r}$, $\gamma_{,{\rm MeV}}$ is the $\gamma$ factor in units of MeV, and $E_{\rm r,MeV}$ is the resonance energy in units of MeV.

When $\omega=1$  as in the reactions considered here, and the radiative decay widths of $^8$B$_X^\ast$ and $^9$B$_X^\ast$, $\Gamma_\gamma$, are 
 much smaller than those for proton emission (as assumed here), the thermal reaction rate is given by
\begin{eqnarray}
N_{\rm A} \langle \sigma v \rangle&=&1.5394\times 10^{11}~{\rm cm}^3 {\rm mol}^{-1} {\rm s}^{-1} A^{-3/2} \Gamma_{\gamma,{\rm MeV}} \nonumber\\
&&\times T_9^{-3/2} \exp(-11.605E_{\rm r,MeV}/T_9)\nonumber\\
&\equiv& C T_9^{-3/2} \exp(-11.605E_{\rm r,MeV}/T_9),
\label{eq14}
\end{eqnarray}
where
$\Gamma_{\gamma,{\rm MeV}}=\Gamma_\gamma/({\rm 1~MeV})$ is the radiative decay width in units of MeV, and
$C$ is a rate coefficient determined from $A$ and $\Gamma_\gamma$.

The rate for a spontaneous emission via an electric dipole (E1) transition is given \citep{Blatt} by
\begin{equation}
\Gamma_\gamma=\frac{16\pi}{9}~e_1^2~E_\gamma^3~\frac{1}{2I_{\rm i}+1} \sum_{M_{\rm
i},~M_{\rm f}} \left|\int r Y_{1\mu}(\hat{r}) \Psi_{\rm f}^\ast \Psi_{\rm i}~d{\bfr}\right|^2,
\label{eq15}
\end{equation}
where
\begin{equation}
e_1= e \frac{Z_1 m_2-Z_2 m_1}{m_1+m_2}
\label{eq16}
\end{equation}
is the effective charge with $m_i$ and $Z_i$ the mass and the charge number of species $i=1$ and 2. $E_\gamma$ is the energy of the emitted photon, $I_{\rm i}$ is the angular momentum of the initial state, $M_{\rm i}$ and $M_{\rm f}$ are magnetic quantum numbers of initial and final states with $\mu=M_{\rm i}-M_{\rm f}$. $\Psi_{\rm i}$ and $\Psi_{\rm f}$ are wave functions of the initial and final states, respectively, and $Y_{1\mu}(\hat{r})$ is the dipole spherical surface harmonic.

We assume that the nuclear states do not significantly change between $^{8,9}$B$_X^{\ast{\rm a}}$ and
$^{8,9}$B$_X$.  For both resonances of $^{8,9}$B$_X^{\ast{\rm a}}$, the quantity $\Gamma_{\gamma,{\rm MeV}}$ is estimated to be
\begin{equation}
\Gamma_{\gamma,{\rm MeV}}=1.26539\times 10^{14} {\rm s}^{-1} e_1^2 (E_{\gamma,{\rm MeV}})^3 (\tau_{\rm if},{\rm fm})^2~,
\label{eq17}
\end{equation}
where
$e_1=e(Z_{\rm B}m_X-Z_Xm_{\rm B})/(m_{\rm B}+m_X)$ is the effective charge with $Z_{\rm B}=5$ and $Z_X=-1$ the charge numbers of $^{8,9}$B and the $X^-$, respectively, and
$\tau_{\rm if} \equiv \int r^2 dr \psi_{\rm f}^\ast r \psi_{\rm i}$ is the radial matrix element.

\begin{figure}
\begin{center}
\includegraphics[width=0.45\textwidth]{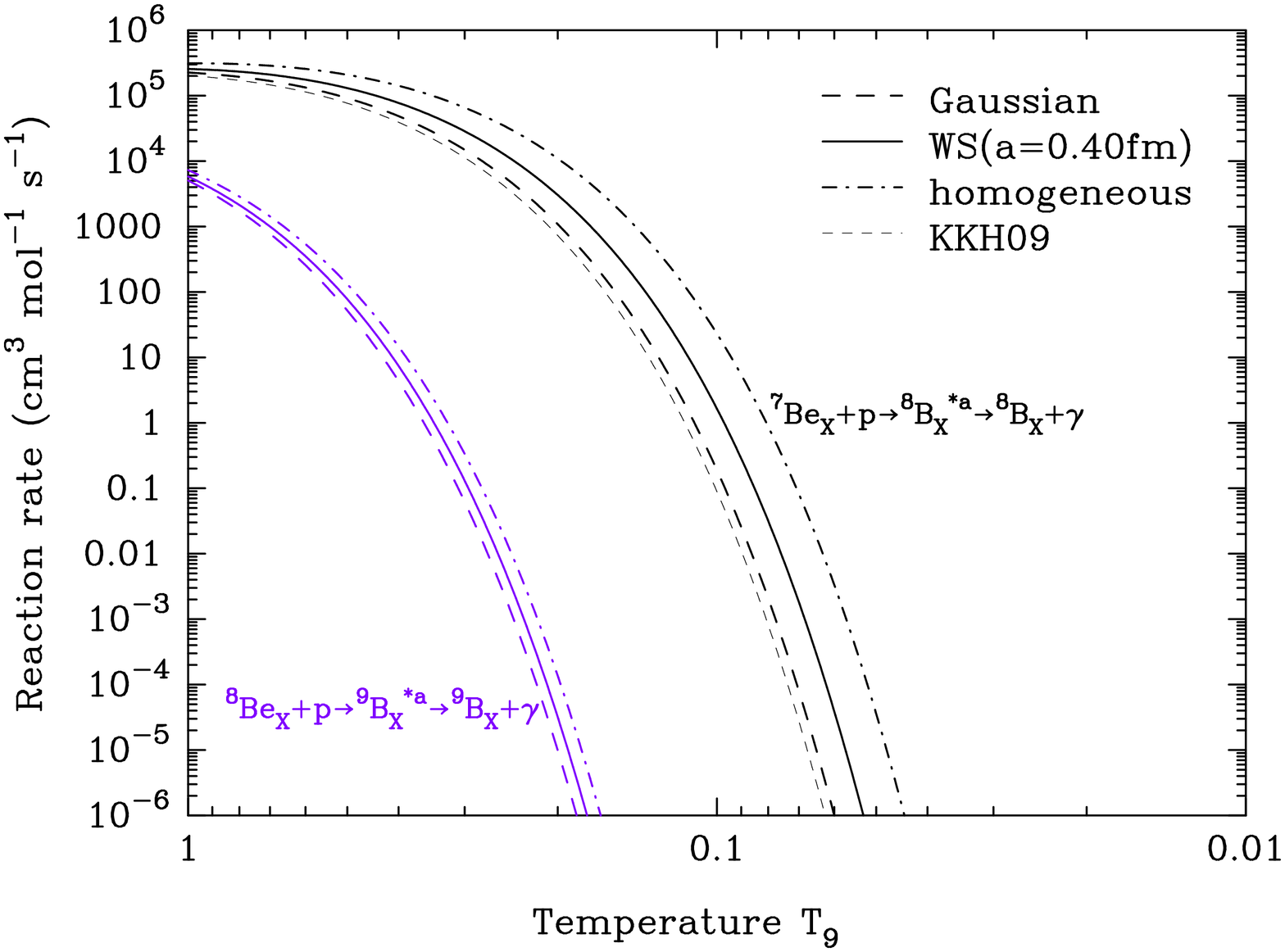}
\end{center}
\caption{Thermonuclear reaction rates for resonant reactions $^7$Be$_X$($p$, $\gamma$)$^8$B$_X$ (black lines) and $^8$Be$_X$($p$, $\gamma$)$^9$B$_X$ (purple lines) as a function of $T_9\equiv T/(10^9~{\rm K})$ for the case of $m_X=1000$ GeV.  Lines are drawn for different nuclear charge distributions  as labeled, i.e.  Gaussian (thick dashed lines), Woods-Saxon type with diffuseness parameter $a=0.40$ fm (solid lines), and a homogeneous well (dot-dashed lines).  The thin dashed line shows the reaction rate for $^7$Be$_X$($p$, $\gamma$)$^8$B$_X$ derived by means of  a quantum many-body model for $m_X=\infty$ \citep{Kamimura:2008fx}. \label{fig5}}
\end{figure}

Figure \ref{fig5} shows thermonuclear reaction rates for resonant reactions $^7$Be$_X$($p$, $\gamma$)$^8$B$_X$ (black lines) and $^8$Be$_X$($p$, $\gamma$)$^9$B$_X$ (purple lines) as a function of $T_9\equiv T/(10^9~{\rm K})$ for the case of $m_X=1000$ GeV.  Thick dashed, solid, and dot-dashed lines correspond to Gaussian type, WS40, and homogeneous type of nuclear charge distributions, respectively.  The thin dashed line corresponds to the reaction rate for $^7$Be$_X$($p$, $\gamma$)$^8$B$_X$ derived by means of  a quantum many-body model calculation for $m_X=\infty$ \citep{Kamimura:2008fx}.  Since the resonant reaction rate is proportional to the Boltzmann suppression factor of $\exp(-E_{\rm r}/T)$, relatively small differences in resonance energies between different charge distribution cases (Fig. \ref{fig2}) can lead to significant differences in the reaction rates.


Tables \ref{tab3} and \ref{tab4} show calculated parameters for the resonant reactions $^7$Be$_X$($p$, $\gamma$)$^8$B$_X$ and $^8$Be$_X$($p$, $\gamma$)$^9$B$_X$, for the three model  charge distributions and  with a fixed mass of $m_X=1$ TeV.  The matrix elements, the resonance energies, the energies of emitted photons, the radiative decay widths of the resonances, the rate coefficients, and the reaction $Q$-values are listed in the second to seventh columns, respectively.

\placetable{tab3}
\placetable{tab4}

The resonance energy, the dipole photon energy, and the reaction $Q$-value are given by
\begin{eqnarray}
E_{\rm r}&=&-E_{\rm B}({\rm B}_X^{\ast{\rm a}})-[E({\rm Be}+p)-E_{\rm B}({\rm Be}_X)], \nonumber\\
E_\gamma&=&E_{\rm B}({\rm B}_X)-E_{\rm B}({\rm B}_X^{\ast{\rm a}}), \nonumber\\
Q&=&E({\rm Be}+p)-E_{\rm B}({\rm Be}_X)+E_{\rm B}({\rm B}_X) \nonumber\\
 &=&E_\gamma-E_{\rm r},
\label{eq18}
\end{eqnarray}
respectively, where the quantities $E(^7$Be+$p)=0.1375$ MeV and $E(^8$Be+$p)=-0.1851$ MeV are binding energies of $^8$B and $^9$B with respect to the energies of the separation channels, respectively.

Tables \ref{tab14} and \ref{tab15} show calculated parameters of the resonant reactions $^7$Be$_X$($p$, $\gamma$)$^8$B$_X$ and $^8$Be$_X$($p$, $\gamma$)$^9$B$_X$, respectively, obtained with the WS40 model for $m_X=1$, 10, 100, and 1000 GeV.  The matrix elements, the resonance energies, the energies of emitted photons, the radiative decay widths of resonances, the rate coefficients, and the reaction $Q$-values are listed in the second to seventh columns, respectively.

\placetable{tab14}
\placetable{tab15}

In our BBN calculation, resonant rates for the proton capture reactions are adopted, while the nonresonant rates are taken from \citet{Kamimura:2008fx}.

\section{RADIATIVE RECOMBINATION WITH $X^-$}\label{sec5}
\subsection{$^7$Be}\label{sec5.1}
\subsubsection{Energy Levels}\label{sec5.1.1}

Table \ref{tab5} shows the binding energies of $^7$Be$_X$ atomic states with main quantum numbers $n$ ranging from one to seven.  Since the $^7$Be nuclear charge distribution has a finite size, the amplitude of the Coulomb potential at small $r$ is less than that for two point charges.  Wave functions at small radii and binding energies of tightly bound states with small $n$ values, therefore, deviate from those of the Bohr model.  Binding energies in the Bohr model are given by $E_{\rm B}^{\rm Bohr}=-Z^2 \alpha^2 \mu/2n^2$, where $\alpha$ is the fine structure constant.  On the other hand, the binding energies of loosely bound states with large $n$ values are similar to those of the Bohr model.

\placetable{tab5}

\subsubsection{$^7$Be($X^-$, $\gamma$)$^7$Be$_X$ Resonant Rate}\label{sec5.1.2}

The resonant rates of the reaction $^7$Be($X^-$, $\gamma$)$^7$Be$_X$ are calculated for $m_X$=1, 10, 100, and 1000 GeV adopting the WS40 model for the nuclear charge distribution.  The normalization of the total charge leads to a radius parameter, $R=2.63$ fm.  Radiative decay widths for E1 transitions are calculated taking into account the change of the E1 effective charge as a function of $m_X$.

In general, the recombination can efficiently proceed via resonant reactions through atomic states ${^7A^\ast}_X^{\ast{\rm a}}$, composed of a nuclear excited state $^7A^\ast$ and an $X^-$ \citep{Bird:2007ge}.  In these reactions, the resonances radiatively decay to lower energy states of ${^7A^\ast}_X^{\ast{\rm a}}$, ${^7A^\ast}_X$, ${^7A}_X^{\ast{\rm a}}$, and ${^7A}_X$ that have  larger binding energies.  Once bound states are produced in the reaction, subsequent transitions via radiative decays to lower energy states occur quickly.  Finally, the GS $^7A_X$ is produced after atomic states are converted to the atomic GS, and the nuclear excited state $^7A^\ast$ inside the atomic states is converted to the nuclear GS \citep{Bird:2007ge}.

Table \ref{tab6} shows calculated parameters of important transitions related to the reaction $^7$Be($X^-$, $\gamma$)$^7$Be$_X$ for $m_X=1$, 10, 100, and 1000 GeV.  There are an infinite number of atomic states of $^7$Be$^\ast_X$, composed of the first nuclear excited state $^7$Be$^\ast$[$\equiv ^7$Be$^\ast$(0.429 MeV, $1/2^-$)] and an $X^-$.   Among them states that satisfy $E_{\rm B}\lesssim 0.4291$ MeV are important resonances in the  recombination.  We take into account atomic resonances with binding energies of $0.23$ MeV $\leq E_{\rm B}\leq 0.43$ MeV.  They are the 1S state for $m_X=1$ GeV, the 2S and the 2P states for $m_X=10$ GeV, and the 3S, the 3P, and the 3D states for $m_X=100$ GeV and 1000 GeV.  The transitions, the matrix elements, the radiative decay widths of the resonance, and the resonance energies are listed in the second to fifth columns, respectively.  

Binding energies of $^7$Be$^\ast_X$ are taken to be the same as those of $^7$Be$_X$.  
This approximation is justified since the quantum three body model \citep{Kamimura:2008fx} for $\alpha$+$^3$He+$X^-$ showed that the RMS charge radii of $^7$Be and $^7$Be$^\ast$ differ by only 0.05 fm.   For the case of $m_X=1$ GeV, there is no important resonance of atomic excited states because of the relatively small binding energies of $^7$Be$_X$.  The most important resonance is then the atomic GS of $^7$Be$^\ast_X$(1S), which can decay only into atomic states of the nuclear GS, i.e., $^7$Be$_X^{\ast{\rm a}}$ and $^7$Be$_X$.  We take the measured rate for the radiative decay of $^7$Be$^\ast$ \citep{Tilley2002} as that for the decay of $^7$Be$^\ast_X$(1S) into the GS $^7$Be$_{X}$(1S).  This rate is listed although this transition is a magnetic dipole transition and, therefore,  relatively weak \citep{Tilley2002}.

\placetable{tab6}

We note that if a final state of the resonance decay is a resonance above the energy threshold of the $A+X^-$ separation channel, the final state instantaneously decays into the separation channel.  The resonant reaction with the final state is, therefore, not an available path to the GS $A_X$.  For example, in the case of $m_X=10$ GeV, the state of $^7$Be$_{X}^{\ast{\rm a}}$(2P) can be produced  via the resonance $^7$Be$_{X}^{\ast{\rm a}}$(2S) with a resonance energy of $E_{\rm r}=0.103$ MeV.  However, the 2P state quickly decays into the separation channel before it can radiatively decay to the GS.

Pathways in the resonant reaction $^7$Be($X^-$, $\gamma$)$^7$Be$_X$ are divided into three types according to the final states in the transitions from  atomic state resonances $^7A_X^\ast$ or ${^7A^\ast}_X^{\ast{\rm a}}$.  Type 1 involves  transitions to atomic states of the same nuclear state ($^7A_X^\ast$ or ${^7A^\ast}_X^{\ast{\rm a}}$).  For Type 1, decay widths for the transitions can be approximately calculated by taking into account only the atomic wave functions.  Type 2 involves  transitions to the nuclear GS of the same atomic state ($^7A_X$ or ${^7A}_X^{\ast{\rm a}}$).  
For Type 2, the decay widths can be approximately calculated by taking into account only the nuclear wave functions.  Type 3 denotes  transitions to different atomic states of the nuclear GS ($^7A_X$ or ${^7A}_X^{\ast{\rm a}}$).  This transition type simultaneously involves both atomic and nuclear transitions, and the number of possible final states can be very large.  In addition, calculations of decay widths for the transitions need both nuclear and atomic wave functions.  
Although a precise calculation of decay widths is beyond the scope of this study, we show in Appendix \ref{app1} that the E1 widths for Type 3 transitions are significantly smaller than those of Type 1.  In Appendix \ref{app1}, we suggest that the E1 width for Type 3 transitions can be interestingly large for exotic atomic systems involving a negatively charged particle with a mass equal to or larger than the nuclear mass.  Most importantly, Type 3 transition widths can be much larger than those of normal atomic systems composed of nuclei and electrons.

We suppose that in Type 1 transitions the  nuclear states do not significantly change and  only their atomic states change.  Then, one can simply  take atomic wave functions expressed as $\Psi_{\rm i}(\bfr)=\psi_{\rm i}(r)Y_{l_{\rm i}m_{\rm i}}(\hat{r})$ and $\Psi_{\rm f}(\bfr)=\psi_{\rm f}(r)Y_{l_{\rm f}m_{\rm f}}(\hat{r})$, where $\psi_{\rm i}(r)$ and $\psi_{\rm f}(r)$ are radial wave functions of initial and final states, respectively, $l_{\rm i}$ and $m_{\rm i}$ are the azimuthal and magnetic quantum numbers, respectively, of the initial state, and $l_{\rm f}$ and $m_{\rm f}$ are those of the final state.  The radiative decay width [Eq. (\ref{eq15})] of the resonance ${^7A^\ast_X}^{\ast{\rm a}}$ is then rewritten in the form 
\begin{eqnarray}
\Gamma_\gamma&=&C(l_{\rm i}, l_{\rm f})~e_1^2~E_\gamma^3 \tau_{\rm if}^2.
\label{eq19}
\end{eqnarray}
where $C(l_{\rm i}, l_{\rm f})$ is a constant  function of angular momenta $l_{\rm i}$ and $l_{\rm f}$.  The values, $c(0,1)=4/3$, $c(1,0)=4/9$, and $c(2,1)=8/15$, are used in deriving the following rates.

The thermal resonant rate is given by Eq. (\ref{eq11}), where in the $^7A$+$X^-$ recombination (for $A$=Li or Be) the reduced mass in amu is $A=A_A A_X /(A_A+A_X)$, and 
the statistical factor is
\begin{eqnarray}
\omega&=&\frac{2J+1}{(2I_A+1)(2I_X+1)} \nonumber\\
&=&\frac{(2l_{\rm res}+1)[2I(A^\ast(1/2^-))+1]}{2I(A(3/2^-))+1}=\frac{(2l_{\rm res}+1)}{2},~~~
\label{eq20}
\end{eqnarray}
where
$l_{\rm res}$ is the azimuthal quantum number of the resonance, and
$I(A(3/2^-))=3/2$ and $I(A(1/2^-))=1/2$ are the spins of the GS and the first nuclear excited state of $^7A$, respectively.

The resonant rates via Types 1 and 2 (for $m_X=1$ GeV) transitions are derived as
\begin{widetext}
\begin{numcases}
{N_{\rm A} \langle \sigma v \rangle_{\rm R} = }
2.94 \times 10^2~{\rm cm}^3 {\rm mol}^{-1} {\rm s}^{-1} T_9^{-3/2} \exp(-1.02/T_9)
& ~~~(for~$m_X=1$~GeV) \label{eq24}
\\ 
7.40 \times 10^4~{\rm cm}^3 {\rm mol}^{-1} {\rm s}^{-1} T_9^{-3/2} \exp(-0.230/T_9)
& ~~~(for~$m_X=10$~GeV) \label{eq21}
\\ 
3.73 \times 10^4~{\rm cm}^3 {\rm mol}^{-1} {\rm s}^{-1} T_9^{-3/2} \exp(-1.62/T_9) & \nonumber\\
+1.49 \times 10^4~{\rm cm}^3 {\rm mol}^{-1} {\rm s}^{-1} T_9^{-3/2} \exp(-1.80/T_9)
& ~~~(for~$m_X=100$~GeV) \label{eq22}
\\ 
3.86 \times 10^4~{\rm cm}^3 {\rm mol}^{-1} {\rm s}^{-1} T_9^{-3/2} \exp(-1.43/T_9) & \nonumber \\
+1.44 \times 10^4~{\rm cm}^3 {\rm mol}^{-1} {\rm s}^{-1} T_9^{-3/2} \exp(-1.64/T_9)
& ~~~(for~$m_X=1000$~GeV). \label{eq23}
\end{numcases}
\end{widetext}
The rate for $m_X=1$ GeV corresponds to the pure nuclear transition from the resonance  $^7$Be$^\ast_X$(1S) to the GS ${^7{\rm Be}}_X$(1S).
The rate for $m_X=10$ GeV corresponds to the atomic transition from the resonance $^7$Be${^\ast_X}^{\ast{\rm a}}$(2P) to the GS $^7$Be$^\ast_X$(1S).  The first terms in the rates for $m_X=100$ and 1000 GeV correspond to the atomic transition from the resonance $^7$Be${^\ast_X}^{\ast{\rm a}}$(3D) to $^7$Be${^\ast_X}^{\ast{\rm a}}$(2P), while the second terms correspond to sums of the atomic transitions from the resonance $^7$Be${^\ast_X}^{\ast{\rm a}}$(2P) to $^7$Be${^\ast_X}^{\ast{\rm a}}$(2S) and $^7$Be$^\ast_X$(1S).

This calculated rate is compared to the previous rate derived in the limit of infinite $m_X$ [Eq. (2.9) of \citet{Bird:2007ge}] \footnote{In \citet{Bird:2007ge}, the effect of direct capture to the state $^7$Be$^\ast_X$(2S) is estimated in the extreme assumption that the 2S state lies above the threshold of $^7$Be+$X^-$ with a resonance energy of 10 keV [Eq. (2.11) of \citet{Bird:2007ge}].  However, a three body calculation for the $\alpha$+$^3$He+$X^-$ system has confirmed that the 2S state is below the energy threshold, and thus not a resonance.  The resonant rate without the effect of the 2S state, i.e., Eq. (2.9) of \citet{Bird:2007ge}, should  therefore be used (M. Kamimura 2008; private communications; Sec. 3.6 in \citet{Kamimura:2008fx}).}.  
We take the rate for $m_X=1000$ GeV for this comparison.  Our first term for the transition 3D $\rightarrow$ 2P is a factor of $\sim 2$ higher than that of \citet{Bird:2007ge}.  Our second term for the transition 3P $\rightarrow$  2S and 1S is roughly the same as that of \citet{Bird:2007ge}.

\subsubsection{$^7$Be($X^-$, $\gamma$)$^7$Be$_X$ Nonresonant Rate}\label{sec5.1.3}

We fitted the function, i.e., $N_{\rm A} \langle \sigma v \rangle=(a+b T_9)/T_9^{1/2}$, to calculated nonresonant rates for the recombination of nuclei and $X^-$ particles in the temperature region of $T_9=[10^{-3}, 1]$, and obtained approximate analytical expressions.

With higher CM energy, the frequencies for the oscillations of continuum-state wave functions increases.  Thus, it takes more computational time to precisely calculate the radial matrix elements or cross sections at larger energy.  In the present study, we derived the cross sections only in the energy range of $10^{-5}$ MeV $<E< 1$ MeV, and the recombination rates are calculated in the temperature range of $T_9 \le 1$ using the derived cross sections and just setting cross sections for $E>1$ MeV to be zero.  Since the nucleosynthesis as well as recombinations of $^4$He and heavier nuclei with $X^-$ proceed after the temperature of the universe decreases down to $T_9<1$, the reaction rates for higher temperatures $T_9>1$ are not necessary in BBN calculations.  Considering that at the relevant temperatures,  the contribution to the thermal rates from reactions at CM energies greater  than the temperature is small, our reaction rates can be safely used in the desired temperature regime.

The nonresonant rate for the reaction $^7$Be($X^-$, $\gamma$)$^7$Be$_X$ is then derived to be
\begin{widetext}
\begin{numcases}
{N_{\rm A} \langle \sigma v \rangle_{\rm NR} = }
2.44 \times 10^5~{\rm cm}^3 {\rm mol}^{-1} {\rm s}^{-1}  \left(1-0.344 T_9 \right) T_9^{-1/2}
& ~~~(for~$m_X=1$~GeV) \label{eq28}
\\ 
5.98 \times 10^4~{\rm cm}^3 {\rm mol}^{-1} {\rm s}^{-1}  \left(1-0.211 T_9 \right) T_9^{-1/2}
& ~~~(for~$m_X=10$~GeV) \label{eq29}
\\ 
4.07 \times 10^4~{\rm cm}^3 {\rm mol}^{-1} {\rm s}^{-1}  \left(1-0.196 T_9 \right) T_9^{-1/2}
& ~~~(for~$m_X=100$~GeV) \label{eq30}
\\ 
3.86 \times 10^4~{\rm cm}^3 {\rm mol}^{-1} {\rm s}^{-1}  \left(1-0.194 T_9 \right) T_9^{-1/2}
& ~~~(for~$m_X=1000$~GeV). \label{eq31}
\end{numcases}
\end{widetext}

Nonresonant cross sections are calculated  with RADCAP taking into account the multiple components of partial waves for scattering states.  We show continuum wave functions at the CM energy $E=0.07$ MeV, which is the average energy corresponding to the temperature of the recombination of $^7$Be+$X^-$ for the case of $m_X=1000$ GeV, i.e., $E=3T/2$ with $T\sim 0.4 \times 10^9$ K.

The total cross section for the absorption of an unpolarized photon with frequency $\nu$ via an E1 transition from a bound state ($n$, $l$) to a continuum state ($E$) is given \citep{Gaunt1930,Karzas1961} by
\begin{eqnarray}
\sigma_{nl\rightarrow E}&=&\frac{16 \pi^2}{3} e_1^2 \mu k \nu \nonumber\\
&&\times \left[\frac{l+1}{2l+1} \left( \tau_{nl}^{E,~l+1}\right)^2 + \frac{l}{2l+1} \left( \tau_{nl}^{E,~l-1}\right)^2 \right],\nonumber\\
\label{eq32}
\end{eqnarray}
where
$k=\sqrt{\mathstrut 2\mu E}$ is the wave number, and
\begin{equation}
\tau_{nl}^{E,~l\pm1}=\int r^2 dr \psi_{E,~l\pm1}(r) r \psi_{nl}(r),
\label{eq33}
\end{equation}
is the radial matrix element for the radius $r$, and wave functions are normalized as
\begin{equation}
\int r^2 dr \left|\psi_{nl}(r)\right|^2=1,
\label{eq34}
\end{equation}
and asymptotically
\begin{equation}
\psi_{E,~l}(r)\sim \frac{\sin\left[kr -\eta \ln (2kr) -l\pi/2+ \sigma_l +\delta_l\right]}{kr}
\label{eq35}
\end{equation}
at large $r$, where $\eta$ is defined by
\begin{equation}
\eta=\frac{Z}{ka_{\rm B}}=\left( \frac{Z^2 \alpha^2 \mu}{2E} \right)^{1/2}.
\label{eq38}
\end{equation}
with $a_{\rm B}=1/(\mu \alpha)$ the Bohr radius,
$\sigma_l$ is the Coulomb phase shift, and
$\delta_l$ is the phase shift due to the difference in Coulomb potential between cases of the point charge and finite size nuclei \citep{Burke2011}.
  The parameter $e_1$ is the effective charge as defined in Eq. (\ref{eq16}).  We note that the precise cross section [Eq. (\ref{eq32})] includes $e_1^2$ instead of $\alpha$ which is usually adopted for hydrogen-like normal atoms.

We compare the calculated cross sections with those for the recombination of two point charges. Wave functions of scattering and bound states, and the bound-free absorption cross section in a pure Coulomb field have been derived analytically.  The bound and continuum state wave functions are given \citep{Karzas1961} by
\begin{eqnarray}
\psi_{nl}(r)&=& \left(\frac{2Z}{na_{\rm B}}\right)^{3/2} \left[\frac{\Gamma(n+l+1)}{\Gamma(n-l)2n}\right]^{1/2} \frac{(2Zr/na_{\rm B})^l}{(2l+1)!} \nonumber \\
&&\nonumber\\
&&\times \mathrm{e}^{-Zr/na_{\rm B}} {}_1F_1\left(l+1-n;~2l+2;~\frac{2Zr}{na_{\rm B}}\right),\nonumber\\
\label{eq36}
\end{eqnarray}
\begin{eqnarray}
\psi_{E,~l}(r)&=& \exp\left(\frac{\eta \pi}{2}\right) \frac{\left|\Gamma(l+1-i\eta)\right|}{(2l+1)!} \left(2kr \right)^l \mathrm{e}^{ikr} \nonumber\\
&&\times {}_1F_1\left(l+1-i\eta;~2l+2;~-2ikr \right),
\label{eq37}
\end{eqnarray}
where
${}_1F_1$ is the regular confluent hypergeometric function.

The cross section for absorption or ionization  is analytically given \citep[Eqs. (36) and (37) of][]{Karzas1961} by
\begin{widetext}
\begin{eqnarray}
\sigma_{nl\rightarrow E,~l-1}&=& \frac{\pi e_1^2}{\mu \nu} \frac{2^{4l}}{3} \frac{l^2 (n+l)! \left\{\left(1^2+\eta^2 \right) \left(2^2+\eta^2 \right) ... \left[(l-1)^2+\eta^2 \right]\right\}}{(2l+1)! (2l-1)! (n-l-1)!}
\frac{\exp(-4\eta \cot^{-1} \rho)}{1-\mathrm{e}^{-2\pi \eta}} \nonumber\\
&&\times \frac{\rho^{2l+2}}{(1+\rho^2)^{2n-2}} \left[G_l(l+1-n;~\eta;~\rho) -(1+\rho^2)^{-2} G_l(l-1-n;~\eta;~\rho)\right]^2,
\label{eq39}
\end{eqnarray}
\end{widetext}
where the quantity in the curly brackets is unity when $l=1$, and
\begin{widetext}
\begin{eqnarray}
\sigma_{nl\rightarrow E,~l+1}&=& \frac{\pi e_1^2}{\mu \nu} \frac{2^{4l+6}}{3} \frac{(l+1)^2 (n+l)! \left(1^2+\eta^2 \right) \left(2^2+\eta^2 \right) ... \left[(l+1)^2+\eta^2 \right]}{(2l+1) (2l+1)! (2l+2)! (n-l-1)!\left[(l+1)^2+\eta^2 \right]^2}
\frac{\exp(-4\eta \cot^{-1} \rho)}{1-\mathrm{e}^{-2\pi \eta}} 
\frac{\rho^{2l+4} \eta^2}{(1+\rho^2)^{2n}}  \nonumber\\
&&\times \left[\left(l+1-n \right)G_{l+1}(l+1-n;~\eta;~\rho) +\frac{l+1+n}{1+\rho^2} G_{l+1}(l-n;~\eta;~\rho)\right]^2.
\label{eq40}
\end{eqnarray}
\end{widetext}
The first and second equations correspond to transitions to the continuum states with angular momenta $l-1$ and $l+1$, respectively.  The parameter $\rho$ is defined $\rho \equiv \eta/n$, and the real polynomial $G_l$ is given by:
\begin{equation}
G_l(-m,~\eta,~\rho)=\sum_{s=0}^{2m} b_s \rho^s,
\label{eq41}
\end{equation}
with coefficients
\begin{equation}
b_0=1,~~~~~b_1=\frac{2m\eta}{l},\nonumber\\
\label{eq42}
\end{equation}
\begin{eqnarray}
b_s&=&-\frac{1}{s(s+2l-1)} \left[ 4\eta (s-1-m) b_{s-1} \right. \nonumber\\
&&\left.~~~~~~~+(2m+2-s) (2m+2l+1-s) b_{s-2} \right].~~~
\label{eq43}
\end{eqnarray}

The recombination cross section can be  derived using the principle of  detailed balance \citep{Blatt,Rybicki1979} \footnote{It appears that Eq. (31) of \citet{Bertulani:2003kr} has an error.}:
\begin{equation}
\frac{\sigma_{E,~l\pm1\rightarrow nl}}{\sigma_{nl\rightarrow E,~l\pm1}}=\frac{[2I(n,l)+1]}{(2I_1+1)(2I_2+1)} \left(\frac{E_\gamma^2}{\mu E}\right),
\label{eq44}
\end{equation}
where
$I_1$ and $I_2$ are spins of particles 1 and 2 constituting the bound state,
$I(n,l)$ is the spin of the bound state $(n,l)$, and
the radiation energy is related to the CM energy and the binding energy by $E_\gamma=E+E_{\rm B}$.

Thermal recombination rate is derived as a function of temperature $T$ by integrating the calculated cross section $\sigma(E)$ over energy [Eq. (\ref{eq10})].  The analytical expression for the  wave function in  the case of a point charge nucleus \citep[Eqs. (31) and (32) of][]{Karzas1961} is derived using  the confluent hypergeometric function calculated with algorithm 707  of \citet{Nardin1992}.

\begin{figure}
\begin{center}
\includegraphics[width=0.45\textwidth]{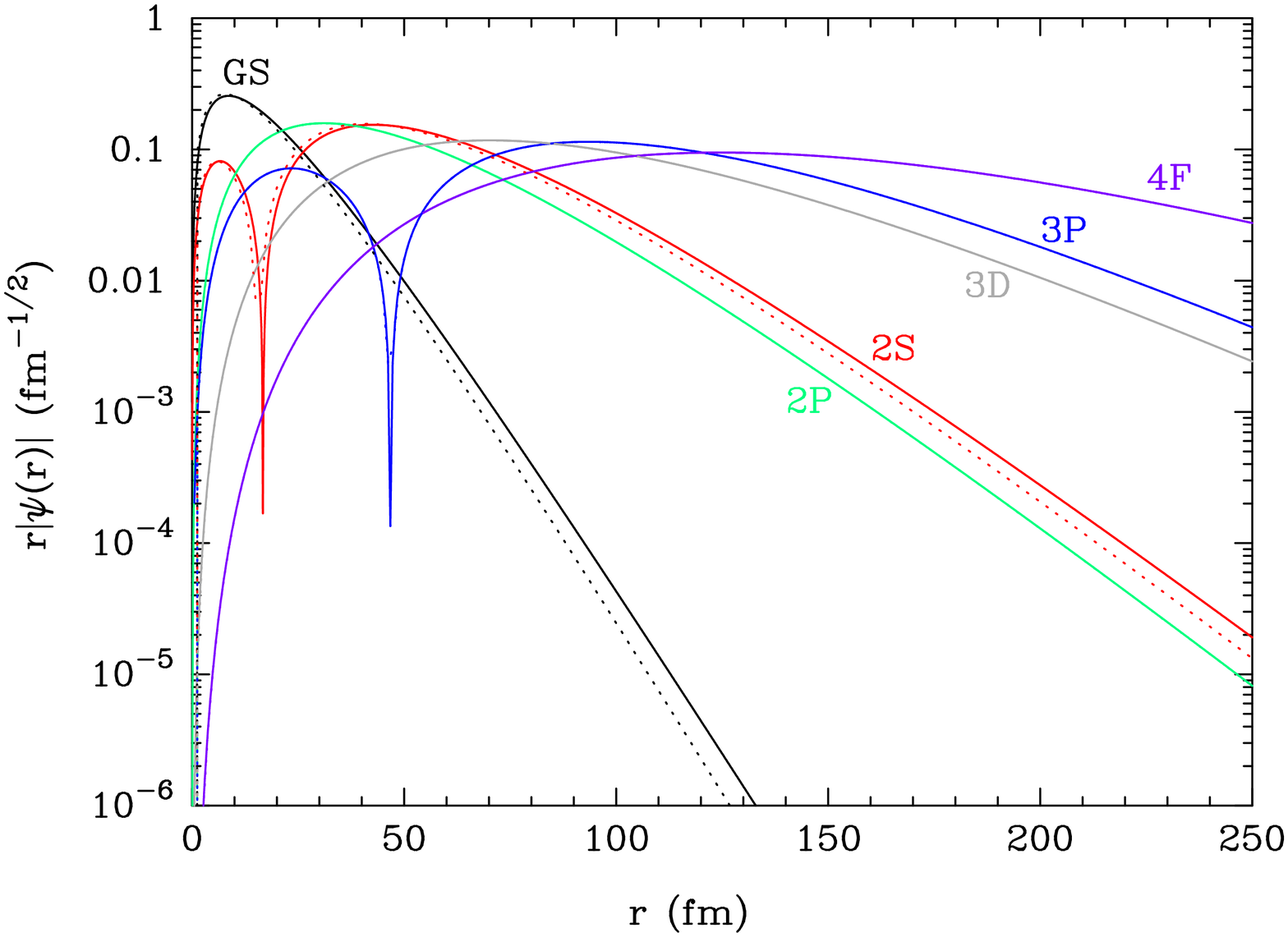}\\
\includegraphics[width=0.45\textwidth]{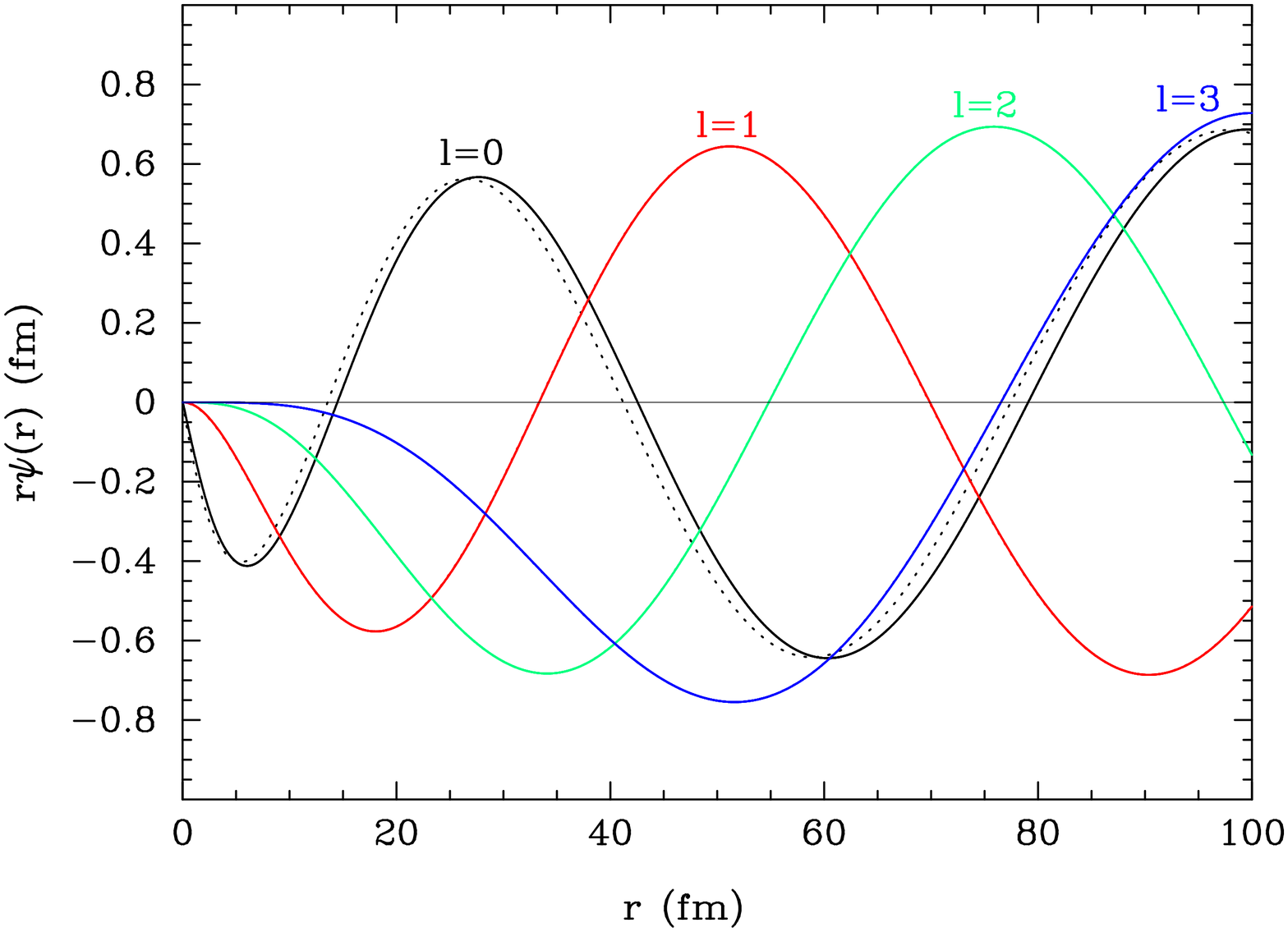}\\
\includegraphics[width=0.45\textwidth]{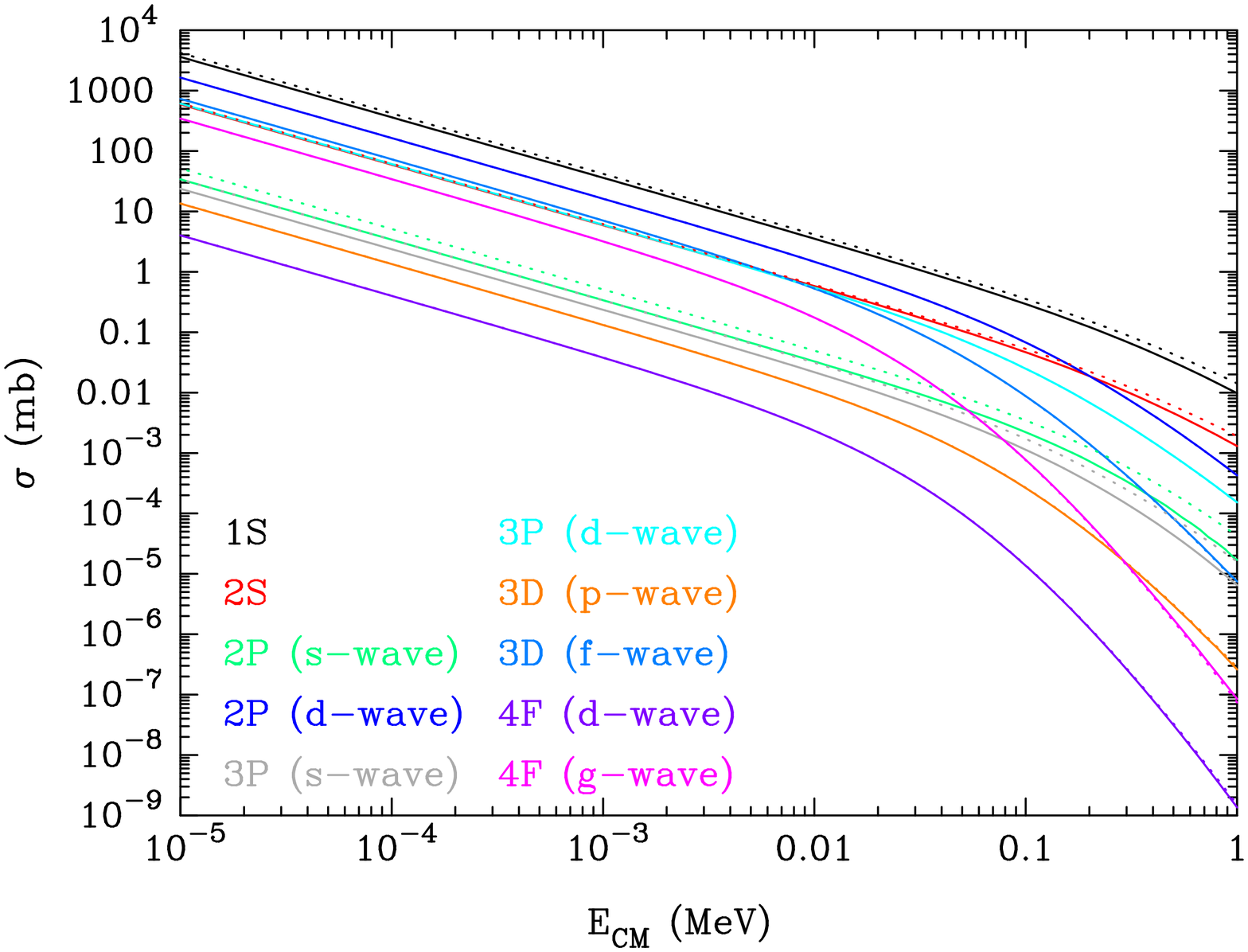}
\end{center}
\caption{Bound-state wave functions (upper panel) and continuum wave functions at $E=3T/2=0.07$ MeV (middle panel) for  the $^7$Be+$X^-$ system as a function of radius $r$ for the case of $m_X=1$ GeV.    The bottom panel shows the recombination cross section as a function of CM energy $E$.  In all panels the solid lines correspond to calculated results while the dotted lines correspond to analytical formulae for hydrogen-like atomic states composed of two point charges.   \label{fig6}}
\end{figure}

Figure \ref{fig6} shows bound-state wave functions (upper panel) and continuum wave functions (middle panel) at $E=0.07$ MeV for the $^7$Be+$X^-$ system as a function of radius $r$ for the case of $m_X=1$ GeV.   Solid lines correspond to calculated wave functions while the dotted lines correspond to the analytical formula for hydrogen-like atomic states composed of two point charges [Eqs. (\ref{eq36}) and (\ref{eq37})].  In the upper panel, wave functions for  the GS (1S state), 2S, 2P, 3P, 3D, and 4F states are plotted.  Here, one can see  that the wave functions for the  GS and 2S state in the finite charge distribution case (solid lines) deviate from those of the point charge case (dotted lines).  The wave functions of other states agree with those for the point charge case.  The scattering wave functions for the  $s$-, $p$-, $d$-, and $f$-waves are plotted in the middle panel.  Note,  that the normalization for the amplitude of the wave function adopted in RADCAP is different from that in \citet{Karzas1961}.
 Hence,  the latter wave functions are normalized to satisfy the former normalization.  In addition, wave functions derived with RADCAP are multiplied by $\exp(i\theta)$, where $\theta$ are arbitrary real constants, and then transformed into real numbers.  Only the wave function of the $l=0$ state for the finite charge distribution case (solid lines) deviates from that of  the point charge case (dotted lines).

The bottom panel shows the recombination cross section as a function of the energy $E$.  Solid lines correspond to the calculated results, while the dotted lines correspond to the analytical solution for the two point charges [Eqs. (\ref{eq39}), (\ref{eq40}), and (\ref{eq44})].  Partial cross sections for the following transitions are drawn:  scattering $p$-wave $\rightarrow$ bound 1S state (black lines); $p$-wave $\rightarrow$ 2S (red); $s$-wave $\rightarrow$ 2P (green); $d$-wave $\rightarrow$ 2P (blue); $s$-wave $\rightarrow$ 3P (gray);  $d$-wave $\rightarrow$ 3P (sky blue); $p$-wave $\rightarrow$ 3D (orange);  $f$-wave $\rightarrow$ 3D (cyan); $d$-wave $\rightarrow$ 4F (violet); and $g$-wave $\rightarrow$ 4F (magenta).  Since the mass $m_X$ is relatively small, the reduced mass is small, and the spatial extent  of the bound-state wave functions is large.  The effect of a finite size charge distribution is only important for a small $r$ and  is, therefore, small.  
Small differences 
 in bound and scattering state wave functions lead to small changes in the cross sections through differences in the binding energies and wave function shapes.  The largest differences in the cross sections are found for  the two transitions starting from an initial $s$-wave, i.e., $s$-wave $\rightarrow$ 2P and $s$-wave $\rightarrow$ 3P.  This is caused by differences in the scattering $s$-wave function.


\begin{figure}
\begin{center}
\includegraphics[width=0.45\textwidth]{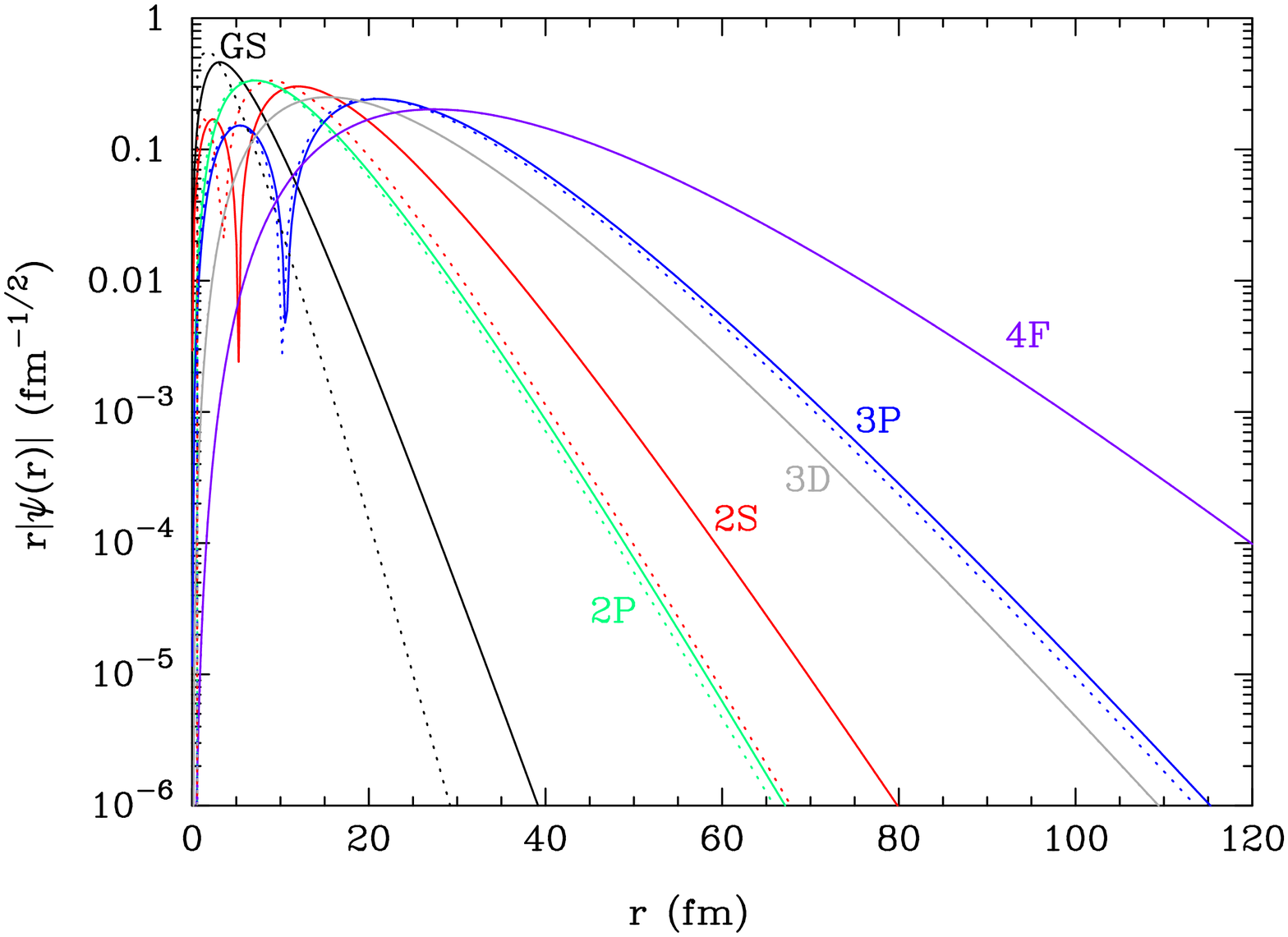}\\
\includegraphics[width=0.45\textwidth]{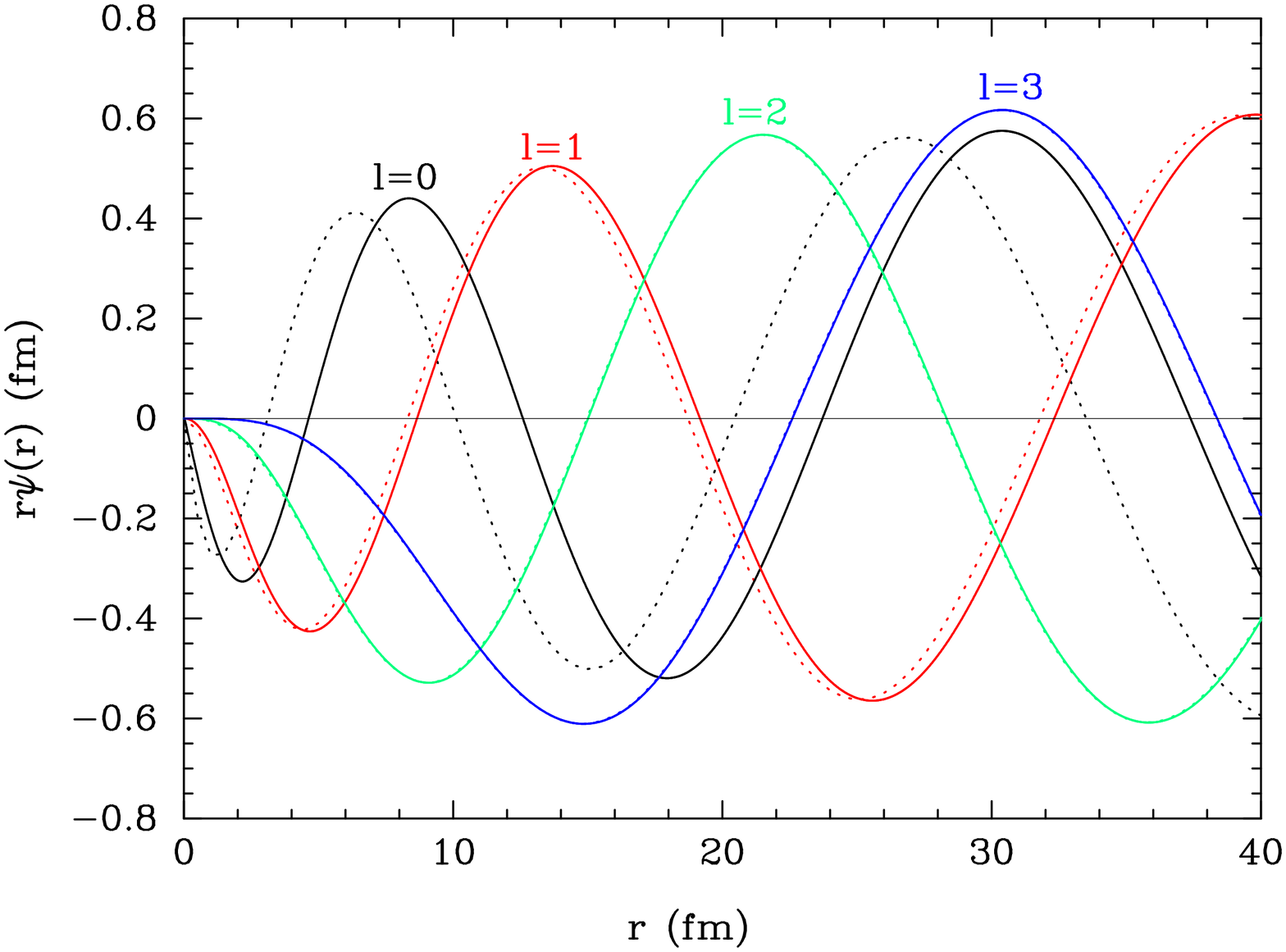}\\
\includegraphics[width=0.45\textwidth]{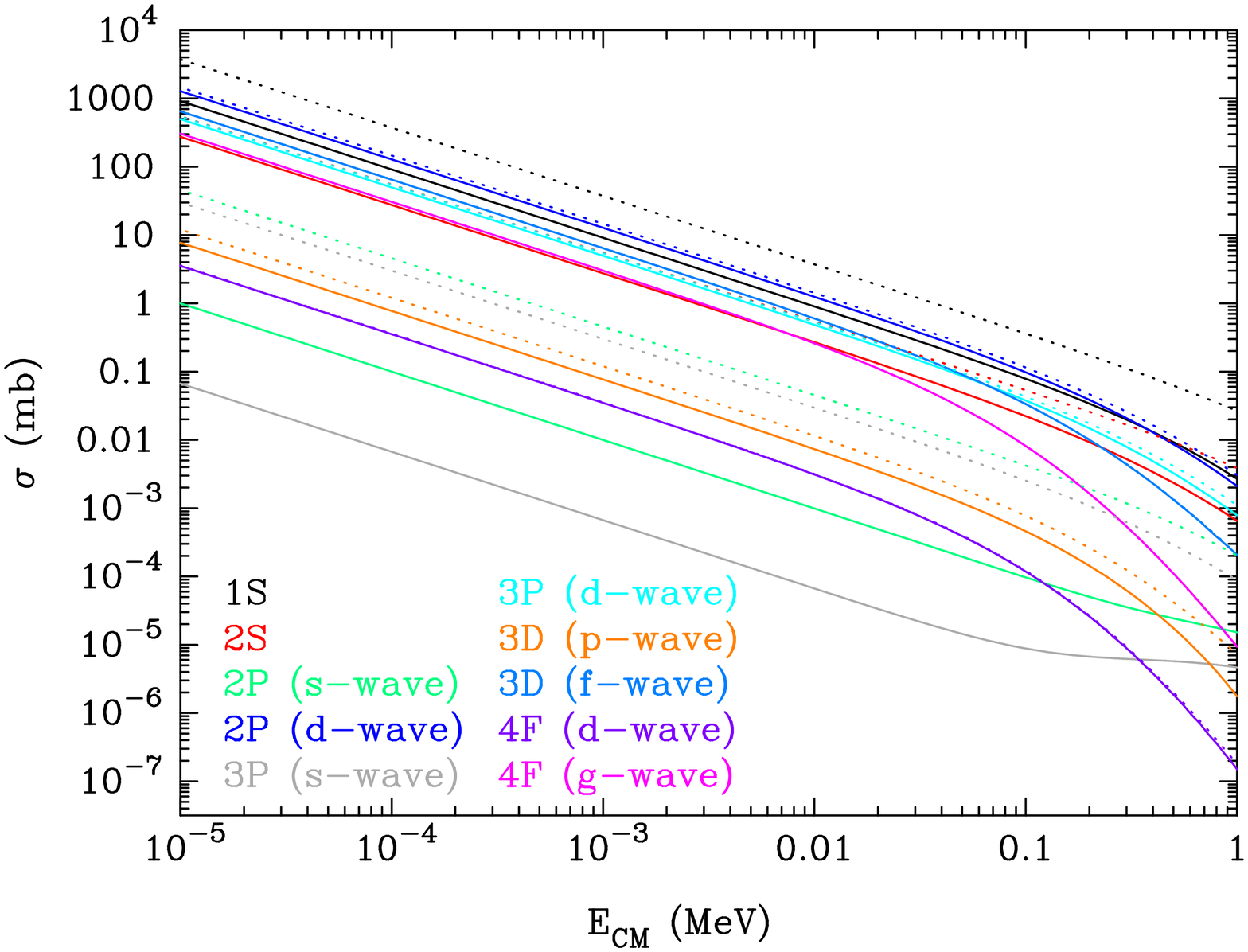}
\end{center}
\caption{Same as Fig. \ref{fig6}, but  for  $m_X=10$ GeV.  \label{fig7}}
\end{figure}

Figure \ref{fig7} shows bound state wave functions (upper panel) and continuum wave functions (middle panel) at $E=0.07$ MeV of the $^7$Be+$X^-$ system as a function of radius $r$ for the case of $m_X=10$ GeV.   Line types indicate the same quantities as in Fig. \ref{fig6}.  In the upper panel, the wave functions for the  GS and 2S state in  the finite charge distribution case (solid lines) deviate significantly  from those for the point-charge case (dotted lines).  Also, the wave functions for the  2P and 3P states  deviate slightly.  In the middle panel, the difference in the wave function for the  $l=0$ state is very large.  A difference in the $l=1$ state exists although it is not large.  The bottom panel shows the recombination cross section as a function of  energy $E$.  Line types indicate the same quantities as in Fig. \ref{fig6}.  
Because of the larger $m_X$ value, the effect of a finite-size charge distribution is more important.  Bound- and scattering-state wave functions, and recombination cross sections are then significantly different from those for the point charge case.  Because of the large difference in the scattering $s$-wave function, the cross sections for transitions from an initial $s$-wave, i.e., $s$-wave $\rightarrow$ 2P and $s$-wave $\rightarrow$ 3P are much  smaller than those in the point charge case.  Partial cross sections for transitions from an initial $p$-wave to bound 1S, 2S and 3D states are also altered  by the finite-size charge distribution.  The cross sections for transitions to 1S and 2S states are also affected by differences in binding energies of the states between the finite- and point-charge cases.


\begin{figure}
\begin{center}
\includegraphics[width=0.45\textwidth]{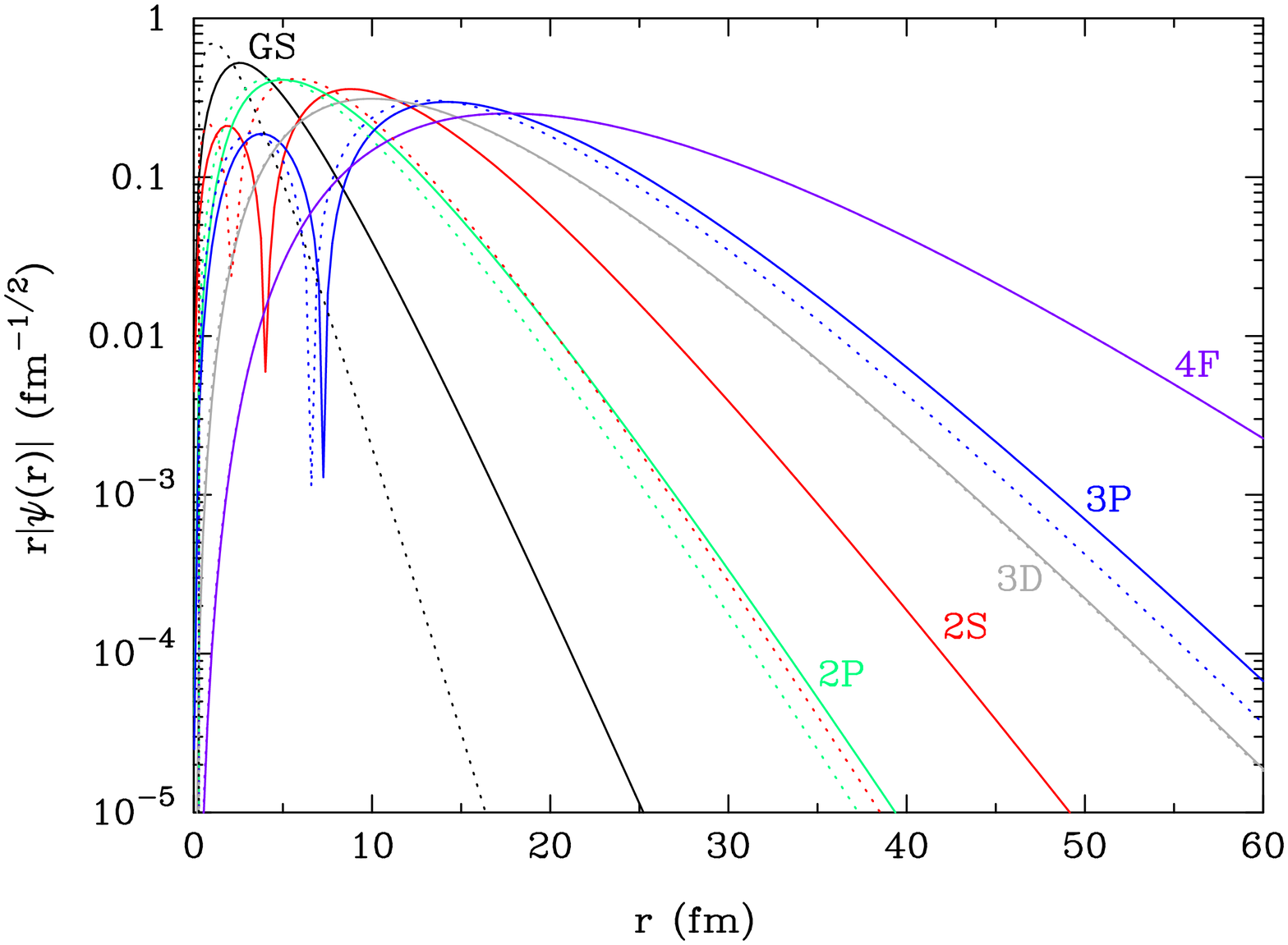}\\
\includegraphics[width=0.45\textwidth]{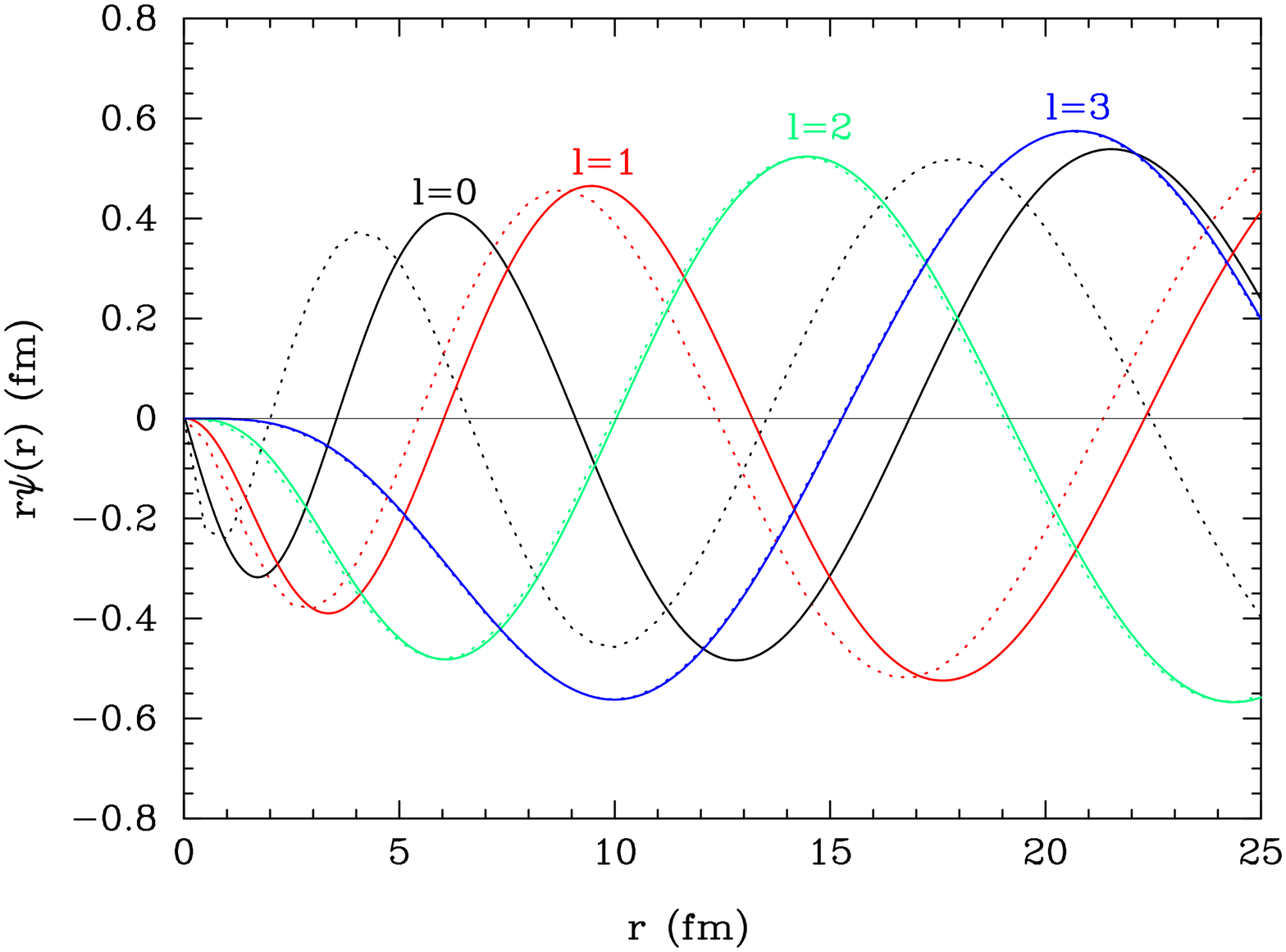}\\
\includegraphics[width=0.45\textwidth]{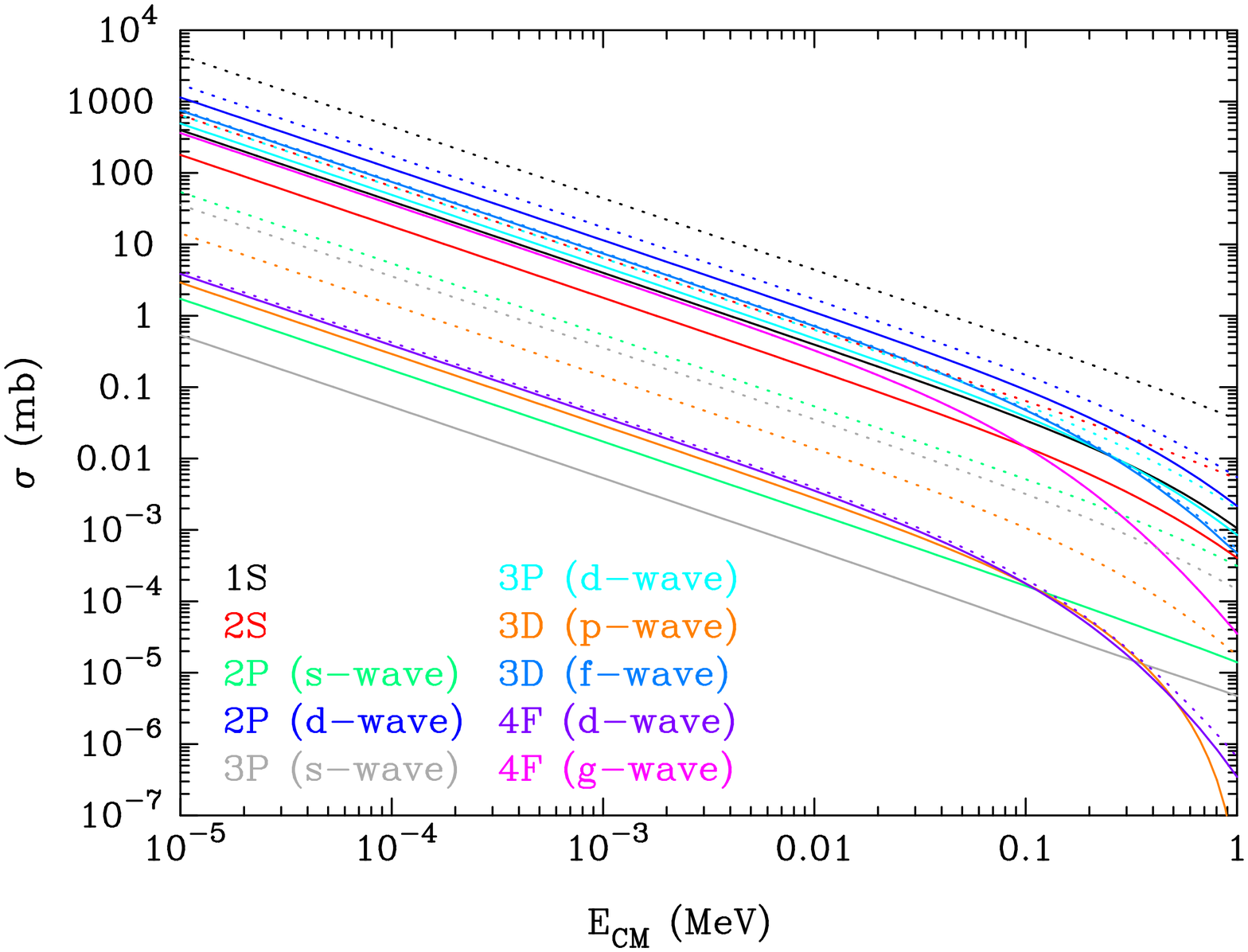}
\end{center}
\caption{Same as Fig. \ref{fig6} for the case of $m_X=100$ GeV.  \label{fig8}}
\end{figure}

Figure \ref{fig8} shows bound-state wave functions (upper panel) and continuum wave functions (middle panel) at $E=0.07$ MeV for the $^7$Be+$X^-$ system as a function of radius $r$ for the case of $m_X=100$ GeV.   Line types indicate the same quantities as in Fig. \ref{fig6}.  It is clear from a comparison of Figs. \ref{fig6}, \ref{fig7}, and \ref{fig8} that deviations of the wave functions from those in the point charge cases become larger as $m_X$ increases.  We can see that deviations of wave functions for bound GS, 2S, 2P and 3P states and scattering wave functions of $l=0$ and $l=1$ states are very large, and that a deviation exists for the $l=1$ state exist, but  it is not large.  The bottom panel shows the recombination cross section as a function of the energy $E$.  Line types indicate the same quantities as in Fig. \ref{fig6}.  Differences in the solid and dotted lines are even larger than in the case of $m_X=10$ GeV (Fig. \ref{fig7}).


\begin{figure}
\begin{center}
\includegraphics[width=0.45\textwidth]{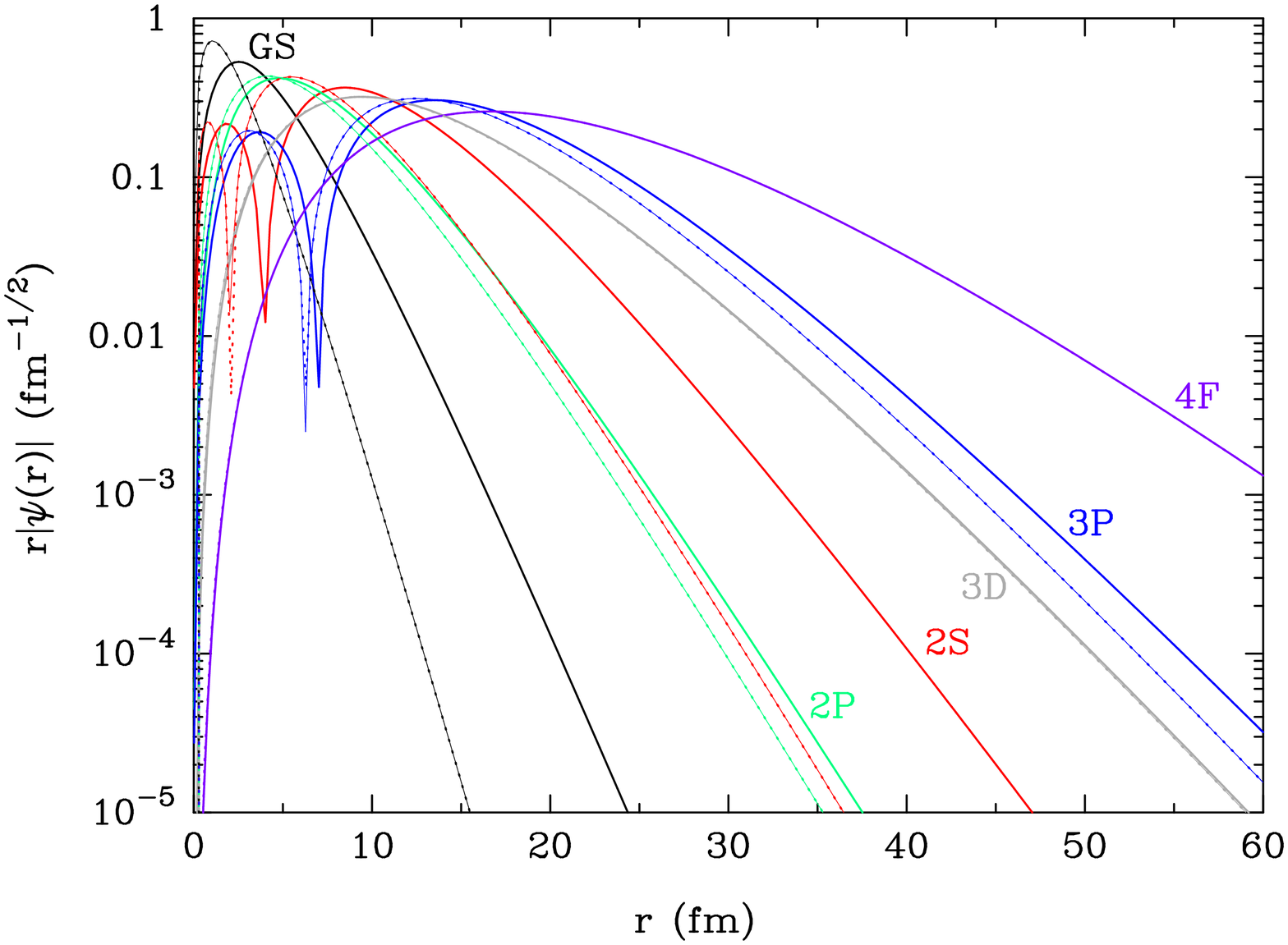}\\
\includegraphics[width=0.45\textwidth]{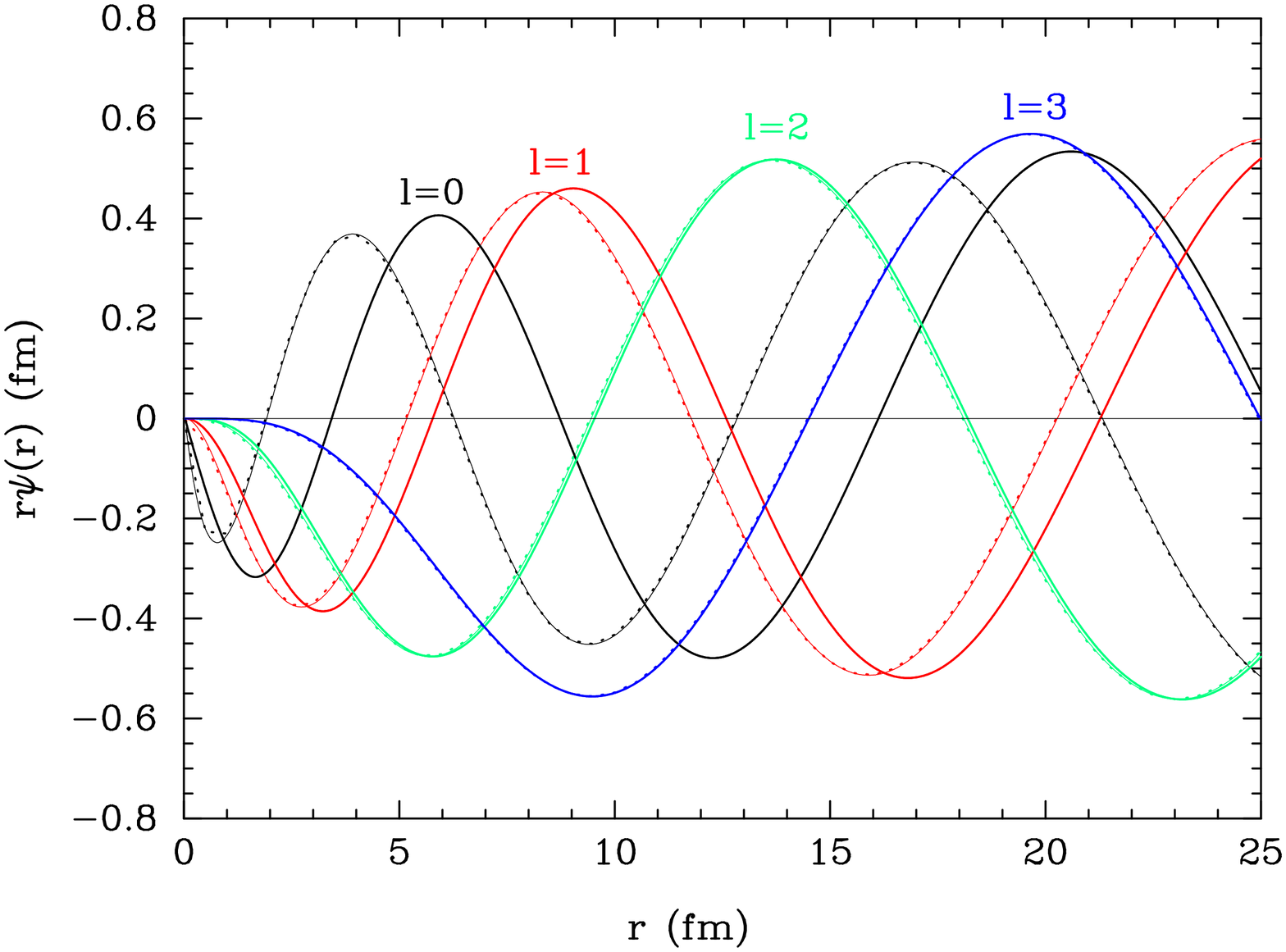}\\
\includegraphics[width=0.45\textwidth]{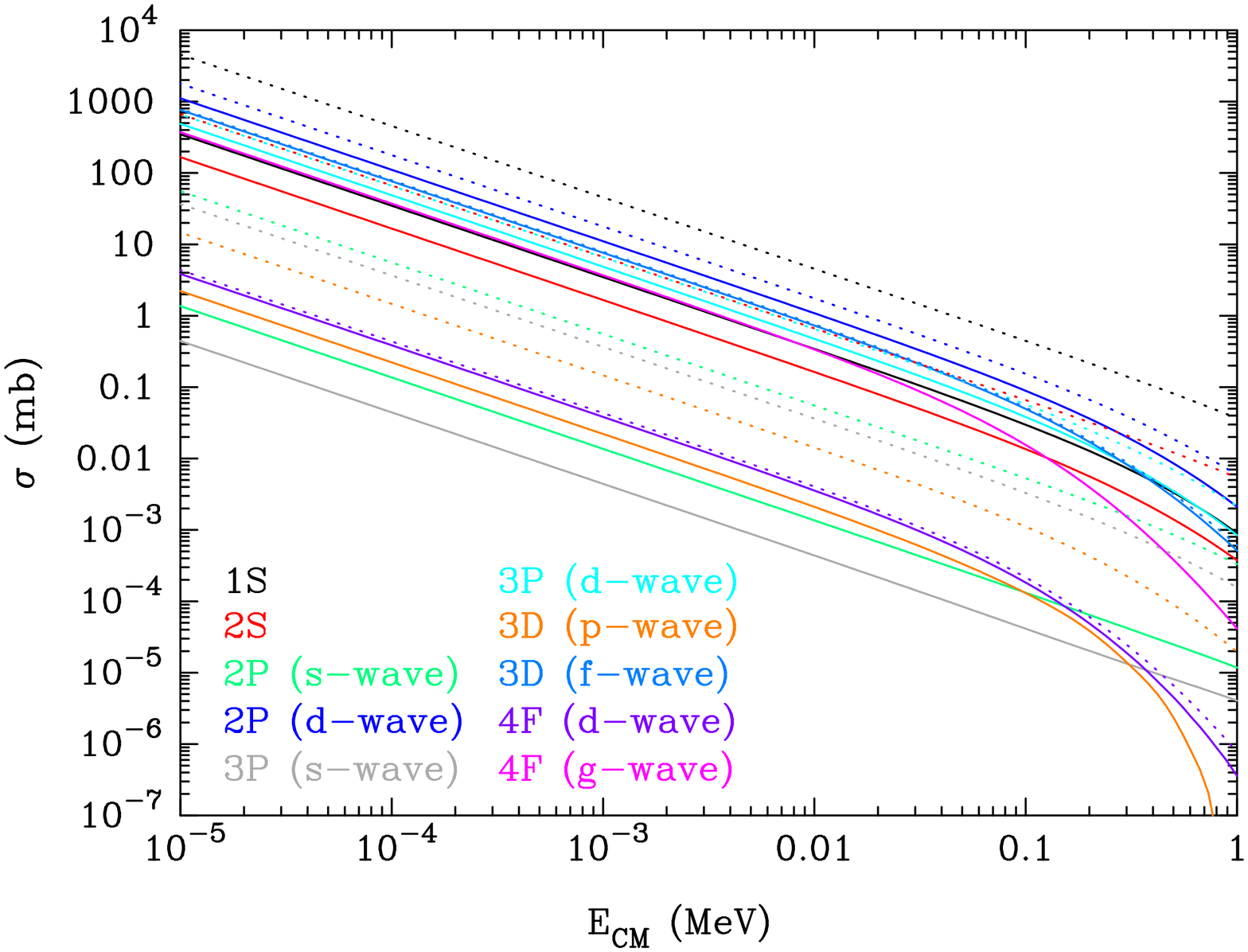}
\end{center}
\caption{Same as Fig. \ref{fig6} for the case of $m_X=1000$ GeV.   Thin solid lines in the upper and middle panels show results calculated under the assumption that $^7$Be has a point charge.  \label{fig9}}
\end{figure}

Figure \ref{fig9} shows bound state wave functions (upper panel) and continuum wave functions (middle panel) at $E=0.07$ MeV as a function of radius $r$ for  the $^7$Be+$X^-$ system in  the case of $m_X=1000$ GeV.  Also shown is the  recombination cross section as a function of the energy $E$ (bottom panel).   Thick solid and dotted lines indicate the same quantities as in Fig. \ref{fig6}.  Since the reduced mass is similar to that in the case of $m_X=100$ GeV, this figure is rather similar to Fig. \ref{fig8}.  In order to check  our calculations, we also calculate the wave functions and the cross sections for case of point-charge nuclei using  the same code (a modified version of RADCAP) as used for the finite charge distribution case.  Thin solid lines in the upper and middle panels show that the calculated results  agree with analytical solutions (dotted lines) quite well.


We found an important characteristic of the $^7$Be+$X^-$ recombination based upon our precise calculation including many transition channels:    In the limit of a heavy $X^-$ particle, i.e., $m_X\gtrsim 100$ GeV, the most important transition in the recombination is the $d$-wave $\rightarrow$ 2P.  This fact does not hold in the case of the point charge model.  In that case the transition $p$-wave $\rightarrow$ 1S is predominant (see dotted lines in Figs. \ref{fig6}--\ref{fig9}).  In the case of a finite size charge distribution, in addition to the main pathway of $d$-wave $\rightarrow$ 2P, cross sections for the transitions $f$-wave $\rightarrow$ 3D and $d$-wave $\rightarrow$ 3P are also larger than that for the GS formation.  It is thus found that estimations of recombination cross sections taking into account only the GS as the final state may not be correct.

We note that our rate for $m_X=1000$ GeV is more than 6 times larger than the previous rate \citep{Bird:2007ge}.  We confirmed that the previous rate \citep{Bird:2007ge} is somewhat close to our rate when only taking into account the transition from the scattering $p$-wave to the bound 1S and 2S states.  In \citet{Bird:2007ge}, it is described that the capture of $^7$Be directly to the GS of $^7$Be$_X$ has the largest cross section, closely followed by the capture to the 2S level.  This is true for hydrogen-like ions composed of point charged particles.   
However, we found that the most important transition is from the scattering $d$-wave to the bound 2P state.  The previous rate \citep{Bird:2007ge} was adopted in most previous  studies on BBN involving the $X^-$ particle,  including studies by part of the present authors \citep{Kusakabe:2007fu,Kusakabe:2007fv,Kusakabe:2010cb}.  The nonresonant recombination rate is  important for the $^7$Be destruction and also for a constraint on the parameter region  for solving the Li problem.  The significant improvement in the rate found in the present work, therefore, makes it possible to derive an improved constraint on the $X^-$ particle as shown in Sec. \ref{sec8}.

\subsection{$^7$Li}\label{sec5.2}
\subsubsection{Energy Levels}\label{sec5.2.1}

Table \ref{tab7} shows binding energies for the  $^7$Li$_X$ atomic states having  main quantum numbers $n$ from one to seven.

\placetable{tab7}

\subsubsection{$^7$Li($X^-$, $\gamma$)$^7$Li$_X$ Resonant Rate}\label{sec5.2.2}

The resonant rates of the reaction $^7$Li($X^-$, $\gamma$)$^7$Li$_X$ were  calculated for $m_X$=1, 10, 100, and 1000 GeV in  the WS40 model.  The radius parameter for the WS40 model is $R=2.48$ fm.

Table \ref{tab8} shows calculated parameters of important transitions related to the reaction $^7$Li($X^-$, $\gamma$)$^7$Li$_X$ for $m_X=1$, 100, and 1000 GeV.  Similarly to the recombination of $^7$Be+$X^-$, the recombination can efficiently proceed via resonant reactions involving  atomic states of ${^7{\rm Li}^\ast_X}^{\ast{\rm a}}$ composed of the first nuclear excited state $^7$Li$^\ast$[$\equiv ^7$Li$^\ast$(0.478 MeV, $1/2^-$)].  Important resonances for ${^7{\rm Li}^\ast_X}^{\ast{\rm a}}$ satisfy $E_{\rm B}\lesssim$ 0.47761 MeV.  We take into account atomic resonances with binding energies of 0.28 MeV $\leq E_{\rm B}\leq$ 0.48 MeV except for  the atomic GS for the case of $m_X=1$ GeV.
The transitions, the matrix elements, the radiative decay widths of the resonances, and the resonance energies are listed in the second to fifth columns, respectively.  For the case of $m_X=1$ GeV, there are no important resonances of atomic excited states because of the relatively small binding energies of $^7$Li$_X$.  The most important resonance in the recombination reaction is then the atomic GS of $^7$Li$^\ast_X$(1S), which can only decay into atomic states of the nuclear GS, i.e., $^7$Li$_X^{\ast{\rm a}}$ and $^7$Li$_X$.  We take the measured rate for the radiative decay of $^7$Li$^\ast$ \citep{Tilley2002} for the decay of $^7$Li$^\ast_X$(1S) into the GS $^7$Li$_{X}$(1S).

\placetable{tab8}

The resonant rates for the reaction $^7$Li($X^-$, $\gamma$)$^7$Li$_X$ via Types 1 and 2 (only for $m_X=1$ GeV) transitions are derived to be
\begin{widetext}
\begin{numcases}
{N_{\rm A} \langle \sigma v \rangle_{\rm R} = }
5.37 \times 10^2~{\rm cm}^3 {\rm mol}^{-1} {\rm s}^{-1} T_9^{-3/2} \exp(-3.24/T_9)
& ~~~(for~$m_X=1$~GeV) \label{eq47}
\\ 
1.72 \times 10^4~{\rm cm}^3 {\rm mol}^{-1} {\rm s}^{-1} T_9^{-3/2} \exp(-1.44/T_9)
& ~~~(for~$m_X=100$~GeV) \label{eq45}
\\ 
1.68 \times 10^4~{\rm cm}^3 {\rm mol}^{-1} {\rm s}^{-1} T_9^{-3/2} \exp(-1.22/T_9)
& ~~~(for~$m_X=1000$~GeV). \label{eq46}
\end{numcases}
\end{widetext}
The rate for $m_X=1$ GeV corresponds to a pure nuclear transition from the resonance  $^7$Li$^\ast_X$(1S) to the GS ${^7{\rm Li}}_X$(1S) \citep[magnetic dipole transition][]{Tilley2002}.
The rate for $m_X=10$ GeV is zero since there are no important resonances operating as a path for the recombination reaction.  The rates for $m_X=100$ and 1000 GeV correspond to the atomic transition from the resonance $^7$Li${^\ast_X}^{\ast{\rm a}}$(2P) to $^7$Li$^\ast_X$(1S).

\subsubsection{$^7$Li($X^-$, $\gamma$)$^7$Li$_X$ Nonresonant Rate}\label{sec5.2.3}

The thermal nonresonant rate for the reaction $^7$Li($X^-$, $\gamma$)$^7$Li$_X$ was derived as a function of temperature $T$ by integrating the calculated cross section $\sigma(E)$ over energy [Eq. (\ref{eq10})].  The derived rates are 
\begin{widetext}
\begin{numcases}
{N_{\rm A} \langle \sigma v \rangle_{\rm NR} = }
1.15 \times 10^5~{\rm cm}^3 {\rm mol}^{-1} {\rm s}^{-1}  \left(1-0.453 T_9 \right)  T_9^{-1/2}
& ~~~(for~$m_X=1$~GeV) \label{eq50}
\\ 
2.50 \times 10^4~{\rm cm}^3 {\rm mol}^{-1} {\rm s}^{-1}  \left(1-0.271 T_9 \right) T_9^{-1/2}
& ~~~(for~$m_X=10$~GeV) \label{eq51}
\\ 
1.70 \times 10^4~{\rm cm}^3 {\rm mol}^{-1} {\rm s}^{-1}  \left(1-0.247 T_9 \right) T_9^{-1/2}
& ~~~(for~$m_X=100$~GeV) \label{eq52}
\\ 
1.62 \times 10^4~{\rm cm}^3 {\rm mol}^{-1} {\rm s}^{-1}  \left(1-0.245 T_9 \right) T_9^{-1/2}
& ~~~(for~$m_X=1000$~GeV). \label{eq53}
\end{numcases}
\end{widetext}

\begin{figure}
\begin{center}
\includegraphics[width=0.45\textwidth]{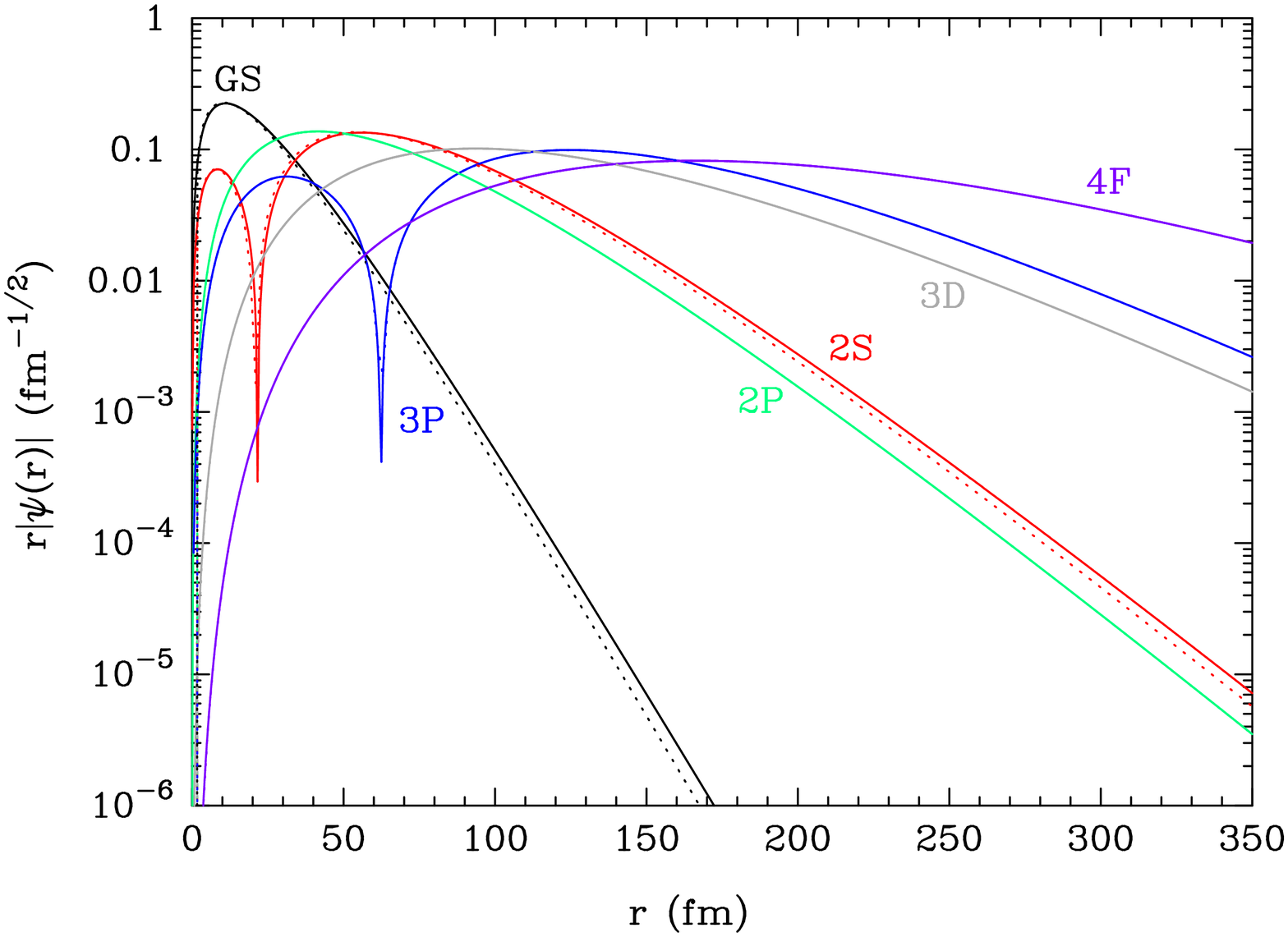}\\
\includegraphics[width=0.45\textwidth]{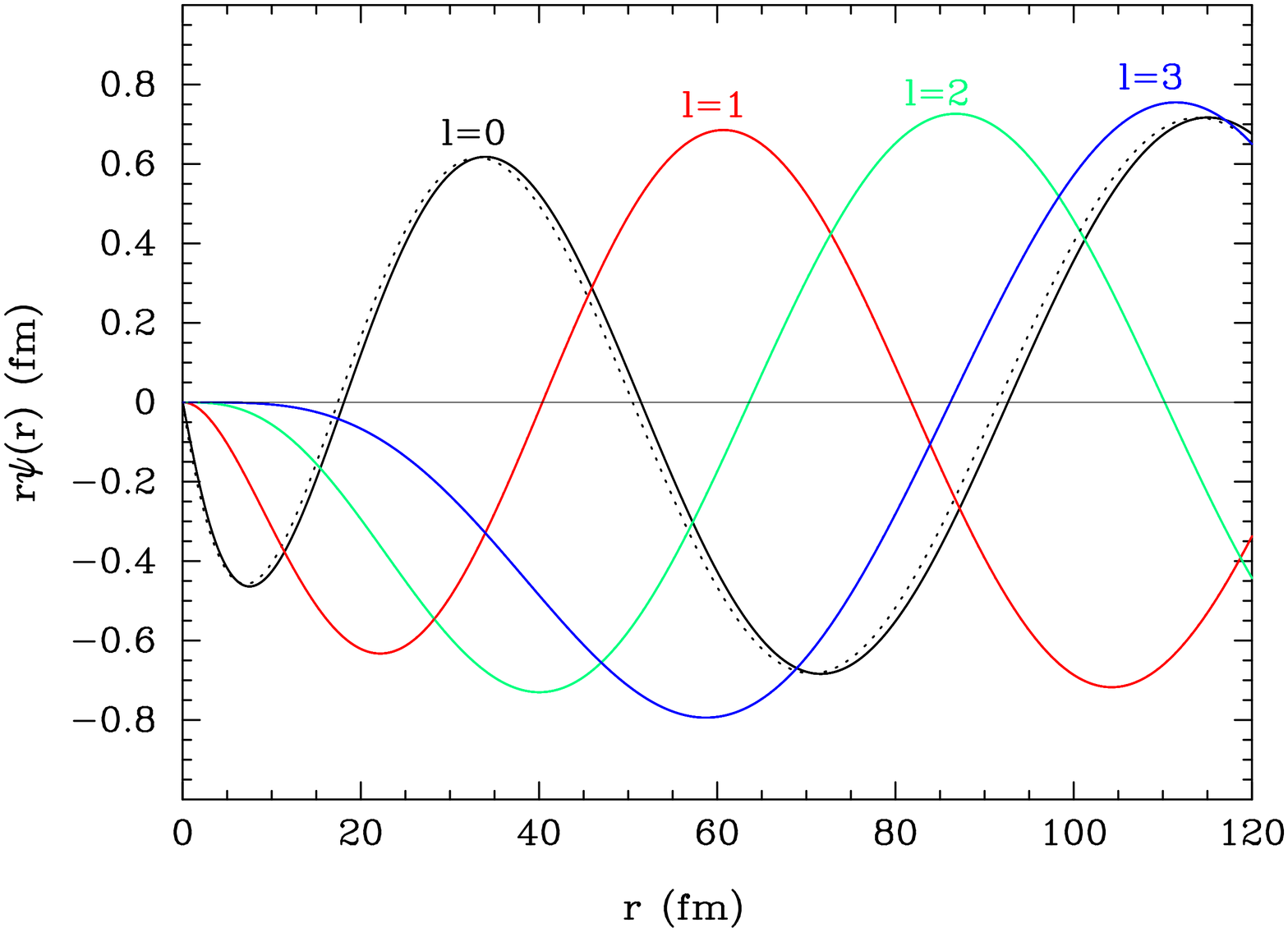}\\
\includegraphics[width=0.45\textwidth]{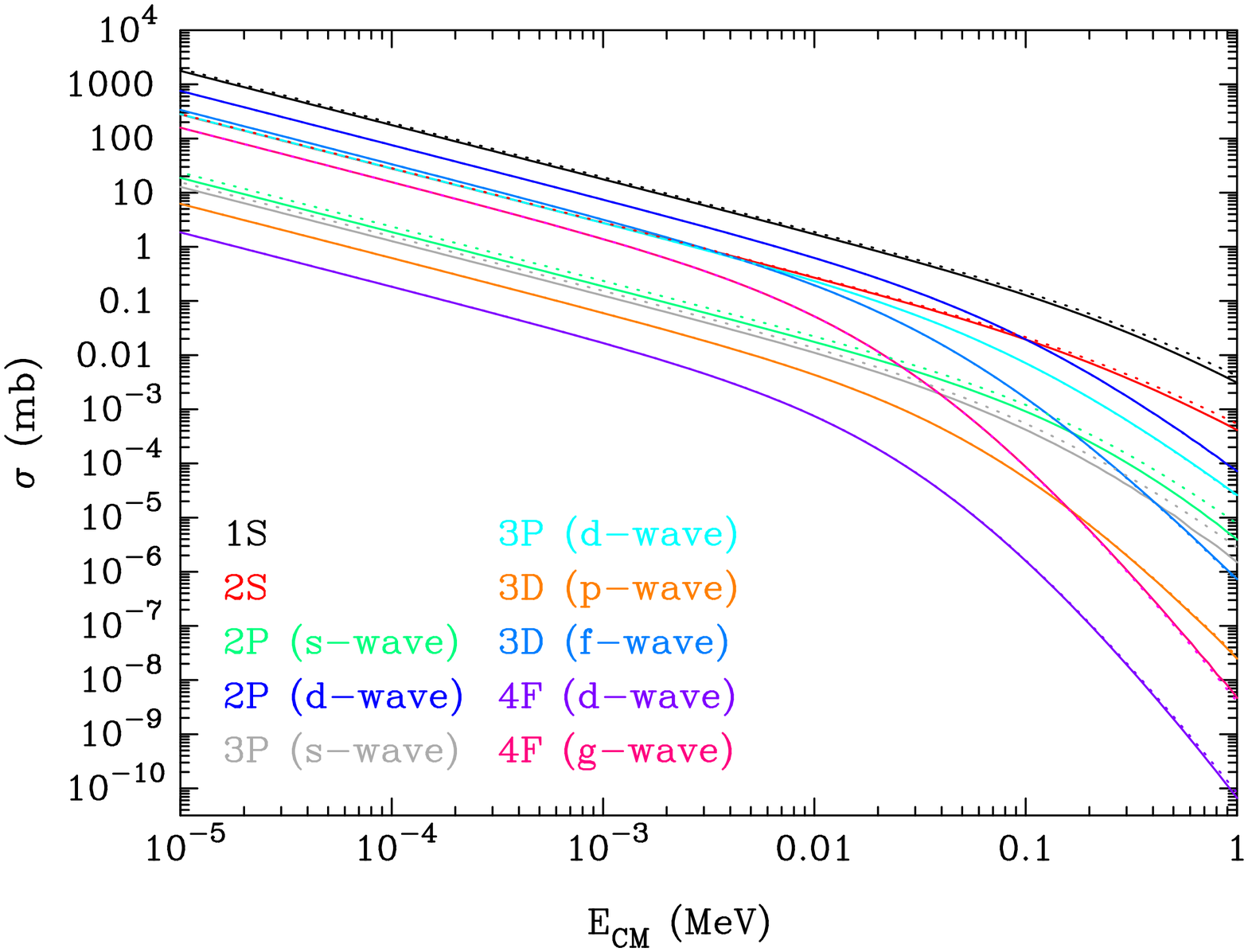}
\end{center}
\caption{Same as Fig. \ref{fig6} for the $^7$Li+$X^-$ system with $m_X=1$ GeV. \label{fig10}}
\end{figure}

Figure \ref{fig10} shows bound-state wave functions (upper panel) and continuum wave functions (middle panel) at $E=0.07$ MeV as a function of radius $r$ for  the $^7$Li+$X^-$ system with $m_X=1$ GeV.  
Also shown is the  recombination cross section as a function of the energy $E$ (bottom panel).   Line types indicate the same quantities as in Fig. \ref{fig6}.  In general, the trends of calculated results are similar to those of the $^7$Be+$X^-$ system (Fig. \ref{fig6}).  However, the Coulomb potential in the $^7$Li+$X^-$ system is smaller than that in the $^7$Be+$X^-$ system.  Therefore, the spatial widths of wave functions in the former system are larger.  The effect of the finite size of the charge distribution as shown by  differences between the solid and dotted lines is then somewhat smaller in the $^7$Li+$X^-$ system than  in the $^7$Be+$X^-$ system.


\begin{figure}
\begin{center}
\includegraphics[width=0.45\textwidth]{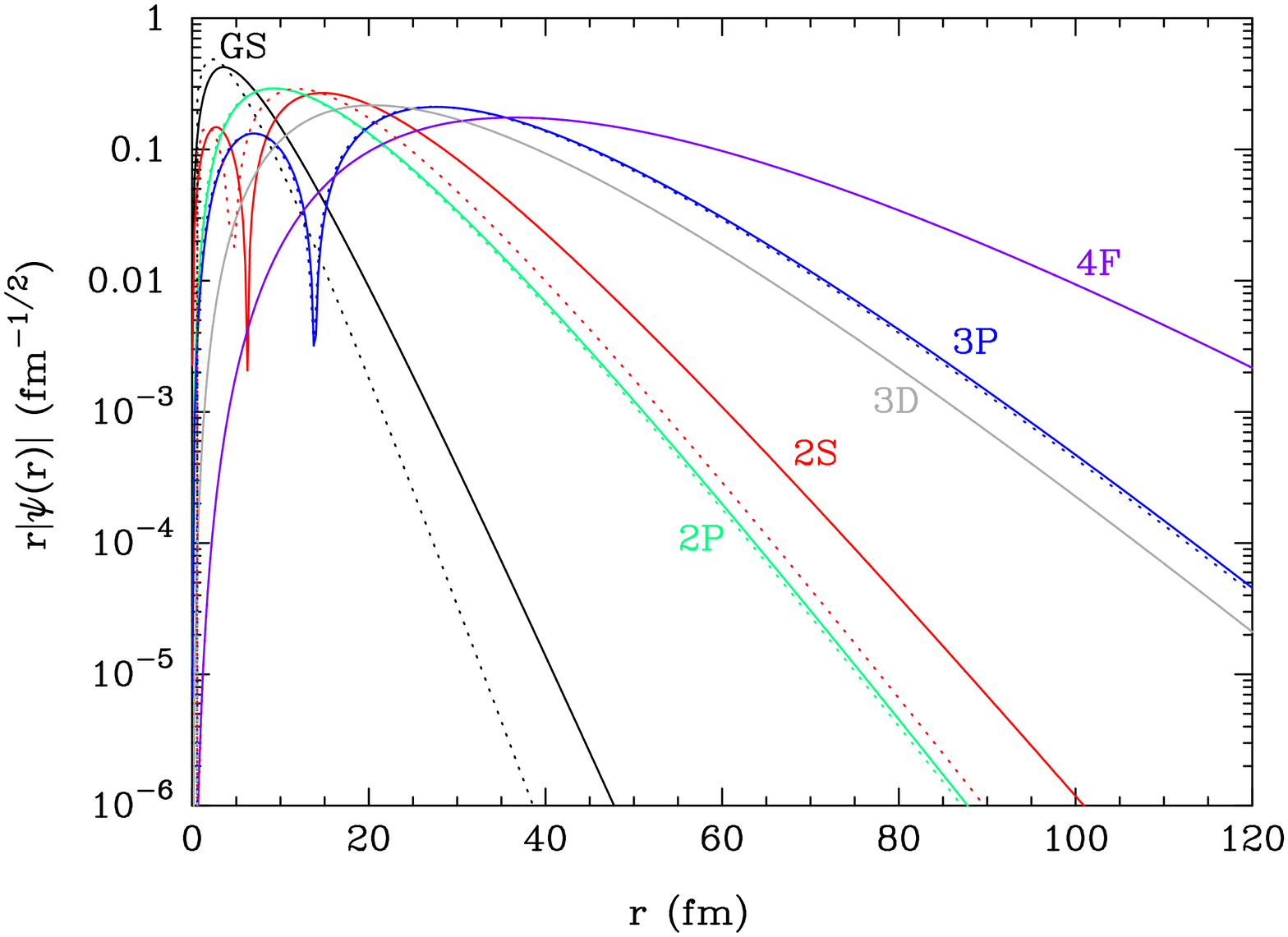}\\
\includegraphics[width=0.45\textwidth]{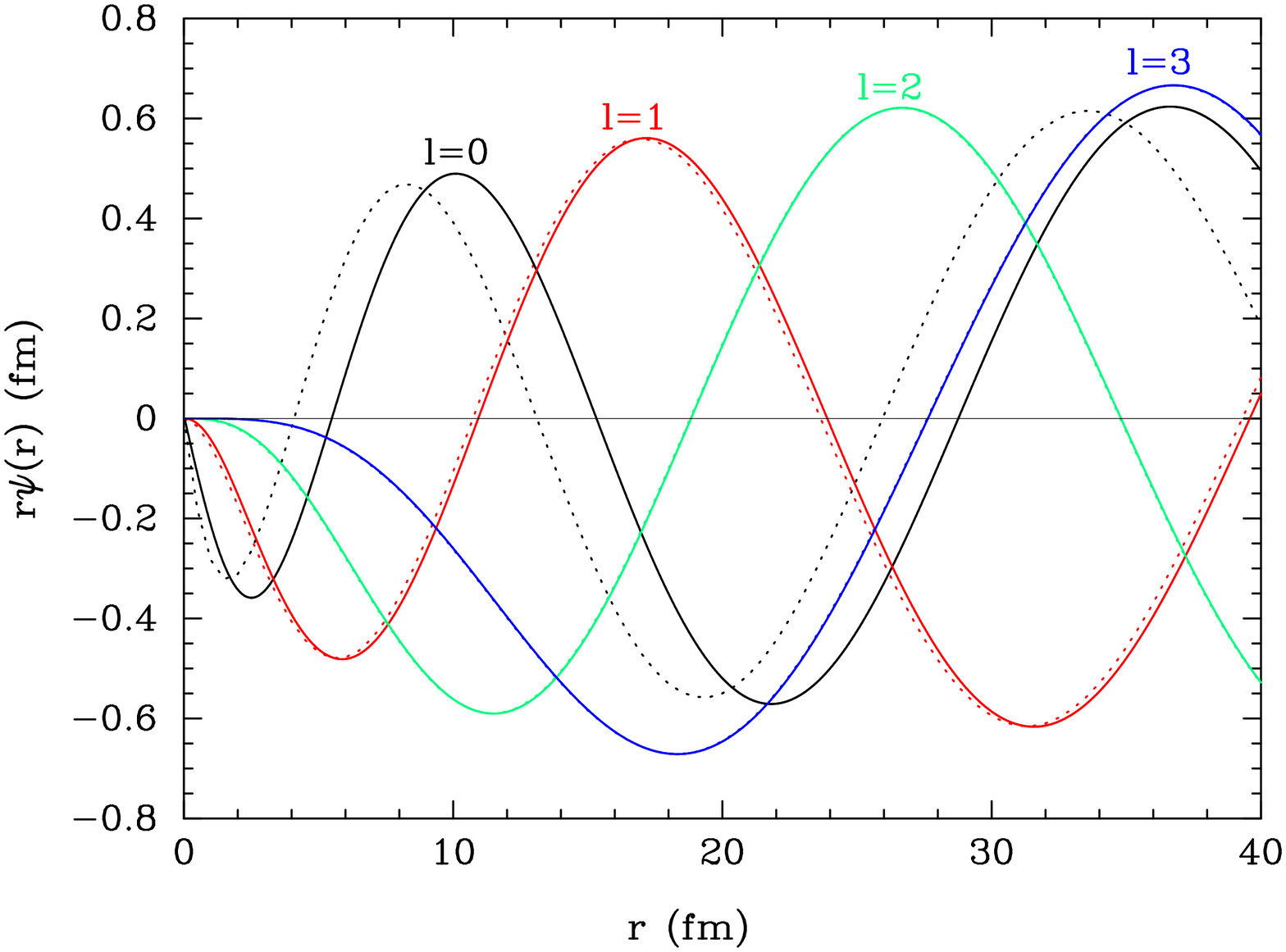}\\
\includegraphics[width=0.45\textwidth]{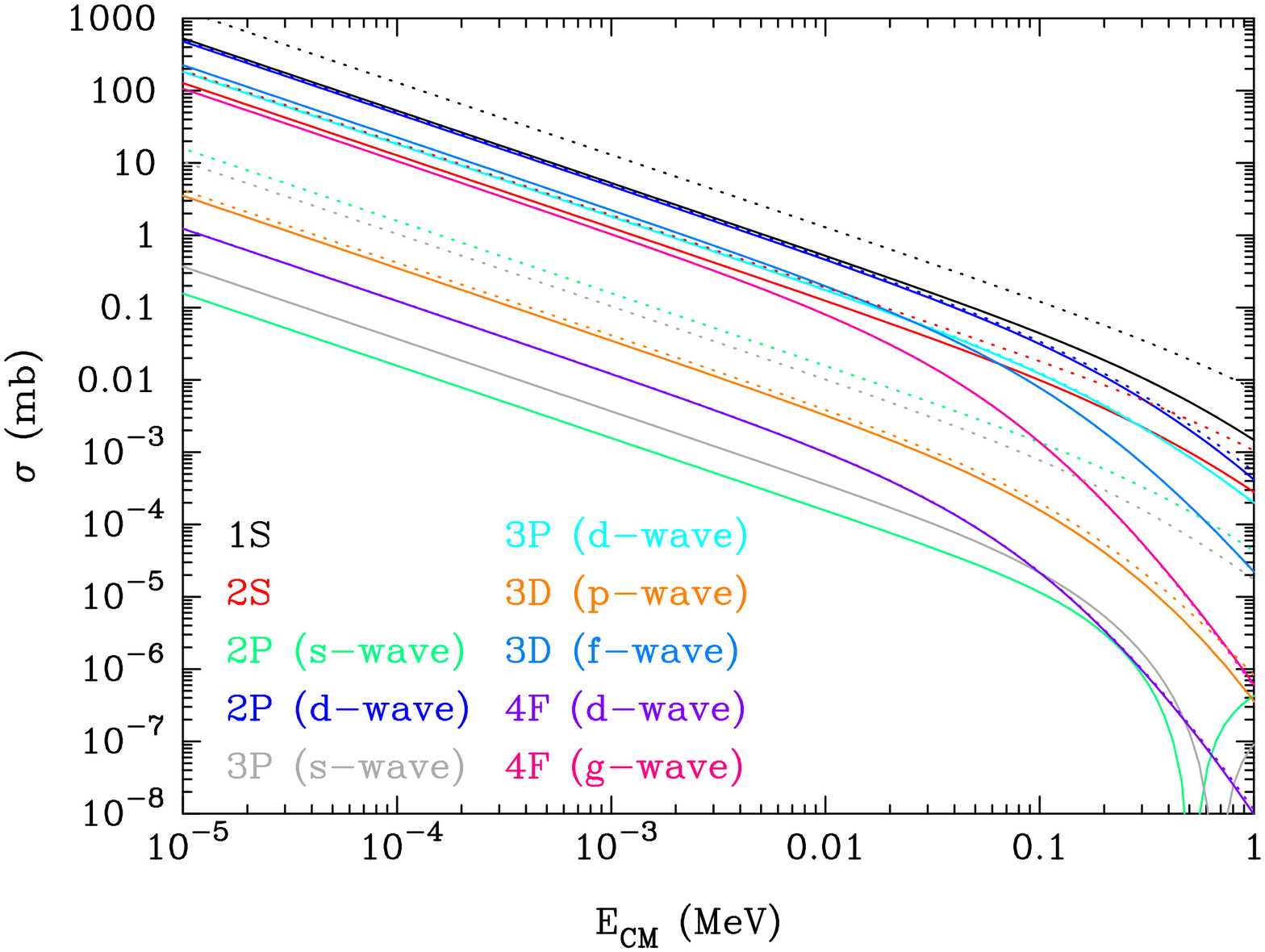}
\end{center}
\caption{Same as Fig. \ref{fig6} for the $^7$Li+$X^-$ system with $m_X=10$ GeV.  \label{fig11}}
\end{figure}

\begin{figure}
\begin{center}
\includegraphics[width=0.45\textwidth]{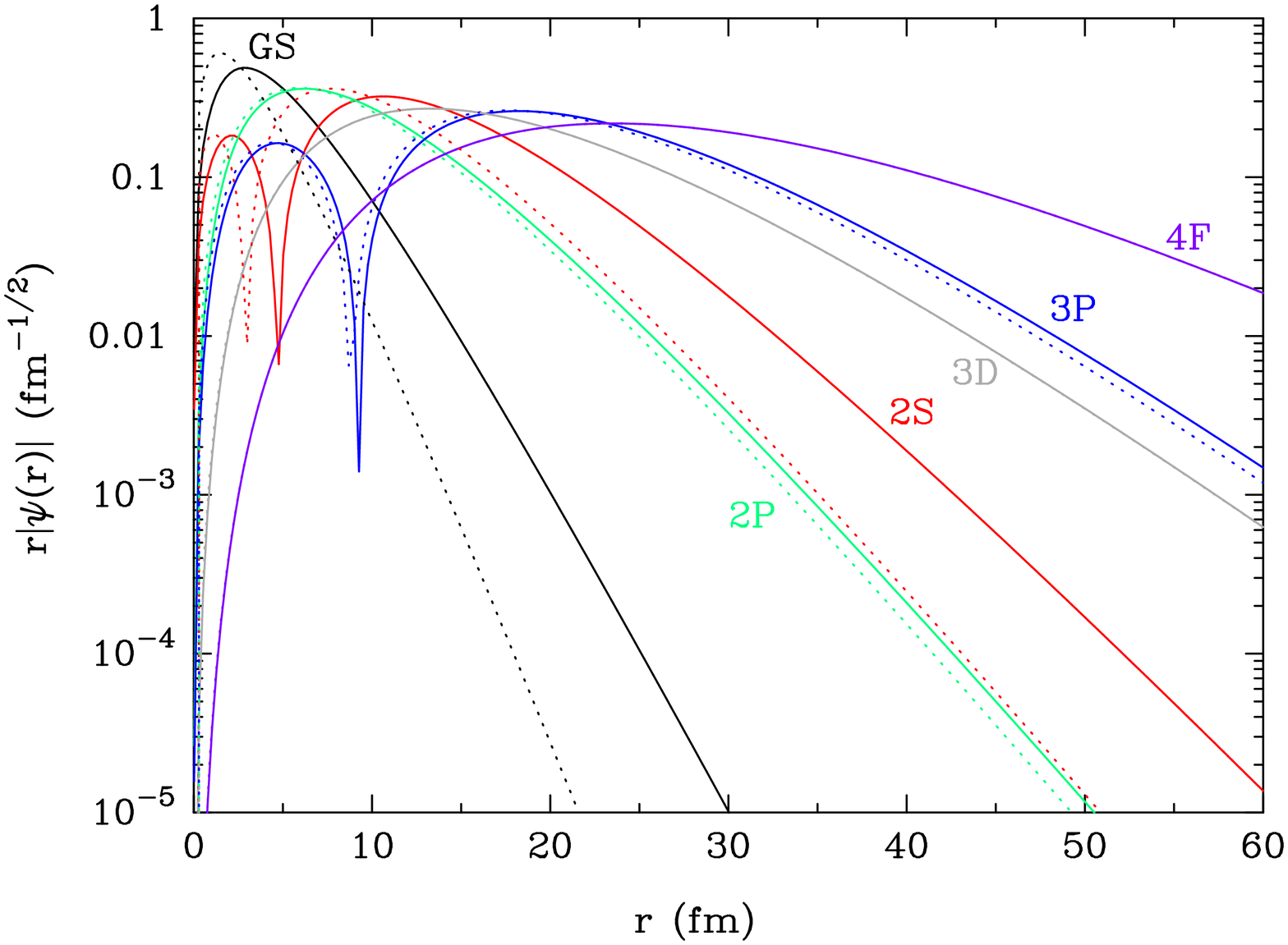}\\
\includegraphics[width=0.45\textwidth]{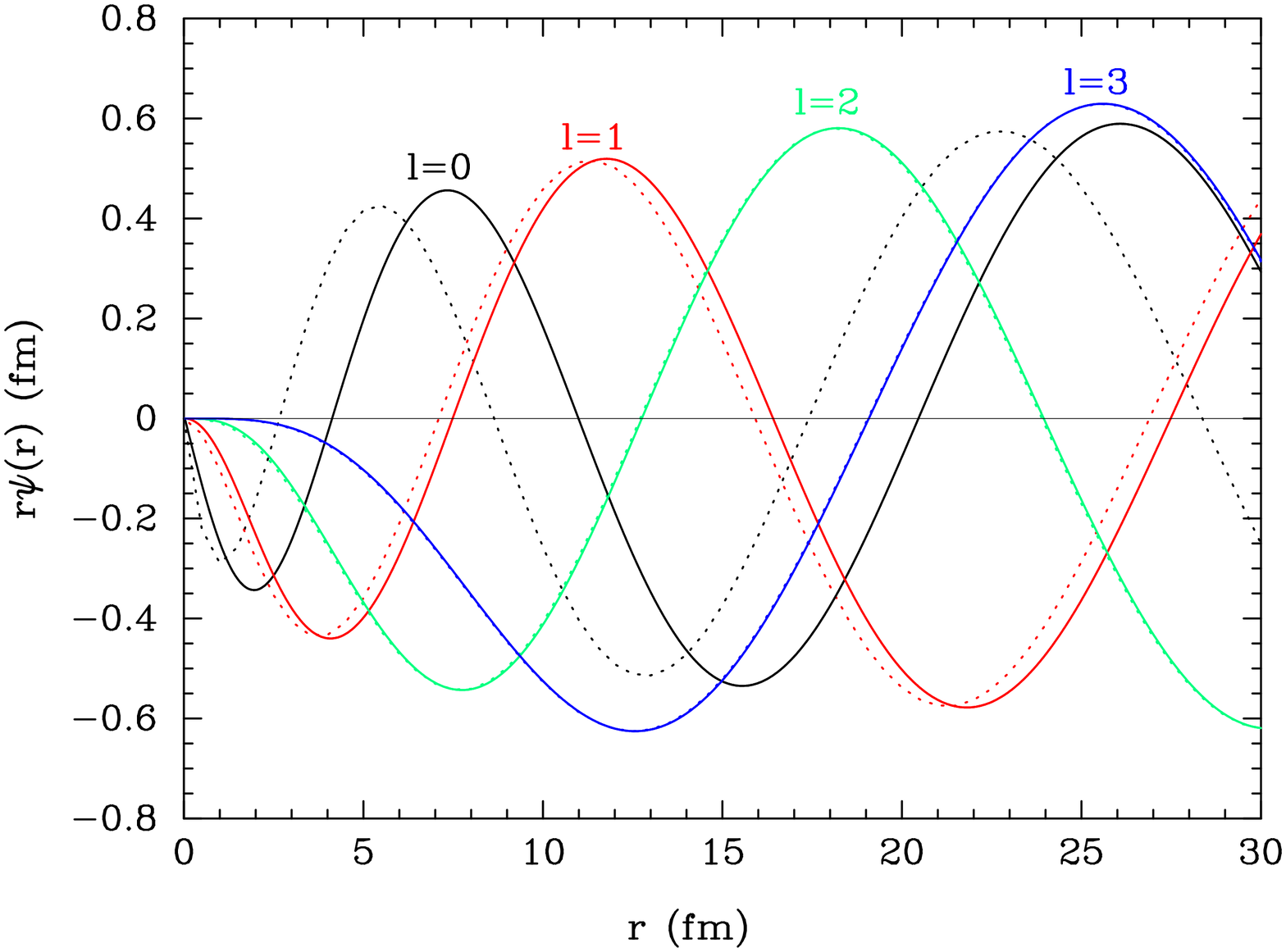}\\
\includegraphics[width=0.45\textwidth]{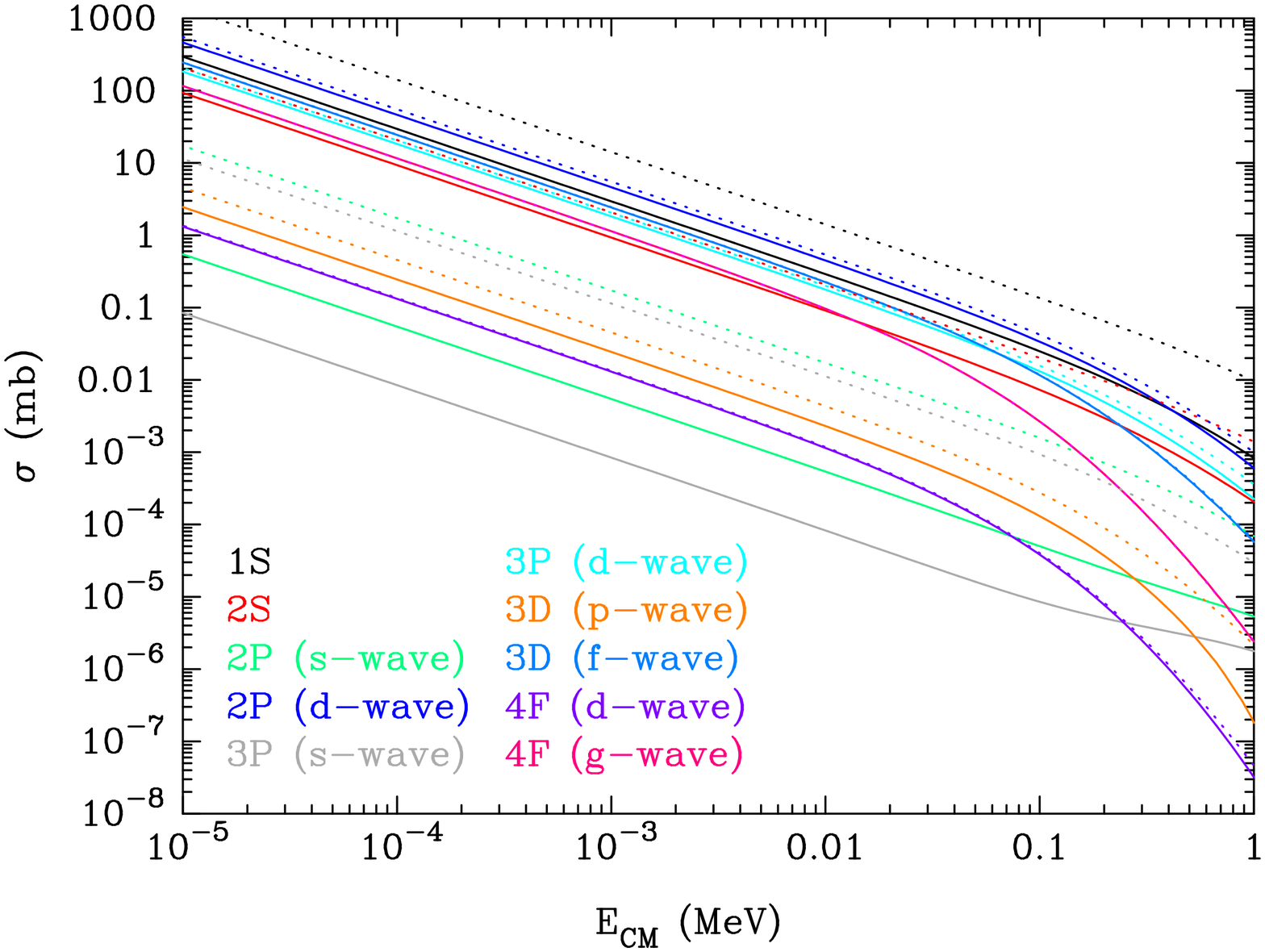}
\end{center}
\caption{Same as Fig. \ref{fig6} for the $^7$Li+$X^-$ system with $m_X=100$ GeV.  \label{fig12}}
\end{figure}

\begin{figure}
\begin{center}
\includegraphics[width=0.45\textwidth]{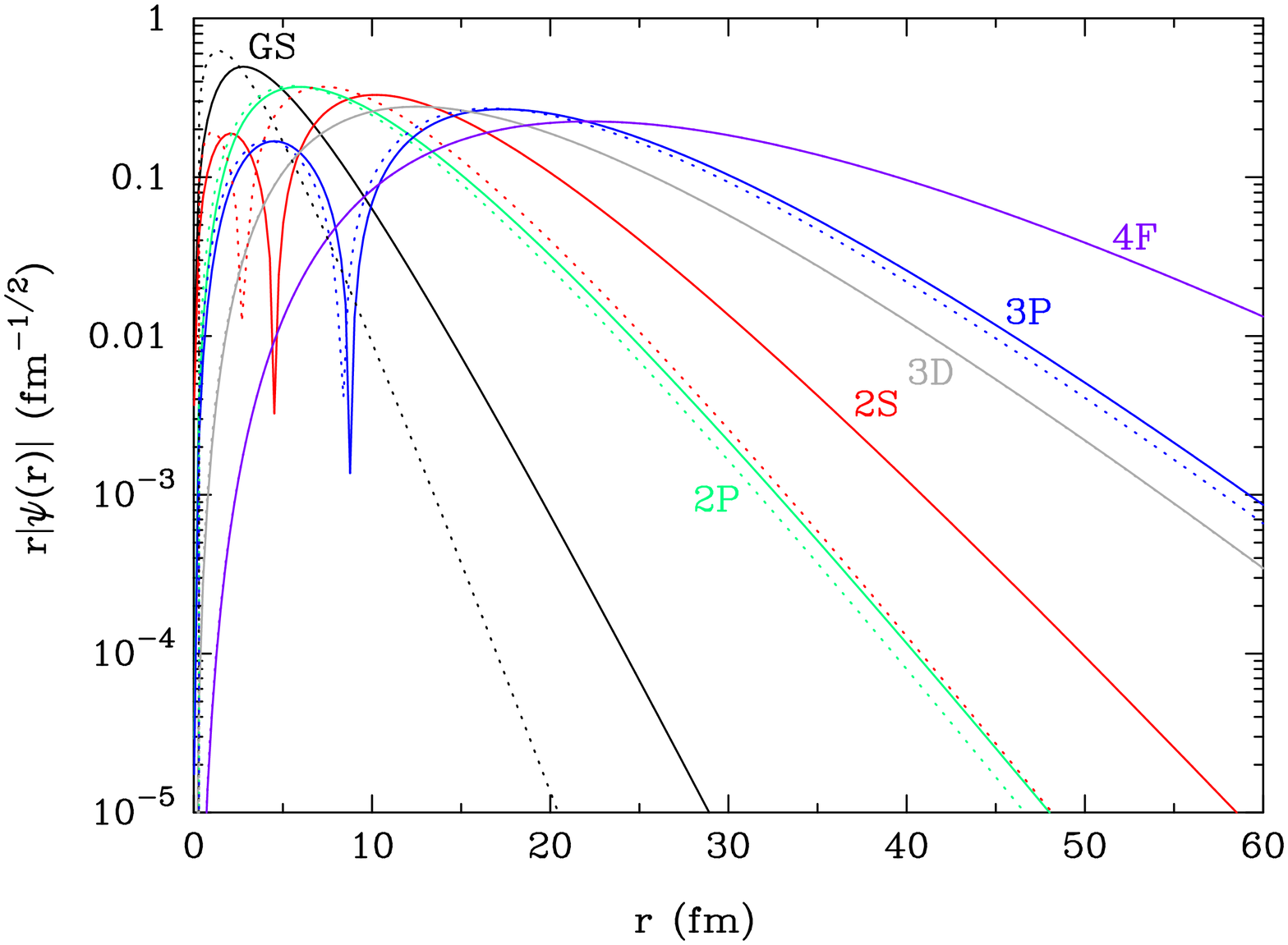}\\
\includegraphics[width=0.45\textwidth]{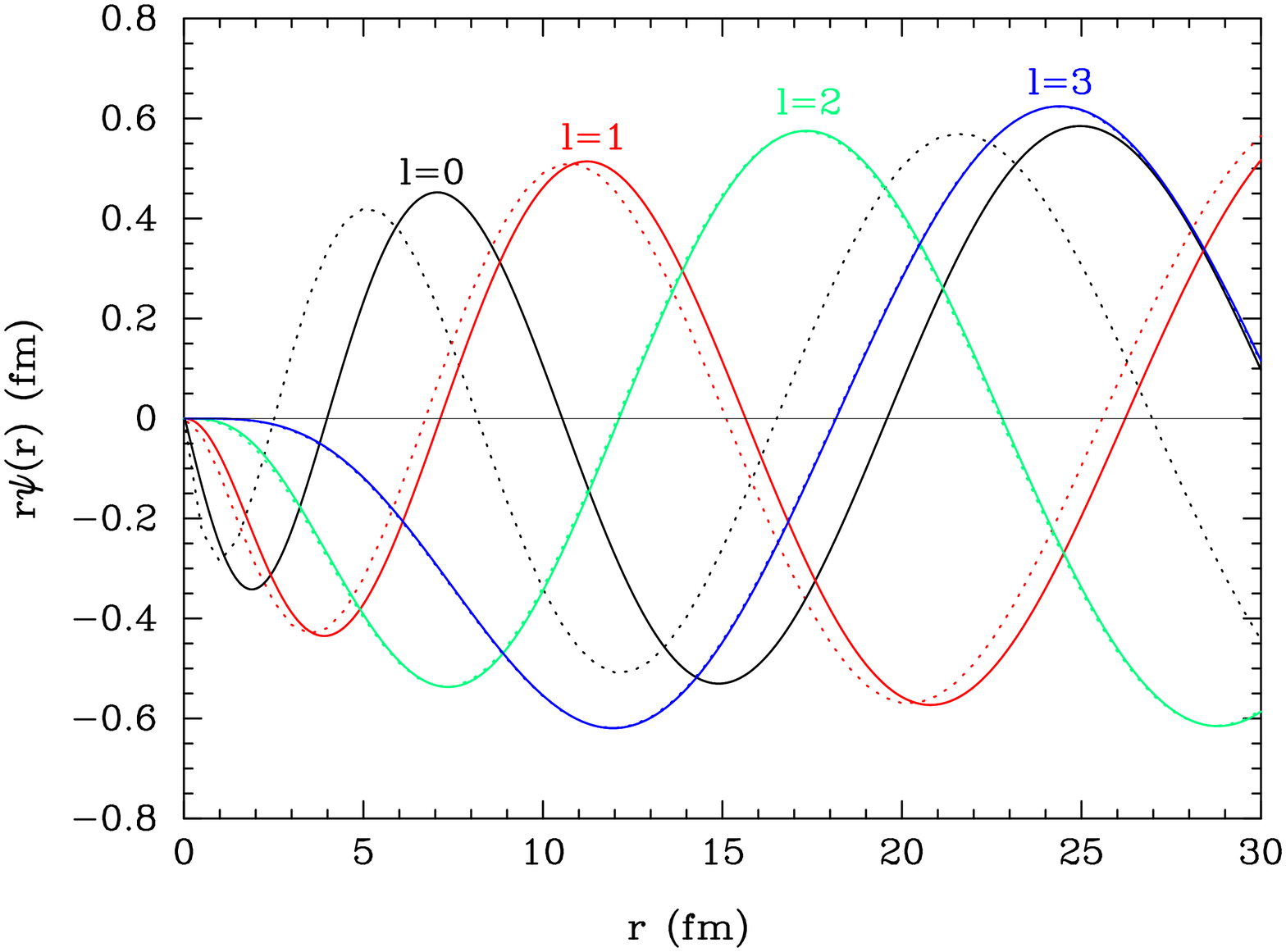}\\
\includegraphics[width=0.45\textwidth]{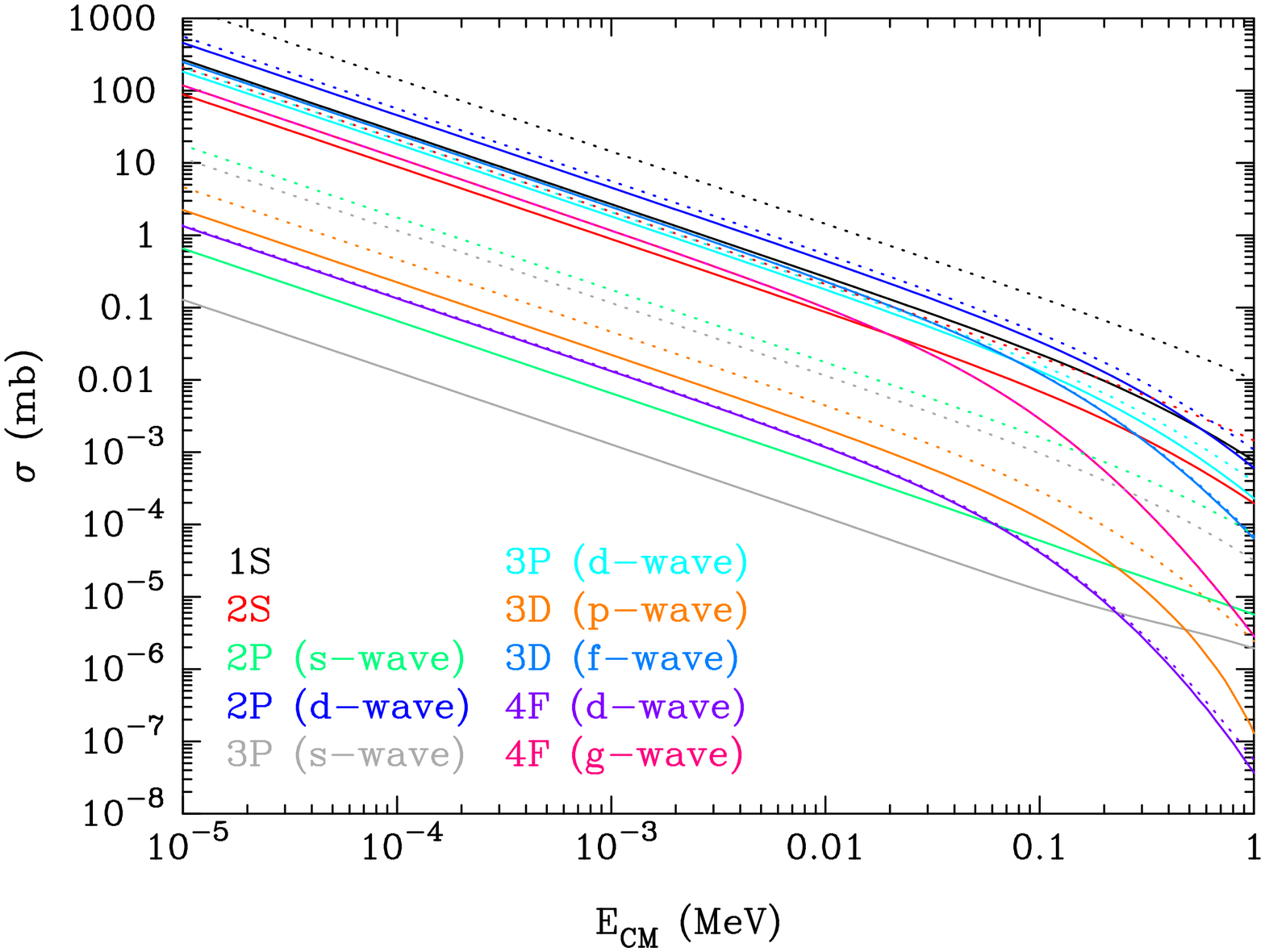}
\end{center}
\caption{Same as Fig. \ref{fig6} for the $^7$Li+$X^-$ system with $m_X=1000$ GeV.  \label{fig13}}
\end{figure}

Figures \ref{fig11}, \ref{fig12}, and \ref{fig13} show bound-state wave functions (upper panel) and continuum wave functions (middle panel) at $E=0.07$ MeV as a function of radius $r$ for the $^7$Li+$X^-$ system in the case of $m_X=10$ GeV, 100 GeV, and 1000 GeV, respectively.  Also shown is the  recombination cross section as a function of the energy $E$ (bottom panel).   Line types indicate the same quantities as in Fig. \ref{fig6}.  Similar to the case of the $^7$Be+$X^-$ system, larger $m_X$ values lead to larger differences in both the wave functions and the recombination cross sections between the finite-size charge and point-charge cases.  We also find that this $^7$Li+$X^-$ system has the important characteristic that the transition $d$-wave $\rightarrow$ 2P is the most important for $m_X\gtrsim 100$ GeV.


\subsection{$^9$Be}\label{sec5.3}
\subsubsection{Energy Levels}\label{sec5.3.1}

Table \ref{tab9} shows the binding energies of $^9$Be$_X$ atomic states that have  main quantum numbers $n$ from one to seven.

\placetable{tab9}

\subsubsection{$^9$Be($X^-$, $\gamma$)$^9$Be$_X$ Nonresonant Rate}\label{sec5.3.2}

The nonresonant reaction rates for $^9$Be($X^-$, $\gamma$)$^9$Be$_X$ were also  calculated for $m_X$=1, 10, 100, and 1000 GeV in  the WS40 model.  The radius parameter for the WS40 model is $R=2.59$ fm.

In the estimation of recombination rate for $^9$Be, the resonant reactions involving atomic states and  nuclear excited states for $^9$Be$^\ast$ were neglected since even the first nuclear excited state has a large excitation energy of 1.684 MeV.  We therefore only calculated the nonresonant rate.

The thermal nonresonant rate was  derived as a function of temperature $T$ by integrating the calculated cross section $\sigma(E)$ over energy [Eq. (\ref{eq10})].  It is then
\begin{widetext}
\begin{numcases}
{N_{\rm A} \langle \sigma v \rangle_{\rm NR} = }
2.07 \times 10^5~{\rm cm}^3 {\rm mol}^{-1} {\rm s}^{-1}  \left(1-0.339 T_9 \right) T_9^{-1/2}
& ~~~(for~$m_X=1$~GeV) \label{eq54}
\\ 
3.89 \times 10^4~{\rm cm}^3 {\rm mol}^{-1} {\rm s}^{-1}  \left(1-0.199 T_9 \right) T_9^{-1/2}
& ~~~(for~$m_X=10$~GeV) \label{eq55}
\\ 
2.32 \times 10^4~{\rm cm}^3 {\rm mol}^{-1} {\rm s}^{-1}  \left(1-0.180 T_9 \right) T_9^{-1/2}
& ~~~(for~$m_X=100$~GeV) \label{eq56}
\\ 
2.14 \times 10^4~{\rm cm}^3 {\rm mol}^{-1} {\rm s}^{-1}  \left(1-0.179 T_9 \right) T_9^{-1/2}
& ~~~(for~$m_X=1000$~GeV). \label{eq57}
\end{numcases}
\end{widetext}

\begin{figure}
\begin{center}
\includegraphics[width=0.45\textwidth]{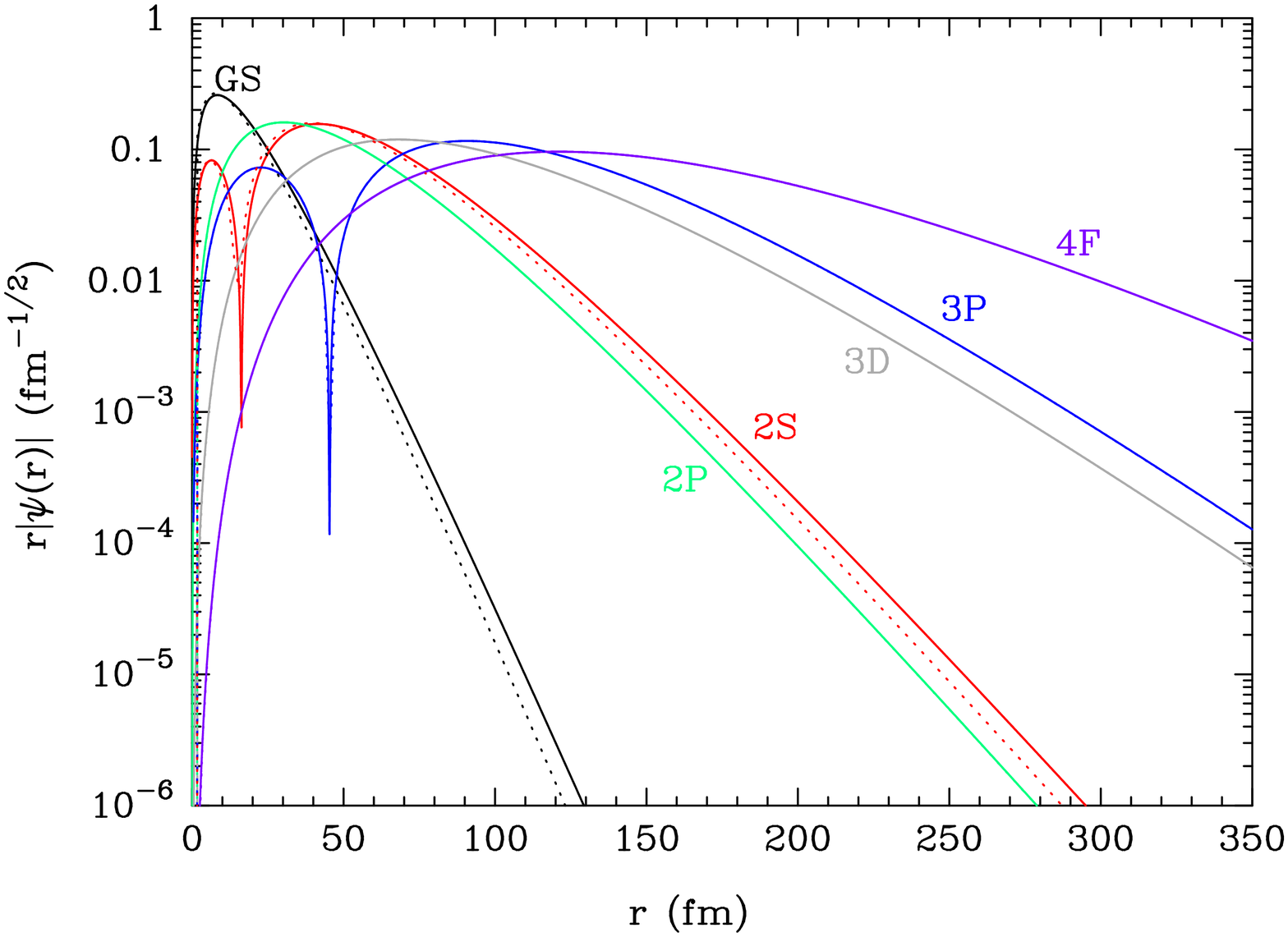}\\
\includegraphics[width=0.45\textwidth]{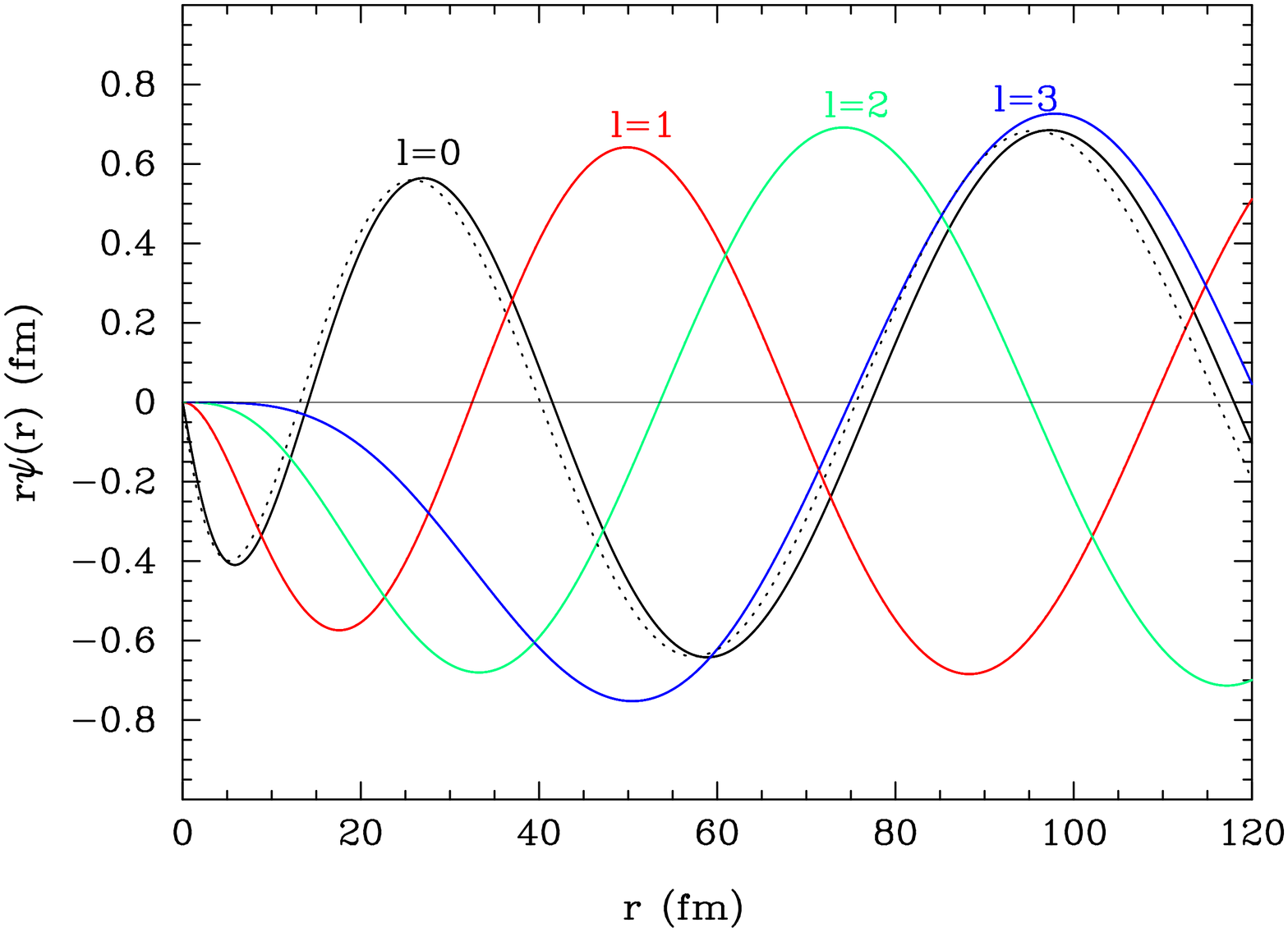}\\
\includegraphics[width=0.45\textwidth]{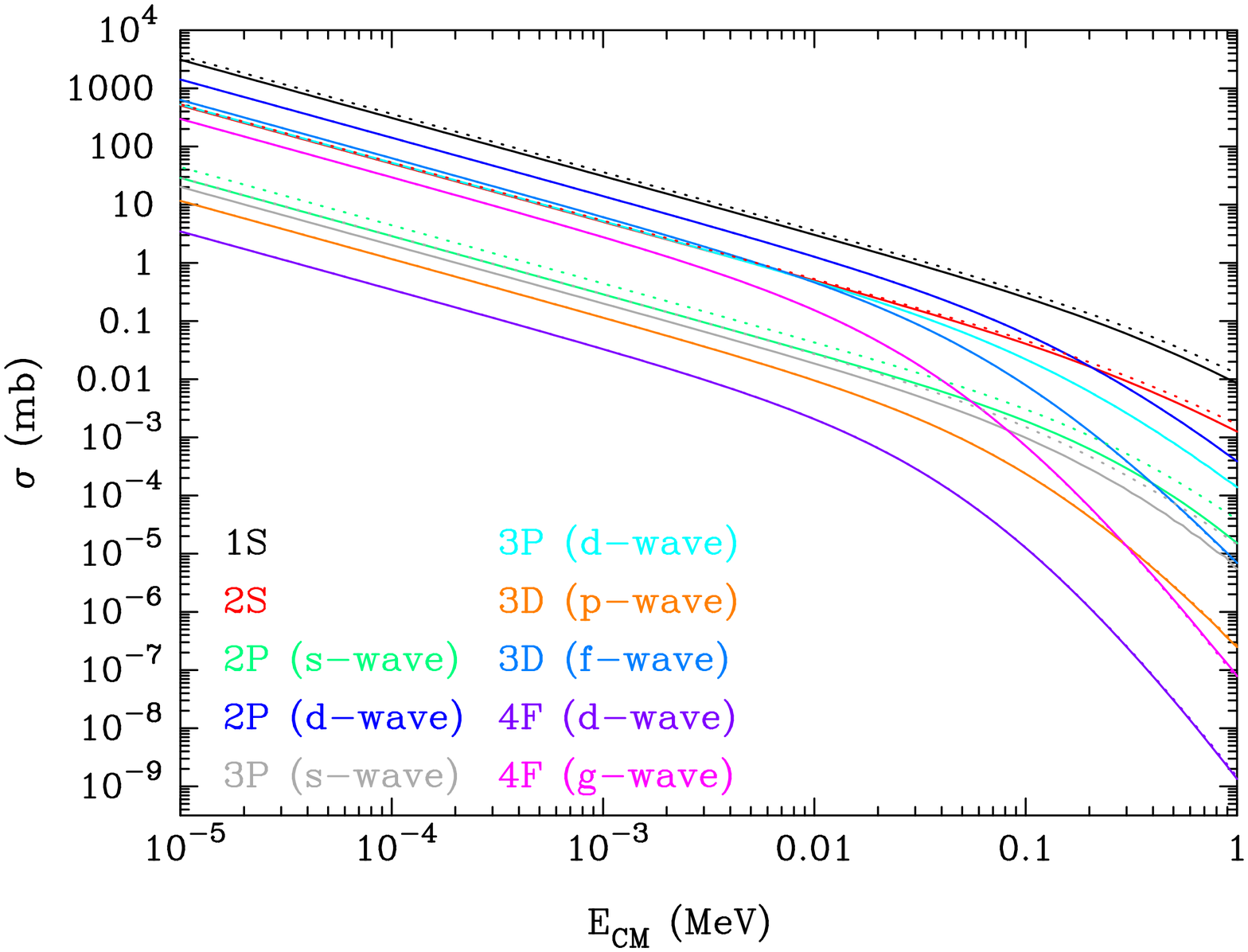}
\end{center}
\caption{Same as Fig. \ref{fig6} for the $^9$Be+$X^-$ system with $m_X=1$ GeV.  \label{fig14}}
\end{figure}

Figure \ref{fig14} shows bound-state wave functions (upper panel) and continuum wave functions (middle panel) at $E=0.07$ MeV as a function of radius $r$ for the $^9$Be+$X^-$ system for the case of $m_X=1$ GeV.  We also show  recombination cross section as a function of the energy $E$ (bottom panel).   Line types indicate the same quantities as in Fig. \ref{fig6}.  Trends of calculated results are similar to those of the $^7$Be+$X^-$ system (Fig. \ref{fig6}).


\begin{figure}
\begin{center}
\includegraphics[width=0.45\textwidth]{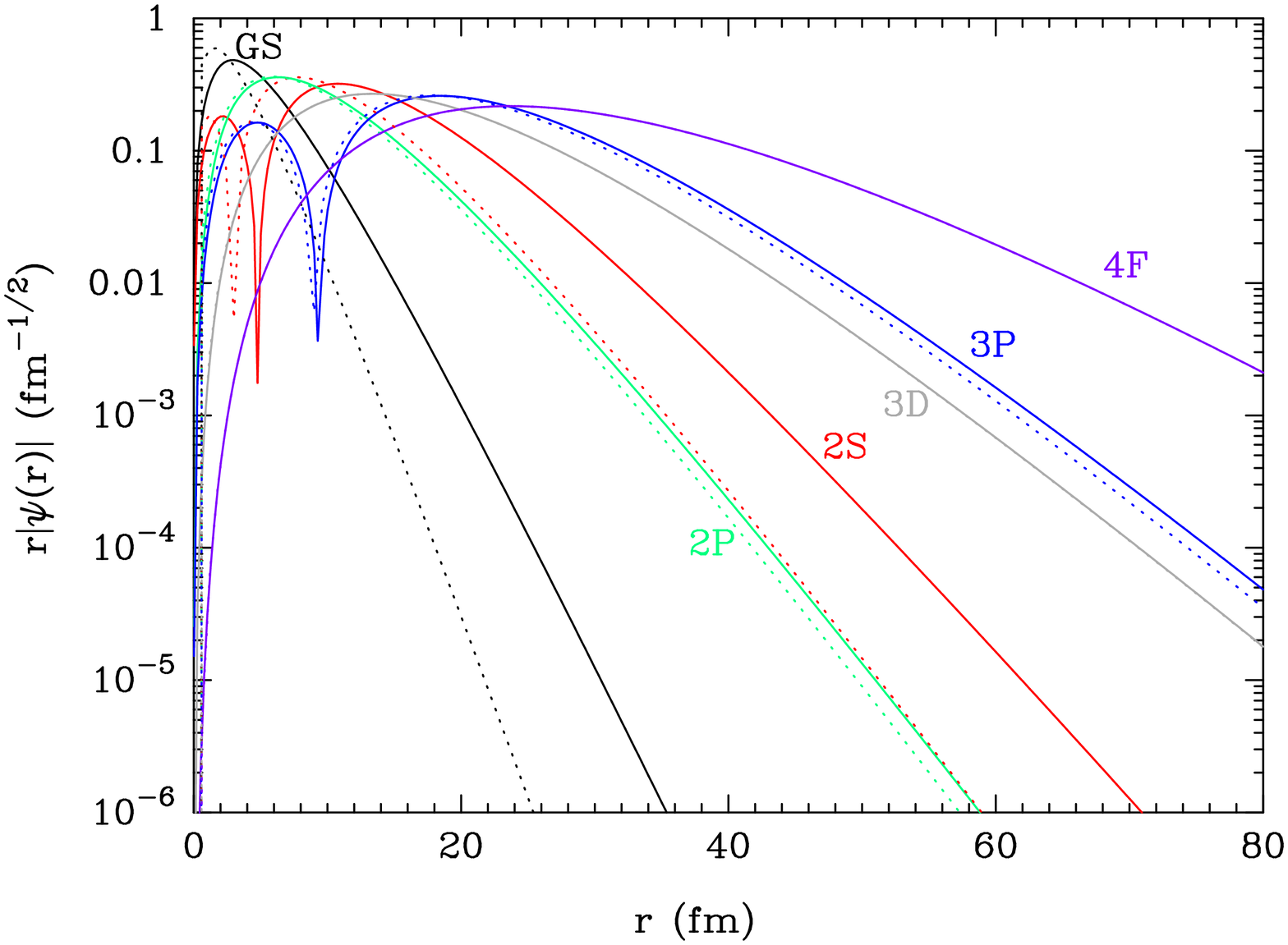}\\
\includegraphics[width=0.45\textwidth]{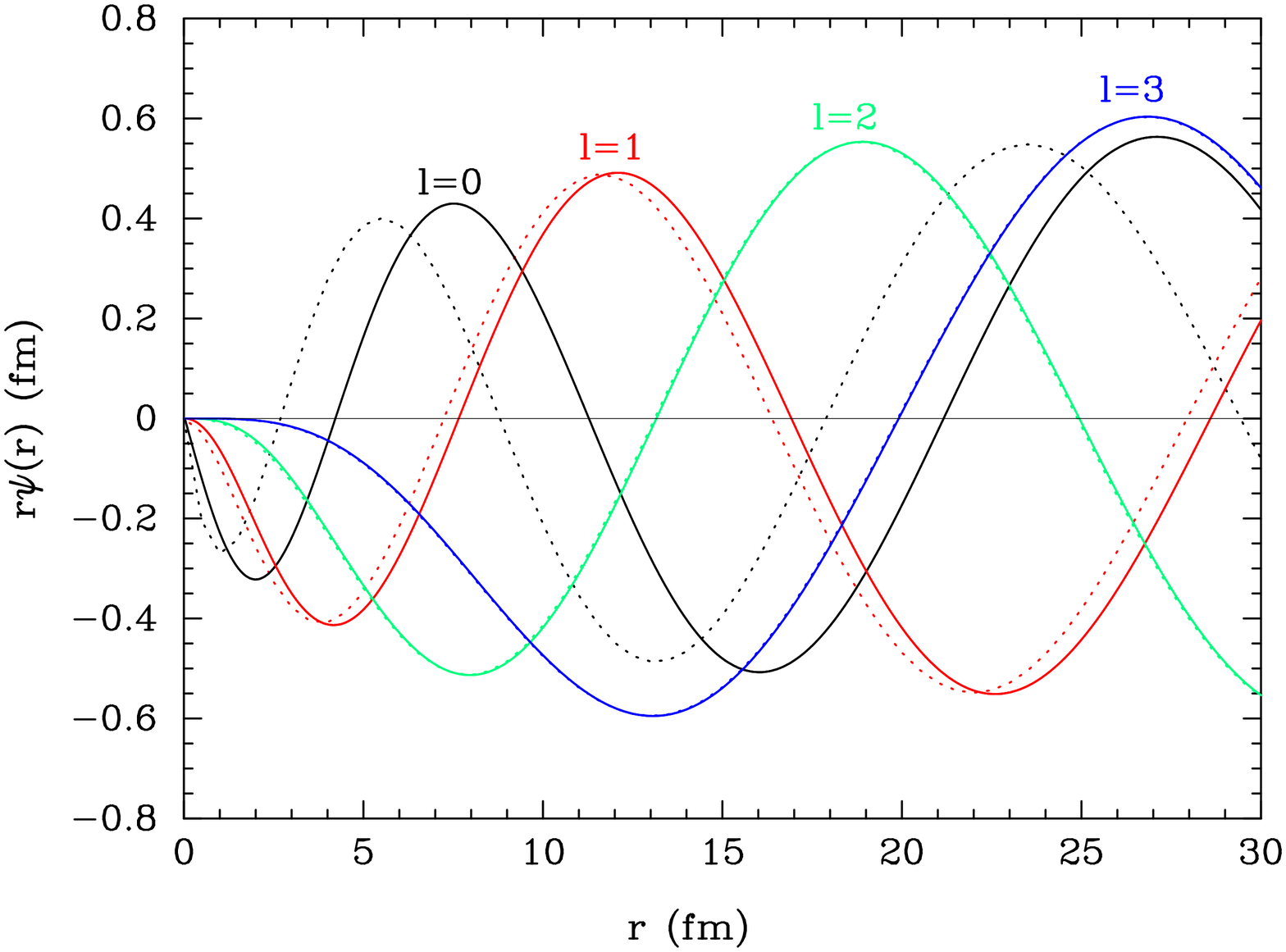}\\
\includegraphics[width=0.45\textwidth]{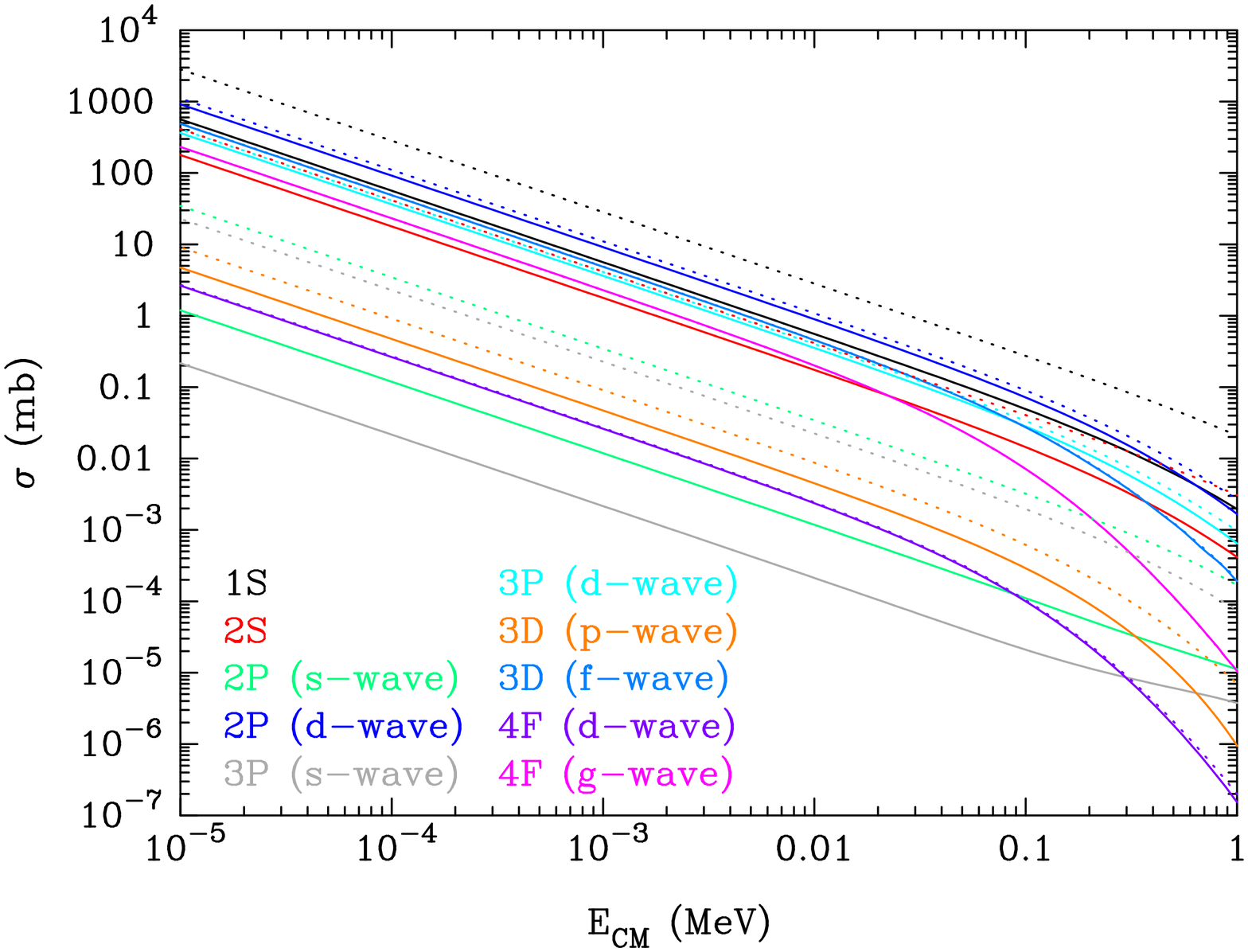}
\end{center}
\caption{Same as Fig. \ref{fig6} for the $^9$Be+$X^-$ system with $m_X=10$ GeV.  \label{fig15}}
\end{figure}

\begin{figure}
\begin{center}
\includegraphics[width=0.45\textwidth]{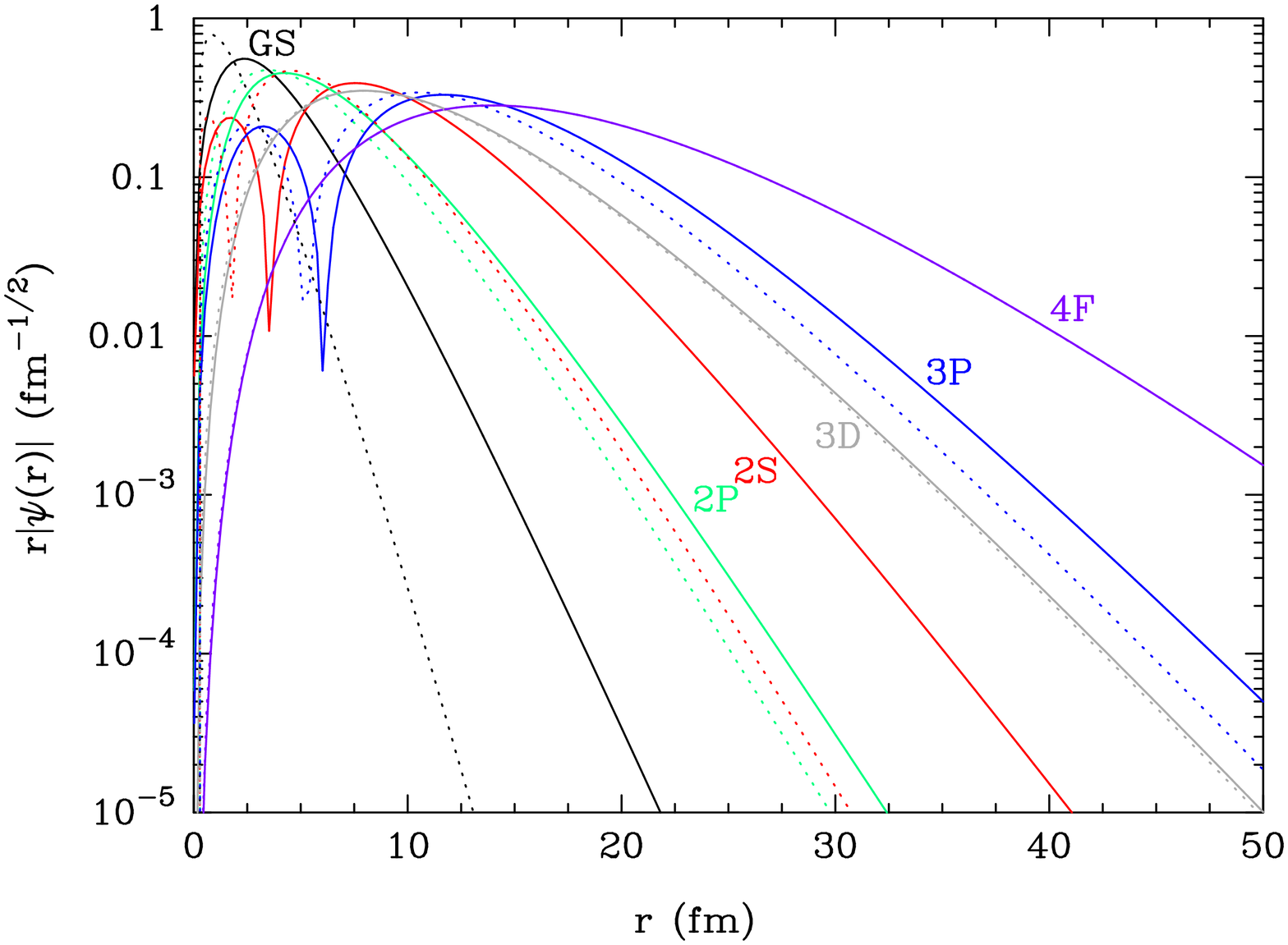}\\
\includegraphics[width=0.45\textwidth]{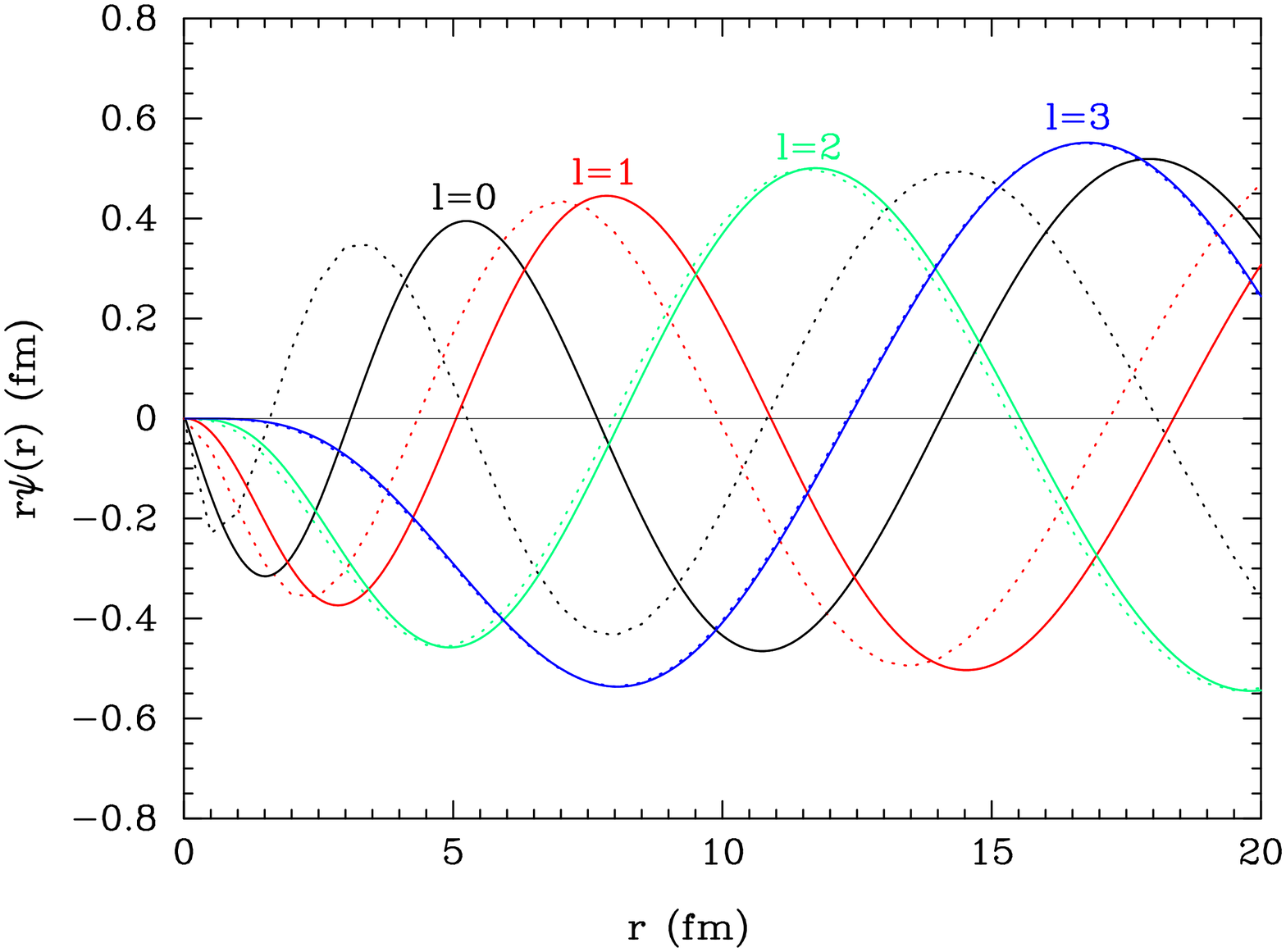}\\
\includegraphics[width=0.45\textwidth]{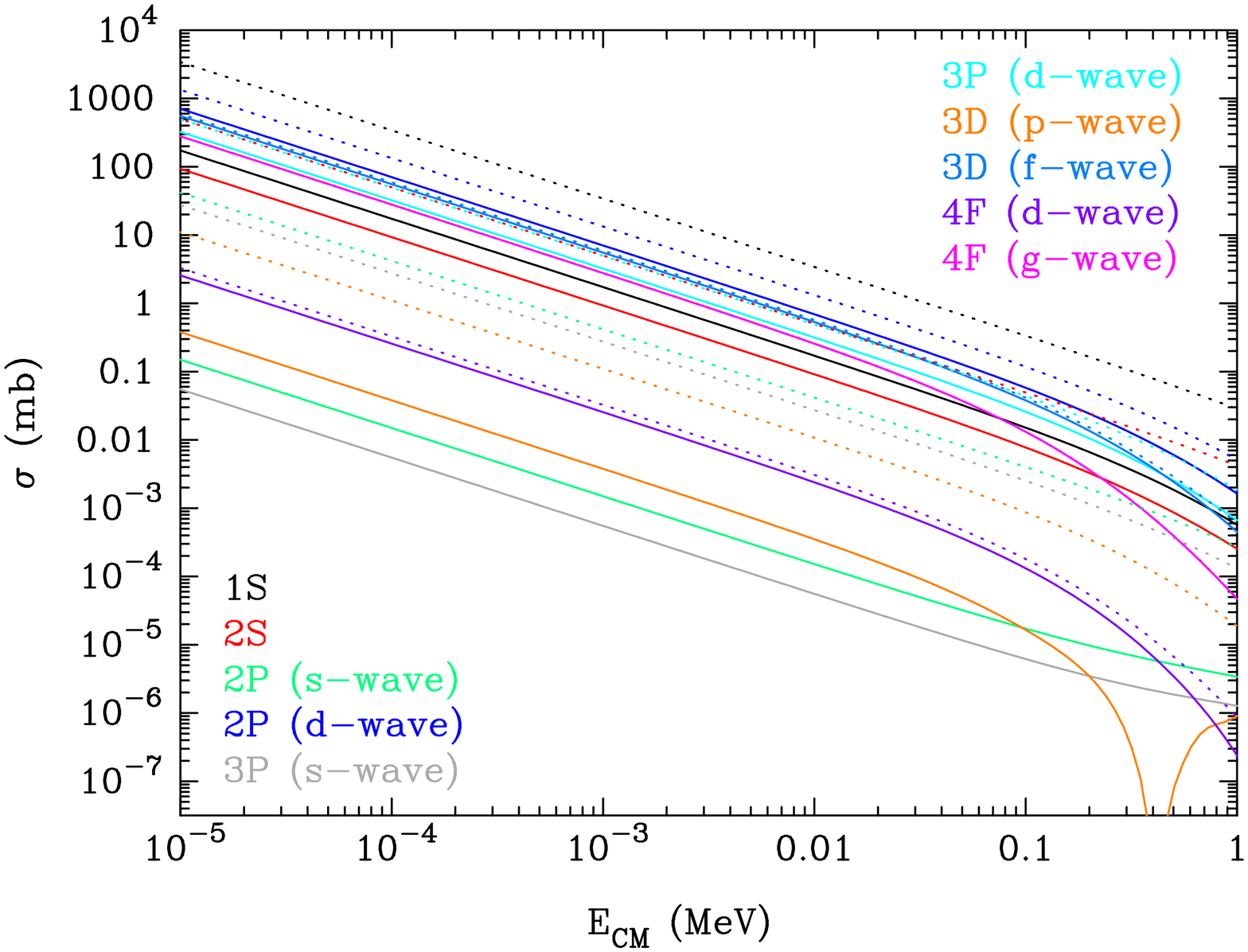}
\end{center}
\caption{Same as Fig. \ref{fig6} for the $^9$Be+$X^-$ system with $m_X=100$ GeV.  \label{fig16}}
\end{figure}

\begin{figure}
\begin{center}
\includegraphics[width=0.45\textwidth]{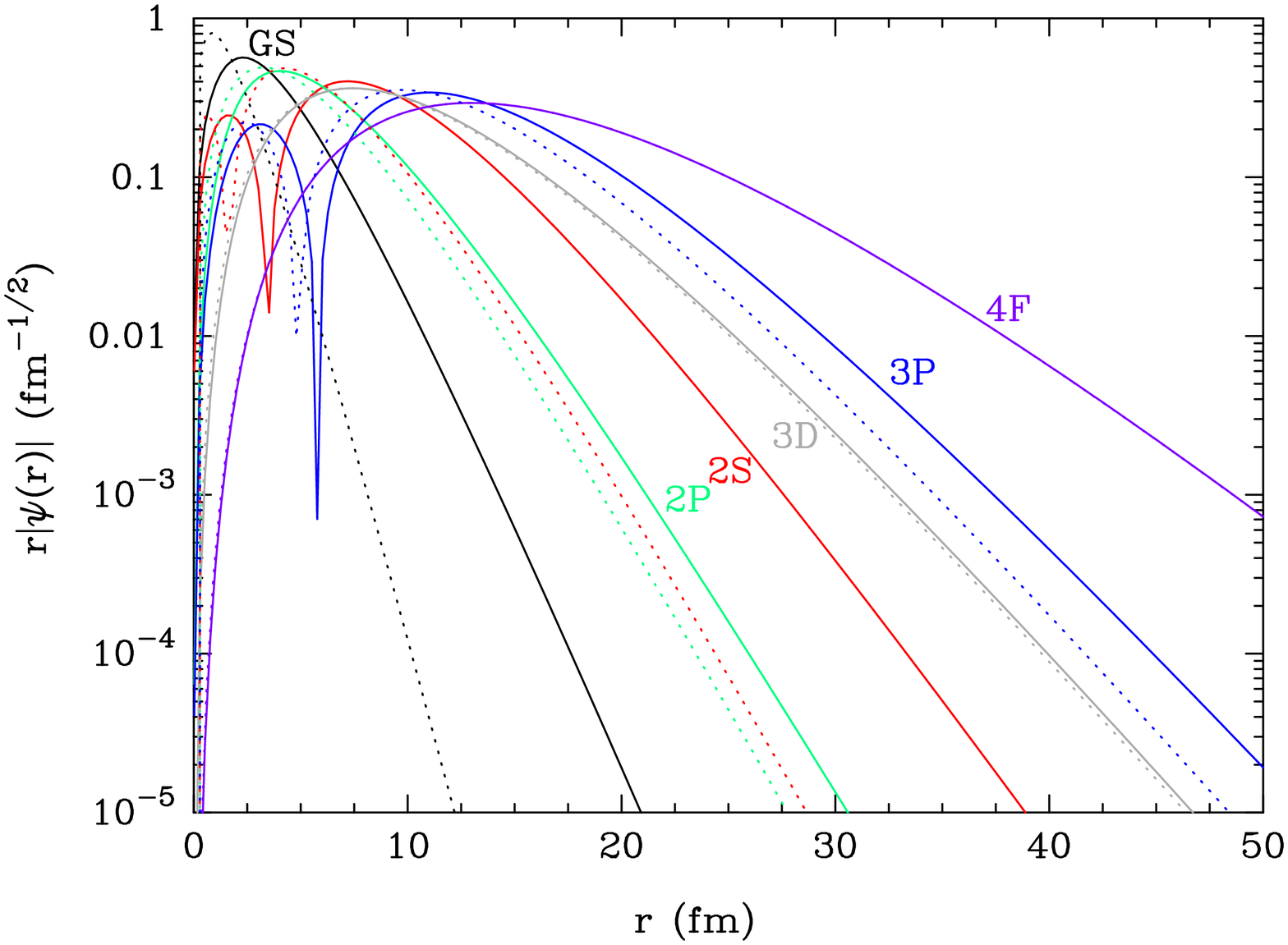}\\
\includegraphics[width=0.45\textwidth]{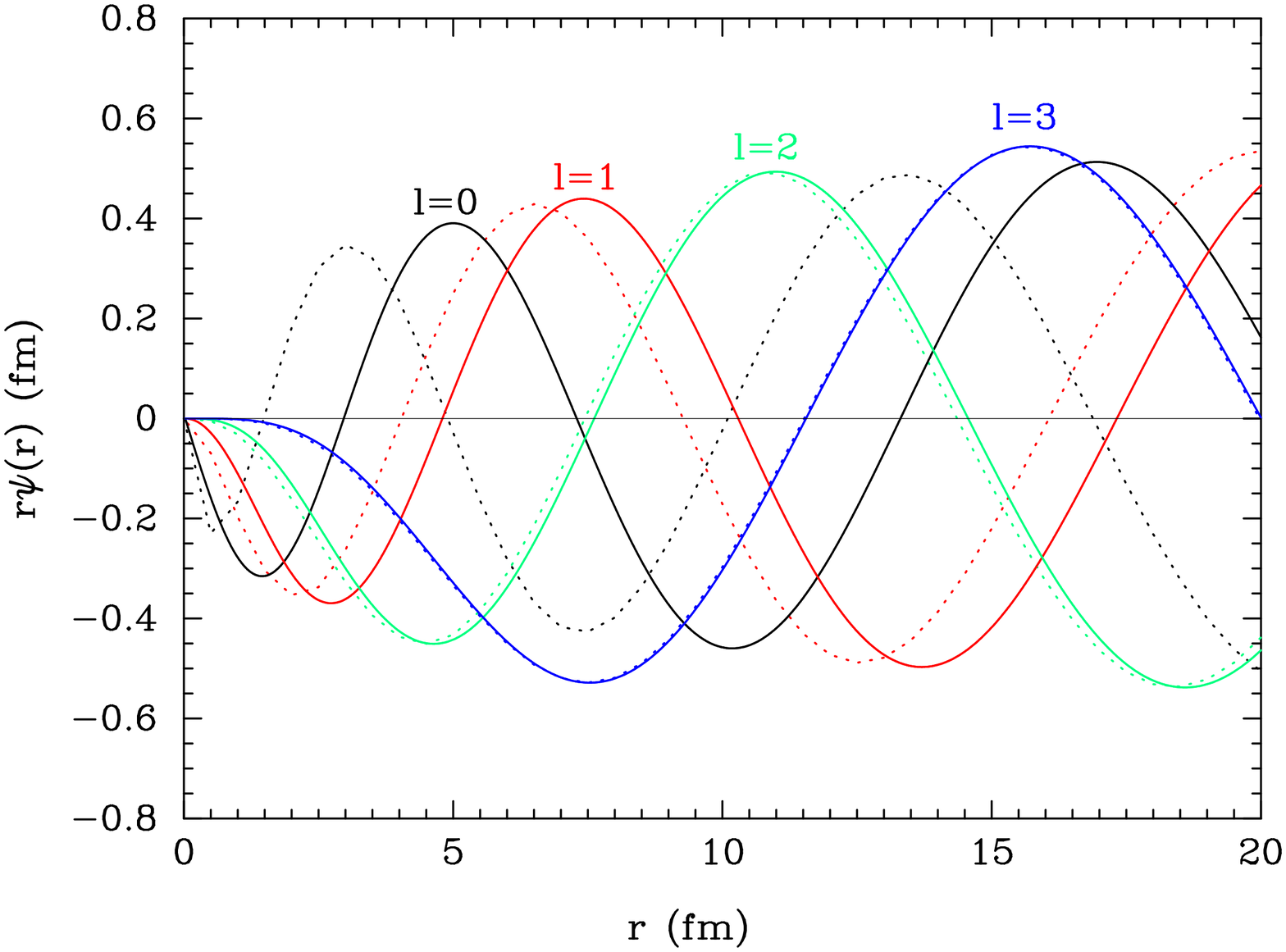}\\
\includegraphics[width=0.45\textwidth]{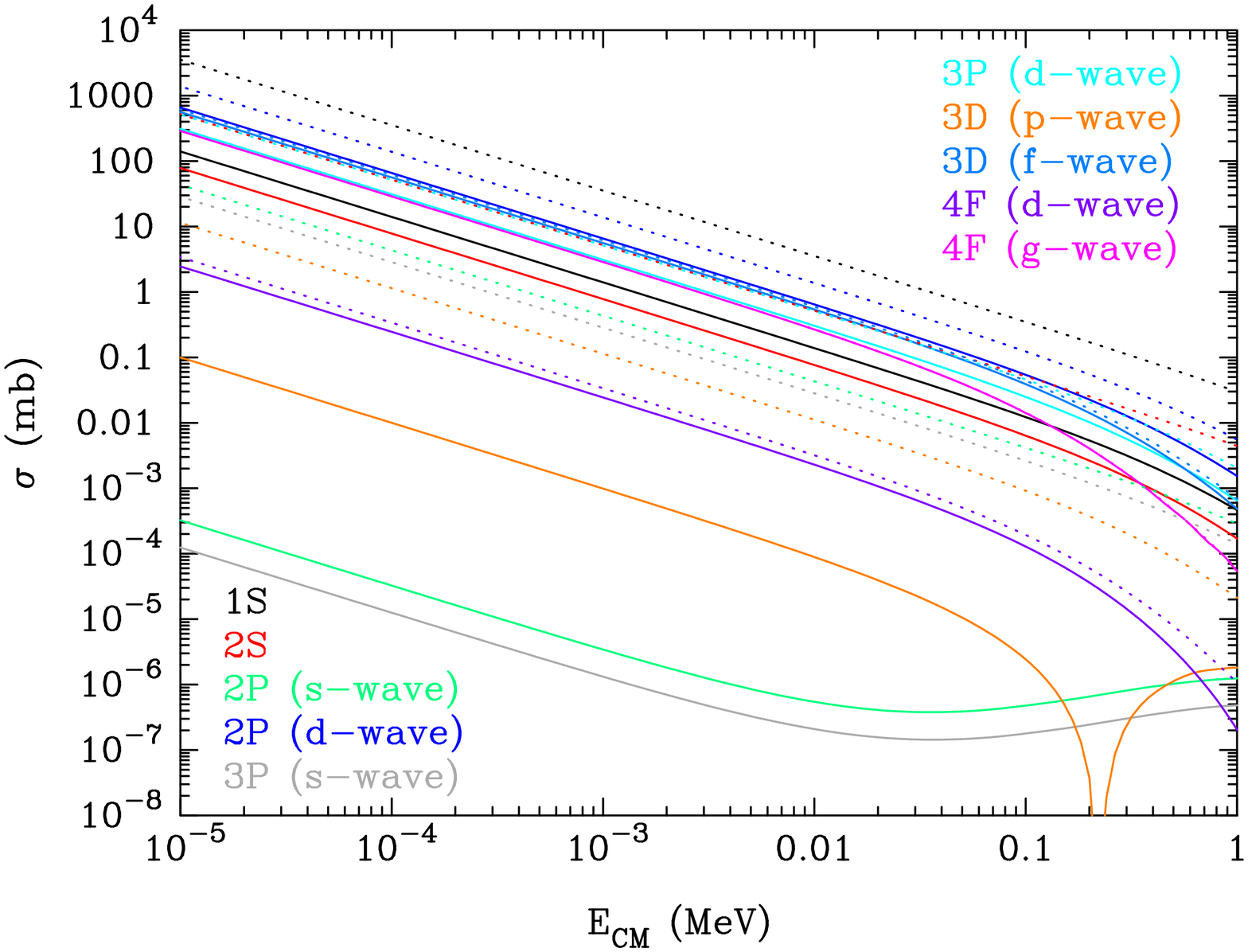}
\end{center}
\caption{Same as Fig. \ref{fig6}, but  for the $^9$Be+$X^-$ system with $m_X=1000$ GeV.  \label{fig17}}
\end{figure}

Figures \ref{fig15}, \ref{fig16}, and \ref{fig17} show bound state wave functions (upper panel) and continuum wave functions (middle panel) at $E=0.07$ MeV as a function of radius $r$, and recombination cross sections as a function of the energy $E$ (bottom panel) of the $^9$Be+$X^-$ system for the cases of $m_X=10$ GeV, 100 GeV, and 1000 GeV, respectively.   Line types indicate the same quantities as in Fig. \ref{fig6}.  Similar to the case of the $^7$Be+$X^-$ system, larger $m_X$ values lead to larger differences in wave functions and recombination cross sections between the finite-size charge and point-charge cases.  We also find  that the transition, $d$-wave $\rightarrow$ 2P, is most important for $m_X\gtrsim 100$ GeV in this $^9$Be+$X^-$ system.


\subsection{$^4$He}\label{sec5.4}
\subsubsection{Energy Levels}\label{sec5.4.1}

Table \ref{tab10} shows the binding energies of $^4$He$_X$ atomic states that have  main quantum numbers $n$ from one to seven.

\placetable{tab10}

\subsubsection{$^4$He($X^-$, $\gamma$)$^4$He$_X$ Nonresonant Rate}\label{sec5.4.2}

The nonresonant rates of the reaction $^4$He($X^-$, $\gamma$)$^4$He$_X$ were calculated for $m_X$=1, 10, 100, and 1000 GeV using  the WS40 model.  For this case, the radius parameter is $R=1.31$ fm.  Since all excited states of $^4$He$^\ast$ have excitation energies larger than 20 MeV, atomic states of nuclear excited states $^4$He$_X^\ast$ are never important resonances in the recombination process.  We then calculate only the nonresonant rate.

The thermal nonresonant rates were derived as a function of temperature $T$ by integrating the calculated cross section $\sigma(E)$ over energy [Eq. (\ref{eq10})].  The resultant rates are
\begin{widetext}
\begin{numcases}
{N_{\rm A} \langle \sigma v \rangle_{\rm NR} = }
5.38 \times 10^4~{\rm cm}^3 {\rm mol}^{-1} {\rm s}^{-1}  \left(1-0.648 T_9 \right) T_9^{-1/2}
& ~~~(for~$m_X=1$~GeV) \label{eq58}
\\ 
1.63 \times 10^4~{\rm cm}^3 {\rm mol}^{-1} {\rm s}^{-1}  \left(1-0.404 T_9 \right) T_9^{-1/2}
& ~~~(for~$m_X=10$~GeV) \label{eq59}
\\ 
1.32 \times 10^4~{\rm cm}^3 {\rm mol}^{-1} {\rm s}^{-1}  \left(1-0.367 T_9 \right) T_9^{-1/2}
& ~~~(for~$m_X=100$~GeV) \label{eq60}
\\ 
1.29 \times 10^4~{\rm cm}^3 {\rm mol}^{-1} {\rm s}^{-1}  \left(1-0.363 T_9 \right) T_9^{-1/2}
& ~~~(for~$m_X=1000$~GeV). \label{eq61}
\end{numcases}
\end{widetext}

\begin{figure}
\begin{center}
\includegraphics[width=0.45\textwidth]{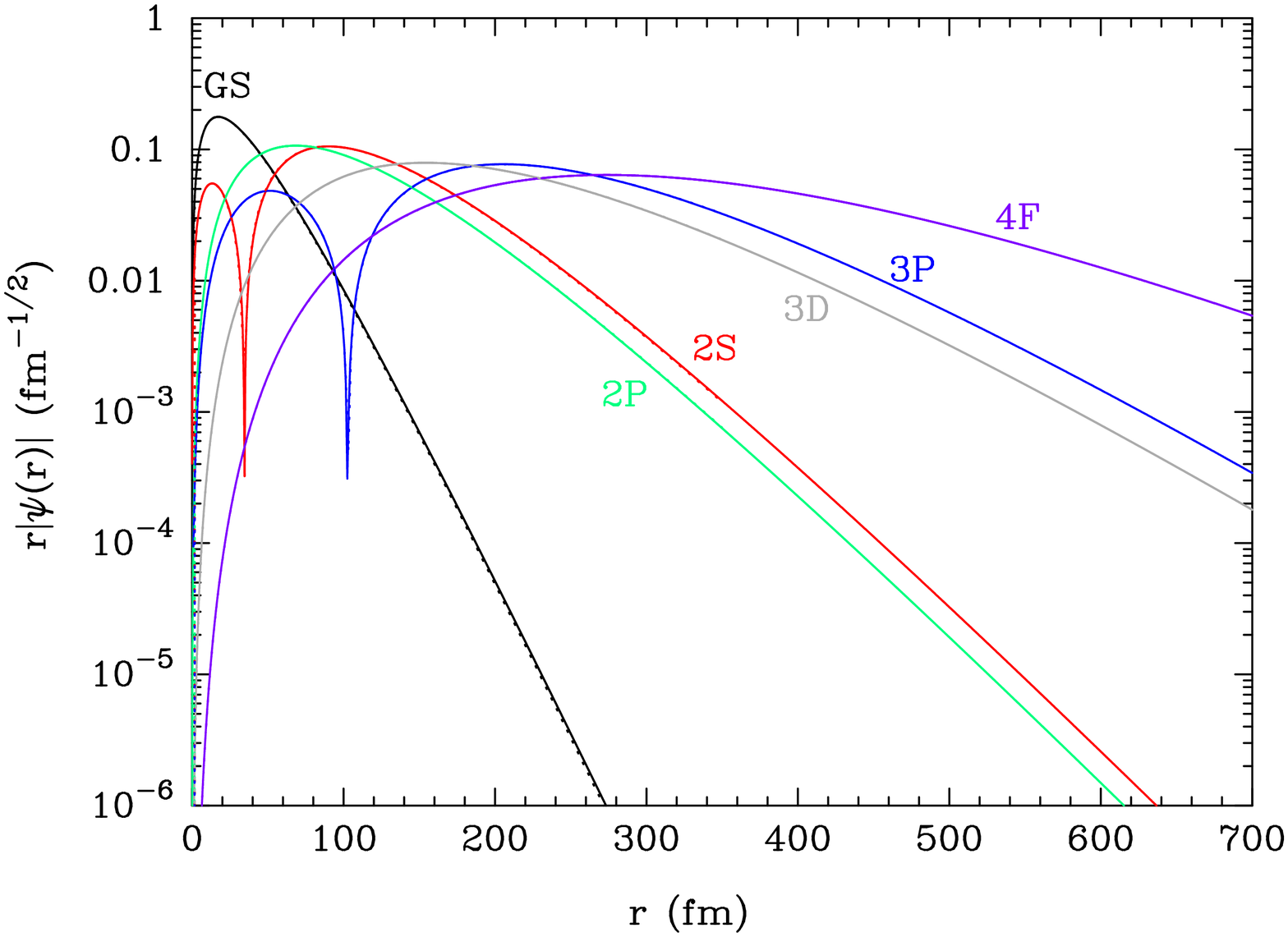}\\
\includegraphics[width=0.45\textwidth]{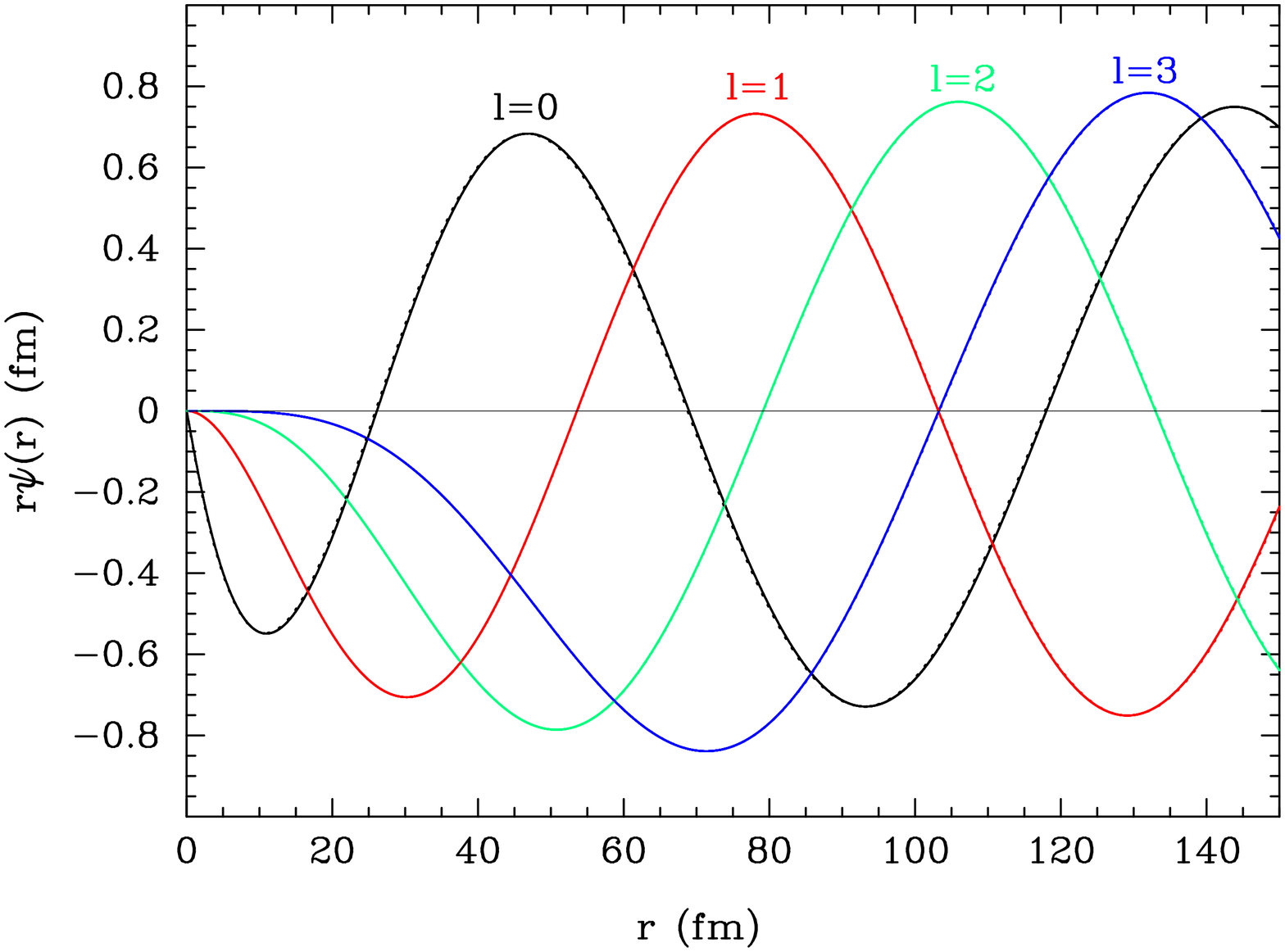}\\
\includegraphics[width=0.45\textwidth]{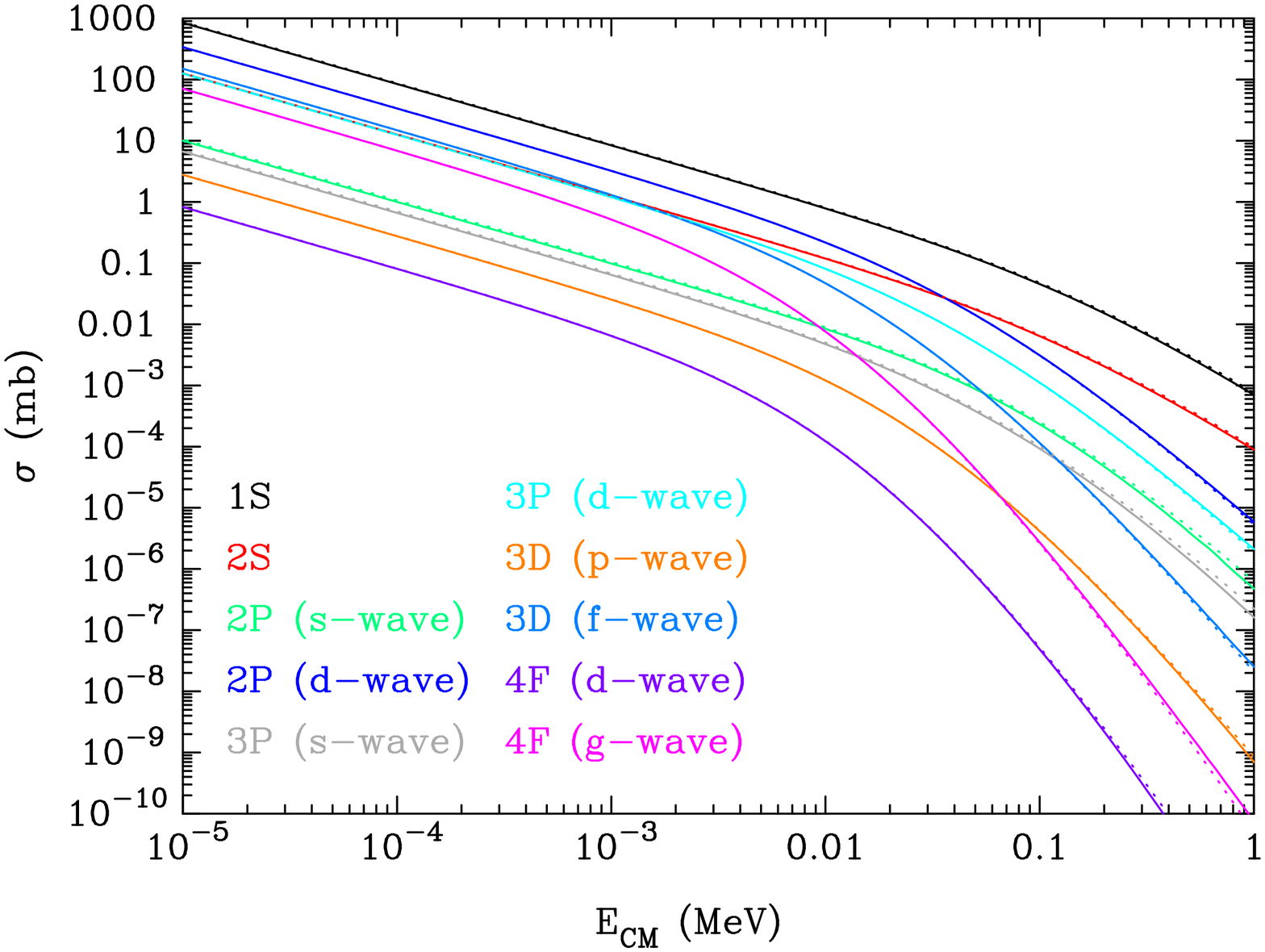}
\end{center}
\caption{Same as Fig. \ref{fig6}, but  for the $^4$He+$X^-$ system with $m_X=1$ GeV.  \label{fig18}}
\end{figure}

Figure \ref{fig18} shows bound-state wave functions (upper panel) and continuum wave functions (middle panel) at $E=0.07$ MeV as a function of radius $r$ for the $^4$He+$X^-$ system in the case of $m_X=1$ GeV.  The recombination cross section is also given as a function of the energy $E$ (bottom panel).   Line types indicate the same quantities as in Fig. \ref{fig6}.  Since the Coulomb potential in the $^4$He+$X^-$ system is small, the wave functions are more extended spatially.  Therefore, the effect of the finite-size charge distribution  is small as evidenced by  the fact that the   solid and dotted lines almost overlap in this figure.


\begin{figure}
\begin{center}
\includegraphics[width=0.45\textwidth]{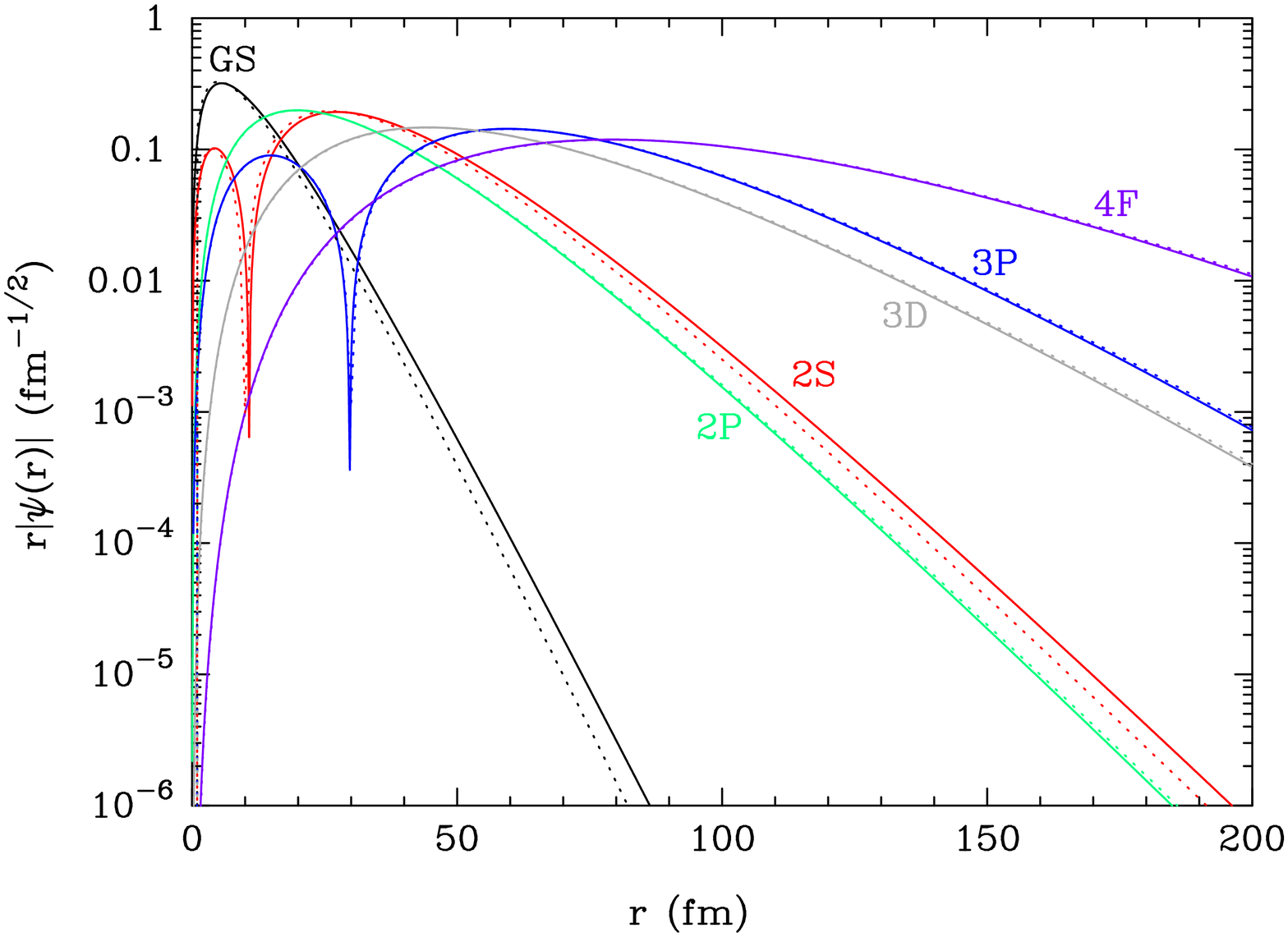}\\
\includegraphics[width=0.45\textwidth]{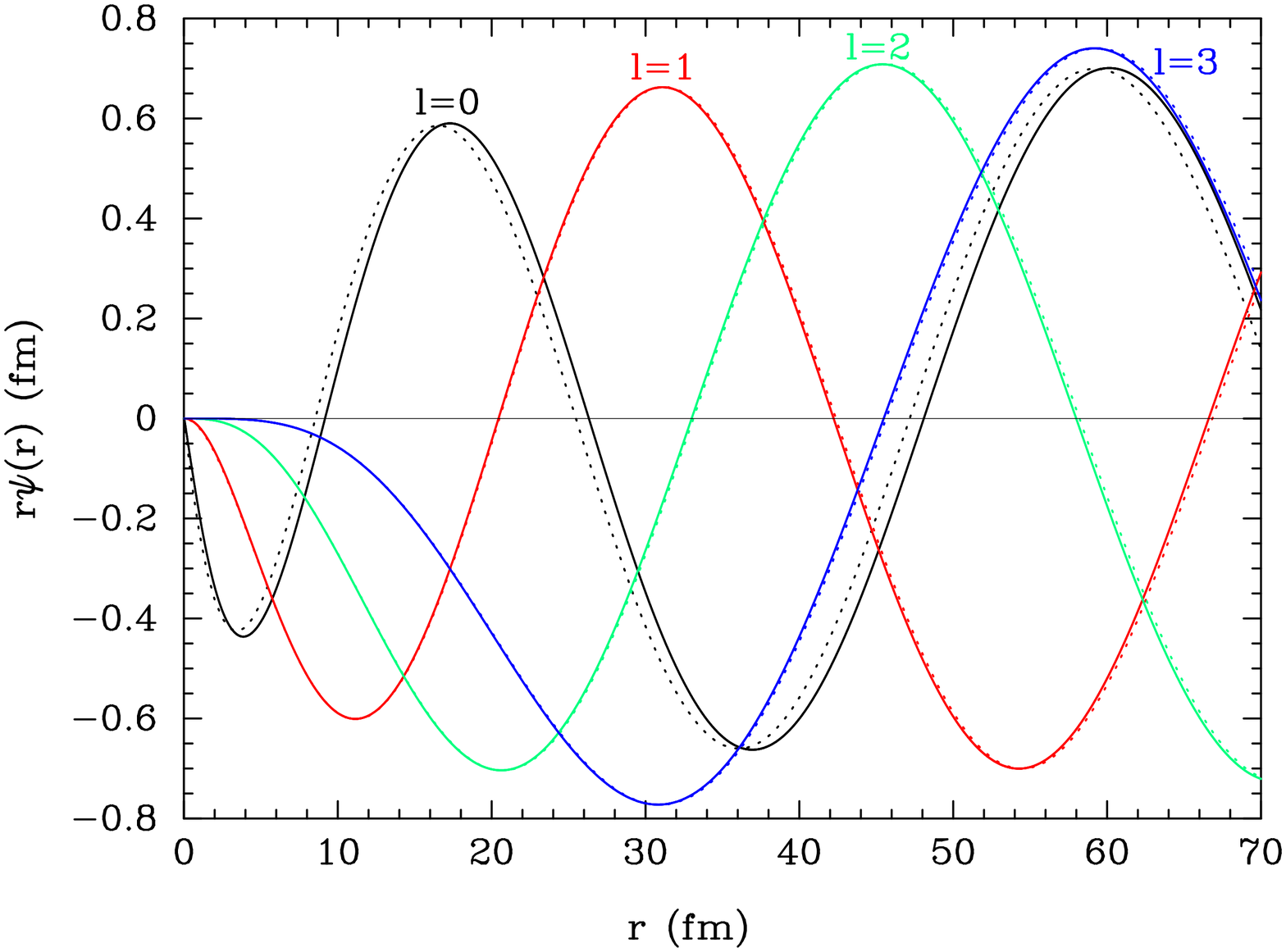}\\
\includegraphics[width=0.45\textwidth]{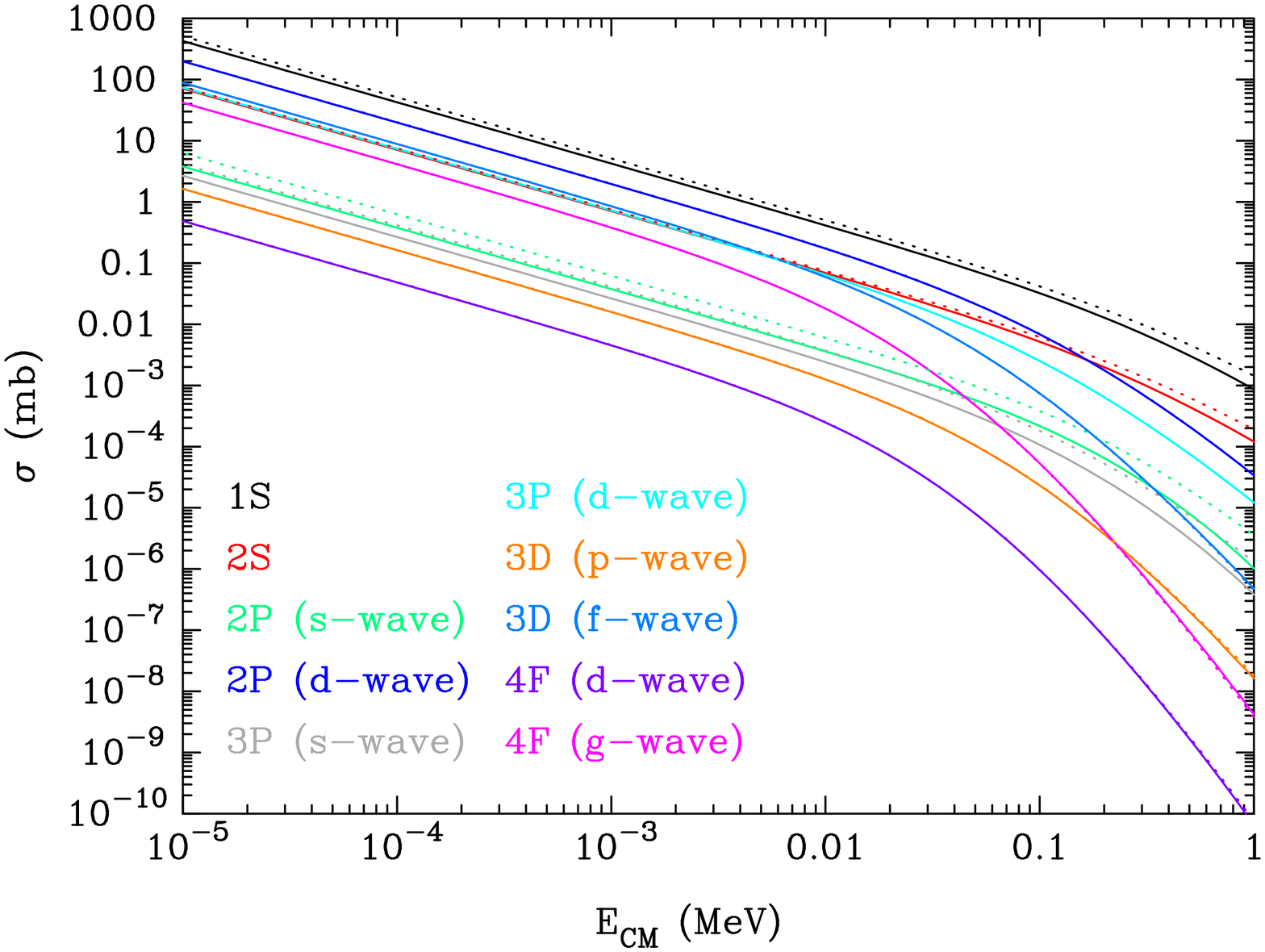}
\end{center}
\caption{Same as Fig. \ref{fig6}, but  for the $^4$He+$X^-$ system with $m_X=10$ GeV.  \label{fig19}}
\end{figure}

\begin{figure}
\begin{center}
\includegraphics[width=0.45\textwidth]{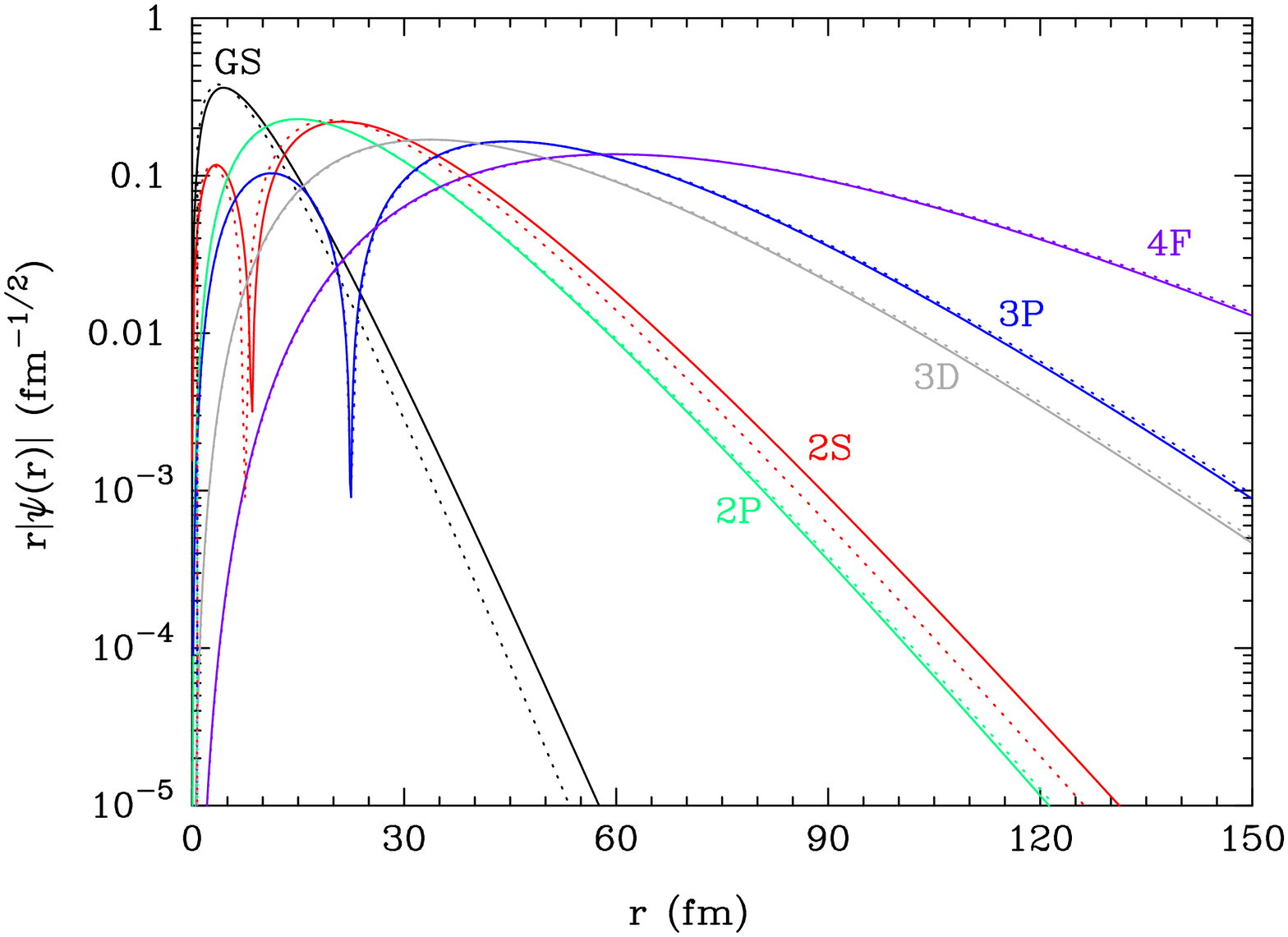}\\
\includegraphics[width=0.45\textwidth]{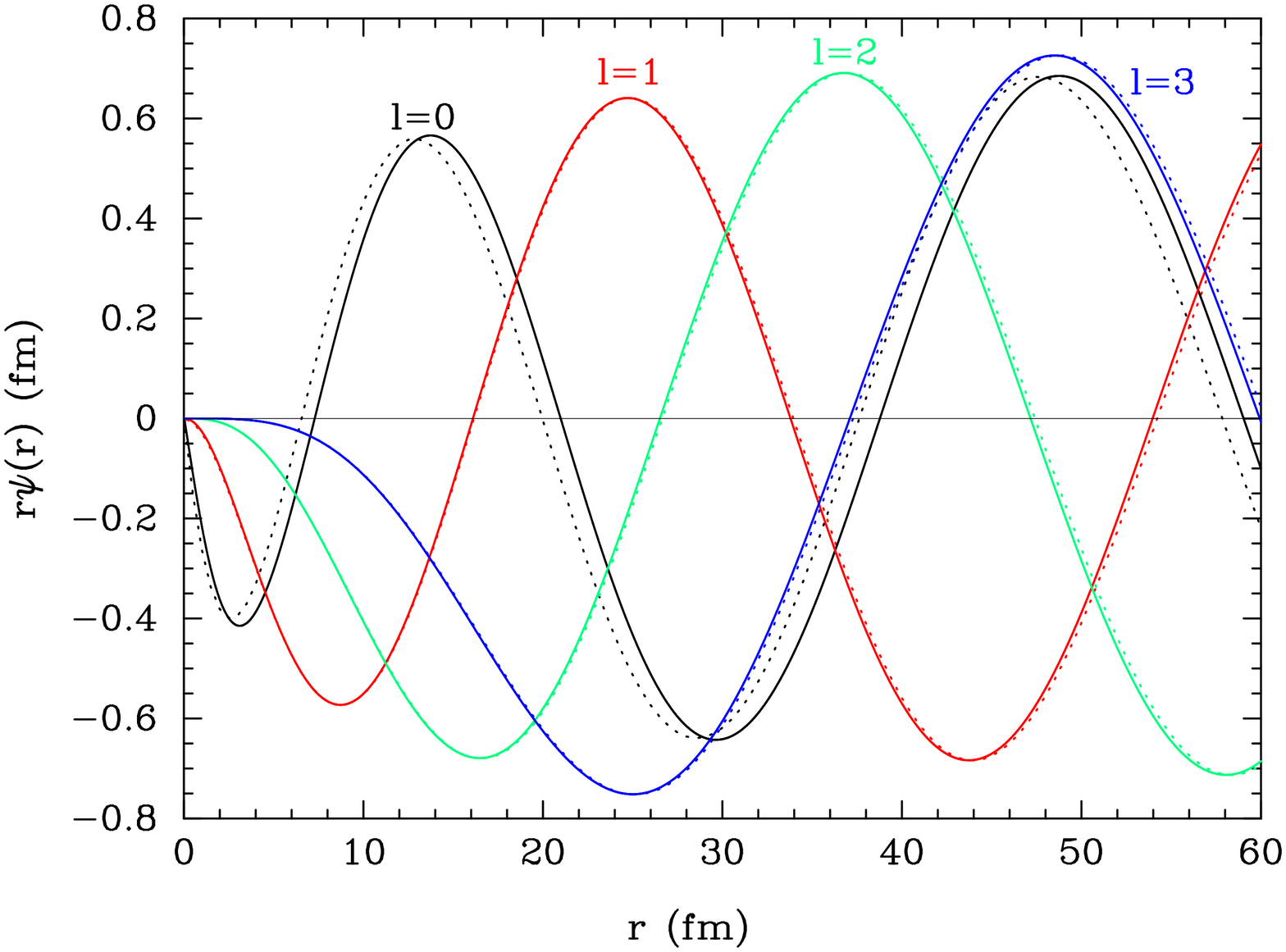}\\
\includegraphics[width=0.45\textwidth]{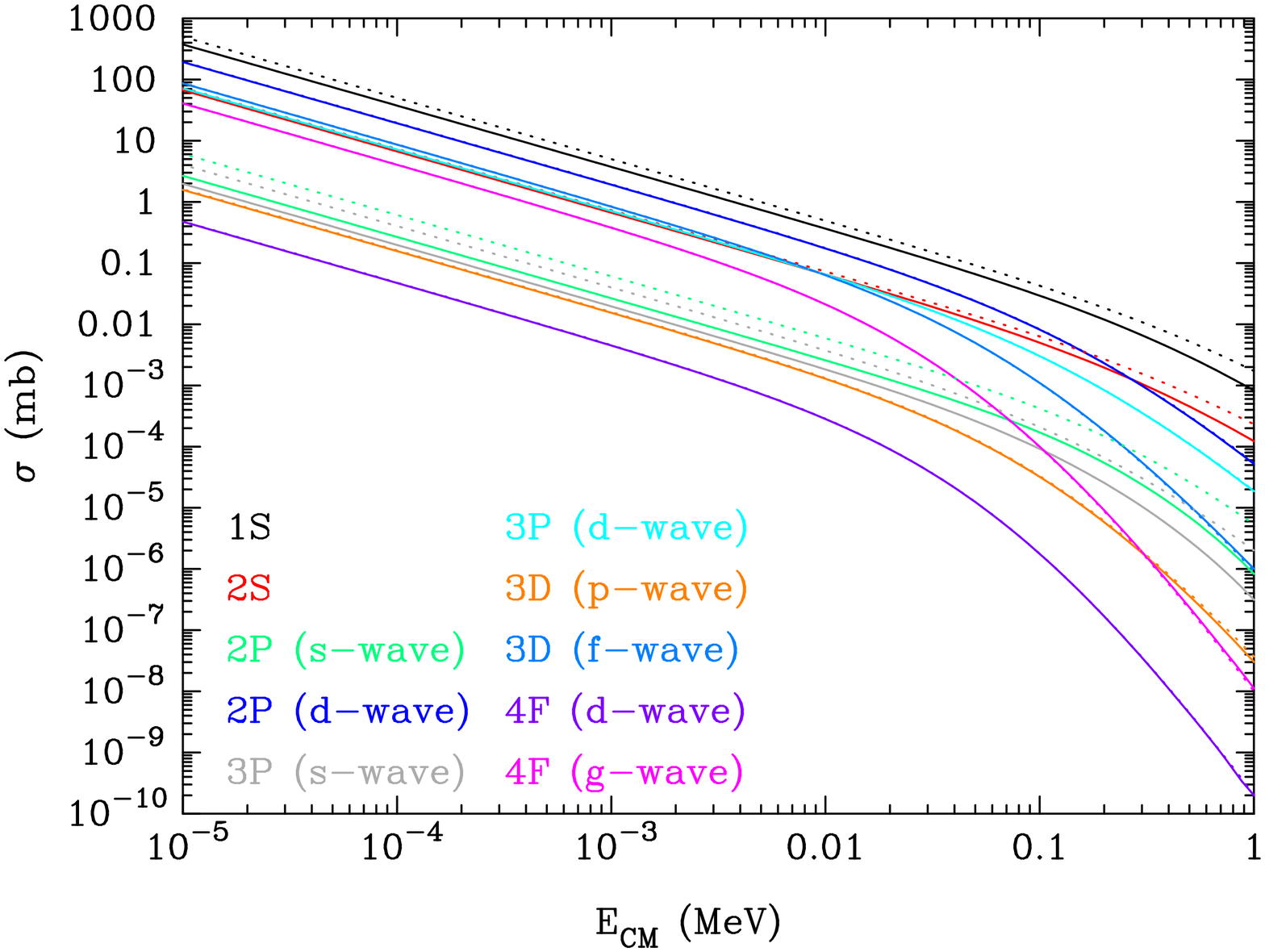}
\end{center}
\caption{Same as Fig. \ref{fig6}, but  for the $^4$He+$X^-$ system with $m_X=100$ GeV.  \label{fig20}}
\end{figure}

\begin{figure}
\begin{center}
\includegraphics[width=0.45\textwidth]{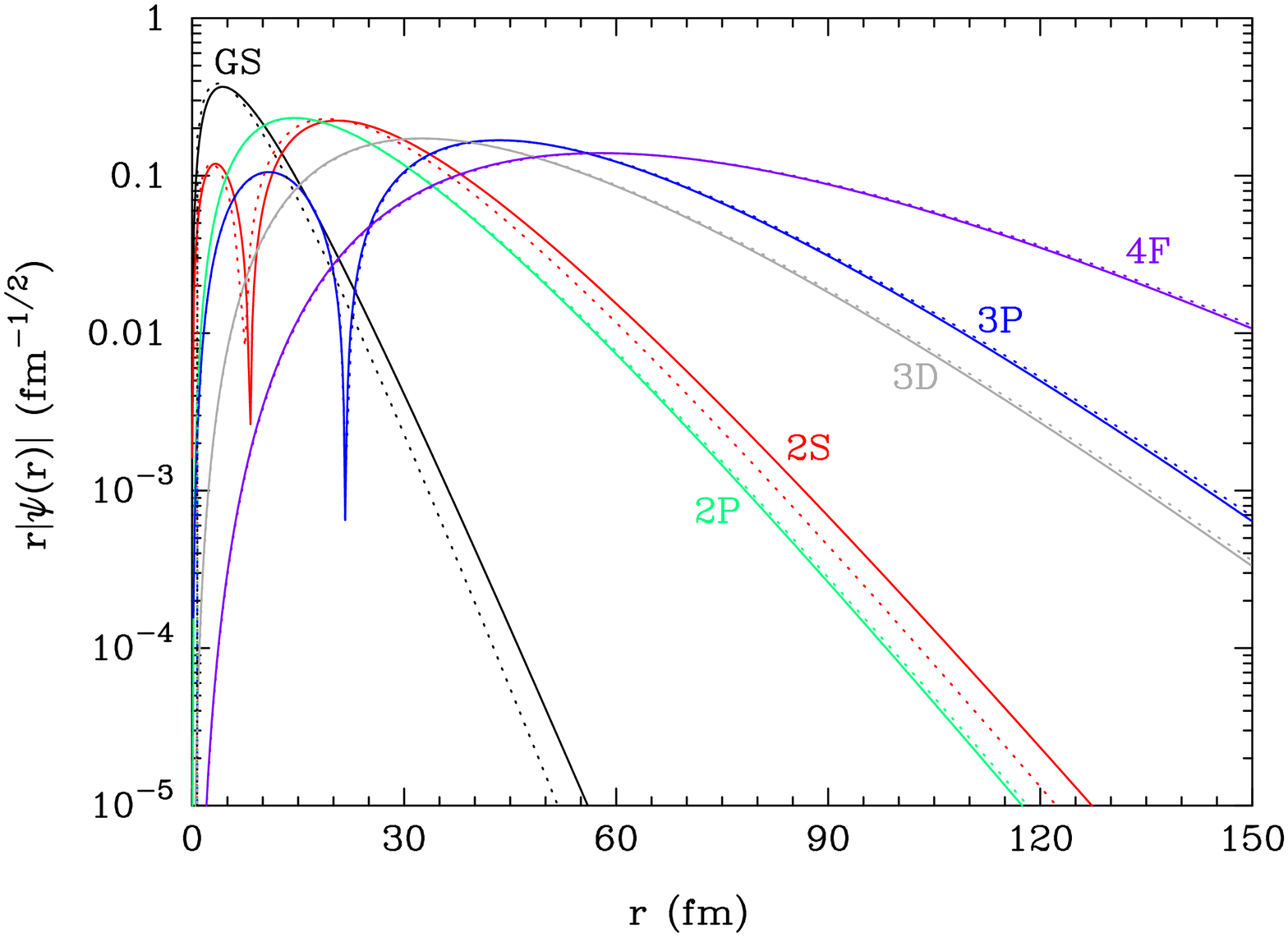}\\
\includegraphics[width=0.45\textwidth]{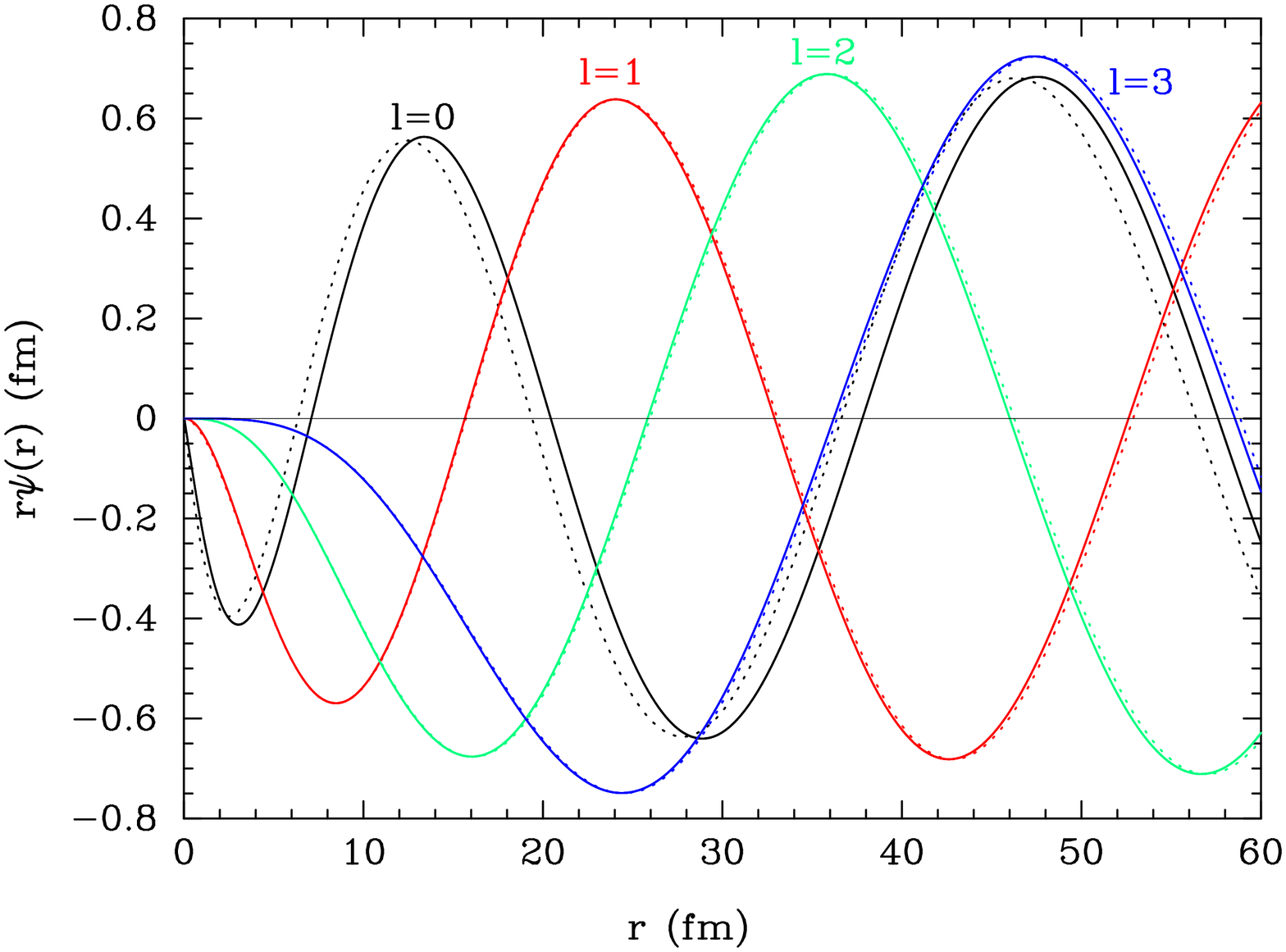}\\
\includegraphics[width=0.45\textwidth]{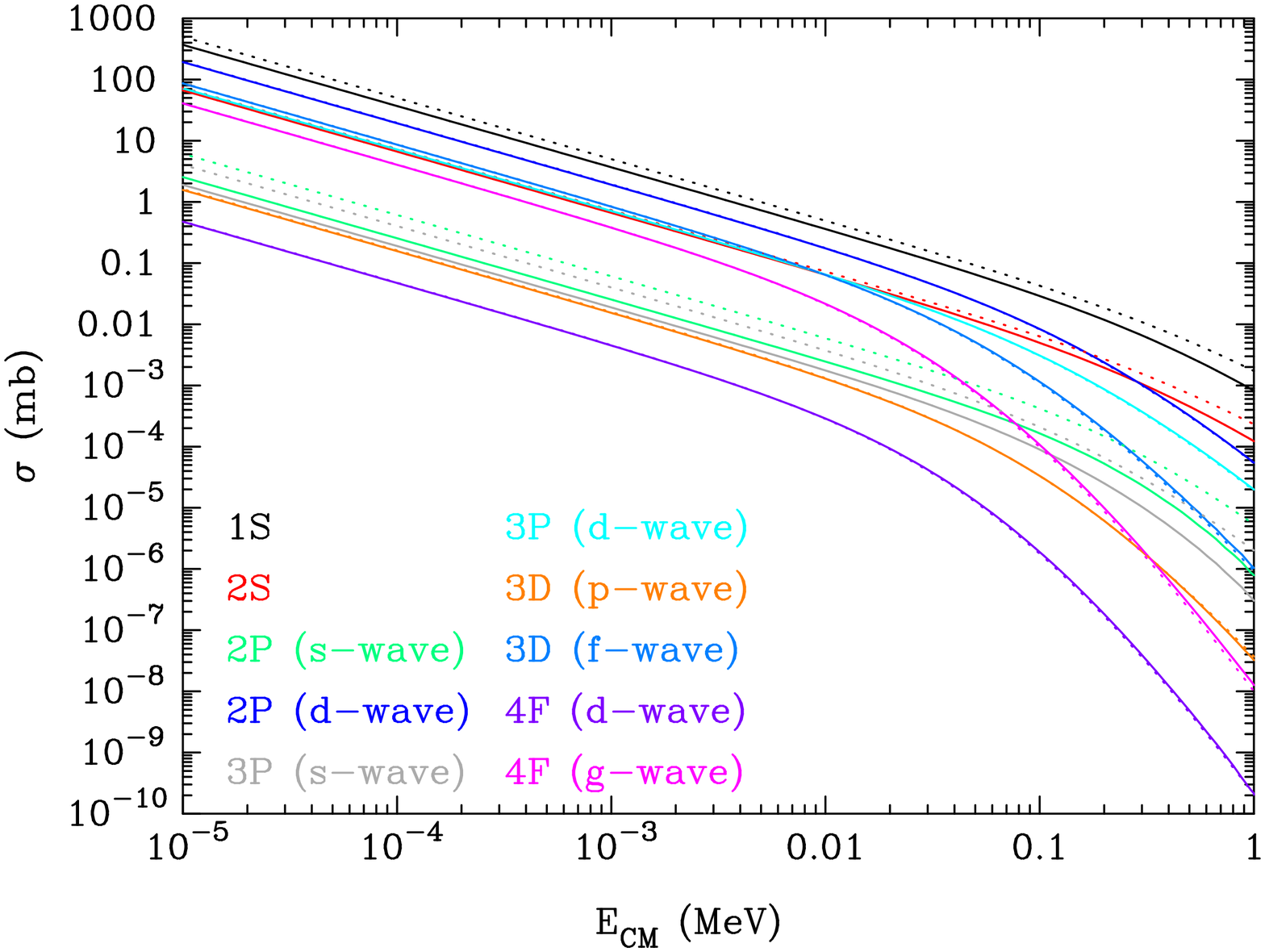}
\end{center}
\caption{Same as Fig. \ref{fig6}, but  for the $^4$He+$X^-$ system with $m_X=1000$ GeV.  \label{fig21}}
\end{figure}

Figures \ref{fig19}, \ref{fig20}, and \ref{fig21} show bound-state wave functions (upper panel) and continuum wave functions (middle panel) at $E=0.07$ MeV as a function of radius $r$ for the $^4$He+$X^-$ system in the case of $m_X=10$ GeV, 100 GeV, and 1000 GeV, respectively. The  recombination cross section is also shown as a function of the energy $E$ (bottom panel).   Line types indicate the same quantities as in Fig. \ref{fig6}.  It is apparent that larger $m_X$ values lead to larger differences in the wave functions and recombination cross sections due to  a finite-size vs. a point charge distribution.  However, because of the small amplitude of the Coulomb potential, the effect of the finite-size nuclear charge does not significantly affect the wave functions and cross sections.  As a result, even in the case of heavy $X^-$ particles ($m_X\gtrsim 100$ GeV), the dominant transition contributing to the recombination is the $p$-wave $\rightarrow$ 1S, similarly to the case of the Coulomb potential for point charges.


\subsection{Other $X$-nuclei}\label{sec5.5}

As seen in the Secs. \ref{sec5.1}--\ref{sec5.4}, realistic wave functions and recombination cross sections for $X$-nuclei can be  significantly different from those derived using  two point charged particles.  Hence, the recombination rates based upon the Bohr atomic model that were utilized in the previous studies \citep[e.g.,][]{Dimopoulos:1989hk,rujula90,Kohri:2006cn,Kusakabe:2007fu,Kusakabe:2007fv} should be considered uncertain by as much as  one order of magnitude.  
Although precise recombination rates for minor nuclei are not derived yet, we assume that the Bohr atom  formula  \citep{Bethe1957} for the minor nuclei is sufficient.  We adopt cross sections in the limit that the CM kinetic energy, $E$, is much smaller than the binding energy, $E_{\rm B}$.  This is justified since the condition, $E=\mu v^2/2=3T/2 \ll E_{\rm B}$ with $v$ the relative velocity of a nucleus $A$ and $X^-$, always holds when the bound state formation is more efficient than its destruction.  The cross sections sections are  thus given by
\begin{equation}
\sigma_{\rm rec}=\frac{2^9 \pi^2 e_1^2}{3\mathrm{e}^4} \frac{E_{\rm B}}{\mu^3 v^2},
\label{eq62}
\end{equation}
where
$\mathrm{e}=2.718$ is the base of the natural logarithm.  The thermal reaction rate is then given by
\begin{eqnarray}
N_{\rm A} \langle \sigma_{\rm rec} v \rangle &=& \frac{2^{19/2} \pi^{3/2} N_{\rm A}e_1^2}{3\mathrm{e}^4} \frac{E_{\rm B}}{\mu^{5/2} T^{1/2}}\nonumber\\
&=&1.37\times 10^4~{\rm cm}^3~{\rm s}^{-1} \frac{(e_1/e)^2 Q_9}{A^{5/2}T_9^{1/2}}\nonumber\\
&\equiv& C_1 T_9^{-1/2}~{\rm cm}^3~{\rm s}^{-1},
\label{eq63}
\end{eqnarray}
where
$Q_9=Q/{\rm MeV}$ is the $Q$-value in units of MeV, and we defined a rate coefficient $C_1=1.37\times 10^4(e_1/e)^2 Q_9/A^{5/2}$.  The $Q$-value for the recombination is equal to the binding energy of the $X$-nucleus $E_{\rm B}$.

The thermal rate for reverse reaction is related to that for the forward reaction through the reciprocity theorem.  Using the relation between the reverse rate $\langle C+D \rangle$ and the forward rate $\langle A+B \rangle$ \citep{Fowler1967,Angulo1999}, the reverse rate coefficient is defined for a non-radiative reaction $A$($B$, $C$)$D$  by
\begin{eqnarray}
C_{\rm r}&\equiv& \frac{\langle C+D \rangle}{\langle A+B \rangle}=\frac{(1+\delta_{CD})}{(1+\delta_{AB})} \frac{g_A g_B}{g_C g_D} \left(\frac{A_A A_B}{A_C A_D}\right)^{3/2} \nonumber\\
&&\times \exp\left(-Q/T\right),
\label{eq64}
\end{eqnarray}
where
$g_i=2I_i+1$ accounts for  the spin degrees of freedom with $I_i$ the nuclear spin of species $i$.  For a radiative reaction $A$($B$, $\gamma$)$C$, on the other hand, the reverse rate coefficient is given by
\begin{eqnarray}
C_{\rm r}&\equiv& 10^{-10}~{\rm cm}^{-3}~\frac{n_\gamma \langle C+\gamma \rangle}{N_{\rm A}\langle A+B \rangle}\exp(Q/T) \left(\frac{T}{10^9~{\rm K}}\right)^{-3/2} \nonumber\\
&=&0.987\frac{g_A g_B}{(1+\delta_{AB})g_C}\left(\frac{A_A A_B}{A_C}\right)^{3/2},
\label{eq65}
\end{eqnarray}
where
$n_\gamma=2\zeta(3)T^3/\pi^2$ is the number density of photon with $\zeta(3)=1.202$ the Riemann zeta function of 3.

Table \ref{tab11} shows approximate recombination rates for nuclei which are not treated in Sec. \ref{sec5}.  The second and third columns correspond to the rate coefficients $C_1$ and reverse rate coefficients $C_{\rm r}$, respectively [Eqs. (\ref{eq63}) and (\ref{eq65})], for the case of $m_X=1$ GeV.  The $C_1$ and $C_{\rm r}$ values for $m_X=10$, 100, and 1000 GeV are listed in fourth to ninth columns.

\placetable{tab11}

\section{$^9$B\lowercase{e} PRODUCTION FROM $^7$L\lowercase{i}}\label{sec6}

We suggest the possibility of a significant production of $^9$Be catalyzed by the negatively charged $X^-$ particle through the deuteron transfer reaction $^7$Li$_X$($d$, $X^-$)$^9$Be.  This reaction rate depends on both resonant and nonresonant components.  Since a realistic theoretical estimate of the rate for this reaction is not currently available, we adopt a simple ansatz that the astrophysical $S$ factor for the reaction can be  taken from the existing data for $^7$Li($d$, $n\alpha$)$^4$He, i.e, $S=30$ MeV b \citep{Caughlan1988}.  We note that the cross section values for $^7$Li($d$, $n\alpha$)$^4$He recommended by the Evaluated Nuclear Data File \citep[ENDF/B-VII.1, 2011; ][]{ENDF-B-VII.1short} corresponds to $S\sim 10$ MeV b for an energy range of 0.1 MeV $\leq E \leq$ 1 MeV.

Realistic theoretical estimates of  the nonresonant cross section for $^7$Li$_X$($d$, $X^-$)$^9$Be would be difficult (M. Kamimura 2013; private communications).  Since the structure of the $^9$Be nucleus is approximately described as $\alpha+\alpha+n$, there is no existing study on the probability that the $^9$Be nucleus is described as $^7$Li$+d$ bound states.  In addition  experiments on the  low energy nuclear scattering of $^7$Li+$d$ are needed to construct    the imaginary potential for $^7$Li$+d$ elastic scattering in the quantum mechanical calculations.  Hence, the cross section for $^7$Li$_X$($d$, $X^-$)$^9$Be assumed in this study is probably uncertain by as much as an order of magnitude, and could be much smaller.

\section{$\beta$-DECAY AND NONRESONANT NUCLEAR REACTIONS}\label{sec7}

Mass excesses of $X$-nuclei, and $Q$-values for possible reactions associated with the $X^-$ particle were  calculated using the  binding energies of $X$-nuclei derived in Sec. \ref{sec3}.  The $\beta$-decay rates of $X$-nuclei ($A_X$), $\Gamma_{\beta X}$, are estimated using experimental values for normal nuclei ($A$), $\Gamma_{\beta}$,  taking into account the momentum phase space factor related to the reaction $Q$-value.  The adopted rates are then given by $\Gamma_{\beta X}=\Gamma_\beta(Q_X/Q)^5$, where $Q_X$ and $Q$ are $Q$-values for the $\beta$-decay of $A_X$ and $A$, respectively.  The decay rate $\Gamma_\beta$ is related to the half life $T_{1/2}$, i.e  $\Gamma_\beta=\ln 2/T_{1/2}$.  An exception to this is  the $\beta$-decay rate of $^6$Be$_X$.  
In this case the $\beta$-decay  rate is estimated
from  that of $^6$He assuming an approximate isospin symmetry.  Adopted data values are as follows:  1) $Q=3.508$ MeV and $\Gamma_\beta=0.859$ s$^{-1}$ for $^6$He(, $e^- \bar{\nu_e}$)$^6$Li \citep{Tilley2002}, 2) $Q=16.005$ MeV and $\Gamma_\beta=0.825$ s$^{-1}$ for $^8$Li(, $e^- \bar{\nu}_e$)$^8$Be \citep{Tilley2004}, and 3) $Q=17.979$ MeV and $\Gamma_\beta=0.900$ s$^{-1}$ for $^8$B(, $e^+ \nu_e$)$^8$Be \citep{Tilley2004}.

Table \ref{tab12} shows the adopted $\beta$-decay rates for $X$-nuclei.  The second and third columns correspond to the $Q$-value and the decay rate $\Gamma_{\beta X}$, respectively, for the case of $m_X=1$ GeV.  The $Q$ and $\Gamma_{\beta X}$ values for $m_X=10$, 100, and 1000 GeV are listed in the fourth to ninth columns.

\placetable{tab12}

For nonresonant thermonuclear reaction rates between two charged nuclei,  the  astrophysical $S$-factors for $X$-nuclear reactions are taken to be as the same as those for the corresponding normal nuclear reactions \citep{Caughlan1988,Smith:1992yy}.  However  changes in the reduced mass and charge numbers are corrected  exactly the same as in \citet{Kusakabe:2007fv}.  For the reactions $^6$Li$_X$($p$, $^3$He)$^4$He$_X$, $^7$Be$_X$($p$, $\gamma$)$^8$B$_X$, $^4$He$_X$($d$, $X^-$)$^6$Li, $^4$He$_X$($t$, $X^-$)$^7$Li, and $^4$He$_X$($^3$He, $X^-$)$^7$Be, the cross sections have been calculated in a quantum mechanical model \citep{Hamaguchi:2007mp,Kamimura:2008fx}.  
We, therefore, take those astrophysical $S$-factors from the published results,  corrected for changes in the reduced mass.  Since both forward and reverse reaction rates depend upon $m_X$, they are different from the reaction rates estimated under the assumption of $m_X \rightarrow \infty$ which have been already published \citep{Kusakabe:2007fv}.

For reactions between a neutron and $X$-nuclei and also those between nuclei and neutral $X$-nuclei, Coulomb repulsion does not exist.  The reactions have, however, already been found to be unimportant within the parameter region for which the predicted light element abundances are consistent with observational constraints \citep{Kusakabe:2010cb}.  We, therefore,  utilize  the same rates as assumed in \citet{Kusakabe:2007fv} for the neutron induced non-radiative reactions, and those published in \citet{Kamimura:2008fx} for reactions of neutral $X$-nuclei.

Table \ref{tab13} shows parameters of nuclear reaction rates for $X$-nuclei.  The second and third columns correspond to the reverse rate coefficient $C_{\rm r}$ [Eqs. (\ref{eq64}) and (\ref{eq65})] and the $Q_9$-value, respectively, for the case of $m_X=1$ GeV.  The $C_{\rm r}$ and $Q_9$ values for $m_X=10$, 100, and 1000 GeV are listed in fourth to ninth columns.  Since the baryon density is low in the early universe, the rates for three-body reactions are small.  Therefore, the reverse reactions of $^6$Li$_X$($p$, $^3$He $\alpha$)$X^-$ and $^7$Be$_X$($n$, $p ^7$Li)$X^-$ are neglected in our calculation, and the reverse rate coefficients are  not shown in this table.

\placetable{tab13}

Although the $Q$-value for the $^8$Be$_X$($p$, $\gamma$)$^9$B$_X$ reaction is negative in the case of $m_X=1$ GeV, its rate is estimated by considering only the reduced mass factor.  Since the $|Q|$ value is very small, it can be regarded as effectively zero in the relevant temperature range.

\section{BBN REACTION NETWORK}\label{sec8}
We utilized a modified   \citep{Kusakabe:2007fv,Kusakabe:2010cb} version of the Kawano reaction network code \citep{Kawano1992,Smith:1992yy}  to calculate nucleosynthesis for four different $X^-$ particle masses, $m_X$.  The nuclear charge distribution was assumed to be given by the WS40 model.  The free $X^-$ particle and bound $X$-nuclei are encoded as new species whose abundances are to be calculated.  Reactions involving the $X^-$ particle are encoded as new reactions.  The mass excesses of $X$-nuclei are input into the code.  In this way  the energy generation through the recombination of normal nuclei and an $X^-$ particle, and the nuclear reactions of $X$-nuclei are precisely taken into account in the thermodynamics of the expanding universe \citep{Kawano1992}.  

Our BBN code includes many reactions associated with the $X^-$ particle.  It then solves the non-equilibrium nuclear and chemical reaction network associated with the $X^-$ with improved reaction rates derived from quantum many-body calculations~\citep{Kamimura:2008fx}. The neutron lifetime was updated to be $878.5 \pm 0.7_{\rm stat} \pm 0.3_{\rm sys}$~s~\citep{Serebrov:2010sg,Mathews:2004kc} based upon improved measurements \citep{Serebrov:2004zf}.  Rates for reactions of normal nuclei with mass numbers $A \le  10$ have been updated with the JINA REACLIB Database V1.0 \citep{Cyburt2010}.  The baryon-to-photon ratio was taken from the determination with the WMAP \citep{Spergel:2003cb,Spergel:2006hy,Larson:2010gs,Hinshaw:2012fq} $\Lambda$CDM model (WMAP9 data only):  i.e.~$\eta=(6.19\pm 0.14) \times 10^{10}$~\citep{Hinshaw:2012fq}.

Reaction rates derived in this paper are included in the code.  We note that
the nonresonant radiative neutron capture reactions of $X$-nuclei considered in the previous study \citep{Kusakabe:2007fv} are switched off for the following reason.  Rates for the reactions generally depend on $m_X$.  When $m_X$ is much larger than the nucleon mass, the radiative neutron capture reactions via electric multipole transitions are strongly hindered because of the very small effective charges \citep{Kusakabe:2009jt}.  In addition, independently of whether the mass of the $m_X$ is large or not, the nucleosynthesis triggered by the $X^-$ particle occurs rather late in the BBN epoch when the neutron abundance is already small.  Thus, neglecting the reactions does not significantly change the time evolutions of the nuclear abundances.

Recombination rates for $^7$Be($X^-$, $\gamma$)$^7$Be$_X$, $^7$Li($X^-$, $\gamma$)$^7$Li$_X$, $^9$Be($X^-$, $\gamma$)$^9$Be$_X$, and $^4$He($X^-$, $\gamma$)$^4$He$_X$ were modified.
$^9$Be$_X$ production through $^8$Be$_X$, i.e., $^4$He$_X$($\alpha$, $\gamma$)$^8$Be$_X$($n$, $\gamma$)$^9$Be, depends significantly on the energy levels of $^8$Be$_X$ and $^9$Be$_X$ \citep{Pospelov:2007js,Kamimura:2008fx,Cyburt:2012kp}, and precise calculations with a quantum four body model by another group is under way \citep{Kamimura2010}.  In this paper, we neglect those reaction series, and leave that discussion as a future work.  The reaction $^4$He$_X$($\alpha$, $\gamma$)$^8$Be$_X$ is thus not included, and the abundance of $^8$Be$_X$ is not shown in the figures below.

\section{RESULTS}\label{sec9}
We show calculated results of BBN for four values of $m_X$.  First, we analyze the time evolution for abundances of normal and $X$-nuclei.  Then we  update constraints on the parameters characterizing  the $X^-$ particle.

The two free parameters in this BBN calculation are the ratio of number abundance of the $X^-$ particles to the  total baryon density, $Y_X=n_X/n_{\rm b}$, and the decay lifetime of the $X^-$ particle, $\tau_X$.  The lifetime is assumed to be much smaller than the age of the present universe, i.e., $<<14$ Gyr \citep{Hinshaw:2012fq}.  The primordial $X$-particles from the early universe are thus by now, long  extinct.  When the $m_X$ value is small, the annihilation cross section for the  $X^-$ and its antiparticle $X^+$ is expected to be large.  Since a large cross section tends to a small freeze-out abundance of $X^-$, it is naturally expected that the abundance $Y_X$ would be very small for  small $m_X$.  However, we also perform calculations for large values of $Y_X$ even in the case of a small $m_X$ value taking into account the possibility that there may be a difference in number abundances of $X^-$ and $X^+$.  If the abundance of $X^-$ had been larger than $X^+$, the freeze-out 
 abundances can be much larger than that for the case of equal abundances of $X^-$ and $X^+$.  In this case, however, charge neutrality still requires the condition of zero net global charge density during the BBN epoch.

As for the fate of $X$-nuclei, it is assumed that the total kinetic energy of products generated from  the decay of $X^-$ is large enough so that all $X$-nuclei can decay into normal nuclei plus the decay products of $X^-$.  The $X$ particle is detached from $X$-nuclei with its rate given by the $X^-$ decay rate.  The lifetime of $X$-nuclei is, therefore, given by the lifetime of the $X^-$ particle itself.

 To identify the  important reactions that affect the abundances of $^6$Li, $^7$Li, $^7$Be, and $^9$Be we tried multiple calculations by switching off respective reactions.  A detailed analysis of the nuclear flow is  described below.

\subsection{Abundance Constraints}\label{sec9.1}
Observational constraints on the deuterium abundance are taken from the mean value of ten QSO absorption line systems, and the abundance corresponding to the best measured damped Lyman alpha system of quasi-stellar object SDSS J1419+0829, i.e., log(D/H)=$-4.58\pm 0.02$ and log(D/H)=$-4.596\pm 0.009$, respectively \citep{Pettini:2012ph}.  Constraints on the primordial $^3$He abundance are taken from the mean value of Galactic HII regions measured through the 8.665~GHz hyperfine transition of $^3$He$^+$, i.e., $^3$He/H=$(1.9\pm 0.6)\times 10^{-5}$~\citep{Bania:2002yj}.  Constraints on the $^4$He abundance are taken from observational values of metal-poor extragalactic HII regions, i.e, $Y_{\rm p}=0.2565\pm 0.0051$~\citep{Izotov:2010ca} and $Y_{\rm p}=0.2561\pm 0.0108$~\citep{Aver:2010wq}.
We take the observational constraint on the $^7$Li abundance from the central  value of log($^7$Li/H)$=-12+(2.199\pm 0.086)$ derived in the  3D NLTE model of 
\citet{Sbordone:2010zi}.  
On the other hand, the constraint on the $^6$Li abundance is chosen more conservatively.  We adopted the least stringent $2~\sigma$ (95\% C.L.) upper limit for all stars reported in \citet{Lind:2013iza}, i.e., $^6$Li/H=$(0.9\pm 4.3)\times 10^{-12}$ for the G64-12 (NLTE model with 5 free parameters).

\subsection{$m_X=1$ GeV}\label{sec9.2}
\subsubsection{Nucleosynthesis}
\begin{figure}
\begin{center}
\includegraphics[width=0.45\textwidth]{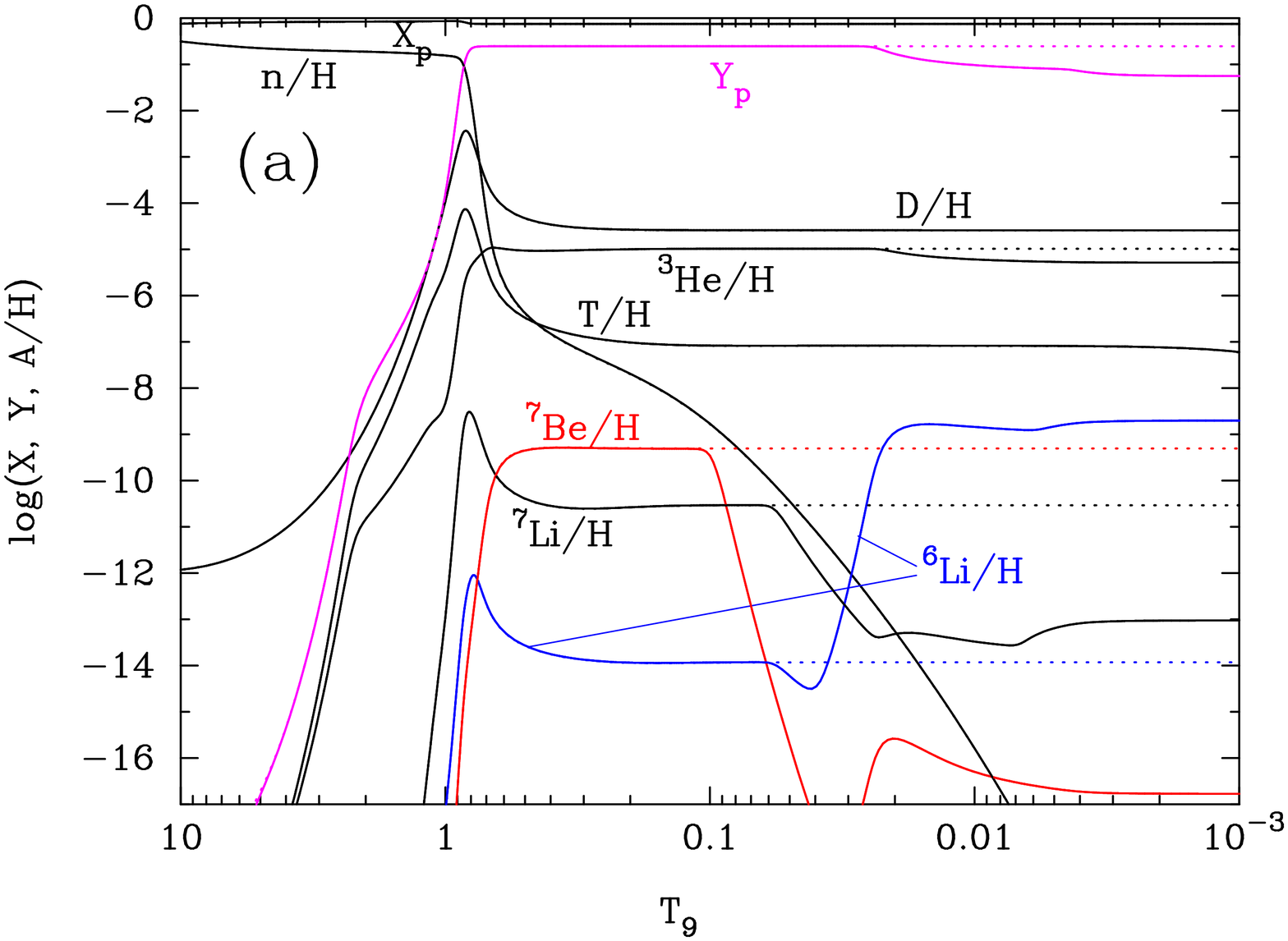}\\
\includegraphics[width=0.45\textwidth]{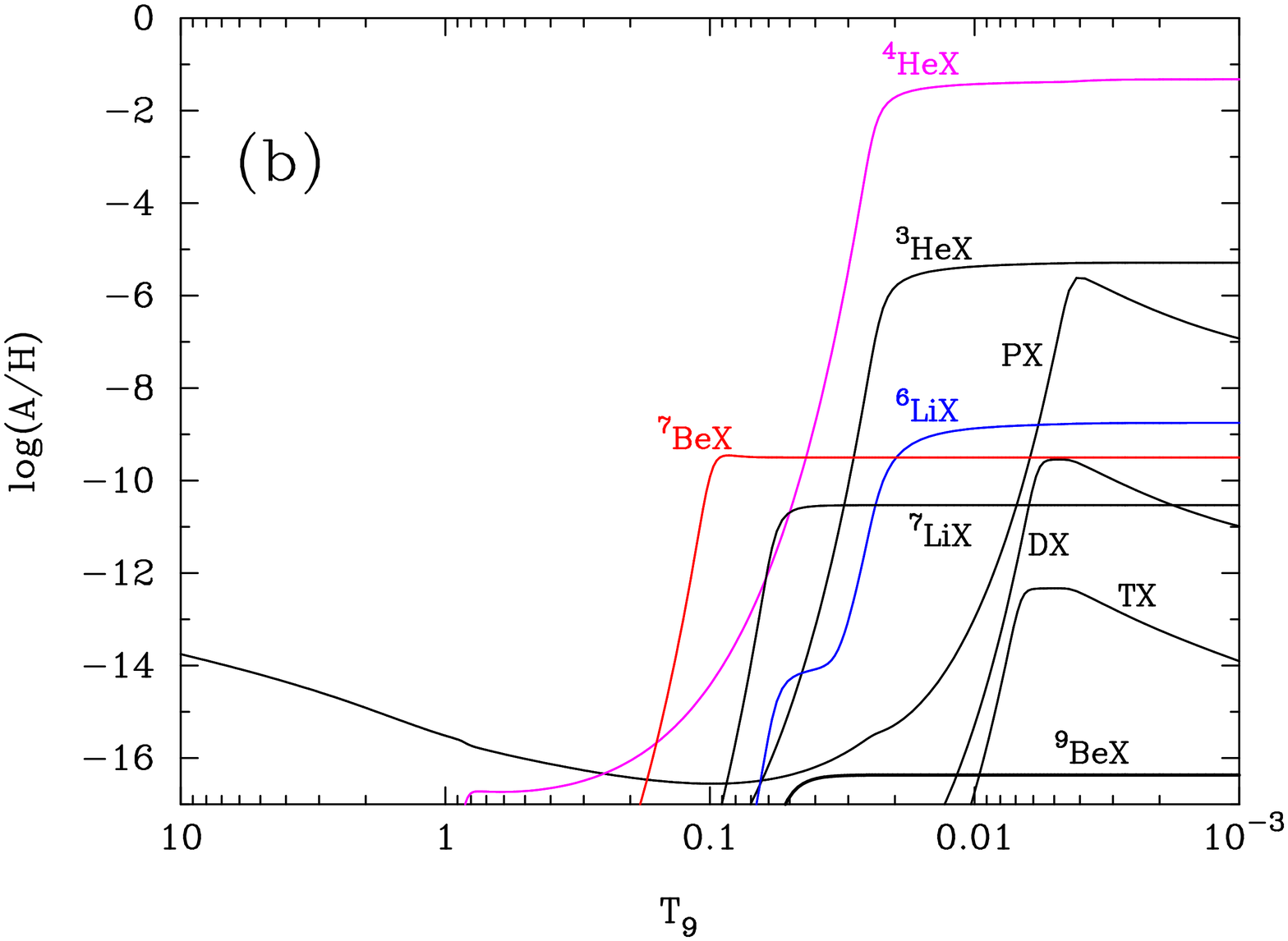}
\end{center}
\caption{Calculated abundances of normal nuclei (a) and $X$-nuclei (b) as a function of $T_9$ for $m_X=1$ GeV.  $X_{\rm p}$ and $Y_{\rm p}$ are the mass fractions of $^1$H and $^4$He, in total baryonic matter, while the other curves correspond to number abundances with respect to that of hydrogen.  The abundance and  lifetime of the $X^-$ particle are taken to be $Y_X=n_X/n_b=0.05$ and $\tau_X=\infty$, respectively.  The dotted lines show the results of the SBBN model.  \label{fig22}}
\end{figure}

Figure \ref{fig22} shows the calculated abundances of normal nuclei (a) and $X$ nuclei (b) as a function of $T_9$ for $m_X=1$ GeV.  Curves for $^1$H and $^4$He correspond to the mass fractions, $X_{\rm p}$ ($^1$H) and $Y_{\rm p}$ ($^4$He) in total baryonic matter, while the other curves correspond to number abundances with respect to that of hydrogen.  The dotted lines show the  result of the SBBN model.  The abundance and the lifetime of the $X^-$ particle are assumed to be $Y_X=0.05$ and $\tau_X=\infty$, respectively, for this example.


Early in the BBN epoch  ($T_9\gtrsim 1$), $p_X$ is the only $X$-nuclide with an abundance larger than $A_X/$H$>10^{-17}$.  Its abundance is the equilibrium value determined by the balance between  the recombination of $p$ and $X^-$ and the photoionization of $p_X$.  When the temperature decreases to $T_9\lesssim 1$, $^4$He is produced as in SBBN (panel a).  Simultaneously, the abundance of $^4$He$_X$ increases through the recombination of $^4$He and $X^-$ (panel b).  As the temperature decreases further, the recombination of nuclei with $X^-$ gradually proceeds in order of decreasing binding energies of $A_X$,  similar to the  recombination of nuclei with electrons.

$^7$Be first recombines with $X^-$ via the $^7$Be($X^-$, $\gamma$)$^7$Be$_X$ reaction at $T_9\lesssim 0.1$.  The produced $^7$Be$_X$ nuclei are then slightly destroyed via the
$^7$Be$_X$($p$, $\gamma$)$^8$B$_X$ reaction.  In the  late epoch, the $^7$Be abundance increases through the reaction $^4$He$_X$($^3$He, $X^-$)$^7$Be at $T_9\sim 0.03-0.02$.

The $^6$Li abundance decreases through the recombination reaction $^6$Li($X^-$, $\gamma$)$^6$Li$_X$ operating at $T_9\lesssim 0.05$.  However, soon after the start of recombination, it is produced through the reaction $^4$He$_X$($d$, $X^-$)$^6$Li at $T_9\sim 0.03-0.02$.  After this production, the $^6$Li$_X$ abundance also increases through recombination.  In the  late epoch, the $^6$Li abundance increases through the reaction $^2$H$_X$($\alpha$, $X^-$)$^6$Li at $T_9\sim 4\times 10^{-3}$.

At $T_9\lesssim 0.05$, the $^7$Li abundance decreases through the recombination reaction  $^7$Li($X^-$, $\gamma$)$^7$Li$_X$.  A small amount of $^7$Li is later produced through the reactions $^4$He$_X$($t$, $X^-$)$^7$Li ($T_9\sim 0.02-0.01$), and $^3$H$_X$($\alpha$, $X^-$)$^7$Li [$T_9\sim (6-5)\times 10^{-3}$].

$^9$Be is predominantly produced through the reaction $^7$Li$_X$($d$, $X^-$)$^9$Be at $T_9\sim 0.06-0.05$.  At $T_9\lesssim 0.05$, the recombination $^9$Be($X^-$, $\gamma$)$^9$Be$_X$ enhances the abundance of $^9$Be$_X $.  The abundance of $^9$Be is small and not seen in this figure since it is converted to $^9$Be$_X$ via the recombination.  We note that the proton capture reaction $^9$Be$_X$($p$, $^6$Li)$^4$He$_X$ does not work efficiently at this temperature of $^9$Be$_X$ production.

The recombination of $^7$Be with an $X^-$ particle and the subsequent radiative proton capture of $^7$Be$_X$ occurs at $T_9\sim 0.1$ although the effect of the latter reaction can not be seen well in this figure.  In this case, the abundances of $^7$Be and $^7$Be$_X$ only change through the recombination at $T_9\sim 0.1$.  The abundance ratio of $^7$Be to $^7$Be$_X$ is then simply described with chemical equilibrium  \citep{Rybicki1979} as
\begin{eqnarray}
\frac{n_A n_X}{n_{A_X}} &=&\frac{g_A g_X}{g_{A_X}} \left(\frac{m_A m_X}{m_{A_X}} \frac{T}{2\pi}\right)^{3/2} \mathrm{e}^{(m_{A_X}-m_A-m_X)/T} \nonumber\\
&\approx& \left(\frac{\mu T}{2\pi}\right)^{3/2} \mathrm{e}^{-E_{\rm B}(A_X)/T},
\label{eq66}
\end{eqnarray}
where
the spin factor of $X^-$ is $g_X=1$, and only the GS of $^7$Be$_X$ is considered so that $g_{A_X}=g_A$.

The baryon number density determined from the CMB WMAP measurement is
\begin{eqnarray}
n_b&\approx& \frac{\rho_b}{m_p}=\frac{\rho_c\Omega_b}{m_p}\left(1+z\right)^3\nonumber\\
&=&1.26
\times 10^{19}~{\rm cm}^{-3}\left(\frac{h}{0.700}\right)^2 \left(\frac{\Omega_b}{0.0463}\right)T_9^3,~~~~
\label{eq67}
\end{eqnarray}
where
$\rho_b$ and $\rho_c$ are the baryon density and the present critical density, respectively.
$\Omega_b=0.0463 \pm 0.0024$ is the baryon density parameter,
$z$ is the redshift of the universe which is related to temperature as $(1+z)=T/T_0$ with $T_0=2.7255$ K the present radiation temperature of the universe \citep{Fixsen:2009ug}, and
$h=H_0$/(100 km s$^{-1}$ Mpc$^{-1}$)$=0.700\pm 0.022$ is the reduced Hubble constant with Hubble constant $H_0$.  The cosmological parameters have been taken from values determined from the WMAP \citep{Spergel:2003cb,Spergel:2006hy,Larson:2010gs,Hinshaw:2012fq} ($\Lambda$CDM model; WMAP9 data only).

We define the recombination temperature $T_{\rm rec}(A)$ at which abundances of the ionized nuclei $A$ and the bound state $A_X$ are equal.  The $T_{\rm rec}(A)$ value is determined as a function of the abundance of $X^-$, $Y_X$, using Eqs. (\ref{eq66}) and (\ref{eq67}).  For example, the recombination temperature of $^7$Be for the case of $m_X=1$ GeV and $Y_X=0.05$ is $T_{\rm rec}(^7{\rm Be})=8.49$ keV (corresponding to $T_9=0.0985$).  Since the recombination proceeds at temperatures lower than in case of larger $m_X$, the number density of protons at the recombination is smaller.  As a result, the rate for $^7$Be$_X$ to experience  radiative proton capture in the temperature range of $T_9\lesssim 1$ is small.  The reduction of the $^7$Be$_X$ abundance through the proton capture is, therefore, less efficient than in the cases with  $m_X=100$ GeV and 1000 GeV.   However,  it is still much more efficient than the case with  $m_X=10$ GeV because of the smaller resonant energy in the resonant reaction $^7$Be$_X$($p$, $\gamma$)$^8$B$_X$ (see Sec. \ref{sec4}).

\subsubsection{Constraints on the $X^-$ Particle}
\begin{figure}
\begin{center}
\includegraphics[width=0.45\textwidth]{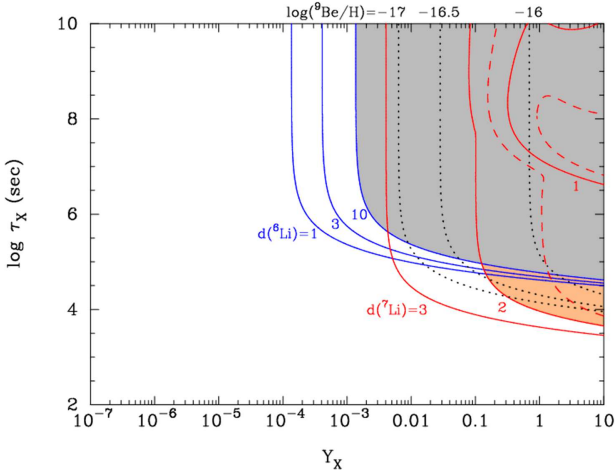}
\end{center}
\caption{Contours of constant lithium abundances relative to the observed values in MPSs, i.e., $d$($^6$Li)=$^6$Li$^{\rm Cal}$/$^6$Li$^{\rm Obs}$ (blue lines) and $d$($^7$Li)=$^7$Li$^{\rm Cal}$/$^7$Li$^{\rm Obs}$ (red lines) for the case of $m_X=1$ GeV.  The adopted observational constraint on the $^7$Li abundance is the central  value of log($^7$Li/H)$=-12+(2.199\pm 0.086)$ derived in the  3D NLTE model of \citet{Sbordone:2010zi}.  The  $^6$Li constraint is taken from the $2 \sigma$  upper limit for the G64-12 \citep[NLTE model with 5 parameters;][]{Lind:2013iza}, of $^6$Li/H=$(0.9\pm 4.3)\times 10^{-12}$.  Dashed lines around the line of $d$($^{7}$Li)=1 correspond to the $2 \sigma$  uncertainty in the observational constraint.  The gray region located to the  right from the contours of $d$($^6$Li)=10 or the 2 sigma lower limit, $d$($^7$Li)=0.67, is  excluded by the overproduction of $^6$Li and underproduction of $^7$Li, respectively.  The orange region is 
the interesting parameter region in which a significant reduction  in $^7$Li is realized without an overproduction of $^6$Li.  Dotted lines are contours of the abundance ratio of $^9$Be/H predicted when the unknown rate for  the reaction $^7$Li$_X$($d$, $X^-$)$^9$Be is adopted  as described in the text.  \label{fig23}}
\end{figure}

Figure \ref{fig23} shows contours of calculated final lithium abundances for the case of $m_X=1$ GeV.  These are normalized to the values observed in MPSs, i.e., $d$($^6$Li)=$^6$Li$^{\rm Cal}$/$^6$Li$^{\rm Obs}$ (blue lines) and $d$($^7$Li)=$^7$Li$^{\rm Cal}$/$^7$Li$^{\rm Obs}$ (red lines).  The final $^7$Li abundance is a sum of the abundances of $^7$Li and $^7$Be produced in BBN.  This is because  $^7$Be is converted to $^7$Li via the electron capture at a later epoch.  The dashed lines around the line of $d$($^{7}$Li)=1 correspond to the 2$\sigma$ uncertainty in the observational constraint.  The gray region  located to the  right of the contours for  $d$($^6$Li)=10 and/or the 2 sigma lower limit, $d$($^7$Li)=0.67, are excluded by the overproduction of $^6$Li and underproduction of $^7$Li, respectively.  The orange region is the interesting parameter region in which a significant $^7$Li reduction occurs without inducing an overproduction of $^6$Li.  Dotted lines are contours of the calculated abundance ratio of $^9$Be/H assuming a rate for the reaction $^7$Li$_X$($d$, $X^-$)$^9$Be as described above in Sec. \ref{sec6}.


If the $X^-$  lifetime is long enough, larger $X^-$ abundances lead to more production of $^6$Li.  This is because the production rate of $^6$Li through the reaction $^4$He$_X$($d$, $X^-$)$^6$Li is proportional to the abundance of $^4$He$_X$ which is proportional to the $X^-$ abundance as long as $n_X\leq n_\alpha$.  On the other hand, for large $Y_X$ values  the amount of $^7$Be destruction is not  proportional to the $X^-$ abundance.  The reason is that the  amount of $^7$Be destruction roughly depends upon the recombination temperature and the conversion rate of $^7$Be$_X$ through the reaction $^7$Be$_X$($p$, $\gamma$)$^8$B$_X$.  However, the recombination temperature is almost independent of $Y_X$.  Therefore  the conversion rate is independent of $Y_X$ although it is dependent on $n_p$.  The destruction is, therefore,  not efficient even if the $Y_X$ values were very high.

The excluded gray region corresponds to $Y_X \gtrsim 10^{-3}$ in the limit of a long $X^-$-particle life time $\tau_X \gtrsim 10^8$ s.  This region is determined from the overproduction of $^6$Li.  The orange region corresponding to  a solution to the $^{7}$Li problem is located at $Y_X \gtrsim 0.1$ and $\tau_X\sim 5\times 10^3$--$10^5$ s.  Within this region, the primordial $^9$Be abundance is predicted to be $^9$Be/H$\lesssim 3\times 10^{-16}$.  This is much larger than the SBBN value of $9.60\times 10^{-19}$  \citep{Coc:2011az}.  Since the abundances of D, $^3$He, and $^4$He are not significantly altered, the adopted constraints on their primordial abundances are satisfied in this region.

\begin{figure}
\begin{center}
\includegraphics[width=0.45\textwidth]{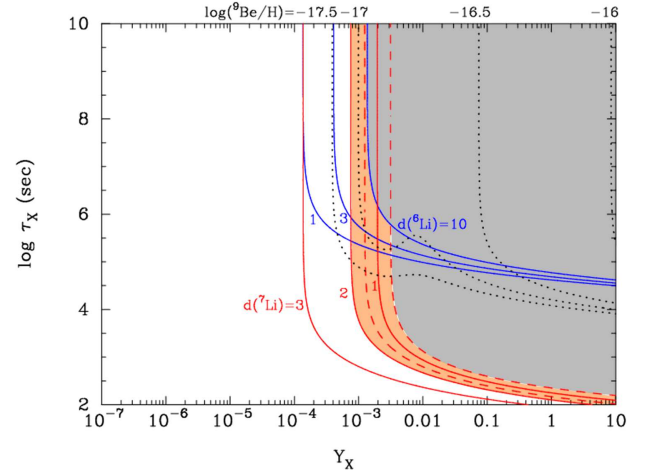}
\end{center}
\caption{Same as in Fig. \ref{fig23},  but the charged-current decay of $^7$Be$_X\rightarrow ^7$Li+$X^0$ is also included.  \label{fig24}}
\end{figure}

Figure \ref{fig24} shows the same contours for calculated abundances of $^{6,7}$Li and $^9$Be as in Fig. \ref{fig23}.  In this case  the instantaneous charged-current decay of $^7$Be$_X\rightarrow ^7$Li+$X^0$ \citep{Jittoh:2007fr,Jittoh:2008eq,Jittoh:2010wh,Bird:2007ge} is also taken into account.  In this  case  the $X^-$ particle interacts not only via its charge but also a  weak interaction  \citep{Jittoh:2007fr,Jittoh:2008eq,Jittoh:2010wh}.  $^7$Be$_X$ can then be converted to $^7$Li plus a neutral particle $X^0$.  Other $X$-nuclei may also  decay depending upon the mass of the $X^0$.  The prohibition of  $^6$Li overproduction, however, limits the length of the  lifetime of the $X^-$ as seen in Fig. \ref{fig23}.  Effects of the weak decay catalyzed by the $X^-$ then appear through the conversion of $X$-nuclei produced just  before the epoch of $^6$Li production.  


Above the recombination temperature for $^4$He$_X$ at which $^6$Li production also proceeds, $^6$Li$_X$, $^7$Li$_X$, and $^7$Be$_X$ can all be produced with large fractions of  bound states (see Fig. \ref{fig22}).  Among these three $X$-nuclei, $^7$Be$_X$ is the most abundant, and its abundance evolution affects the parameter region for a solution to the Li problem.  Therefore, for simplicity we only consider the decay of $^7$Be$_X$  here.

The contours for the $^6$Li abundance is similar to those in Fig. \ref{fig23}.  On the other hand, the $^7$Li abundance is much different from that in Fig. \ref{fig23} because of the different process for $^7$Be destruction. Including the charged-current decay of $^7$Be$_X$, the destruction rate of $^7$Be in this model is the same as the recombination rate of $^7$Be itself.  In the model without the decay, the destruction rate requires  that $^7$Be$_X$ nuclei produced via the recombination then  experience a proton capture reaction without being re-ionized to a $^7$Be+$X^-$ free state.
The different processes of $^7$Be$_X$ destruction, therefore, cause a difference in the efficiency for the final $^7$Li reduction.  In this model with the decay, the amount of $^7$Be destruction roughly scales as $Y_X$ unlike the model without the decay.

The excluded region is wider than in Fig. \ref{fig23}.  This region is determined from the $^7$Li underproduction.  This region also involves lower values of $\tau_X$ than in Fig. \ref{fig23}.  The solution to the $^{7}$Li problem is at $Y_X \gtrsim 8\times 10^{-4}$ and $\tau_X\gtrsim 10^2$ s (orange region).  In this region, the $^9$Be abundance is calculated to be $^9$Be/H$\lesssim 3\times 10^{-17}$.

\subsection{$m_X=10$ GeV}\label{sec9.3}
\subsubsection{Nucleosynthesis}
\begin{figure}
\begin{center}
\includegraphics[width=0.45\textwidth]{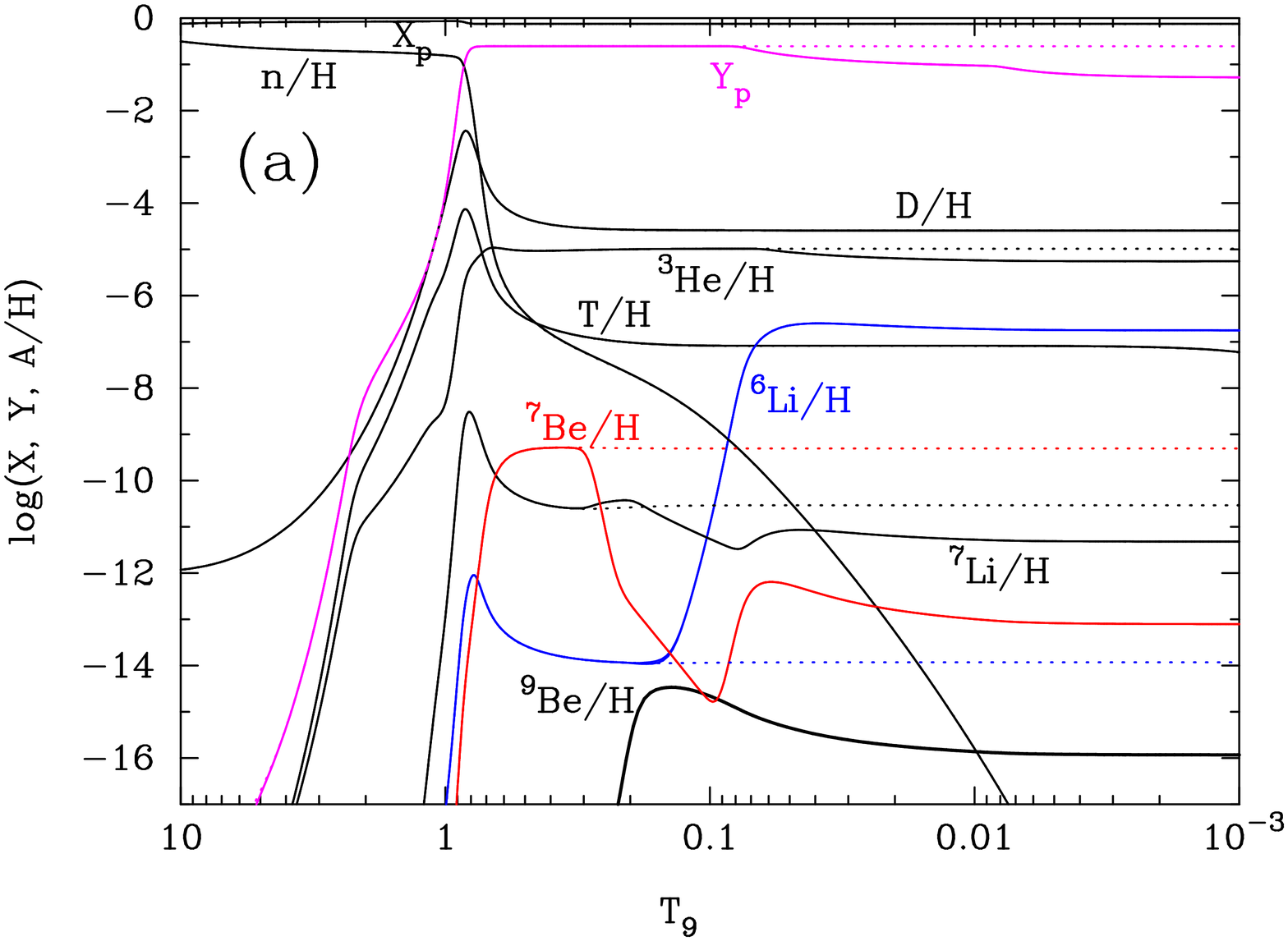}\\
\includegraphics[width=0.45\textwidth]{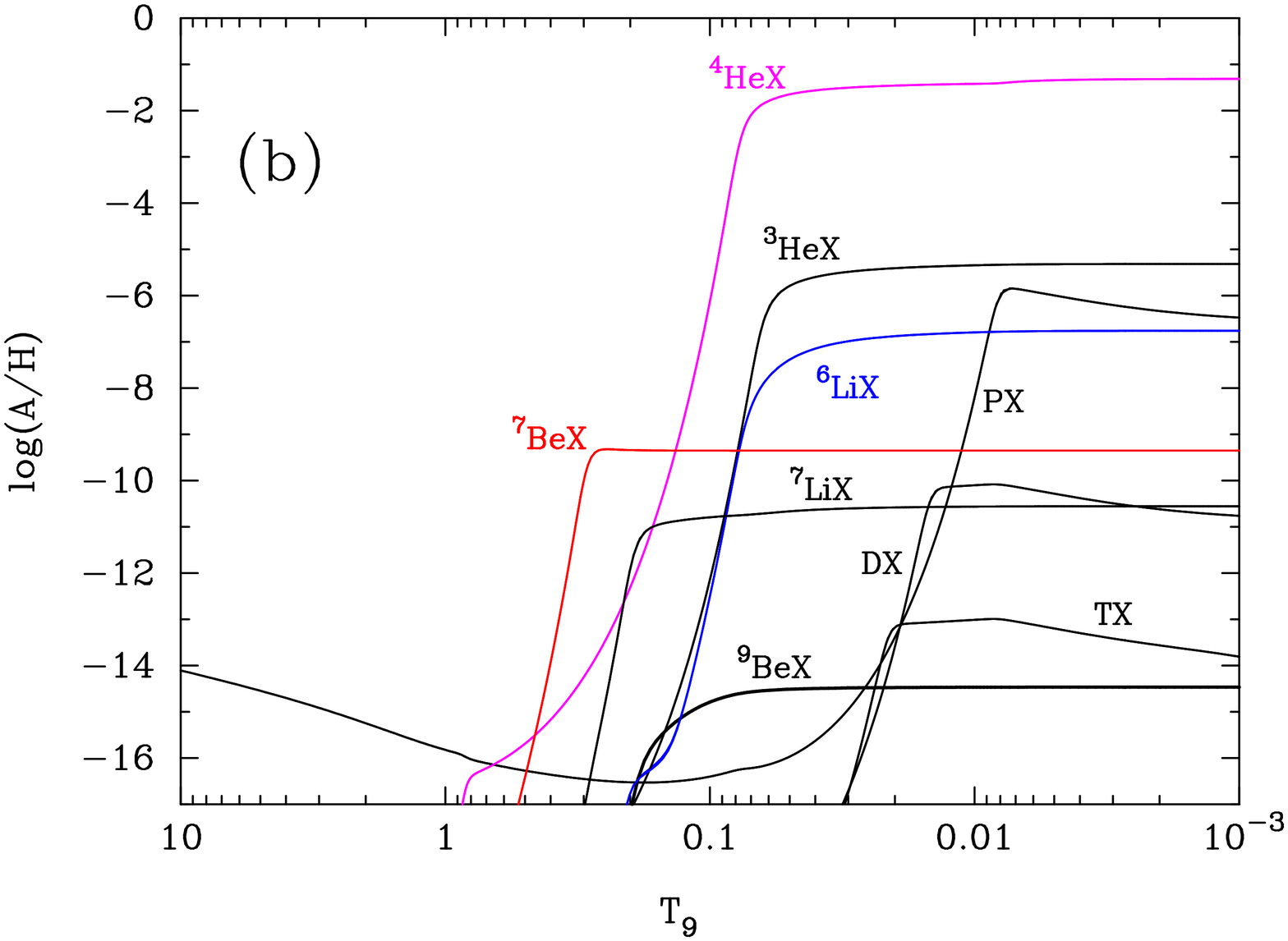}
\end{center}
\caption{Same as in Fig. \ref{fig22}, but  for the case of $m_X=10$ GeV.  \label{fig25}}
\end{figure}

Figure \ref{fig25} shows the same abundances as a function of $T_9$ as in Fig. \ref{fig22}, but  for the case of $m_X=10$ GeV and without the decay of $^7$Be$_X$.


The $^7$Be nuclide recombines with $X^-$ at $T_{\rm rec}(^7{\rm Be})=25.1$ keV ($T_9=0.291$).  Although this temperature is higher than in the case of $m_X=1$ GeV, the resonant peak in the $^7$Be$_X$($p$, $\gamma$)$^8$B$_X$ reaction is higher.  The efficiency for  $^7$Be$_X$ destruction is then smaller than that for $m_X=1$ GeV.  During a  late epoch, the $^7$Be abundance increases mainly through the reaction $^4$He$_X$($^3$He, $X^-$)$^7$Be at $T_9\lesssim 0.1$.  In the same epoch, the $^7$Be abundance increases also through the reaction $^6$Li($p$, $\gamma$)$^7$Be.  However, in this case the production rate is much smaller than that via $^4$He$_X$($^3$He, $X^-$)$^7$Be.  It is thus found that $^7$Be is produced by the  $^6$Li($p$, $\gamma$)$^7$Be reaction if the abundance of $^6$Li during BBN is much larger than in SBBN as realized in this model by including the $X^-$ particle.

$^6$Li is produced through the reaction $^4$He$_X$($d$, $X^-$)$^6$Li at $T_9\sim 0.06$.  $^6$Li$_X$ is then produced through the recombination $^6$Li($X^-$, $\gamma$)$^6$Li$_X$.

At first the $^7$Li abundance increases through the two  reaction pathways of $^7$Be$_X$($n$, $p$$^7$Li)$X^-$ and $^7$Be$_X$($n$, $p$)$^7$Li$_X$($\gamma$, $X^-$)$^7$Li at $T_9\sim 0.3-0.2$.   This is seen as a bump in the abundance curve.  The existence of this bump depends upon the reaction rates of $^7$Be$_X$($n$, $p$$^7$Li)$X^-$ and $^7$Be$_X$($n$, $p$)$^7$Li$_X$, which are assumed to be the same as that of $^7$Be($n$, $p$)$^7$Li in this paper.  This possible bump appears during the epoch when the recombination of $^7$Be has started but that of $^7$Li has not.   Then, the $^7$Li abundance decreases through the recombination  reaction $^7$Li($X^-$, $\gamma$)$^7$Li$_X$ at $T_9\sim 0.2-0.1$.  The proton capture reaction $^7$Li$_X$($p$, 2$\alpha$)$X^-$ also partly destroys the $^7$Li nuclei  produced via the recombination.  Finally, $^7$Li is produced through the reaction $^4$He$_X$($t$, $X^-$)$^7$Li at $T_9\sim 0.07-0.06$.

$^9$Be is produced through the reaction $^7$Li$_X$($d$, $X^-$)$^9$Be at $T_9\sim 0.2-0.1$.  The recombination of $^9$Be increases the abundance of $^9$Be$_X $ at $T_9\sim 0.2-0.1$.  When the  proton-capture reaction $^9$Be$_X$($p$, $^6$Li)$^4$He$_X$ is operative at $T_9\gtrsim 0.07$,
 it decreases the abundance of $^9$Be$_X$.
  
\subsubsection{Constraints on the $X^-$ Particle}
\begin{figure}
\begin{center}
\includegraphics[width=0.45\textwidth]{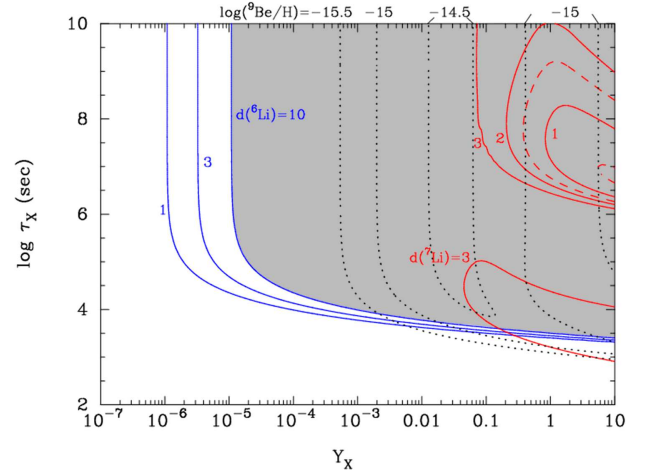}
\end{center}
\caption{Same as in Fig. \ref{fig23}, but  for the case of $m_X=10$ GeV.  Note that there is no interesting parameter region in which a $^7$Li reduction occurs without an  overproduction of $^6$Li.  \label{fig26}}
\end{figure}

Figure \ref{fig26} shows the same contours for calculated abundances of $^{6,7}$Li and $^9$Be as in Fig. \ref{fig23} without the decay of $^7$Be$_X$, but  for $m_X=10$ GeV.  The excluded gray region is larger than that in Fig. \ref{fig23} because of the enhanced production rate of $^6$Li.  In addition, there is no parameter region for the solution to the $^{7}$Li problem because of the smaller destruction rate for $^7$Be.


\begin{figure}
\begin{center}
\includegraphics[width=0.45\textwidth]{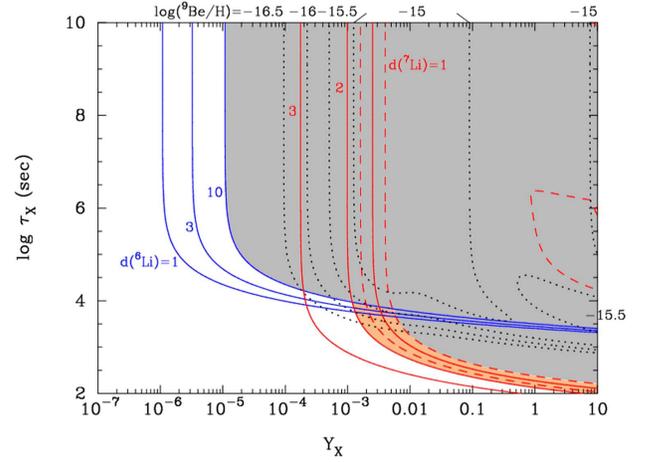}
\end{center}
\caption{Same as in Fig. \ref{fig23}, but  for the case of $m_X=10$ GeV when the charged-current decay of $^7$Be$_X\rightarrow ^7$Li+$X^0$ is included.  \label{fig27}}
\end{figure}

Figure \ref{fig27} shows the same contours for calculated abundances of $^{6,7}$Li and $^9$Be as in Fig. \ref{fig23}, but  for $m_X=10$ GeV and with the decay of $^7$Be$_X$.  The contours for the $^6$Li abundance are similar to those in Fig. \ref{fig26}.  The $^7$Li abundance is different from that in Fig. \ref{fig25} for the same reason described above for Fig. \ref{fig24}.  The excluded region is determined from the combination of  $^7$Li underproduction and  $^6$Li overproduction.  It  is wider than in Fig. \ref{fig26}.  The region for the $^{7}$Li problem is at $Y_X \gtrsim 10^{-3}$ and $\tau_X \sim 10^2$--$10^4$ s.  The $^9$Be abundance in this region is $^9$Be/H$\lesssim 10^{-15}$.


\subsection{$m_X=100$ GeV}\label{sec9.4}
\subsubsection{Nucleosynthesis}
\begin{figure}
\begin{center}
\includegraphics[width=0.45\textwidth]{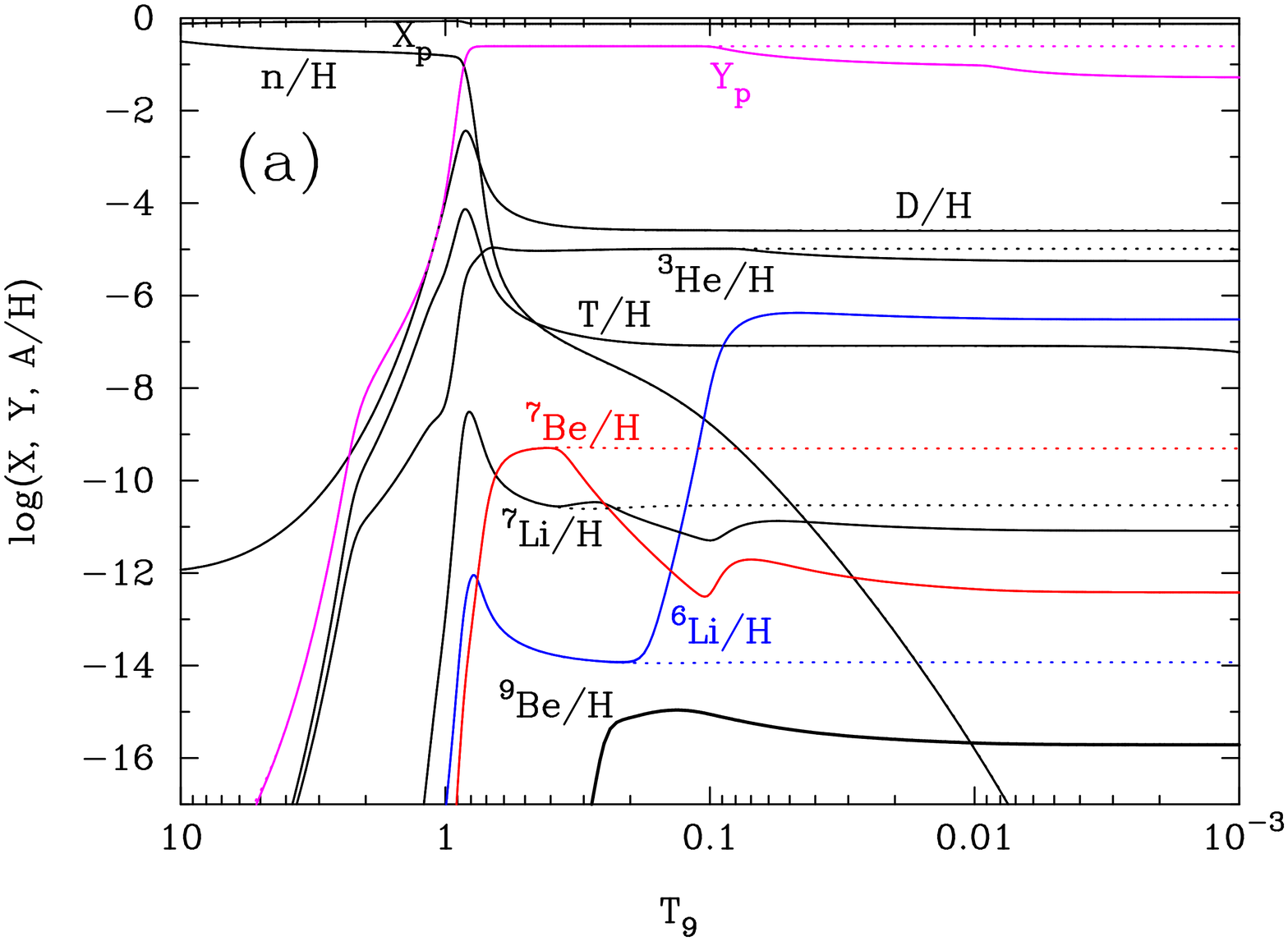}\\
\includegraphics[width=0.45\textwidth]{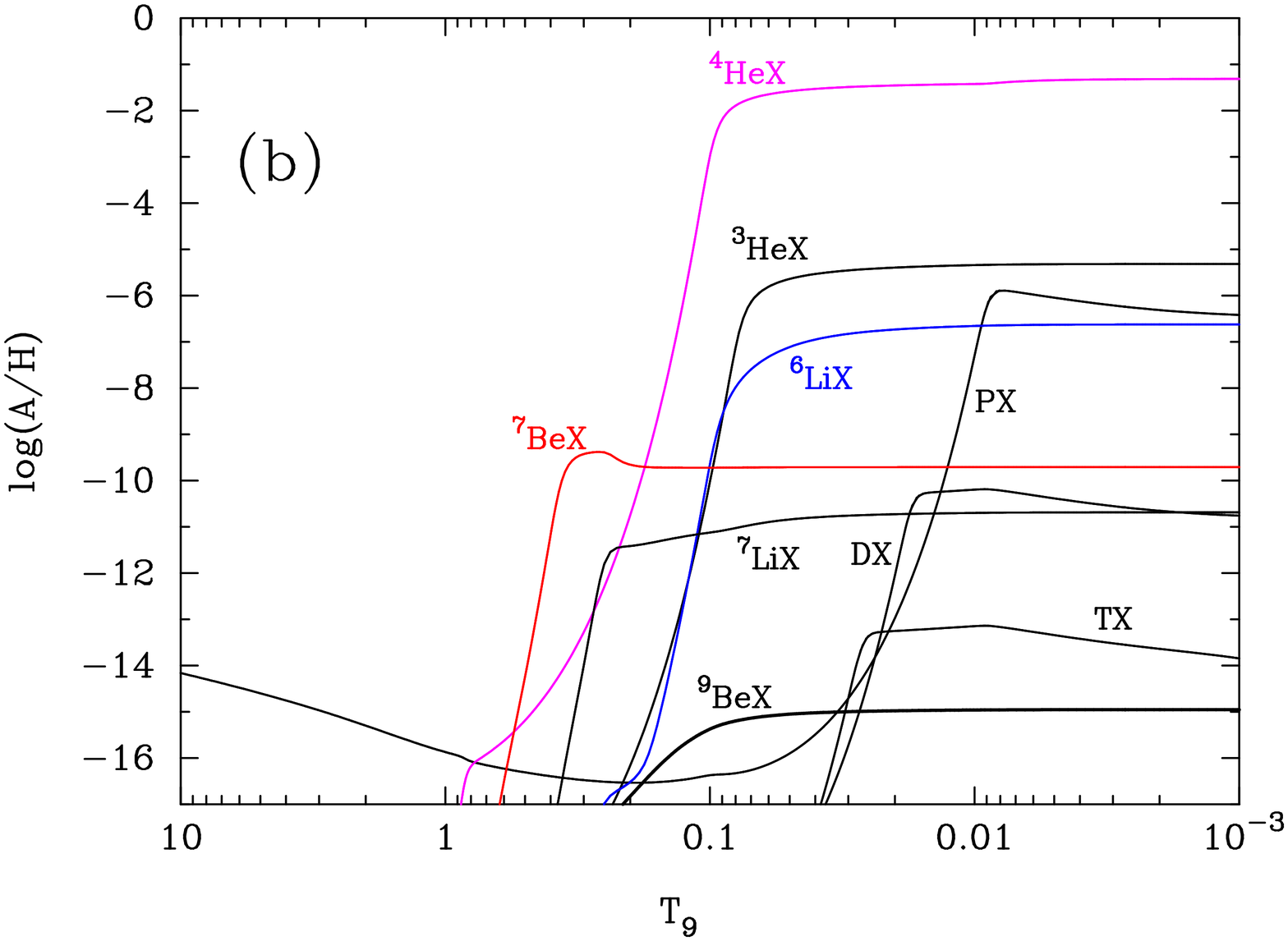}
\end{center}
\caption{Same as in Fig. \ref{fig22}, but  for the case of $m_X=100$ GeV.  \label{fig28}}
\end{figure}

Figure \ref{fig28} shows the same abundances as a function of $T_9$ as in Fig. \ref{fig22} without the decay of $^7$Be$_X$, but  for the case of $m_X=100$ GeV .


The $^7$Be nuclide recombines with $X^-$ at $T_{\rm rec}(^7{\rm Be})=30.9$ keV ($T_9=0.359$).  The efficiency of $^7$Be$_X$ destruction through the reaction
$^7$Be$_X$($p$, $\gamma$)$^8$B$_X$ at $T_9=0.3-0.2$ is larger than in the cases with  $m_X=1$ GeV and 10 GeV as seen in this figure.  This high efficiency is because of the higher recombination temperature and the relatively smaller peak of the resonant reaction $^7$Be$_X$($p$, $\gamma$)$^8$B$_X$.  During  later epoch, the $^7$Be abundance increases mainly through the reaction $^4$He$_X$($^3$He, $X^-$)$^7$Be and somewhat less  through the reaction $^6$Li($p$, $\gamma$)$^7$Be.  The production rate via the former reaction is somewhat larger than that via the latter.

$^6$Li is produced through the reaction $^4$He$_X$($d$, $X^-$)$^6$Li at $T_9\sim 0.1$.  The abundance of $^6$Li$_X$ increases through the recombination reaction 
$^6$Li($X^-$, $\gamma$)$^6$Li$_X$.  Some  of the $^6$Li$_X$ nuclei are then destroyed through  proton capture via the $^6$Li$_X$($p$, $^3$He$\alpha$)$X^-$ 
reaction in the temperature range of $T_9\gtrsim 0.05$.

In the interval of recombination temperatures for $^7$Be and $^7$Li, i.e., $T_9\sim 0.3$--$0.2$, the $^7$Li abundance at first increases  through the  neutron-induced   reactions on  $^7$Be$_X$ as in the case of $m_X=10$ GeV.   Then, the $^7$Li abundance decreases through recombination with $X^-$ at $T_9\lesssim 0.2$.  At $T_9\gtrsim 0.05$, the proton capture reaction $^7$Li$_X$($p$, 2$\alpha$)$X^-$ partly destroy $^7$Li nuclei produced via the recombination.  Finally, $^7$Li is produced through the reaction $^4$He$_X$($t$, $X^-$)$^7$Li at $T_9\lesssim 0.1$.

$^9$Be is produced through the reaction $^7$Li$_X$($d$, $X^-$)$^9$Be at $T_9\sim 0.3$--$0.1$.  The recombination $^9$Be($X^-$, $\gamma$)$^9$Be$_X$ reaction enhances the abundance of $^9$Be$_X $at $T_9\sim 0.2$--$0.1$.  The proton capture reaction $^9$Be$_X$($p$, $^6$Li)$^4$He$_X$ then decreases the abundance of $^9$Be$_X$ at $T_9\gtrsim 0.1$.

\subsubsection{Constraints on the $X^-$ Particle}
\begin{figure}
\begin{center}
\includegraphics[width=0.45\textwidth]{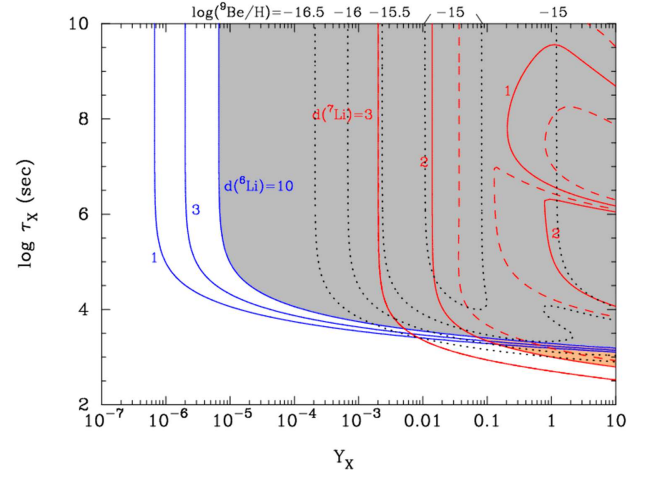}
\end{center}
\caption{Same as in Fig. \ref{fig23}, but  for the case of $m_X=100$ GeV.  \label{fig29}}
\end{figure}

Figure \ref{fig29} shows the same contours for calculated abundances of $^{6,7}$Li and $^9$Be as in Fig. \ref{fig23} without the decay of $^7$Be$_X$, but  for $m_X=100$ GeV.  The excluded gray region is even  larger than that for $m_X=10$ GeV because of the enhanced production rate of $^6$Li.  The parameter region for the solution to the $^{7}$Li problem is at $Y_X \gtrsim 0.07$ and $\tau_X \sim (0.6$--$3) \times 10^3$ s.  The $^9$Be abundance in this region is $^9$Be/H$\lesssim 3\times 10^{-16}$.


\begin{figure}
\begin{center}
\includegraphics[width=0.45\textwidth]{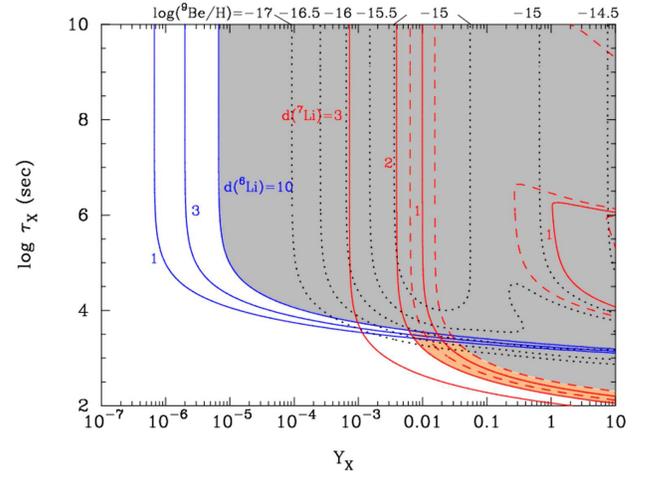}
\end{center}
\caption{Same as in Fig. \ref{fig23}, but  for the case of $m_X=100$ GeV and  the charged-current decay of $^7$Be$_X\rightarrow ^7$Li+$X^0$ is included.  \label{fig30}}
\end{figure}

Figure \ref{fig30} shows the same contours for calculated abundances of $^{6,7}$Li and $^9$Be as in Fig. \ref{fig23}, but  for $m_X=100$ GeV and with the decay of $^7$Be$_X$.  The excluded region is determined from the combination of the $^7$Li underproduction and the $^6$Li overproduction.  The region for the $^{7}$Li problem is at $Y_X \gtrsim 6\times 10^{-3}$ and $\tau_X \sim 10^2$--$4\times 10^3$ s.  In this region, the $^9$Be abundance is $^9$Be/H$\lesssim 3\times 10^{-16}$.


\subsection{$m_X=1000$ GeV}\label{sec9.5}
\subsubsection{Nucleosynthesis}\label{sec9.5.1}
\begin{figure}
\begin{center}
\includegraphics[width=0.45\textwidth]{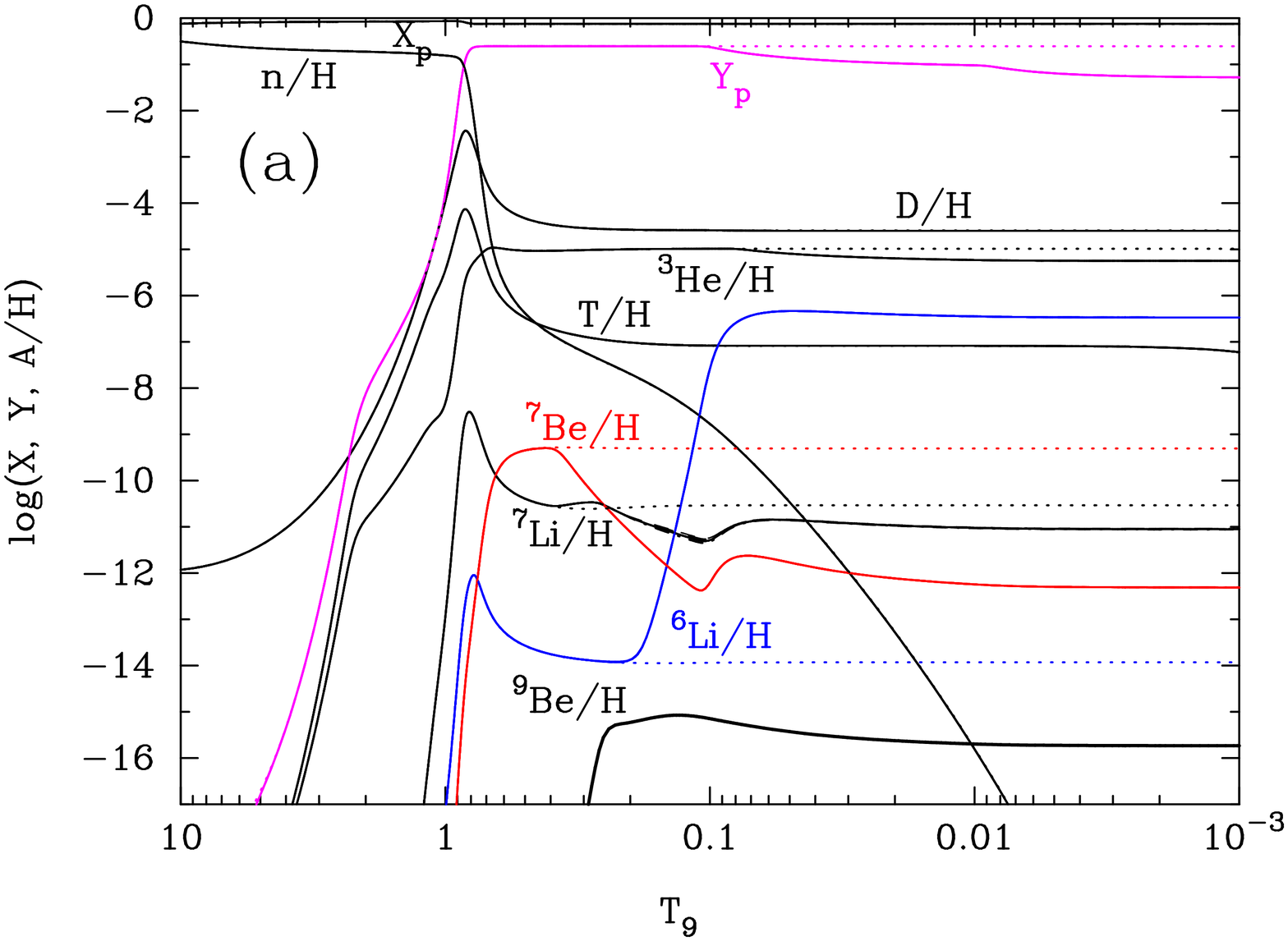}\\
\includegraphics[width=0.45\textwidth]{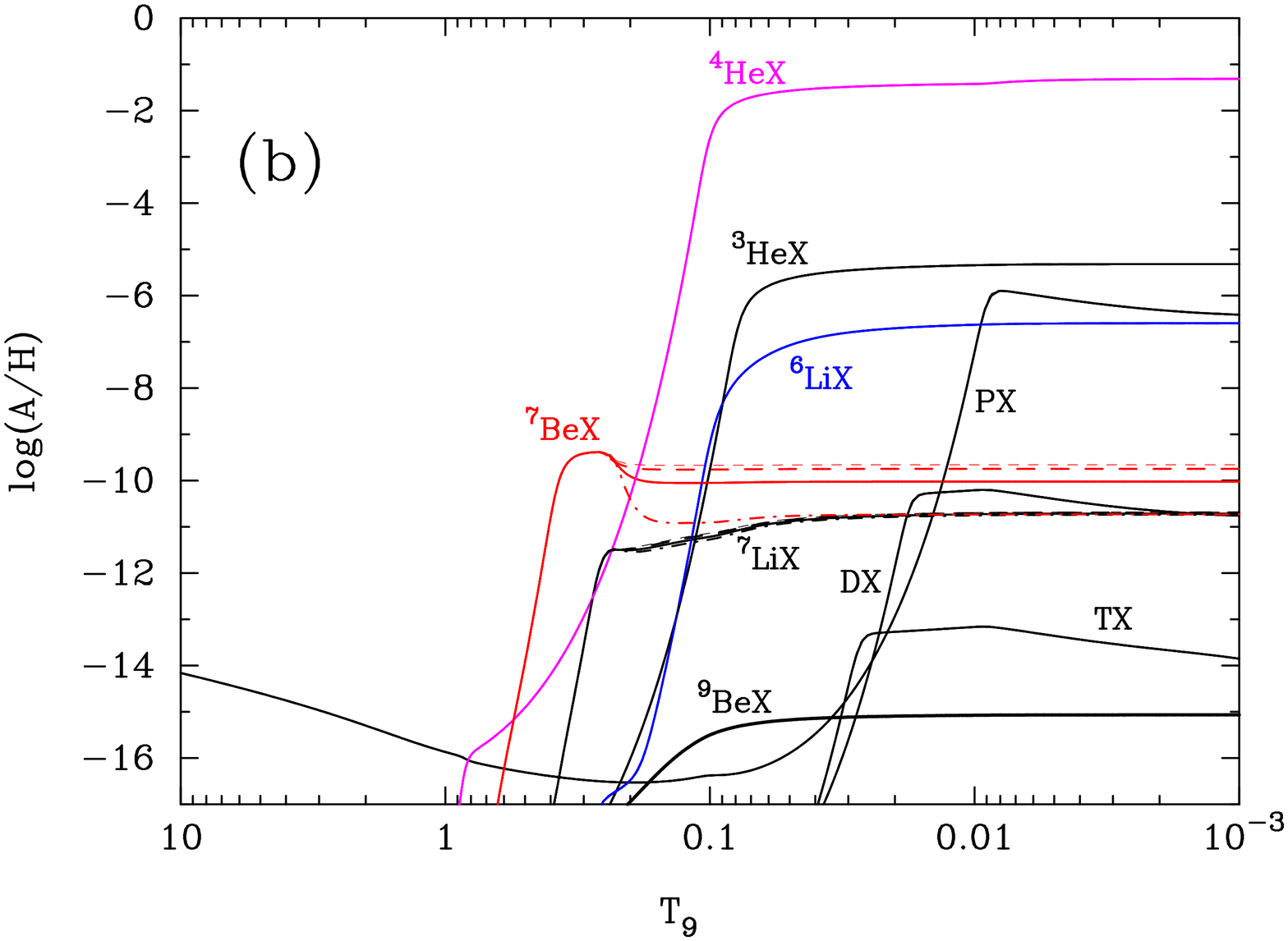}
\end{center}
\caption{Same as in Fig. \ref{fig22}, but  for the case of $m_X=1000$ GeV.  \label{fig31}}
\end{figure}

Figure \ref{fig31} shows the same abundances as a function of $T_9$ as in Fig. \ref{fig22} without the decay of $^7$Be$_X$, but  for the case of $m_X=1000$ GeV.  This result is very similar to that for $m_X=100$ GeV except for  the abundance of $^7$Be$_X$.


The recombination temperature of $^7$Be is the same as in the case of $m_X=100$ GeV, i.e., $T_{\rm rec}(^7{\rm Be})=30.9$ keV ($T_9=0.359$).  The efficiency of $^7$Be$_X$ destruction through the reaction $^7$Be$_X$($p$, $\gamma$)$^8$B$_X$ is slightly larger than that for $m_X=100$ GeV mainly because of the smaller resonant height of the reaction $^7$Be$_X$($p$, $\gamma$)$^8$B$_X$.

BBN calculations for $m_X=1000$ GeV are performed for four cases of reaction rates for $^7$Be$_X$($p$, $\gamma$)$^8$B$_X$ and $^8$Be$_X$($p$, $\gamma$)$^9$B$_X$.  Three cases correspond to Gaussian (thick dashed lines), WS40 (solid lines), and well (dot-dashed lines) models for nuclear charge distributions studied in this paper (Sec. \ref{sec4}), while one case corresponds to the previous calculation \citep{Kusakabe:2007fv} in which the reaction rate for $^7$Be$_X$($p$, $\gamma$)$^8$B$_X$ derived with quantum many-body model \citep{Kamimura:2008fx} was adopted.  It is found that amounts of $^7$Be$_X$ destruction vary significantly when the nuclear charge distributions are changed.  The result for our Gaussian charge distribution model (thick dashed line) is close to that for the quantum many-body model in which charge distribution of cluster components has been assumed to be Gaussian shape also. The differences in the curves for $^7$Be$_X$ thus indicate the effect of uncertainties in charge density, which are estimated from measurements of RMS charge radii.

\subsubsection{Constraints on the $X^-$ Particle}\label{sec9.5.2}
\begin{figure}
\begin{center}
\includegraphics[width=0.45\textwidth]{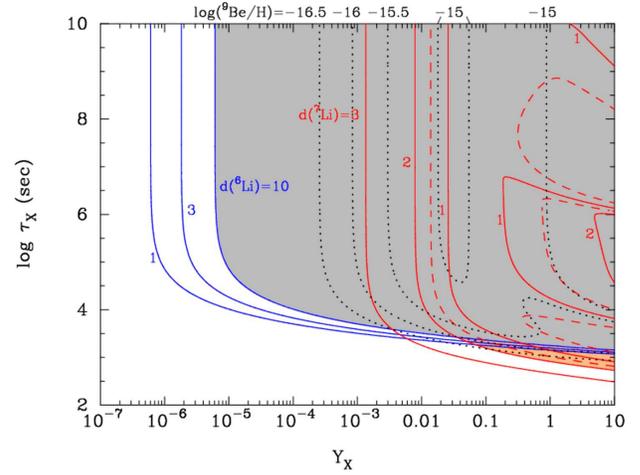}
\end{center}
\caption{Same as in Fig. \ref{fig23}, but for the case of $m_X=1000$ GeV.  \label{fig32}}
\end{figure}

Figure \ref{fig32} shows the same contours for calculated abundances of $^{6,7}$Li and $^9$Be as in Fig. \ref{fig23}  without the decay of $^7$Be$_X$, but  for $m_X=1000$ GeV.  The parameter region for the solution to the $^{7}$Li problem is at $Y_X \gtrsim 0.04$ and $\tau_X \sim (0.6$--$3) \times 10^3$ s.  The $^9$Be abundance in this region is $^9$Be/H$\lesssim 3\times 10^{-16}$.


\begin{figure}
\begin{center}
\includegraphics[width=0.45\textwidth]{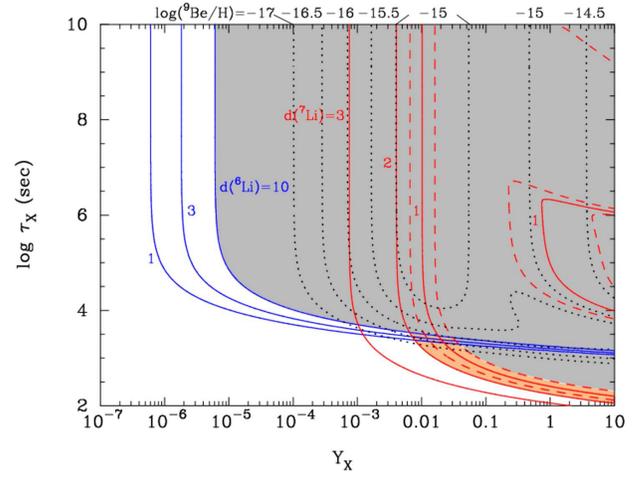}
\end{center}
\caption{Same as in Fig. \ref{fig23}, but  for the case of $m_X=1000$ GeV and the charged-current decay of $^7$Be$_X\rightarrow ^7$Li+$X^0$ is also included.  \label{fig33}}
\end{figure}

Figure \ref{fig33} shows the same contours for calculated abundances of $^{6,7}$Li and $^9$Be as in Fig. \ref{fig23}, but  for $m_X=1000$ GeV and with the decay of $^7$Be$_X$ also included.  The region for the solution of the $^{7}$Li problem does not significantly differ from that for $m_X=100$ GeV, and is at $Y_X \gtrsim 6\times 10^{-3}$ and $\tau_X \sim 10^2$--$4\times 10^3$ s.  The $^9$Be abundance in this region is $^9$Be/H$\lesssim 3\times 10^{-16}$.


\subsection{Comparison with Previous Constraints}
Previous constraints are all derived in the limit of $m_X \rightarrow \infty$.  It is, therefore, appropriate to compare them to the new constraints for the largest mass case, i.e., $m_X=1000$ GeV.  We compare rates of nuclear recombination with $X^-$ first, and parameter regions for the $^7$Li reduction second.

\subsubsection{Recombination Rates}
\begin{figure}
\begin{center}
\includegraphics[width=0.45\textwidth]{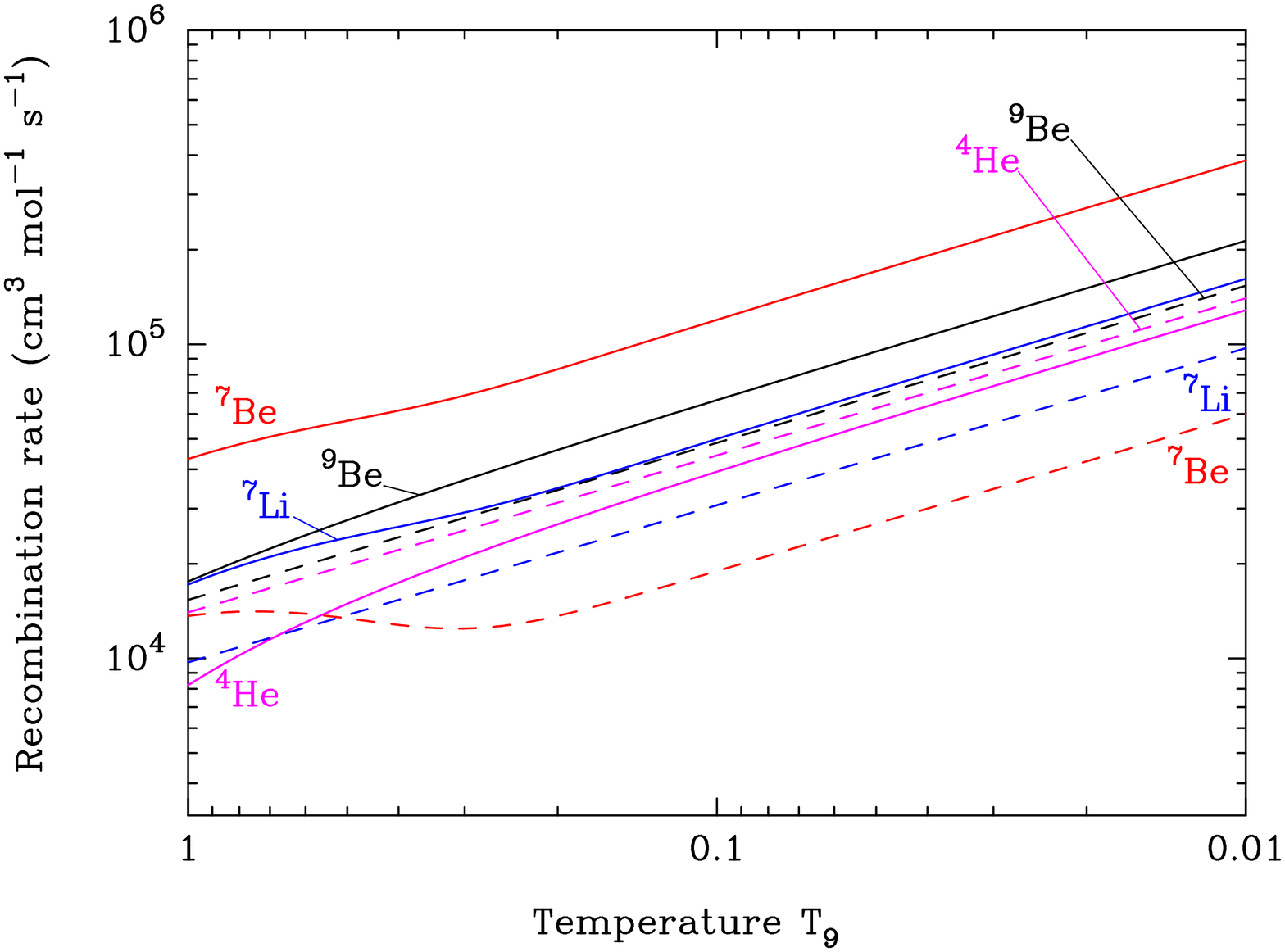}
\end{center}
\caption{Rates for recombination of $^7$Be, $^7$Li, $^9$Be, and $^4$He with $X^-$ in the case of $m_X=1000$ GeV as a function of temperature.  Solid lines show the recombination rates derived in this paper, while dashed lines show the rates adopted in the previous studies (see text).  \label{fig34}}
\end{figure}

Figure \ref{fig34} shows rates for the recombination of $^7$Be, $^7$Li, $^9$Be, and $^4$He with $X^-$ as a function of temperature $T_9$ in the case of $m_X=1000$ GeV.  Solid lines correspond to the recombination rates derived in this paper: Eqs. (\ref{eq23}) and (\ref{eq31}) for $^7$Be, Eqs. (\ref{eq46}) and (\ref{eq53}) for $^7$Li, Eq. (\ref{eq57}) for $^9$Be, and Eq. (\ref{eq61}) for $^4$He.  Dashed lines, on the other hand, correspond to the rates adopted in the previous studies \citep[e.g.][]{Kusakabe:2007fv,Kusakabe:2010cb}: Eq. (2.9) of \citet{Bird:2007ge} for $^7$Be, and Eq. (\ref{eq63}) with $m_X\rightarrow \infty$ for other nuclides.  The $^7$Be rate in the present study is much larger than the previous rate.  The present rates for $^7$Li and $^9$Be are also significantly larger than the previous rates.  The present $^4$He rate is, on the other hand, not significantly different for temperatures $T_9 \lesssim 0.1$ where the recombination effectively proceeds.  The new precise rates for $^7$Be, $^7$Li, and $^9$Be are larger than the previous rates, while that for $^4$He is smaller than the previous rate.


\subsubsection{Case without $^7$Be$_X$ decay}
When the charged-current decay of $^7$Be$_X\rightarrow ^7$Li+$X^0$ is absent, the following reactions predominantly determine the abundance evolution of $^7$Be, $^7$Li, and $^6$Li:  (1) $^7$Be($X^-$, $\gamma$)$^7$Be$_X$,  (2) $^7$Be$_X$($p$, $\gamma$)$^8$B$_X$, (3) $^7$Li($X^-$, $\gamma$)$^7$Li$_X$, (4) $^7$Li$_X$($p$, 2$\alpha$)$X^-$, (5) $^4$He($X^-$, $\gamma$)$^4$He$_X$, and (6) $^4$He$_X$($d$, $X^-$)$^6$Li.  We updated the recombination rates for (1), (3), (5), and also the resonant proton capture rate for (2).  The rates for (4) and (6) are taken from the same reference \citep{Kamimura:2008fx} as adopted in the previous studies.

In the theoretical calculation of $^7$Li/H, the most important reactions are (1) and (2).  The new recombination rate for (1) is about 5 times larger than the previous rate (Fig. \ref{fig34}) at the recombination temperature of $T_9 \sim 0.36$ (Sec. \ref{sec9.5.1}).  On the other hand, the adopted destruction rate for (2) for the WS40 model is about 2.7 times larger than the previous rate (Fig. \ref{fig5}) at the destruction temperature $T_9\sim 0.3$ (Fig. \ref{fig31}).  Because of increases in the two reaction rates, the effective rate for $^7$Be destruction (or reduction in the final $^7$Li/H value) becomes higher than the previous one.  

Second, the observational constraint on the abundance $^7$Li/H has been updated from $^7$Li/H$=(1.23^{+0.68}_{-0.32}) \times 10^{-10}$ \citep[95 \% confidence limits;][]{Ryan:1999vr} to log($^7$Li/H)$=-12+(2.199\pm 0.086)$ \citep{Sbordone:2010zi}.  As a result, curves of $d$($^7$Li)=2 correspond to different abundances: $^7$Li/H$=3.16 \times 10^{-10}$ in this study, while $^7$Li/H$=2.46 \times 10^{-10}$ in the previous studies.  In this study, therefore, we need less destruction of $^7$Be to realize $d$($^7$Li)=2.  

Both the theoretical and observational improvements indicate that it is easier to reduce the primordial $^7$Li abundance to the level of $d$($^7$Li)=2.  The contours of $d(^7$Li$)$, therefore, move left by a factor of about 20 in the parameter plane of Fig. \ref{fig32}.  In the parameter region around $d$($^7$Li)=2, a partial destruction of $^7$Be is realized.  In this region, the destruction rate of $^7$Be is proportional to the product of reaction rates (1) and (2).  The factor $\sim$20 can be explained by this proportionality and the difference in the destruction fraction of $^7$Be required from observations.

The $^6$Li/H abundance, on the other hand, is not much changed from that in the previous studies both theoretically and observationally.  In the theoretical part, the most important reactions are (5) and (6).  The new recombination rate of (5) is lower than the previous rate by only about 10 \% (Fig. \ref{fig34}) at the recombination temperature $T_9 \sim 0.1$ (Fig. \ref{fig31}).  In the observational part, the constraint has been updated from $^6$Li/H$=(7.1\pm 0.7) \times 10^{-12}$ \citep[the average of stars with $^6$Li detections in][]{Asplund:2005yt} to $^6$Li/H=$(0.9\pm 4.3)\times 10^{-12}$ \citep{Lind:2013iza}.  Curves of $d$($^6$Li)=10 then correspond to $^6$Li/H$=9.5 \times 10^{-11}$ in this study, and $^6$Li/H$=7.1 \times 10^{-11}$ in the previous studies.  These slight changes do not move the contour of $d$($^6$Li)=10 much.  The contour then moves left only by a factor of about 1.4.

The interesting parameter region for the $^7$Li reduction subsequently moves upper left.  The constraint on the $^7$Li abundance is significantly changed while that on the $^6$Li abundance is not changed.  The parameter region is, therefore, exclusively affected by the change of the $^7$Li contour.  The minimum $X^-$ abundance required for the effective $^7$Li reduction is $Y_X=0.04$ in this study.  This value is only about a twenty-fifth of the previous estimate $Y_X\sim 1$ \citep{Kusakabe:2010cb}.

\subsubsection{Case with $^7$Be$_X$ decay}
When the charged-current decay of $^7$Be$_X\rightarrow ^7$Li+$X^0$ is operative, the $^7$Be destruction rate is determined only by (1) $^7$Be($X^-$, $\gamma$)$^7$Be$_X$ since the $^7$Be$_X$ nucleus is assumed to be instantaneously destroyed.

In the theoretical calculation of $^7$Li/H, the new larger rate of (1) moves the contours of $d(^7$Li$)$ to the left by a factor of about 5.  The constraint on the $^6$Li/H abundance is the same as in the case without $^7$Be$_X$ decay.  The interesting parameter region then moves to upper left because of the changes of the theoretical result and the observational constraint on the $^7$Li abundance.  The minimum $X^-$ abundance required for the effective $^7$Li reduction is $Y_X=6\times 10^{-3}$ in this study.  This value is about a factor of 7 below that of the previous estimate $Y_X\sim 0.04$ \citep{Kusakabe:2007fv}.

\section{SUMMARY}\label{sec10}
We have completed a new detailed study of the  effects of a long-lived negatively charged massive particle, i.e., $X^-$, on BBN.  The BBN model including the $X^-$ particle is motivated by the  discrepancy between the $^7$Li abundances predicted in SBBN model and those inferred from spectroscopic observations of MPSs.  
In the BBN model including the $X^-$, $^7$Be is destroyed via a recombination reaction with the $X^-$ followed by a radiative proton capture reaction, i.e., $^7$Be($X^-$, $\gamma$)$^7$Be$_X$($p$, $\gamma$)$^8$B$_X$.   Since the primordial $^7$Li abundance is mainly from the abundance of $^7$Be produced during BBN, this $^7$Be destruction leads to a reduction of the primordial $^7$Li abundance, and it can explain the observed abundances.  In addition, $^6$Li is produced via the recombination of $^4$He and $X^-$ followed by a deuteron capture reaction, i.e., $^4$He($X^-$, $\gamma$)$^4$He$_X$($d$, $X^-$)$^6$Li.  Although the effects of many possible reactions have been studied, the $^9$Be abundance is not significantly enhanced in this BBN model.

In this paper, we have also made a new study of the  effects of uncertainties in the nuclear charge distributions on the binding energies of nuclei and $X^-$ particles, the reaction rates, and the resultant BBN.  We also calculated new radiative recombination rates for $^7$Be, $^7$Li, $^9$Be, and $^4$He with $X^-$ taking into account the contributions from many partial waves of the scattering states.  We also suggest a new reaction of $^9$Be production that enhances  the primordial $^9$Be abundance to a level that might be detectable   in future observations of MPSs.

In detail, this work can be  summarized as follows.
\begin{enumerate}
\item We assumed three shapes for  the nuclear charge density, i.e., Woods-Saxon, Gaussian, and homogeneous sphere types which were  parameterized  to reproduce the experimentally measured RMS charge radii.  The potentials between the  $X^-$ and nuclei were  then derived by folding the Coulomb potential and the nuclear charge densities (Sec. \ref{sec2}).  Binding energies for  nuclei plus $X^-$ were  calculated for the different nuclear charge densities and different masses of the $X^-$, $m_X$.  Along with the binding energies of the  GS $X$-nuclei, those of the first atomic excited states of $^8$B$_X^{\ast{\rm a}}$ and $^9$B$_X^{\ast{\rm a}}$ were  derived since these states provide important resonances in the $^7$Be($p$, $\gamma$)$^8$B$_X$ and $^8$Be($p$, $\gamma$)$^9$B$_X$ reactions (Sec. \ref{sec3}).  Resonant rates for the radiative proton capture were then calculated.  We found that the different charge distributions result in reaction rates that can  differ by significant factors depending upon the temperature.  
This is because the rates depend on the resonance energy heights that are sensitive to relatively small changes in binding energies of $X$-nuclei caused by the different nuclear charge distributions (Sec. \ref{sec4}).

\item We also calculated new precise rates for the radiative recombinations of $^7$Be, $^7$Li, $^9$Be, and $^4$He with $X^-$ for four cases of $m_X$.  For that purpose, binding energies and wave functions of the respective $X$-nuclei were derived for several bound states.  
In the recombination process for $^7$Be and $^7$Li, bound states of the nuclear first excited states, $^7$Be$^\ast$ and $^7$Li$^\ast$, with $X^-$ can operate as effective resonances.  These resonant reaction rates as well as transition matrices, radiative decay widths of the resonances, and resonance energies were  calculated using derived wave functions.  For $^9$Be and $^4$He, however, there are no important resonances in the recombination processes since the resonance energies are much higher than the typical temperatures corresponding to the recombination epoch. (Sec. \ref{sec5})

\item For the four nuclei $^7$Be, $^7$Li, $^9$Be, and $^4$He, we calculated continuum-state wave functions for $l=0$ to 4, and nonresonant recombination rates for the respective partial waves of scattering states and  bound states.  It was found that the finite sizes of the nuclear charge distributions causes deviations in the bound and continuum wave functions compared to  those derived assuming that nuclei are point charges.  These deviations are larger for larger $m_X$ and for heavier nuclei with a larger charge.  In addition, the effect of the finite charge distribution predominantly affects the wave functions for tightly bound states and those for scattering states with small angular momenta $l$.  We found the important characteristics of the $^7$Be+$X^-$ recombination.  That is,  for the heavy $X^-$, $m_X\gtrsim 100$ GeV, the most important transition in the recombination is the $d$-wave $\rightarrow$ 2P.  Transitions $f$-wave $\rightarrow$ 3D and $d$-wave $\rightarrow$ 3P are also more efficient than that for the GS formation.  
This fact is completely different from the formation of hydrogen-like electronic ions described by the point-charge distribution.  In this case the transition $p$-wave $\rightarrow$ 1S is predominant.  The same characteristics that the transition $d$-wave $\rightarrow$ 2P is most important  was found for the recombinations of $^7$Li and $^9$Be.  Since $^4$He is lighter and its charge is smaller than $^7$Li and $^{7,9}$Be, the effect of  a finite charge distribution is smaller.  In the $^4$He recombination, therefore, the transition $p$-wave $\rightarrow$ 1S is predominant similar to the case of a point charge nucleus.  Recombination rates for other nuclei were  estimated using a simple Bohr atomic model formula  (Sec. \ref{sec5}).

\item Our nonresonant rate for  the  $^7$Be($X^-$, $\gamma$)$^7$Be$_X$ reaction with $m_X=1000$ GeV is more than 6 times larger than the previously estimated  rate \citep{Bird:2007ge}.  This difference is caused by our treatment of many bound states and many partial waves for the  scattering states (Sec. \ref{sec5}).  This improvement in the rate provides an improved constraint on the $X^-$ particle properties (Sec. \ref{sec9}).

\item We have also suggested a new reaction for $^9$Be production, i.e, $^7$Li$_X$($d$, $X^-$)$^9$Be.  We adopted an example reaction rate using the astrophysical $S$-factor for the reaction $^7$Li($d$, $n\alpha$)$^4$He as a starting point (Sec. \ref{sec6}).  This reaction was  found to significantly enhance the  primordial $^9$Be abundance from our BBN network calculation (Sec. \ref{sec9}).

\item Using the binding energies of $X$-nuclei calculated in Sec. \ref{sec3}, mass excesses of $X$-nuclei along with  rates and $Q$-values for reactions involving the $X^-$ particle were  calculated for four cases of $m_X$.  
The reaction network  included the $\beta$-decays of $X$-nuclei, nuclear reactions of $X$-nuclei and their inverse reactions.  $Q$-values and reverse reaction coefficients were found to be  heavily dependent on $m_X$  (Sec. \ref{sec7}).  The $X^-$-particle mass dependence of the $Q$-value is especially important for the resonant reaction $^7$Be$_X$($p$, $\gamma$)$^8$B$_X$ (Sec. \ref{sec9}).

\item We constructed an updated BBN code that  includes the new reaction rates derived in this paper (Sec. \ref{sec8}).  BBN calculations based on this code were  then shown for four cases of $m_X$.  It was found that the amounts of $^7$Be destruction depend significantly on the assumed charge distribution form of the $^7$Be nucleus for the $m_X=1000$ GeV case.  Finally, we derived new  most realistic constraints on the initial abundance and the lifetime of the $X^-$ particle.  Parameter regions for the solution to the $^7$Li problem were  identified for the respective $m_X$ cases.  We also derived the expected primordial $^9$Be abundances predicted in the allowed  parameter regions.  The predicted $^9$Be abundances are larger than in the SBBN model, but  smaller than the present observational upper limit from MPSs (Sec. \ref{sec9}).

\item Some discussion was  also  given for E1 transitions that simultaneously change both  nuclear and atomic states of $^7$Be$_X$ and $^7$Li$_X$.  These are hindered because of the near orthogonality of the  atomic  and nuclear wave functions.  It was suggested, however,  that for exotic atoms composed of nuclei and an $X^-$ with mass much larger than the nucleon mass, this orthogonality in the atomic and nuclear wave functions can be somewhat broken.  Such exotic atoms may, therefore, have large rates for E1 transitions that simultaneously change  nuclear and atomic states (Appendix \ref{app1}).

\end{enumerate}

\appendix

\section{TRANSITIONS OF EXOTIC ATOMS THAT SIMULTANEOUSLY CHANGE BOTH NUCLEAR AND ATOMIC STATES}\label{app1}

Here we discuss in detail the type 3 transitions in the $^7$Be($X^-$, $\gamma$)$^7$Be$_X$ reaction that was addressed in Sec. \ref{sec5.1}.

\subsection{Electric Dipole Transition Rate}\label{appa11}
The reduced probability for a transition from an initial state (i) to a final state (f) is given by
\begin{equation}
B(I_{\rm i} \rightarrow I_{\rm f}) =\frac{1}{2I_{\rm i}+1} \sum_{M_{\rm i}, M_{\rm f}} \left|\langle \Psi_{I_{\rm f}}^{M_{\rm f}} \left| {\mathcal O} ({\rm E1},~\mu)\right| \Psi_{I_{\rm i}}^{M_{\rm i}} \rangle \right|^2,
\label{eq1}
\end{equation}
where
$I_k$, $M_k$ and $\Psi_{I_{\rm k}}^{M_{\rm k}}$ are the spins, the magnetic quantum numbers and the wave functions, respectively, of state $k$ for initial (i) and final (f) states, with $\mu=M_{\rm i}-M_{\rm f}$.
We consider the three-body system of $\alpha$, $^3$He, and $X^-$ located at the position vectors $\bfx_i$ for $i$=1 ($\alpha$), 2 ($^3$He), and 3 ($X^-$), respectively.  This system has bound states of $^7$Be$_X$.  The system of $^7$Li$_X$ can be considered similarly to this system.  The electric dipole ($E1$) operator is given by ${\mathcal O}$(E1, $\mu$)$=\sum_{i=1}^3 q_i x_i Y_{1 \mu}(\hat{x_i})$ where  $q_i$ is the electric charge, $x_i=|\bfx_i|$ is the distance from the origin to the position of particle $i$, and $Y_{1 \mu}(\hat{x_i})$ are the spherical surface harmonics.

\subsection{Hindrance of the Matrix Element}\label{appa12}
The wave function describing atoms composed of a nucleus $^7A$ and a negatively charged massive particle $X^-$ is approximately given by a product of functions of a $^7A$ nuclear state and a $^7A_X$ atomic state, i.e.,
\begin{eqnarray}
\Psi_{I_{\rm i}}^{M_{\rm i}}(\bfr,~\bfr^\prime)&=&\sum_{m_1,m_{\rm i}}(j_1 m_1 l_{\rm i} m_{\rm i}|I_{\rm i} M_{\rm i}) {\Psi^{\rm n}}_{j_1}^{m_1}(\bfr) {\Psi^{\rm a}}_{n_{\rm i}l_{\rm i}m_{\rm i}}(\bfr^\prime) \nonumber \\
\Psi_{I_{\rm f}}^{M_{\rm f}}(\bfr,~\bfr^\prime)&=&\sum_{m_2,m_{\rm f}}(j_2 m_2 l_{\rm f} m_{\rm f}|I_{\rm f} M_{\rm f}) {\Psi^{\rm n}}_{j_2}^{m_2}(\bfr) {\Psi^{\rm a}}_{n_{\rm f}l_{\rm f}m_{\rm f}}(\bfr^\prime), \nonumber \\
\label{eq_a2}
\end{eqnarray}
where
${\Psi^{\rm n}}_{j_\beta}^{m_\beta}(\bfr)$ are atomic wave functions for the two body system of particles 1 and 2, with the spin $j_\beta$ and  magnetic quantum numbers $m_\beta$ for $\beta=1$ (for state i) and 2 (for state f).
${\Psi^{\rm a}}_{n_k l_k m_k}(\bfr^\prime)$ is the atomic wave function for the two body system of particles (1+2)+3, with $n_k$, $l_k$, and $m_k$ the main, azimuthal, and magnetic quantum numbers, respectively.
$(j_\beta m_\beta l_k m_k|I_k M_k)$ is the Clebsch-Gordan coefficient for $(\beta,~k)=$(1,~i) and (2, f).
\begin{eqnarray}
\bfr&=&\bfx_1-\bfx_2 \nonumber \\
\bfr^\prime&=& \frac{{\cal M}_1 \bfx_1 + {\cal M}_2 \bfx_2}{{\cal M}_1 + {\cal M}_2} -\bfx_3
\label{eq_a3}
\end{eqnarray}
are Jacobi coordinates where  ${\cal M}_i$ the mass of particle $i$.
The atomic wave function is then simply given by ${\Psi^{\rm a}}_{n_k l_k m_k}(\bfr^\prime)=\psi^{\rm a}_{n_k l_k}(r^\prime) Y^{\rm a}_{l_k m_k}(\hat{r^\prime})$ for $k=$i and f.

One can consider transitions which change the atomic and nuclear states simultaneously.  This type of transition proceeds from states ($^7$Be$^\ast_X$)$^{\ast{\rm a}}$ or $^7$Be$^\ast_X$ to ($^7$Be$_X$)$^{\ast{\rm a}}$ or $^7$Be$_X$, where the initial states are atomic excited or ground states composed of the first nuclear excited state $^7$Be$^\ast(1/2^-)$,  and the final states are atomic excited or ground states of the nuclear ground state $^7$Be$(3/2^-)$.  We show that the E1 rates for such transitions are smaller than those for typical E1 allowed nuclear transitions.  For simplicity, we approximately neglect the finite-size charge distributions of $\alpha$ and $^3$He, and assume that all three particles are  point charges.  Then, the electric dipole moment is given by
\begin{eqnarray}
\bfd(\bfx_1,~\bfx_2,~\bfx_3)&=&\frac{(q_1+q_2)m_3 - q_3({\cal M}_1+{\cal M}_2)}{{\cal M}_1 +{\cal M}_2 +{\cal M}_3} \bfr^\prime \nonumber\\
&& + \frac{{\cal M}_2 q_1 - {\cal M}_1 q_2}{{\cal M}_1 +{\cal M}_2} \bfr \nonumber\\
&\equiv&q_{r^\prime} \bfr^\prime + q_r \bfr,
\label{eq_a6}
\end{eqnarray}
where $q_{r^\prime}$ and $q_{r}$ are  defined as coefficients of $\bfr^\prime$ and $\bfr$, respectively.

Using Eqs. (\ref{eq2}) and (\ref{eq6}), the matrix element in Eq. (\ref{eq1}) can be rewritten as
\begin{widetext}
\begin{eqnarray}
\langle \Psi_{I_{\rm f}}^{M_{\rm f}} \left| {\mathcal O}(E1,\mu)\right| \Psi_{I_{\rm i}}^{M_{\rm i}} \rangle
&=& \sum_{m_1,~m_{\rm i}} \sum_{m_2,~m_{\rm f}} (j_1 m_1 l_{\rm i} m_{\rm i}|I_{\rm i} M_{\rm i}) (j_2 m_2 l_{\rm f} m_{\rm f}|I_{\rm f} M_{\rm f})
\int d\bfr  \int d\bfr^\prime \nonumber\\
&&\times
{{\Psi^{\rm n}}_{j_2}^{m_2}}^\ast(\bfr) {\Psi^{\rm a}}_{n_{\rm f}l_{\rm f}m_{\rm f}}^\ast(\bfr^\prime)
\left[q_{r^\prime} r^\prime Y_{1\mu}(\hat{r^\prime})+ q_r r Y_{1\mu}(\hat{r})\right]
{{\Psi^{\rm n}}_{j_1}^{m_1}}(\bfr) {\Psi^{\rm a}}_{n_{\rm i}l_{\rm i}m_{\rm i}}(\bfr^\prime)\nonumber\\
&=& \sum_{m_1,~m_{\rm i}} \sum_{m_2,~m_{\rm f}} (j_1 m_1 l_{\rm i} m_{\rm i}|I_{\rm i} M_{\rm i}) (j_2 m_2 l_{\rm f} m_{\rm f}|I_{\rm f} M_{\rm f}) \nonumber\\
&&\times \left\{
\langle {\Psi^{\rm n}}_{j_2}^{m_2} \left| \right. {{\Psi^{\rm n}}_{j_1}^{m_1}} \rangle
\int d\bfr^\prime
{\Psi^{\rm a}}_{n_{\rm f}l_{\rm f}m_{\rm f}}^\ast(\bfr^\prime)
\left[q_{r^\prime} r^\prime Y_{1\mu}(\hat{r^\prime}) \right]
{\Psi^{\rm a}}_{n_{\rm i}l_{\rm i}m_{\rm i}}(\bfr^\prime) \right. \nonumber\\
&&+ \left. \langle {\Psi^{\rm a}}_{n_{\rm f}l_{\rm f}m_{\rm f}} \left|\right. {\Psi^{\rm a}}_{n_{\rm i}l_{\rm i}m_{\rm i}} \rangle
\int d\bfr
{{\Psi^{\rm n}}_{j_2}^{m_2}}^\ast(\bfr)
\left[ q_r r Y_{1\mu}(\hat{r}) \right]
{{\Psi^{\rm n}}_{j_1}^{m_1}}(\bfr)
\right\}~.
\label{eq_a7}
\end{eqnarray}
\end{widetext}
The orthogonality of the wave functions satisfies the conditions of $\langle {\Psi^{\rm n}}_{j_2}^{m_2} \left| \right. {{\Psi^{\rm n}}_{j_1}^{m_1}} \rangle =0$ and $\langle {\Psi^{\rm a}}_{n_{\rm f}l_{\rm f}m_{\rm f}} \left|\right. {\Psi^{\rm a}}_{n_{\rm i}l_{\rm i}m_{\rm i}} \rangle$=0 if both the nuclear and atomic states change in the reaction.  This E1 matrix element is thus found to be zero.

\subsection{E1 Rate Enhanced by a Heavy $X^-$ Particle}\label{appa13}
Contrary  to the approximate estimation described above, the E1 transition rate is not expected to vanish,   although it is hindered compared to the E1 rate for allowed nuclear transitions.  This is because particles can have charge distributions of finite size.  In the present case, $\alpha$ and $^3$He have a finite charge distribution.  We explain this effect by comparing the electronic ion, $^7$Be$^{3+}$, and the exotic ion of the  massive $X^-$ particle, $^7$Be$_X$.

Average radii of electronic ions composed of an electron and light nuclei are $\sim \mathcal{O}(10^{-8}~{\rm cm})$ while the average radii of nuclear wave functions for light nuclei are $\sim \mathcal{O}(10^{-13}~{\rm cm})$.  Since the two radius scales are different from each other by a large factor, the atomic and nuclear wave functions can be separately considered for the following reason:  1) nuclear wave functions are not affected by the existence of the electrons which are far away from the nuclei; and 2) atomic wave functions are not affected by the nuclear charge distribution since the Coulomb potential between an electron and the  nucleus does not depend on the nuclear charge distribution except at  very small atomic radii $r^\prime$ comparable to the nuclear charge radius.

When the mass of the $X^-$ is larger than $\sim 1$ GeV, however, the average radii of $^7$Be$_X$ atomic states approach $\mathcal{O}(1~{\rm fm})$.  This is roughly the same order of magnitude as the charge radius of the $^7$Be nucleus.  At large nuclear radii, therefore, effects of the Coulomb forces by the $X^-$ particle are not completely negligible in nuclear wave functions.  Nuclear wave functions then depend  not only on the nuclear radii but also on atomic radii.  In addition, at small atomic radii the effects of the finite nuclear charge distribution reflecting a nuclear cluster structure are not completely negligible in atomic wave functions.  Atomic wave functions then depend not only on atomic radii but also on nuclear radii.  Therefore, nuclear and atomic wave functions are not strictly orthogonal, and the E1 matrix element is finite.

It is physically interesting that rates for E1 transitions simultaneously changing nuclear and atomic states can be larger if the $X^-$ particle is heavier.  The rates are expected to be large for not only the hypothetical $X^-$ particle predicted in beyond the standard model physics,  but also known negatively charged heavy particles such as $\mu^-$, $\pi^-$, ${\bar p}^-$, and so on.  For example, ordinary and radiative muon captures on a proton, in which the latter just corresponds to the recombination process in this work, were performed in TRIUMF \citep{Jonkmans}, but the  theoretical interpretation is   still under discussions \citep{Ch}.  

In addition to the pure Coulomb force, spin-dependent interactions can exist between an $X^-$ particle and nuclear clusters if the $X^-$ particle has a spin.  In this paper, we assumed a spinless $X^-$ particle.  In general, however, spin dependent interactions can mix states of $A+X^-$ and $A^\ast+X^-$ so that the overlap integrals can be non-zero (M. Kamimura 2013; private communications).

\acknowledgments
We are grateful to Professor Masayasu Kamimura for discussion on the reaction cross sections.  This work was supported by the National Research Foundation of Korea (Grant Nos. 2012R1A1A2041974, 2011-0015467, 2012M7A1A2055605), and in part by Grants-in-Aid for Scientific Research of JSPS (24340060), and Scientific Research on Innovative Areas of MEXT (20105004).  Work at the University of Notre Dame (GJM) supported by   
the U.S. Department of Energy under Nuclear Theory Grant DE-FG02-95-ER40934.

\clearpage

\begin{deluxetable*}{ccclllll}
\tablecaption{\label{tab1} Binding Energies of $A_X$ (MeV) for $m_X=100$ TeV}
\tablewidth{0pt}
\tablehead{
\colhead{Nuclei} & \colhead{$\langle r^2 \rangle_{\rm C}^{1/2}$~(fm) } &
 \colhead{Ref. } & \colhead{Gaussian} & \colhead{Homogeneous} & \colhead{WS(0.45~fm)} & \colhead{WS(0.4~fm)} & \colhead{WS(0.35~fm)}}
\startdata
$^1$H                 & 0.875 $\pm$ 0.007                 & 1 & 0.0250 &   0.0250 & 0.0250 & 0.0250 & 0.0250 \\
$^2$H                 & 2.116 $\pm$ 0.006                 & 2 & 0.0489 &   0.0488 & 0.0489 & 0.0489 & 0.0488 \\
$^3$H                 & 1.755 $\pm$ 0.086                 & 3 & 0.0724 &   0.0724 & 0.0725 & 0.0725 & 0.0724 \\
$^3$He                & 1.959 $\pm$ 0.030                 & 3 & 0.268  &   0.267  & 0.268  & 0.268  & 0.267  \\
$^4$He                & 1.80  $\pm$ 0.04                  & 4 & 0.343  &   0.342  & 0.344  & 0.343  & 0.343  \\
$^6$Li                & 2.48  $\pm$ 0.03                  & 4 & 0.806  &   0.790  & 0.802  & 0.799  & 0.797  \\
$^7$Li                & 2.43  $\pm$ 0.02                  & 4 & 0.882  &   0.862  & 0.878  & 0.874  & 0.871  \\
$^8$Li                & 2.42  $\pm$ 0.02                  & 4 & 0.945  &   0.921  & 0.940  & 0.936  & 0.932  \\
$^6$Be                & 2.52  $\pm$ 0.02\tablenotemark{a} & 4 & 1.234  &   1.201  & 1.225  & 1.220  & 1.215  \\
$^7$Be                & 2.52  $\pm$ 0.02                  & 4 & 1.324  &   1.284  & 1.313  & 1.306  & 1.300  \\
$^8$Be                & 2.52  $\pm$ 0.02\tablenotemark{a} & 4 & 1.401  &   1.353  & 1.387  & 1.379  & 1.373  \\
$^9$Be                & 2.50  $\pm$ 0.01                  & 4 & 1.477  &   1.422  & 1.462  & 1.452  & 1.445  \\
$^{10}$Be             & 2.40  $\pm$ 0.02                  & 4 & 1.577  &   1.516  & 1.564  & 1.553  & 1.544  \\
$^7$B                 & 2.68  $\pm$ 0.12\tablenotemark{b} & 5 & 1.752  &   1.684  & 1.726  & 1.717  & 1.709  \\
$^8$B                 & 2.68  $\pm$ 0.12                  & 5 & 1.840  &   1.762  & 1.810  & 1.799  & 1.790  \\
$^9$B                 & 2.68  $\pm$ 0.12\tablenotemark{b} & 5 & 1.917  &   1.829  & 1.883  & 1.871  & 1.860  \\
$^{10}$B              & 2.58  $\pm$ 0.07                  & 6 & 2.036  &   1.939  & 2.004  & 1.989  & 1.976  \\
$^{11}$B              & 2.58  $\pm$ 0.07\tablenotemark{c} & 6 & 2.099  &   1.993  & 2.063  & 2.047  & 2.034  \\
$^{12}$B              & 2.51  $\pm$ 0.02                  & 4 & 2.198  &   2.082  & 2.164  & 2.145  & 2.129  \\
$^9$C                 & 2.51  $\pm$ 0.02\tablenotemark{d} & 4 & 2.554  &   2.428  & 2.517  & 2.496  & 2.479  \\
$^{10}$C              & 2.51  $\pm$ 0.02\tablenotemark{d} & 4 & 2.638  &   2.499  & 2.597  & 2.574  & 2.556  \\
$^{11}$C              & 2.51  $\pm$ 0.02\tablenotemark{d} & 4 & 2.713  &   2.562  & 2.668  & 2.644  & 2.623  \\
$^{12}$C              & 2.51  $\pm$ 0.02                  & 4 & 2.780  &   2.618  & 2.731  & 2.705  & 2.683  \\
\hline							
$^8$B$^{\ast{\rm a}}$ & 2.68  $\pm$ 0.12                  & 5 & 1.021  &   1.024  & 1.022  & 1.022  & 1.023  \\
$^9$B$^{\ast{\rm a}}$ & 2.68  $\pm$ 0.12\tablenotemark{b} & 5 & 1.104  &   1.105  & 1.105  & 1.104  & 1.104  
\enddata
\tablenotetext{a}{Taken from $^7$Be radius}
\tablenotetext{b}{Taken from $^8$B radius}
\tablenotetext{c}{Taken from $^{10}$B radius}
\tablenotetext{d}{Taken from $^{12}$C radius}
\tablecomments{References: 1= \citet{yao06}; 2= \citet{sim81};
 3=TUNL Nuclear Data, http://www.tunl.duke.edu/NuclData;
 4= \citet{tan88}; 5= \citet{fuk99}; 6= \citet{1995PhRvC..51.2406C}.}
\end{deluxetable*}


\begin{deluxetable*}{cllll}
\tablecaption{\label{tab2} Binding Energies of $A_X$ for a Woods-Saxon Charge Density with $a=0.40$ fm (MeV)}
\tablewidth{0pt}
\tablehead{
\colhead{Nuclei} & \colhead{$m_X$=1 GeV} & \colhead{10 GeV} & \colhead{100 GeV} & \colhead{1000 GeV}}
\startdata
$^1$H        & 0.0127    & 0.0228 & 0.0247 & 0.0249 \\
$^2$H        & 0.0173    & 0.0414 & 0.0480 & 0.0488 \\
$^3$H        & 0.0196    & 0.0572 & 0.0706 & 0.0723 \\
$^3$He       & 0.0776    & 0.216  & 0.261  & 0.267  \\
$^4$He       & 0.0830    & 0.263  & 0.333  & 0.342  \\
$^6$Li       & 0.194     & 0.615  & 0.776  & 0.797  \\
$^7$Li       & 0.198     & 0.659  & 0.847  & 0.872  \\
$^8$Li       & 0.201     & 0.693  & 0.904  & 0.932  \\
$^6$Be       & 0.335     & 0.970  & 1.189  & 1.216  \\
$^7$Be       & 0.341     & 1.023  & 1.270  & 1.302  \\
$^8$Be       & 0.346     & 1.066  & 1.340  & 1.375  \\
$^9$Be       & 0.350     & 1.108  & 1.408  & 1.448  \\
$^{10}$Be    & 0.355     & 1.164  & 1.502  & 1.548  \\
$^7$B        & 0.511     & 1.389  & 1.676  & 1.712  \\
$^8$B        & 0.518     & 1.440  & 1.755  & 1.795  \\
$^9$B        & 0.524     & 1.483  & 1.821  & 1.866  \\
$^{10}$B     & 0.532     & 1.554  & 1.933  & 1.983  \\
$^{11}$B     & 0.536     & 1.587  & 1.987  & 2.041  \\
$^{12}$B     & 0.542     & 1.644  & 2.079  & 2.138  \\
$^9$C        & 0.739     & 2.004  & 2.435  & 2.490  \\
$^{10}$C     & 0.745     & 2.050  & 2.508  & 2.568  \\
$^{11}$C     & 0.750     & 2.090  & 2.572  & 2.636  \\
$^{12}$C     & 0.755     & 2.125  & 2.629  & 2.697  \\
\hline				  	
$^8$B$^{\ast{\rm a}}$ & 0.147     & 0.665  & 0.973  & 1.017  \\
$^9$B$^{\ast{\rm a}}$ & 0.149     & 0.703  & 1.047  & 1.099  
\enddata
\end{deluxetable*}


\begin{deluxetable*}{ccccccc}
\tablecaption{\label{tab3} Calculated Parameters for $^7$Be$_X$($p$, $\gamma$)$^8$B$_X$ with $m_X=1$ TeV.}
\tablewidth{0pt}
\tablehead{
\colhead{Model} & \colhead{$\tau_{\rm if}$ (fm)} & \colhead{$E_{\rm r}$ (MeV)} & \colhead{$E_\gamma$ (MeV)} & \colhead{$\Gamma_\gamma$ (eV)} & \colhead{$C$ ($10^{6}$ cm$^3$ mol$^{-1}$ s$^{-1}$)} & \colhead{$Q$-value (MeV)}}
\startdata
Gaussian    & 2.98 & 0.167 & 0.820 & 10.0\phantom{0} & 1.55 & 0.653 \\
homogeneous & 3.18 & 0.124 & 0.740 & \phantom{0}8.43 & 1.30 & 0.615 \\
WS40        & 3.08 & 0.148 & 0.778 & \phantom{0}9.19 & 1.42 & 0.630 
\enddata
\end{deluxetable*}


\begin{deluxetable*}{ccccccc}
\tablecaption{\label{tab4} Calculated Parameters for $^8$Be$_X$($p$, $\gamma$)$^9$B$_X$ with $m_X=1$ TeV.}
\tablewidth{0pt}
\tablehead{
\colhead{Model} & \colhead{$\tau_{\rm if}$ (fm)} & \colhead{$E_{\rm r}$ (MeV)} & \colhead{$E_\gamma$ (MeV)} & \colhead{$\Gamma_\gamma$ (eV)} & \colhead{$C$ ($10^{6}$ cm$^3$ mol$^{-1}$ s$^{-1}$)} & \colhead{$Q$-value (MeV)}}
\startdata
Gaussian    & 2.84 & 0.484 & 0.814 & 8.91 & 1.37 & 0.330 \\
homogeneous & 3.05 & 0.435 & 0.725 & 7.30 & 1.12 & 0.290 \\
WS40        & 2.95 & 0.462 & 0.767 & 8.06 & 1.24 & 0.305 
\enddata
\end{deluxetable*}


\begin{deluxetable*}{ccccccc}
\tablecaption{\label{tab14} Calculated Parameters for  $^7$Be$_X$($p$, $\gamma$)$^8$B$_X$ Obtained with the WS40 Model.}
\tablewidth{0pt}
\tablehead{
\colhead{$m_X$ (GeV)} & \colhead{$\tau_{\rm if}$ (fm)} & \colhead{$E_{\rm r}$ (MeV)} & \colhead{$E_\gamma$ (MeV)} & \colhead{$\Gamma_\gamma$ (eV)} & \colhead{$C$ ($10^{6}$ cm$^3$ mol$^{-1}$ s$^{-1}$)} & \colhead{$Q$-value (MeV)}}
\startdata
\phantom{000} 1  &  9.50   & \phantom{0}0.0568 & 0.372 & \phantom{0}0.838 & \phantom{0}0.154 & 0.315 \\
\phantom{00} 10  &  3.87   &            0.220  & 0.775 &            6.29  &            1.05  & 0.555 \\
\phantom{0} 100  &  3.16   &            0.160  & 0.782 &            8.88  &            1.39  & 0.622 \\
           1000  &  3.08   &            0.148  & 0.778 &            9.19  &            1.42  & 0.630 
\enddata
\end{deluxetable*}


\begin{deluxetable*}{ccccccc}
\tablecaption{\label{tab15} Calculated Parameters for $^8$Be$_X$($p$, $\gamma$)$^9$B$_X$ Obtained with the WS40 Model.}
\tablewidth{0pt}
\tablehead{
\colhead{$m_X$ (GeV)} & \colhead{$\tau_{\rm if}$ (fm)} & \colhead{$E_{\rm r}$ (MeV)} & \colhead{$E_\gamma$ (MeV)} & \colhead{$\Gamma_\gamma$ (eV)} & \colhead{$C$ ($10^{6}$ cm$^3$ mol$^{-1}$ s$^{-1}$)} & \colhead{$Q$-value (MeV)}}
\startdata
\phantom{000} 1  &  9.41   & 0.382 & 0.375 & \phantom{0}0.794 & \phantom{0}0.143 & $-0.00699$ \\
\phantom{00} 10  &  3.76   & 0.549 & 0.780 &            5.65  & \phantom{0}0.940 &  0.232     \\
\phantom{0} 100  &  3.03   & 0.477 & 0.774 &            7.81  &            1.22  &  0.297     \\
           1000  &  2.95   & 0.462 & 0.767 &            8.06  &            1.24  &  0.305     
\enddata
\end{deluxetable*}


\begin{deluxetable*}{c|ccccccc}
\tablecaption{\label{tab5} Binding Energies of $^7$Be$_X$ Atomic States with Main Quantum Numbers $n=1$--$7$ (keV).}
\tablewidth{0pt}
\tablehead{
\colhead{$m_X=1$ GeV} & \colhead{$l=0$} & \colhead{$l=1$} & \colhead{$l=2$} & \colhead{$l=3$} & \colhead{$l=4$} & \colhead{$l=5$} & \colhead{$l=6$}}
\startdata
$n=1$ & $341$\phantom{.00}  &        &        &        &        &        &        \\
$n=2$ & $88.7$ & $92.3$ &        &        &        &        &        \\
$n=3$ & $40.0$ & $41.0$ & $41.1$ &        &        &        &        \\
$n=4$ & $22.6$ & $23.1$ & $23.1$ & $23.1$ &        &        &        \\
$n=5$ & $14.5$ & $14.8$ & $14.8$ & $14.8$ & $14.8$ &        &        \\
$n=6$ & $10.1$ & $10.3$ & $10.3$ & $10.3$ & $10.3$ & $10.3$ &        \\
$n=7$ & \phantom{00}$7.45$ & \phantom{00}$7.54$ & \phantom{00}$7.54$ & \phantom{00}$7.54$ & \phantom{00}$7.54$ & \phantom{00}$7.54$ & \phantom{00}$7.54$ \\
\hline
$m_X=10$ GeV & $l=0$ & $l=1$ & $l=2$ & $l=3$ & $l=4$ & $l=5$ & $l=6$ \\
\hline
$n=1$ & $1023$\phantom{.000} &        &        &        &        &        &        \\
$n=2$ & $326$\phantom{.00}  & $409$\phantom{.00}  &        &        &        &        &        \\
$n=3$ & $158$\phantom{.00}  & $183$\phantom{.00}  & $187$\phantom{.00}  &        &        &        &        \\
$n=4$ & $92.4$ & $104$\phantom{.00}  & $105$\phantom{.00}  & $105$\phantom{.00}  &        &        &        \\
$n=5$ & $60.7$ & $66.4$ & $67.3$ & $67.3$ & $67.3$ &        &        \\
$n=6$ & $42.9$ & $46.2$ & $46.8$ & $46.8$ & $46.8$ & $46.8$ &        \\
$n=7$ & $31.9$ & $34.0$ & $34.4$ & $34.4$ & $34.4$ & $34.4$ & $34.4$ \\
\hline
$m_X=100$ GeV & $l=0$ & $l=1$ & $l=2$ & $l=3$ & $l=4$ & $l=5$ & $l=6$ \\
\hline
$n=1$ & $1270$\phantom{.000} &        &        &        &        &        &        \\
$n=2$ & $451$\phantom{.00}  & $603$\phantom{.00}  &        &        &        &        &        \\
$n=3$ & $226$\phantom{.00}  & $274$\phantom{.00}  & $290$\phantom{.00}  &        &        &        &        \\
$n=4$ & $135$\phantom{.00}  & $156$\phantom{.00}  & $163$\phantom{.00}  & $163$\phantom{.00}  &        &        &        \\
$n=5$ & $89.7$ & $101$\phantom{.00}  & $104$\phantom{.00}  & $105$\phantom{.00}  & $105$\phantom{.00}  &        &        \\
$n=6$ & $63.9$ & $70.4$ & $72.4$ & $72.6$ & $72.6$ & $72.6$ &        \\
$n=7$ & $47.8$ & $51.9$ & $53.2$ & $53.3$ & $53.3$ & $53.3$ & $53.3$ \\
\hline
$m_X=1000$ GeV & $l=0$ & $l=1$ & $l=2$ & $l=3$ & $l=4$ & $l=5$ & $l=6$ \\
\hline
$n=1$ & $1302$\phantom{000} &        &        &        &        &        &        \\
$n=2$ & $469$\phantom{00}  & $632$\phantom{00}  &        &        &        &        &        \\
$n=3$ & $236$\phantom{00}  & $288$\phantom{00}  & $306$\phantom{00}  &        &        &        &        \\
$n=4$ & $142$\phantom{00}  & $164$\phantom{00}  & $172$\phantom{00}  & $173$\phantom{00}  &        &        &        \\
$n=5$ & $94.3$ & $106$\phantom{00}  & $110$\phantom{00}  & $111$\phantom{00}  & $111$\phantom{00}  &        &        \\
$n=6$ & $67.2$ & $74.2$ & $76.6$ & $76.8$ & $76.8$ & $76.8$ &        \\
$n=7$ & $50.3$ & $54.8$ & $56.3$ & $56.4$ & $56.4$ & $56.4$ & $56.4$ 
\enddata
\end{deluxetable*}


\begin{deluxetable*}{rclll}
\tablecaption{\label{tab6} Calculated Parameters for $^7$Be($X^-$, $\gamma$)$^7$Be$_X$ in the WS40 Model.}
\tablewidth{0pt}
\tablehead{
\colhead{$m_X$ (GeV)} & \colhead{Transition} & \colhead{$\tau_{\rm if}$ (fm)} & \colhead{$\Gamma_\gamma$ (eV)} & \colhead{$E_{\rm r}$ (MeV)}}
\startdata
1           & $^7$Be$^\ast_X$(1S)$\rightarrow ^7$Be$_X$(1S)      & ---                  & 0.00343\tablenotemark{a} & 0.0881 \\
10          & $^7$Be$^\ast_X$(2P)$\rightarrow ^7$Be$^\ast_X$(1S) & 4.29                 & 2.80                    & 0.0198 \\
100         & $^7$Be$^\ast_X$(3D)$\rightarrow ^7$Be$^\ast_X$(2P) & 6.04                 & 1.64                    & 0.140 \\
100         & $^7$Be$^\ast_X$(3P)$\rightarrow ^7$Be$^\ast_X$(2S) & 8.09                 & 0.438                   & 0.155 \\
100         & $^7$Be$^\ast_X$(3P)$\rightarrow ^7$Be$^\ast_X$(1S) & 0.738                & 0.653                   & 0.155 \\
1000        & $^7$Be$^\ast_X$(3D)$\rightarrow ^7$Be$^\ast_X$(2P) & 5.80                 & 1.84                    & 0.123 \\
1000        & $^7$Be$^\ast_X$(3P)$\rightarrow ^7$Be$^\ast_X$(2S) & 7.84                 & 0.481                   & 0.141 \\
1000        & $^7$Be$^\ast_X$(3P)$\rightarrow ^7$Be$^\ast_X$(1S) & 0.693                & 0.662                   & 0.141 
\enddata
\tablenotetext{a}{Given by $\Gamma_\gamma = \tau_\gamma^{-1}$ with a lifetime of 192 fs taken from that of the first excited $1/2^-$ state in $^7$Be \citep{Tilley2002}.}
\end{deluxetable*}


\begin{deluxetable*}{c|ccccccc}
\tablecaption{\label{tab7} Binding Energies of $^7$Li$_X$ Atomic States with Main Quantum Numbers $n=1$--$7$ (keV).}
\tablewidth{0pt}
\tablehead{
\colhead{$m_X=1$ GeV} & \colhead{$l=0$} & \colhead{$l=1$} & \colhead{$l=2$} & \colhead{$l=3$} & \colhead{$l=4$} & \colhead{$l=5$} & \colhead{$l=6$}}
\startdata
$n=1$ & $198$\phantom{.00}  &        &        &        &        &        &        \\
$n=2$ & $50.7$ & $52.0$ &        &        &        &        &        \\
$n=3$ & $22.7$ & $23.1$ & $23.1$ &        &        &        &        \\
$n=4$ & $12.8$ & $13.0$ & $13.0$ & $13.0$ &        &        &        \\
$n=5$ & \phantom{00}$8.23$ & \phantom{00}$8.31$ & \phantom{00}$8.31$ & \phantom{00}$8.31$ & \phantom{00}$8.31$ &        &        \\
$n=6$ & \phantom{00}$5.73$ & \phantom{00}$5.77$ & \phantom{00}$5.77$ & \phantom{00}$5.77$ & \phantom{00}$5.77$ & \phantom{00}$5.77$ &        \\
$n=7$ & \phantom{00}$4.21$ & \phantom{00}$4.24$ & \phantom{00}$4.24$ & \phantom{00}$4.24$ & \phantom{00}$4.24$ & \phantom{00}$4.24$ & \phantom{00}$4.24$ \\
\hline
$m_X=10$ GeV & $l=0$ & $l=1$ & $l=2$ & $l=3$ & $l=4$ & $l=5$ & $l=6$ \\
\hline
$n=1$ & $659$\phantom{.00}  &        &        &        &        &        &        \\
$n=2$ & $197$\phantom{.00}  & $234$\phantom{.00}  &        &        &        &        &        \\
$n=3$ & $92.9$ & $104$\phantom{.00}  & $105$\phantom{.00}  &        &        &        &        \\
$n=4$ & $53.9$ & $58.8$ & $59.2$ & $59.2$ &        &        &        \\
$n=5$ & $35.1$ & $37.7$ & $37.9$ & $37.9$ & $37.9$ &        &        \\
$n=6$ & $24.7$ & $26.2$ & $26.3$ & $26.3$ & $26.3$ & $26.3$ &        \\
$n=7$ & $18.3$ & $19.2$ & $19.3$ & $19.3$ & $19.3$ & $19.3$ & $19.3$ \\
\hline
$m_X=100$ GeV & $l=0$ & $l=1$ & $l=2$ & $l=3$ & $l=4$ & $l=5$ & $l=6$ \\
\hline
$n=1$ & $847$\phantom{.00}  &        &        &        &        &        &        \\
$n=2$ & $277$\phantom{.00}  & $354$\phantom{.00}  &        &        &        &        &        \\
$n=3$ & $135$\phantom{.00}  & $159$\phantom{.00}  & $163$\phantom{.00}  &        &        &        &        \\
$n=4$ & $79.5$ & $89.9$ & $91.8$ & $91.9$ &        &        &        \\
$n=5$ & $52.4$ & $57.7$ & $58.8$ & $58.8$ & $58.8$ &        &        \\
$n=6$ & $37.1$ & $40.2$ & $40.8$ & $40.8$ & $40.8$ & $40.8$ &        \\
$n=7$ & $27.6$ & $29.6$ & $30.0$ & $30.0$ & $30.0$ & $30.0$ & $30.0$ \\
\hline
$m_X=1000$ GeV & $l=0$ & $l=1$ & $l=2$ & $l=3$ & $l=4$ & $l=5$ & $l=6$ \\
\hline
$n=1$ & $872$\phantom{.00}  &        &        &        &        &        &        \\
$n=2$ & $289$\phantom{.00}  & $372$\phantom{.00}  &        &        &        &        &        \\
$n=3$ & $141$\phantom{.00}  & $167$\phantom{.00}  & $173$\phantom{.00}  &        &        &        &        \\
$n=4$ & $83.6$ & $94.8$ & $97.1$ & $97.2$ &        &        &        \\
$n=5$ & $55.1$ & $61.0$ & $62.2$ & $62.2$ & $62.2$ &        &        \\
$n=6$ & $39.0$ & $42.5$ & $43.2$ & $43.2$ & $43.2$ & $43.2$ &        \\
$n=7$ & $29.1$ & $31.3$ & $31.7$ & $31.7$ & $31.7$ & $31.7$ & $31.7$ 
\enddata
\end{deluxetable*}


\begin{deluxetable*}{rcclc}
\tablecaption{\label{tab8} Calculated Parameters for $^7$Li($X^-$, $\gamma$)$^7$Li$_X$ in the WS40 Model.}
\tablewidth{0pt}
\tablehead{
\colhead{$m_X$ (GeV)} & \colhead{Transition} & \colhead{$\tau_{\rm if}$ (fm)} & \colhead{$\Gamma_\gamma$ (eV)} & \colhead{$E_{\rm r}$ (MeV)}}
\startdata
   1        & $^7$Li$^\ast_X$(1S)$\rightarrow ^7$Li$_X$(1S) &  ---                 & 0.00627\tablenotemark{a} & 0.280 \\
 100        & $^7$Li$^\ast_X$(2P)$\rightarrow ^7$Li$^\ast_X$(1S) &  3.91                & 1.26                    & 0.124 \\
1000        & $^7$Li$^\ast_X$(2P)$\rightarrow ^7$Li$^\ast_X$(1S) &  3.81                & 1.34                    & 0.105 
\enddata
\tablenotetext{a}{Given by $\Gamma_\gamma = \tau_\gamma^{-1}$ with the life time 105 fs taken from that of the first excited $1/2^-$ state of $^7$Li \citep{Tilley2002}.}
\end{deluxetable*}


\begin{deluxetable*}{c|ccccccc}
\tablecaption{\label{tab9} Binding Energies of $^9$Be$_X$ Atomic States with Main Quantum Numbers $n=1$--$7$ (keV).}
\tablewidth{0pt}
\tablehead{
\colhead{$m_X=1$ GeV} & \colhead{$l=0$} & \colhead{$l=1$} & \colhead{$l=2$} & \colhead{$l=3$} & \colhead{$l=4$} & \colhead{$l=5$} & \colhead{$l=6$}}
\startdata
$n=1$ & $350$\phantom{.00}  &        &        &        &        &        &        \\
$n=2$ & $91.3$ & $95.1$ &        &        &        &        &        \\
$n=3$ & $41.1$ & $42.3$ & $42.3$ &        &        &        &        \\
$n=4$ & $23.3$ & $23.8$ & $23.8$ & $23.8$ &        &        &        \\
$n=5$ & $15.0$ & $15.2$ & $15.2$ & $15.2$ & $15.2$ &        &        \\
$n=6$ & $10.4$ & $10.6$ & $10.6$ & $10.6$ & $10.6$ & $10.6$ &        \\
$n=7$ & \phantom{00}$7.68$ & \phantom{00}$7.77$ & \phantom{00}$7.77$ & \phantom{00}$7.77$ & \phantom{00}$7.77$ & \phantom{00}$7.77$ & \phantom{00}$7.77$ \\
\hline
$m_X=10$ GeV & $l=0$ & $l=1$ & $l=2$ & $l=3$ & $l=4$ & $l=5$ & $l=6$ \\
\hline
$n=1$ & $1108$\phantom{.000} &        &        &        &        &        &        \\
$n=2$ & $364$\phantom{.00}  & $467$\phantom{.00}  &        &        &        &        &        \\
$n=3$ & $178$\phantom{.00}  & $210$\phantom{.00}  & $216$\phantom{.00}  &        &        &        &        \\
$n=4$ & $105$\phantom{.00}  & $119$\phantom{.00}  & $121$\phantom{.00}  & $121$\phantom{.00}  &        &        &        \\
$n=5$ & $69.1$ & $76.3$ & $77.7$ & $77.8$ & $77.8$ &        &        \\
$n=6$ & $48.9$ & $53.2$ & $54.0$ & $54.0$ & $54.0$ & $54.0$ &        \\
$n=7$ & $36.4$ & $39.1$ & $39.7$ & $39.7$ & $39.7$ & $39.7$ & $39.7$ \\
\hline
$m_X=100$ GeV & $l=0$ & $l=1$ & $l=2$ & $l=3$ & $l=4$ & $l=5$ & $l=6$ \\
\hline
$n=1$ & $1408$\phantom{.000} &        &        &        &        &        &        \\
$n=2$ & $531$\phantom{.00}  & $728$\phantom{.00}  &        &        &        &        &        \\
$n=3$ & $272$\phantom{.00}  & $335$\phantom{.00}  & $364$\phantom{.00}  &        &        &        &        \\
$n=4$ & $164$\phantom{.00}  & $193$\phantom{.00}  & $205$\phantom{.00}  & $206$\phantom{.00}  &        &        &        \\
$n=5$ & $110$\phantom{.00}  & $125$\phantom{.00}  & $131$\phantom{.00}  & $132$\phantom{.00}  & $132$\phantom{.00}  &        &        \\
$n=6$ & $78.7$ & $87.5$ & $91.2$ & $91.6$ & $91.6$ & $91.6$ &        \\
$n=7$ & $59.0$ & $64.7$ & $67.0$ & $67.3$ & $67.3$ & $67.3$ & $67.3$ \\
\hline
$m_X=1000$ GeV & $l=0$ & $l=1$ & $l=2$ & $l=3$ & $l=4$ & $l=5$ & $l=6$ \\
\hline
$n=1$ & $1448$\phantom{.000} &        &        &        &        &        &        \\
$n=2$ & $558$\phantom{.00}  & $768$\phantom{.00}  &        &        &        &        &        \\
$n=3$ & $288$\phantom{.00}  & $356$\phantom{.00}  & $391$\phantom{.00}  &        &        &        &        \\
$n=4$ & $175$\phantom{.00}  & $205$\phantom{.00}  & $220$\phantom{.00}  & $222$\phantom{.00}  &        &        &        \\
$n=5$ & $117$\phantom{.00}  & $133$\phantom{.00}  & $141$\phantom{.00}  & $142$\phantom{.00}  & $142$\phantom{.00}  &        &        \\
$n=6$ & $83.9$ & $93.4$ & $97.8$ & $98.5$ & $98.5$ & $98.5$ &        \\
$n=7$ & $63.0$ & $69.2$ & $71.9$ & $72.3$ & $72.4$ & $72.4$ & $72.4$ 
\enddata
\end{deluxetable*}


\begin{deluxetable*}{c|ccccccc}
\tablecaption{\label{tab10} Binding Energies of $^4$He$_X$ Atomic States with Main Quantum Numbers $n=1$--$7$ (keV).}
\tablewidth{0pt}
\tablehead{
\colhead{$m_X=1$ GeV} & \colhead{$l=0$} & \colhead{$l=1$} & \colhead{$l=2$} & \colhead{$l=3$} & \colhead{$l=4$} & \colhead{$l=5$} & \colhead{$l=6$}}
\startdata
$n=1$ & $83.0$ &        &        &        &        &        &        \\
$n=2$ & $20.9$ & $21.0$ &        &        &        &        &        \\
$n=3$ & \phantom{00}$9.29$ & \phantom{00}$9.33$ & \phantom{00}$9.33$ &        &        &        &        \\
$n=4$ & \phantom{00}$5.23$ & \phantom{00}$5.25$ & \phantom{00}$5.25$ & \phantom{00}$5.25$ &        &        &        \\
$n=5$ & \phantom{00}$3.35$ & \phantom{00}$3.36$ & \phantom{00}$3.36$ & \phantom{00}$3.36$ & \phantom{00}$3.36$ &        &        \\
$n=6$ & \phantom{00}$2.33$ & \phantom{00}$2.33$ & \phantom{00}$2.33$ & \phantom{00}$2.33$ & \phantom{00}$2.33$ & \phantom{00}$2.33$ &        \\
$n=7$ & \phantom{00}$1.71$ & \phantom{00}$1.71$ & \phantom{00}$1.71$ & \phantom{00}$1.71$ & \phantom{00}$1.71$ & \phantom{00}$1.71$ & \phantom{00}$1.71$ \\
\hline
$m_X=10$ GeV & $l=0$ & $l=1$ & $l=2$ & $l=3$ & $l=4$ & $l=5$ & $l=6$ \\
\hline
$n=1$ & $263$\phantom{.00}  &        &        &        &        &        &        \\
$n=2$ & $69.0$ & $72.3$ &        &        &        &        &        \\
$n=3$ & $31.1$ & $32.1$ & $32.1$ &        &        &        &        \\
$n=4$ & $17.7$ & $18.1$ & $18.1$ & $18.1$ &        &        &        \\
$n=5$ & $11.4$ & $11.6$ & $11.6$ & $11.6$ & $11.6$ &        &        \\
$n=6$ & \phantom{00}$7.92$ & \phantom{00}$8.03$ & \phantom{00}$8.03$ & \phantom{00}$8.03$ & \phantom{00}$8.03$ & \phantom{00}$8.03$ &        \\
$n=7$ & \phantom{00}$5.82$ & \phantom{00}$5.90$ & \phantom{00}$5.90$ & \phantom{00}$5.90$ & \phantom{00}$5.90$ & \phantom{00}$5.90$ & \phantom{00}$5.90$ \\
\hline
$m_X=100$ GeV & $l=0$ & $l=1$ & $l=2$ & $l=3$ & $l=4$ & $l=5$ & $l=6$ \\
\hline
$n=1$ & $333$\phantom{.00}  &        &        &        &        &        &        \\
$n=2$ & $89.3$ & $95.6$ &        &        &        &        &        \\
$n=3$ & $40.6$ & $42.5$ & $42.5$ &        &        &        &        \\
$n=4$ & $23.1$ & $23.9$ & $23.9$ & $23.9$ &        &        &        \\
$n=5$ & $14.9$ & $15.3$ & $15.3$ & $15.3$ & $15.3$ &        &        \\
$n=6$ & $10.4$ & $10.6$ & $10.6$ & $10.6$ & $10.6$ & $10.6$ &        \\
$n=7$ & \phantom{00}$7.66$ & \phantom{00}$7.81$ & \phantom{00}$7.81$ & \phantom{00}$7.81$ & \phantom{00}$7.81$ & \phantom{00}$7.81$ & \phantom{00}$7.81$ \\
\hline
$m_X=1000$ GeV & $l=0$ & $l=1$ & $l=2$ & $l=3$ & $l=4$ & $l=5$ & $l=6$ \\
\hline
$n=1$ & $342$\phantom{.00}  &        &        &        &        &        &        \\
$n=2$ & $92.0$ & $98.8$ &        &        &        &        &        \\
$n=3$ & $41.9$ & $43.9$ & $43.9$ &        &        &        &        \\
$n=4$ & $23.8$ & $24.7$ & $24.7$ & $24.7$ &        &        &        \\
$n=5$ & $15.4$ & $15.8$ & $15.8$ & $15.8$ & $15.8$ &        &        \\
$n=6$ & $10.7$ & $11.0$ & $11.0$ & $11.0$ & $11.0$ & $11.0$ &        \\
$n=7$ & \phantom{00}$7.91$ & \phantom{00}$8.07$ & \phantom{00}$8.07$ & \phantom{00}$8.07$ & \phantom{00}$8.07$ & \phantom{00}$8.07$ & \phantom{00}$8.07$ 
\enddata
\end{deluxetable*}


\begin{deluxetable*}{ccccccccc}
\tablecaption{\label{tab11} Approximate Recombination Rates}
\tablewidth{0pt}
\tablehead{
\colhead{Reaction} & \multicolumn{2}{c}{$m_X$=1 GeV} & \multicolumn{2}{c}{10 GeV} & \multicolumn{2}{c}{100 GeV} & \multicolumn{2}{c}{1000 GeV}}
\startdata
                                      & $C_1$  & $C_{\rm r}$ & $C_1$  & $C_{\rm r}$ & $C_1$  & $C_{\rm r}$ & $C_1$  & $C_{\rm r}$ \\
$^1$H($X^-$, $\gamma$)$^1$H$_X$       & 1.05$\times 10^4$ & 0.368       & 4.55$\times 10^3$ & 0.863       & 4.04$\times 10^3$ & 0.985       & 3.98$\times 10^3$ & 0.985       \\
$^2$H($X^-$, $\gamma$)$^2$H$_X$       & 6.77$\times 10^3$ & 0.576       & 1.79$\times 10^3$ & 2.160	    & 1.42$\times 10^3$ & 2.783       & 1.38$\times 10^3$ & 2.783       \\
$^3$H($X^-$, $\gamma$)$^3$H$_X$       & 5.61$\times 10^3$ & 0.694       & 1.08$\times 10^3$ & 3.543	    & 7.74$\times 10^2$ & 5.106       & 7.45$\times 10^2$ & 5.106       \\
$^3$He($X^-$, $\gamma$)$^3$He$_X$     & 3.55$\times 10^4$ & 0.694       & 1.30$\times 10^4$ & 3.543	    & 1.12$\times 10^4$ & 5.106       & 1.10$\times 10^4$ & 5.106       \\
$^6$Li($X^-$, $\gamma$)$^6$Li$_X$     & 6.64$\times 10^4$ & 0.857       & 1.76$\times 10^4$ & 7.451	    & 1.35$\times 10^4$ & 14.38\phantom{00}       & 1.31$\times 10^4$ & 14.38\phantom{00}       \\
$^8$Li($X^-$, $\gamma$)$^8$Li$_X$     & 5.63$\times 10^4$ & 0.909       & 1.13$\times 10^4$ & 9.684	    & 7.80$\times 10^3$ & 22.08\phantom{00}       & 7.46$\times 10^3$ & 22.08\phantom{00}       \\
$^8$B($X^-$, $\gamma$)$^8$B$_X$       & 2.06$\times 10^5$ & 0.909       & 5.53$\times 10^4$ & 9.684	    & 4.13$\times 10^4$ & 22.08\phantom{00}       & 3.98$\times 10^4$ & 22.08\phantom{00}       \\
$^{10}$B($X^-$, $\gamma$)$^{10}$B$_X$ & 1.77$\times 10^5$ & 0.942       & 3.83$\times 10^4$ & 11.62\phantom{00}	    & 2.64$\times 10^4$ & 30.77\phantom{00}       & 2.52$\times 10^4$ & 30.77\phantom{00}       \\
$^{11}$B($X^-$, $\gamma$)$^{11}$B$_X$ & 1.66$\times 10^5$ & 0.954       & 3.26$\times 10^4$ & 12.50\phantom{00}	    & 2.16$\times 10^4$ & 35.45\phantom{00}       & 2.05$\times 10^4$ & 35.45\phantom{00}       \\
$^{12}$B($X^-$, $\gamma$)$^{12}$B$_X$ & 1.58$\times 10^5$ & 0.965       & 2.86$\times 10^4$ & 13.31\phantom{00}	    & 1.83$\times 10^4$ & 40.34\phantom{00}       & 1.73$\times 10^4$ & 40.34\phantom{00}       \\
$^{11}$C($X^-$, $\gamma$)$^{11}$C$_X$ & 2.64$\times 10^5$ & 0.954       & 5.83$\times 10^4$ & 12.50\phantom{00}	    & 4.00$\times 10^4$ & 35.45\phantom{00}       & 3.80$\times 10^4$ & 35.45\phantom{00}       \\
$^{12}$C($X^-$, $\gamma$)$^{12}$C$_X$ & 2.49$\times 10^5$ & 0.965       & 5.01$\times 10^4$ & 13.31\phantom{00}	    & 3.31$\times 10^4$ & 40.34\phantom{00}       & 3.13$\times 10^4$ & 40.34\phantom{00}       
\enddata
\end{deluxetable*}


\begin{deluxetable*}{crcrcrcrcrc}
\tablecaption{\label{tab12} $\beta$-Decay Rates}
\tablewidth{0pt}
\tablehead{
\colhead{Reaction} & \multicolumn{2}{c}{$m_X$=1 GeV} & \multicolumn{2}{c}{10 GeV} & \multicolumn{2}{c}{100 GeV} & \multicolumn{2}{c}{1000 GeV}}
\startdata
                                          & $Q_X$ (MeV) & $\Gamma_{\beta X}$ (s$^{-1}$) & $Q_X$ (MeV) & $\Gamma_{\beta X}$ (s$^{-1}$) & $Q_X$ (MeV) & $\Gamma_{\beta X}$ (s$^{-1}$) & $Q_X$ (MeV) & $\Gamma_{\beta X}$ (s$^{-1}$) \\
$^6$Be$_X$(, $e^+ \nu_e$)$^6$Li$_X$       &  3.636 & 1.027 &  3.422 & 0.759 &  3.364 & 0.697 &  3.357 & 0.689 \\
$^8$Li$_X$(, $e^- \bar{\nu}_e$)$^8$Be$_X$ & 16.407 & 0.934 & 15.915 & 0.802 & 15.704 & 0.750 & 15.676 & 0.744 \\
$^8$B$_X$(, $e^+ \nu_e$)$^8$Be$_X$        & 17.296 & 0.742 & 17.095 & 0.699 & 17.054 & 0.691 & 17.049 & 0.690 
\enddata
\end{deluxetable*}

\clearpage

\begin{deluxetable*}{ccrcrcrcr}
\tablecaption{\label{tab13} Reverse Reaction Coefficients and $Q$-values for Nuclear Reactions in the WS40 Model}
\tablewidth{0pt}
\tablehead{
\colhead{Reaction} & \multicolumn{2}{c}{$m_X$=1 GeV} & \multicolumn{2}{c}{10 GeV} & \multicolumn{2}{c}{100 GeV} & \multicolumn{2}{c}{1000 GeV}}
\startdata
                                           & $C_{\rm r}$ & $Q_9$ & $C_{\rm r}$ & $Q_9$ & $C_{\rm r}$ & $Q_9$ & $C_{\rm r}$ & $Q_9$ \\
$^3$He$_X$($d$, $p$)$^4$He$_X$             & 6.108 & 213.050 & 7.641 & 213.536 & 8.376 & 213.822 & 8.478 & 213.862 \\
$^3$He$_X$($\alpha$, $\gamma$)$^7$Be$_X$   & 1.415 &  21.464 & 2.690 &  27.768 & 3.742 &  30.115 & 3.925 &  30.421 \\
$^4$He$_X$($d$, $\gamma$)$^6$Li$_X$        & 1.695 &  18.390 & 2.305 &  21.182 & 2.716 &  22.238 & 2.782 &  22.375 \\
$^4$He$_X$($d$, $X^-$)$^6$Li               & 1.973 &  16.141 & 0.309 &  14.047 & 0.203 &  13.235 & 0.193 &  13.130 \\
$^4$He$_X$($t$, $\gamma$)$^7$Li$_X$        & 1.277 &  29.960 & 1.942 &  33.210 & 2.464 &  34.581 & 2.553 &  34.766 \\
$^4$He$_X$($t$, $X^-$)$^7$Li               & 1.438 &  27.662 & 0.225 &  25.567 & 0.148 &  24.756 & 0.141 &  24.651 \\
$^4$He$_X$($^3$He, $\gamma$)$^7$Be$_X$     & 1.277 &  21.401 & 1.942 &  27.219 & 2.464 &  29.281 & 2.553 &  29.547 \\
$^4$He$_X$($^3$He, $X^-$)$^7$Be            & 1.438 &  17.444 & 0.225 &  15.349 & 0.148 &  14.538 & 0.141 &  14.433 \\
$^4$He$_X$($\alpha$, $\gamma$)$^8$Be$_X$   & 3.299 &   1.986 & 5.503 &   8.250 & 7.483 &  10.610 & 7.846 &  10.922 \\
$^4$He$_X$($^6$Li, $\gamma$)$^{10}$B$_X$   & 1.927 &  56.985 & 3.723 &  66.752 & 5.745 &  70.332 & 6.163 &  70.811 \\
$^6$Li$_X$($n$, $t$)$^4$He$_X$             & 0.950 &  53.263 & 0.698 &  48.376 & 0.593 &  46.508 & 0.579 &  46.266 \\
$^6$Li$_X$($p$, $\gamma$)$^7$Be$_X$        & 1.214 &  66.763 & 1.357 &  69.789 & 1.461 &  70.795 & 1.478 &  70.923 \\
$^6$Li$_X$($p$, $^3$He $\alpha$)$X^-$      & ---   &  44.399 & ---   &  39.513 & ---   &  37.645 & ---   &  37.403 \\
$^6$Li$_X$($\alpha$, $\gamma$)$^{10}$B$_X$ & 1.727 &  55.699 & 2.453 &  62.675 & 3.212 &  65.199 & 3.364 &  65.539 \\
$^7$Li$_X$($p$, $\gamma$)$^8$Be$_X$        & 6.622 & 201.967 & 7.267 & 204.981 & 7.788 & 205.969 & 7.879 & 206.096 \\
$^7$Li$_X$($p$, 2$\alpha$)$X^-$      & ---   & 199.017 & ---   & 193.673 & ---   & 191.490 & ---   & 191.200 \\
$^7$Li$_X$($\alpha$, $\gamma$)$^{11}$B$_X$ & 4.317 & 104.484 & 5.818 & 111.338 & 7.497 & 113.795 & 7.850 & 114.130 \\
$^8$Li$_X$($p$, $\gamma$)$^9$Be$_X$        & 2.111 & 197.719 & 2.285 & 200.809 & 2.438 & 201.835 & 2.467 & 201.968 \\
$^6$Be$_X$($n$, $p$)$^6$Li$_X$             & 0.333 &  57.210 & 0.333 &  54.724 & 0.333 &  54.051 & 0.333 &  53.971 \\
$^7$Be$_X$($n$, $p$)$^7$Li$_X$             & 1.000 &  17.422 & 1.000 &  14.854 & 1.000 &  14.164 & 1.000 &  14.083 \\
$^7$Be$_X$($n$, $p$ $^7$Li)$X^-$           & ---   &  15.124 & ---   &   7.212 & ---   &   4.338 & ---   &   3.968 \\
$^7$Be$_X$($p$, $\gamma$)$^8$B$_X$         & 1.326 &   3.655 & 1.455 &   6.440 & 1.559 &   7.215 & 1.578 &   7.312 \\
$^7$Be$_X$($d$, $p$)$^8$Be$_X$             & 14.24\phantom{00} & 193.572 & 15.63\phantom{00} & 194.019 & 16.75\phantom{00} & 194.317 & 16.95\phantom{00} & 194.363 \\
$^7$Be$_X$($\alpha$, $\gamma$)$^{11}$C$_X$ & 4.317 &  92.305 & 5.818 &  99.941 & 7.497 & 102.664 & 7.850 & 103.038 \\
$^8$Be$_X$($p$, $\gamma$)$^9$B$_X$         & 2.111 &$-$0.081 & 2.285 &   2.688 & 2.438 &   3.446 & 2.467 &   3.541 \\
$^9$Be$_X$($p$, $\gamma$)$^{10}$B$_X$      & 0.980 &  78.540 & 1.049 &  81.611 & 1.116 &  82.524 & 1.128 &  82.641 \\
$^9$Be$_X$($p$, $^6$Li)$^4$He$_X$          & 0.506 &  21.556 & 0.281 &  14.858 & 0.193 &  12.191 & 0.182 &  11.831 \\
$^{10}$Be$_X$($p$, $\gamma$)$^{11}$B$_X$   & 0.433 & 132.395 & 0.459 & 135.205 & 0.487 & 135.932 & 0.492 & 136.023 \\
$^9$B$_X$($p$, $\gamma$)$^{10}$C$_X$       & 6.846 &  49.053 & 7.326 &  53.070 & 7.789 &  54.451 & 7.879 &  54.637 \\
$^{10}$B$_X$($p$, $\gamma$)$^{11}$C$_X$    & 3.030 & 103.370 & 3.215 & 107.055 & 3.406 & 108.260 & 3.445 & 108.422 \\
$^{11}$B$_X$($p$, $\gamma$)$^{12}$C$_X$    & 7.002 & 187.715 & 7.375 & 191.414 & 7.791 & 192.632 & 7.879 & 192.797 \\
\hline
$^1$H$_X$($\alpha$, $p$)$^4$He$_X$         & 2.090 &   0.815 & 5.686 &   2.793 & 7.679 &   3.582 & 7.967 &   3.684 \\
$^1$H$_X$($^7$Li, 2$\alpha$)$X^-$    & ---   & 201.168 & ---   & 201.050 & ---   & 201.028 & ---   & 201.026 \\
$^1$H$_X$($^7$Be, $X^-$)$^8$B              & 3.517 &   1.448 & 1.497 &   1.331 & 1.328 &   1.309 & 1.312 &   1.306 \\
$^2$H$_X$($\alpha$, $d$)$^4$He$_X$         & 1.333 &   0.762 & 2.272 &   2.577 & 2.752 &   3.312 & 2.820 &   3.408 \\
$^2$H$_X$($\alpha$, $X^-$)$^6$Li           & 2.637 &  16.903 & 0.703 &  16.624 & 0.560 &  16.547 & 0.546 &  16.538 \\
$^3$H$_X$($\alpha$, $t$)$^4$He$_X$         & 1.108 &   0.736 & 1.386 &   2.394 & 1.519 &   3.050 & 1.538 &   3.135 \\
$^3$H$_X$($\alpha$, $X^-$)$^7$Li           & 1.597 &  28.397 & 0.313 &  27.961 & 0.225 &  27.806 & 0.217 &  27.786 
\enddata
\end{deluxetable*}


\begin{thebibliography}{999}

\bibitem[Angulo et al.(1999)]{Angulo1999} Angulo, C., Arnould, M., 
Rayet, M., et al.\ 1999, Nuclear Physics A, 656, 3 


\bibitem[Aoki(2012)]{Aoki2012b} Aoki, W.\ 2012, Memorie della 
Societa Astronomica Italiana Supplementi, 22, 35 


\bibitem[Aoki et al.(2009)]{Aoki:2009ce} Aoki, W., Barklem, P.~S., 
Beers, T.~C., et al.\ 2009, \apj, 698, 1803 


\bibitem[Aoki et al.(2012)Aoki, Ito, \& Tajitsu]{Aoki:2012wb} Aoki, W., Ito, H., 
\& Tajitsu, A.\ 2012, \apjl, 751, L6 


\bibitem[Asplund et al.(2006)]{Asplund:2005yt} Asplund, M., Lambert, 
D.~L., Nissen, P.~E., Primas, F., \& Smith, V.~V.\ 2006, \apj, 644, 229 


\bibitem[Asplund 
\& Mel{\'e}ndez(2008)]{asp2008} Asplund, M., \& Mel{\'e}ndez, J.\ 2008, First Stars III, 990, 342 



\bibitem[Aver et al.(2010)]{Aver:2010wq} Aver, E., Olive, K.~A., 
\& Skillman, E.~D.\ 2010, \jcap, 5, 3 


\bibitem[Bailly et al.(2009)Bailly, Jedamzik, \& Moultaka]{Bailly:2008yy} Bailly, S., Jedamzik, 
K., \& Moultaka, G.\ 2009, \prd, 80, 063509 


\bibitem[Bania et al.(2002)]{Bania:2002yj} Bania, T.~M., Rood, 
R.~T., \& Balser, D.~S.\ 2002, \nat, 415, 54 


\bibitem[Bertulani(2003)]{Bertulani:2003kr} Bertulani, C.~A.\ 2003, 
Computer Physics Communications, 156, 123 


\bibitem[Bethe \& Salpeter(1957)]{Bethe1957}
 Bethe, H.~A., \& Salpeter, E.~E.\ 1957,
  Quantum Mechanics of One- and Two-Electron Atoms,
  (Academic Press, NY)


\bibitem[Bird et al.(2008)Bird, Koopmans, \& Pospelov]{Bird:2007ge} Bird, C., Koopmans, K., 
\& Pospelov, M.\ 2008, \prd, 78, 083010 


\bibitem[Blatt \& Weisskopf(1991)]{Blatt}
  Blatt, J. M., \&  Weisskopf, V. F. 1991,
  Theoretical nuclear physics
  (Dover, Mineola, NY)


\bibitem[Boesgaard et al.(1999)]{boe1999} Boesgaard, A.~M., 
Deliyannis, C.~P., King, J.~R., et al.\ 1999, \aj, 117, 1549 


\bibitem[Bonifacio et 
al.(2007)]{bon2007} Bonifacio, P., Molaro, P., Sivarani, T., et al.\ 2007, \aap, 462, 851 


\bibitem[Burke(2011)]{Burke2011} Burke, P.~G.\ 2011, R-Matrix 
Theory of Atomic Collisions: Application to Atomic, Molecular and Optical 
Processes, Springer Series on Atomic, Optical, and Plasma Physics, Volume 
61.~ISBN 978-3-642-15930-5.~Springer-Verlag Berlin Heidelberg, 2011,  


\bibitem[Cahn 
\& Glashow(1981)]{cahn:1981} Cahn, R.~N., \& Glashow, S.~L.\ 1981, Science, 213, 607 


\bibitem[Caughlan 
\& Fowler(1988)]{Caughlan1988} Caughlan, G.~R., \& Fowler, W.~A.\ 1988, Atomic Data and Nuclear Data Tables, 40, 283 


\bibitem[Cayrel et 
al.(1999)]{Cayrel:1999kx} Cayrel, R., Spite, M., Spite, F., et al.\ 1999, \aap, 343, 923 


\bibitem[Cayrel et 
al.(2007)]{Cayrel:2007te} Cayrel, R., Steffen, M., Chand, H., et al.\ 2007, \aap, 473, L37 


\bibitem[Chadwick et al.(2011)]{ENDF-B-VII.1short} Chadwick, M.~B., 
Herman, M., Oblo{\v z}insk{\'y}, P., et al.\ 2011, Nuclear Data Sheets, 
112, 2887 


\bibitem[Chatrchyan et al.(2013)]{CMS:2012xi} Chatrchyan, S., 
Khachatryan, V., Sirunyan, A.~M., et al.\ 2013, \prd, 87, 092008 


\bibitem[Chatrchyan et al.(2013)]{CMS2013JHEP} Chatrchyan, S., 
Khachatryan, V., Sirunyan, A.~M., et al.\ 2013, Journal of High Energy 
Physics, 7, 122 


\bibitem[Cheoun et al.(2003)Cheoun, Kim, \& Choi]{Ch} Cheoun, M.~K., Kim, 
K.~S., \& Choi, T.~K.\ 2003, Journal of Physics G Nuclear Physics, 29, 2099 


\bibitem[Cichocki et al.(1995)]{1995PhRvC..51.2406C} Cichocki, A., Dubach, 
J., Hicks, R.~S., et al.\ 1995, \prc, 51, 2406 


\bibitem[Coc et al.(2012)]{Coc:2011az} Coc, A., Goriely, S., Xu, 
Y., Saimpert, M., \& Vangioni, E.\ 2012, \apj, 744, 158 


\bibitem[Coc et al.(2013)Coc, Uzan, \& Vangioni]{Coc:2013eea} Coc, A., Uzan, J.-P., 
\& Vangioni, E.\ 2013, arXiv:1307.6955 


\bibitem[Cunha et al.(2000)]{cun2000} Cunha, K., Smith, V.~V., 
Boesgaard, A.~M., \& Lambert, D.~L.\ 2000, \apj, 530, 939 


\bibitem[Cyburt et al.(2010)]{Cyburt2010} Cyburt, R.~H., Amthor, 
A.~M., Ferguson, R., et al.\ 2010, \apjs, 189, 240 


\bibitem[Cyburt et al.(2012)]{Cyburt:2012kp} Cyburt, R.~H., Ellis, 
J., Fields, B.~D., et al.\ 2012, \jcap, 12, 37 


\bibitem[Cyburt et al.(2006)]{Cyburt:2006uv} Cyburt, R.~H., Ellis, 
J., Fields, B.~D., Olive, K.~A., \& Spanos, V.~C.\ 2006, \jcap, 11, 14 


\bibitem[{\DJ}apo et al.(2012)]{Dapo2012} {\DJ}apo, H., Boztosun, 
I., Kocak, G., \& Balantekin, A.~B.\ 2012, \prc, 85, 044602 


\bibitem[de R{\'u}jula et al.(1990)de R{\'u}jula, Glashow, \& Sarid]{rujula90} de R{\'u}jula, 
A., Glashow, S.~L., \& Sarid, U.\ 1990, Nuclear Physics B, 333, 173 

\bibitem[Dick et al.(1986)Dick, Greenlees, \& Kaufman]{Dick:1985wk} Dick, W.~J., Greenlees, 
G.~W., \& Kaufman, S.~L.\ 1986, \prd, 33, 32 


\bibitem[Dimopoulos et al.(1990)]{Dimopoulos:1989hk} Dimopoulos, S., 
Eichler, D., Esmailzadeh, R., \& Starkman, G.~D.\ 1990, \prd, 41, 2388 


\bibitem[Duncan et al.(1997)]{dun1997} Duncan, D.~K., Primas, 
F., Rebull, L.~M., et al.\ 1997, \apj, 488, 338 


\bibitem[Fixsen(2009)]{Fixsen:2009ug} Fixsen, D.~J.\ 2009, \apj, 707, 
916 


\bibitem[Fowler et 
al.(1967)Fowler, Caughlan, \& Zimmerman]{Fowler1967} Fowler, W.~A., Caughlan, G.~R., \& Zimmerman, B.~A.\ 1967, \araa, 5, 525 


\bibitem[Fukuda et al.(1999)]{fuk99} Fukuda, M., et al.\ 
1999, Nuclear Physics A, 656, 209


\bibitem[Garc{\'{\i}}a P{\'e}rez et 
al.(2009)]{gar2009} Garc{\'{\i}}a P{\'e}rez, A.~E., Aoki, W., Inoue, S., et al.\ 2009, \aap, 504, 213 


\bibitem[Garcia Lopez et al.(1998)]{gar1998} Garcia Lopez, 
R.~J., Lambert, D.~L., Edvardsson, B., et al.\ 1998, \apj, 500, 241 


\bibitem[Gaunt(1930)]{Gaunt1930} Gaunt, J.~A.\ 1930, Royal 
Society of London Philosophical Transactions Series A, 229, 163 


\bibitem[Gonz{\'a}lez Hern{\'a}ndez et 
al.(2009)]{Hernandez:2009gn} Gonz{\'a}lez Hern{\'a}ndez, J.~I., Bonifacio, P., Caffau, E., et al.\ 2009, \aap, 505, L13 


\bibitem[Hamaguchi et al.(2007)]{Hamaguchi:2007mp} Hamaguchi, K., 
Hatsuda, T., Kamimura, M., Kino, Y., 
\& Yanagida, T.~T.\ 2007, Physics Letters B, 650, 268 


\bibitem[Hemmick et al.(1990)]{Hemmick:1989ns} Hemmick, T.~K., Elmore, 
D., Gentile, T., et al.\ 1990, \prd, 41, 2074 


\bibitem[Hinshaw et al.(2013)]{Hinshaw:2012fq} Hinshaw, G., Larson, 
D., Komatsu, E., et al.\ 2013, \apjs, 208, 19 


\bibitem[Hiyama et al.(2003)Hiyama, Kino, \& Kamimura]{Hiyama:2003cu} Hiyama, E., Kino, Y., 
\& Kamimura, M.\ 2003, Progress in Particle and Nuclear Physics, 51, 223 


\bibitem[Inoue et al.(2005)]{ino05} Inoue, S., Aoki, W., 
Suzuki, T.~K., et al.\ 2005, From Lithium to Uranium: Elemental Tracers of 
Early Cosmic Evolution, 228, 59 


\bibitem[Ito et al.(2009)]{Ito:2009uv} Ito, H., Aoki, W., Honda, 
S., \& Beers, T.~C.\ 2009, \apjl, 698, L37 


\bibitem[Izotov 
\& Thuan(2010)]{Izotov:2010ca} Izotov, Y.~I., \& Thuan, T.~X.\ 2010, \apjl, 710, L67 


\bibitem[Jedamzik(2008a)]{Jedamzik:2007qk} Jedamzik, K.\ 2008, \jcap, 3, 
8 


\bibitem[Jedamzik(2008b)]{Jedamzik:2007cp} Jedamzik, K.\ 2008, \prd, 77, 
063524 


\bibitem[Jedamzik 
\& Pospelov(2009)]{Jedamzik:2009uy} Jedamzik, K., \& Pospelov, M.\ 2009, New Journal of Physics, 11, 105028 


\bibitem[Jittoh et al.(2010)]{Jittoh:2010wh} Jittoh, T., Kohri, K., 
Koike, M., et al.\ 2010, \prd, 82, 115030 


\bibitem[Jittoh et al.(2008)]{Jittoh:2008eq} Jittoh, T., Kohri, K., 
Koike, M., et al.\ 2008, \prd, 78, 055007 


\bibitem[Jittoh et al.(2007)]{Jittoh:2007fr} Jittoh, T., Kohri, K., 
Koike, M., et al.\ 2007, \prd, 76, 125023 


\bibitem[Jonkmans et al.(1996)]{Jonkmans} Jonkmans, G., Ahmad, 
S., Armstrong, D.~S., et al.\ 1996, Physical Review Letters, 77, 4512 


\bibitem[Kamimura et al.(2009)Kamimura, Kino, \& Hiyama]{Kamimura:2008fx} Kamimura, M., Kino, 
Y., \& Hiyama, E.\ 2009, Progress of Theoretical Physics, 121, 1059 


\bibitem[Kamimura et al.(2010)Kamimura, Kino, \& Hiyama]{Kamimura2010} Kamimura, M., Kino, 
Y., 
\& Hiyama, E.\ 2010, American Institute of Physics Conference Series, 1238, 139 


\bibitem[Karzas 
\& Latter(1961)]{Karzas1961} Karzas, W.~J., \& Latter, R.\ 1961, \apjs, 6, 167 


\bibitem[Kawano(1992)]{Kawano1992} Kawano, L.\ 1992, NASA 
STI/Recon Technical Report N, 92, 25163 


\bibitem[Kawasaki et al.(2007)Kawasaki, Kohri, \& Moroi]{Kawasaki:2007xb} Kawasaki, M., Kohri, 
K., \& Moroi, T.\ 2007, Physics Letters B, 649, 436 


\bibitem[Kawasaki et al.(2008)]{Kawasaki:2008qe} Kawasaki, M., Kohri, 
K., Moroi, T., \& Yotsuyanagi, A.\ 2008, \prd, 78, 065011 


\bibitem[Khlopov 
\& Kouvaris(2008)]{Khlopov:2007ic} Khlopov, M.~Y., \& Kouvaris, C.\ 2008, \prd, 77, 065002 


\bibitem[Kohri et al.(2012)]{Kohri:2012gc} Kohri, K., Ohta, S., 
Sato, J., Shimomura, T., \& Yamanaka, M.\ 2012, \prd, 86, 095024 


\bibitem[Kohri 
\& Takayama(2007)]{Kohri:2006cn} Kohri, K., \& Takayama, F.\ 2007, \prd, 76, 063507 


\bibitem[Korn et al.(2007)]{Korn:2007cx} Korn, A.~J., Grundahl, F., 
Richard, O., et al.\ 2007, \apj, 671, 402 


\bibitem[Kusakabe et al.(2007)]{Kusakabe:2007fu} Kusakabe, M., Kajino, 
T., Boyd, R.~N., Yoshida, T., \& Mathews, G.~J.\ 2007, \prd, 76, 121302 


\bibitem[Kusakabe et al.(2008)]{Kusakabe:2007fv} Kusakabe, M., Kajino, 
T., Boyd, R.~N., Yoshida, T., \& Mathews, G.~J.\ 2008, \apj, 680, 846 


\bibitem[Kusakabe et al.(2010)]{Kusakabe:2010cb} Kusakabe, M., Kajino, 
T., Yoshida, T., \& Mathews, G.~J.\ 2010, \prd, 81, 083521 


\bibitem[Kusakabe et al.(2009)]{Kusakabe:2009jt} Kusakabe, M., Kajino, 
T., Yoshida, T., \& Mathews, G.~J.\ 2009, \prd, 80, 103501 


\bibitem[Kusakabe et al.(2013a)]{Kusakabe:2013tra} Kusakabe, M., Kim, 
K.~S., Cheoun, M.-K., Kajino, T., \& Kino, Y.\ 2013, \prd, 88, 063514 


\bibitem[Kusakabe et al.(2013b)]{2013PhRvD..88h9904K} Kusakabe, M., Kim, 
K.~S., Cheoun, M.-K., Kajino, T., \& Kino, Y.\ 2013, \prd, 88, 089904 


\bibitem[Langacker 
\& Steigman(2011)]{Langacker:2011db} Langacker, P., \& Steigman, G.\ 2011, \prd, 84, 065040 


\bibitem[Larson et al.(2011)]{Larson:2010gs} Larson, D., Dunkley, J., 
Hinshaw, G., et al.\ 2011, \apjs, 192, 16 


\bibitem[Lind et 
al.(2013)]{Lind:2013iza} Lind, K., Melendez, J., Asplund, M., Collet, R., \& Magic, Z.\ 2013, \aap, 554, A96 

\bibitem[Lind et 
al.(2009)]{Lind:2009ta} Lind, K., Primas, F., Charbonnel, C., Grundahl, F., \& Asplund, M.\ 2009, \aap, 503, 545 


\bibitem[Nardin et al.(1992)]{Nardin1992}
 Nardin, M., Perger, W.~F. \& Bhalla, A.\ 1992, J. ACM Transactions on Mathematical Software, 18, 345


\bibitem[Mathews et al.(2005)Mathews, Kajino, \& Shima]{Mathews:2004kc} Mathews, G.~J., Kajino, 
T., \& Shima, T.\ 2005, \prd, 71, 021302 


\bibitem[Mel{\'e}ndez 
\& Ram{\'{\i}}rez(2004)]{Melendez:2004ni} Mel{\'e}ndez, J., \& Ram{\'{\i}}rez, I.\ 2004, \apjl, 615, L33 


\bibitem[Meneguzzi et 
al.(1971)Meneguzzi, Audouze, \& Reeves]{men1971} Meneguzzi, M., Audouze, J., \& Reeves, H.\ 1971, \aap, 15, 337 


\bibitem[Monaco et 
al.(2010)]{Monaco:2010mm} Monaco, L., Bonifacio, P., Sbordone, L., Villanova, S., \& Pancino, E.\ 2010, \aap, 519, L3 


\bibitem[Monaco et 
al.(2012)]{Monaco:2011sd} Monaco, L., Villanova, S., Bonifacio, P., et al.\ 2012, \aap, 539, A157 


\bibitem[Mucciarelli et al.(2012)Mucciarelli, Salaris, \& Bonifacio]{Mucciarelli:2011ts} Mucciarelli, A., 
Salaris, M., \& Bonifacio, P.\ 2012, \mnras, 419, 2195 

\bibitem[Nissen et 
al.(1999)]{Nissen:1999iq} Nissen, P.~E., Lambert, D.~L., Primas, F., \& Smith, V.~V.\ 1999, \aap, 348, 211 


\bibitem[Norman et al.(1989)]{Norman:1988fd} Norman, E.~B., Chadwick, 
R.~B., Lesko, K.~T., Larimer, R.-M., 
\& Hoffman, D.~C.\ 1989, \prd, 39, 2499 


\bibitem[Pettini 
\& Cooke(2012)]{Pettini:2012ph} Pettini, M., \& Cooke, R.\ 2012, \mnras, 425, 2477 


\bibitem[Pinsonneault et al.(2002)]{Pinsonneault:2001ub} Pinsonneault, 
M.~H., Steigman, G., Walker, T.~P., 
\& Narayanan, V.~K.\ 2002, \apj, 574, 398 


\bibitem[Pinsonneault et al.(1999)]{Pinsonneault:1998nf} Pinsonneault, 
M.~H., Walker, T.~P., Steigman, G., 
\& Narayanan, V.~K.\ 1999, \apj, 527, 180 


\bibitem[Pospelov(2007)]{Pospelov:2007js} Pospelov, M.\ 2007, 
arXiv:0712.0647 


\bibitem[Pospelov(2007)]{Pospelov:2006sc} Pospelov, M.\ 2007, Physical 
Review Letters, 98, 231301 


\bibitem[Pospelov 
\& Pradler(2010)]{Pospelov:2010hj} Pospelov, M., \& Pradler, J.\ 2010, Annual Review of Nuclear and Particle Science, 60, 539 


\bibitem[Pospelov et al.(2008)Pospelov, Pradler, \& Steffen]{Pospelov:2008ta} Pospelov, M., Pradler, 
J., \& Steffen, F.~D.\ 2008, \jcap, 11, 20 


\bibitem[Pradler 
\& Steffen(2008a)]{Pradler:2007ar} Pradler, J., \& Steffen, F.~D.\ 2008, European Physical Journal C, 56, 287 


\bibitem[Pradler 
\& Steffen(2008b)]{Pradler:2007is} Pradler, J., \& Steffen, F.~D.\ 2008, Physics Letters B, 666, 181 


\bibitem[Prantzos(2012)]{pra2012} Prantzos, N.\ 2012, \aap, 542, A67 


\bibitem[Prantzos(2006)]{pra2006} Prantzos, N.\ 2006, \aap, 448, 665 


\bibitem[Primas et 
al.(1999)]{Primas:1998gp} Primas, F., Duncan, D.~K., Peterson, R.~C., \& Thorburn, J.~A.\ 1999, \aap, 343, 545 


\bibitem[Primas et 
al.(2000)]{Primas:2000ee} Primas, F., Molaro, P., Bonifacio, P., \& Hill, V.\ 2000, \aap, 362, 666 


\bibitem[Reeves(1970)]{ree1970} Reeves, H.\ 1970, \nat, 226, 
727 


\bibitem[Reeves(1974)]{ree1974} Reeves, R.\ 1974, \araa, 12, 437 


\bibitem[Rybicki 
\& Lightman(1979)]{Rybicki1979} Rybicki, G.~B., \& Lightman, A.~P.\ 1979, New York, Wiley-Interscience, 1979.~393 p.,  


\bibitem[Rich 
\& Boesgaard(2009)]{Rich:2009gj} Rich, J.~A., \& Boesgaard, A.~M.\ 2009, \apj, 701, 1519 


\bibitem[Richard et al.(2005)Richard, Michaud, \& Richer]{Richard:2004pj} Richard, O., Michaud, 
G., \& Richer, J.\ 2005, \apj, 619, 538 


\bibitem[Ryan et al.(2000)]{Ryan:1999vr} Ryan, S.~G., Beers, T.~C., 
Olive, K.~A., Fields, B.~D., \& Norris, J.~E.\ 2000, \apjl, 530, L57 


\bibitem[Sbordone et 
al.(2010)]{Sbordone:2010zi} Sbordone, L., Bonifacio, P., Caffau, E., et al.\ 2010, \aap, 522, A26 


\bibitem[Serebrov et al.(2005)]{Serebrov:2004zf} Serebrov, A., 
Varlamov, V., Kharitonov, A., et al.\ 2005, Physics Letters B, 605, 72 


\bibitem[Serebrov 
\& Fomin(2010)]{Serebrov:2010sg} Serebrov, A.~P., \& Fomin, A.~K.\ 2010, \prc, 82, 035501 


\bibitem[Shi et 
al.(2007)]{shi2007} Shi, J.~R., Gehren, T., Zhang, H.~W., Zeng, J.~L., \& Zhao, G.\ 2007, \aap, 465, 587 


\bibitem[Simon et al.(1981)]{sim81} Simon, G.~G., Schmitt, 
C., \& Walther, V.~H.\ 1981, Nuclear Physics A, 364, 285 


\bibitem[Smiljanic et 
al.(2009)]{Smiljanic:2009dt} Smiljanic, R., Pasquini, L., Bonifacio, P., et al.\ 2009, \aap, 499, 103 


\bibitem[Smith et al.(1993)Smith, Kawano, \& Malaney]{Smith:1992yy} Smith, M.~S., Kawano, 
L.~H., \& Malaney, R.~A.\ 1993, \apjs, 85, 219 


\bibitem[Smith et al.(1982)]{Smith1982} Smith, P.~F., Bennett, 
J.~R.~J., Homer, G.~J., et al.\ 1982, Nuclear Physics B, 206, 333 


\bibitem[Smith et al.(1998)Smith, Lambert, \& Nissen]{smith98} Smith, V.~V., Lambert, 
D.~L., \& Nissen, P.~E.\ 1998, \apj, 506, 405 


\bibitem[Smith et al.(1993)Smith, Lambert, \& Nissen]{smith93} Smith, V.~V., Lambert, 
D.~L., \& Nissen, P.~E.\ 1993, \apj, 408, 262 


\bibitem[Smith et al.(2001)]{smith2001} Smith, V.~V., 
Vargas-Ferro, O., Lambert, D.~L., \& Olgin, J.~G.\ 2001, \aj, 121, 453 


\bibitem[Spergel et al.(2007)]{Spergel:2006hy} Spergel, D.~N., Bean, 
R., Dor{\'e}, O., et al.\ 2007, \apjs, 170, 377 


\bibitem[Spergel et al.(2003)]{Spergel:2003cb} Spergel, D.~N., Verde, 
L., Peiris, H.~V., et al.\ 2003, \apjs, 148, 175 


\bibitem[Spite 
\& Spite(1982)]{Spite:1982dd} Spite, F., \& Spite, M.\ 1982, \aap, 115, 357 


\bibitem[Steffen et al.(2010)]{ste2010} Steffen, M., Cayrel, 
R., Bonifacio, P., Ludwig, H.-G., 
\& Caffau, E.\ 2010, IAU Symposium, 265, 23 


\bibitem[Steffen et al.(2012)]{ste2012} Steffen, M., Cayrel, 
R., Caffau, E., et al.\ 2012, Memorie della Societa Astronomica Italiana 
Supplementi, 22, 152 


\bibitem[Tan et al.(2009)Tan, Shi, \& Zhao]{Tan:2008md} Tan, K.~F., Shi, J.~R., 
\& Zhao, G.\ 2009, \mnras, 392, 205 


\bibitem[Tanihata et al.(1988)]{tan88} Tanihata, I., Kobayashi, T., 
Yamakawa, O., Shimoura, S., Ekuni, K., Sugimoto, K., Takahashi, N.,
Shimoda, T., \& Sato, H.\ 1988, Physics Letters B, 106, 592 


\bibitem[Tilley et al.(2002)]{Tilley2002} Tilley, D.~R., Cheves, 
C.~M., Godwin, J.~L., et al.\ 2002, Nuclear Physics A, 708, 3 


\bibitem[Tilley et al.(2004)]{Tilley2004} Tilley, D.~R., Kelley, 
J.~H., Godwin, J.~L., et al.\ 2004, Nuclear Physics A, 745, 155 


\bibitem[Turkevich et al.(1984)Turkevich, Wielgoz, \& Economou]{Tur1984} Turkevich, A., 
Wielgoz, K., \& Economou, T.~E.\ 1984, \prd, 30, 1876 


\bibitem[Verkerk et al.(1992)]{Verkerk:1991jf} Verkerk, P., Grynberg, 
G., Pichard, B., et al.\ 1992, Physical Review Letters, 68, 1116 


\bibitem[Yamagata et al.(1993)Yamagata, Takamori, \& Utsunomiya]{Yamagata:1993jq} Yamagata, T., 
Takamori, Y., \& Utsunomiya, H.\ 1993, \prd, 47, 1231 


\bibitem[Yao \& et al.(2006)]{yao06} Yao, W.-M., \& et al.\ 
2006, Journal of Physics G Nuclear Physics, 33, 1 
\end{thebibliography}
\end{document}